\documentclass[dissertation]{uathesis}

\usepackage[colorlinks=true,linkcolor=blue,citecolor=magenta]{hyperref}
\usepackage{natbib}
\usepackage{pdfpages}
\usepackage{authblk}
\usepackage{caption}
\usepackage{lscape}
\usepackage[tableposition=t]{caption} 
\usepackage{mathtools}
\usepackage{bm}
\usepackage{bbm}
\usepackage{enumitem}
\usepackage{units}
\usepackage{graphicx}
\usepackage{amsmath}
\usepackage{amssymb}
\usepackage{epstopdf}
\usepackage{geometry}
\usepackage{comment}
\usepackage{xcolor}
\usepackage{float}
\usepackage{doi}
\usepackage{mathrsfs}
\usepackage{multirow}
\usepackage[utf8]{inputenc}
\usepackage{booktabs}
\usepackage{fancyhdr}
\usepackage{csquotes}
\usepackage{setspace}
\usepackage[normalem]{ulem}

\newcommand{\orcidicon}{\includegraphics[width=0.32cm]{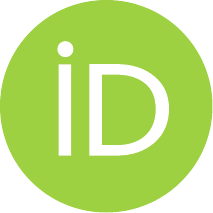}}
\newcommand{\orc}[1]{\href{https://orcid.org/#1}{\orcidicon}}

\newcommand{\orcA}{0000-0001-8217-1484}
\newcommand{\orcB}{0000-0001-5038-8427}
\newcommand{\orcC}{0000-0001-5474-2649}
\newcommand{\orcD}{0000-0003-2704-6474}
\newcommand{\orcE}{0000-0002-2289-4856}
\newcommand{\orcF}{0000-0001-5884-5047}

\newcommand*{\keV}{\text{ keV}}
\newcommand*{\eV}{\text{ eV}}

\newcommand*{\bb}{\boldsymbol}
\newcommand*{\beqn}{\begin{equation}}
\newcommand*{\eeqn}{\end{equation}}
\newcommand{\req}[1]{Eq.~(\ref{#1})}
\newcommand{\rf}[1]{Figure~{\ref{#1}}}
\newcommand{\rt}[1]{Table~{\ref{#1}}}
\newcommand{\rsec}[1]{Section~{\ref{#1}}}
\newcommand{\rchap}[1]{Chapter~{\ref{#1}}}
\newcommand{\rapp}[1]{Appendix~{\ref{#1}}}

\newcommand\Tstrut{\rule{0pt}{2.6ex}}
\newcommand\Bstrut{\rule[-0.9ex]{0pt}{0pt}}
\newcommand{\TBstrut}{\Tstrut\Bstrut}

\newcounter{daggerfootnote}

\completetitle{Modern topics in relativistic spin dynamics and magnetism}
\fullname{Andrew James Steinmetz}
\degreename{Doctor of Philosophy}
\degreemajor{Physics} 
\begin{document}
\maketitlepage
{DEPARTMENT OF PHYSICS}
{2023}							

\approval
{October 27th, 2023}		
{Johann Rafelski}	
{John Rutherfoord}	
{Shufang Su}		
{Sean Fleming}		
{Stefan Meinel}		
{}		

\statementbyauthor

\incacknowledgements{subtex/acknowledgements}
\incdedication{subtex/dedication}
\tableofcontents
\listoffigures
\listoftables
\newgeometry{left=1.25in,right=1.25in,top=1.5in,bottom=1.0in}
\incabstract{subtex/abstract}


\chapter*{Publications and author contributions}
\label{sec:pubs}
\addcontentsline{toc}{chapter}{PUBLICATIONS AND AUTHOR CONTRIBUTIONS}
I, Andrew James Steinmetz\orc{\orcC}, in the course of satisfying the University of Arizona Department of Physics's requirements for a Ph.D. doctoral dissertation, prepared the following publications which are reprinted in full in the appendices. These articles are not ordered chronologically, but in the contextual order of presentation in this document. My contribution to each work is described under each item.
\begin{itemize}
    \item \rapp{appendixA} - ``Magnetic dipole moment in relativistic quantum mechanics'' by~\citet*{Steinmetz:2018ryf} is a study and comparison of DP and KGP wave equations for homogeneous magnetic fields and hydrogen-like atoms. I performed all computation, writing, and figure making in preparation of the first draft and approved the final draft before submission. I acknowledge the help and consultation of Martin Formanek\orc{\orcD} (MF) and Johann Rafelski\orc{\orcA} (JR) in research, writing and editing.
    \item \rapp{appendixB} - ``Strong fields and neutral particle magnetic moment dynamics'' by~\citet*{Formanek:2017mbv} is an overview of our research group's efforts in studying neutral particle dynamics in electromagnetic fields. I wrote Section 2.1 in collaboration with MF. I consulted and helped lead author MF and co-authors Stefan Evans\orc{\orcF} (SE) and Cheng Tao Yang\orc{\orcB} (CTY) in editing and revising the overall manuscript.
    \item \rapp{appendixC} - ``Relativistic dynamics of point magnetic moment'' by~\citet*{Rafelski:2017hce} introduces a new covariant formulation of classical spin dynamics and unifies Gilbertian and Amp{\`e}rian dipoles. I wrote Section 3 in collaboration with JR and MF and aided in the computation in Section 5.1. I otherwise consulted in the research, writing, and editing process of this publication.
    \item \rapp{appendixD} - ``Dynamic fermion flavor mixing through transition dipole moments'' by \citet*{Rafelski:2023zgp} is a study of Majorana neutrino flavor mixing in electromagnetic fields and proposes a novel dynamical EM-mass basis for propagating neutrinos. The article was written originally via invitation of JR by Gerhard Buchalla, Dieter L\"ust and Zhi-Zhong Xing as a memorial chapter in a book dedicated to Harald Fritzsch. I performed all computation and writing in preparation of the first draft and approved the final draft before submission. I acknowledge the help and consultation of JR and CTY in research, writing and editing.
    \item \rapp{appendixE} - ``A Short Survey of Matter-Antimatter Evolution in the Primordial Universe'' by~\citet*{Rafelski:2023emw} is a 50 page long review with many novel results describing the role of antimatter in the early universe. I supervised (in collaboration with CTY) the document creation, combining the writing contributions of all authors (including myself, Jeremiah Birrell\orc{\orcE} (JB), CTY, and JR) into one coherent presentation. I also coordinated with all authors in formatting and editing the technical figures in this review by JB, CTY, and JR.
    \item \rapp{appendixF} - ``Matter-antimatter origin of cosmic magnetism'' by~\citet*{Steinmetz:2023nsc} proposes a model of para-magnetization driven by the large matter-antimatter (electron-positron) content of the early universe. I carried out all writing in preparation of the first draft and approved the final draft before submission. Computation and figure making was done in collaboration with CTY who contributed key results and five technical figures. I acknowledge the help and consultation of CTY and JR in research, writing and editing.
\end{itemize}

This is not a total catalogue of my research efforts, but lists the works that form the foundation of \rchap{chap:moment}, \rchap{chap:neutrino} and \rchap{chap:cosmo} of this dissertation. Where noted, these chapters also contain sections of complete yet unpublished work. \rchap{chap:future} contains brief discussions of still-in-progress research efforts to be completed after submission of this dissertation.

I was also co-author on the following publications which are not used extensively in this dissertation and are not reprinted as appendices. They are listed in chronological order below. In these three works I consulted with MF, CTY and JR in research and editing making content clarifying contributions to these manuscripts:
\begin{itemize}
    \item ``Classical neutral point particle in linearly polarized EM plane wave field'' by~\citet*{Formanek:2019cga} explores the dynamical equations presented in \rapp{appendixC} for neutral particles with magnetic moment.
    \item ``Radiation reaction friction: Resistive material medium'' by~\citet*{Formanek:2020zwc} introduces a novel model of relativistic covariant friction within a medium.
    \item ``Motion of classical charged particles with magnetic moment in external plane-wave electromagnetic fields'' by~\citet*{Formanek:2021mcp} is a followup to the above 2019 work and \rapp{appendixC} for charged particles with magnetic moment.
    \item ``Decomposition of Fermi gas into zero and finite temperature distributions with examples'' by~\citet*{Yang:2023aaa} is a mathematical methods paper detailing a novel analytic form of the finite temperature behavior of the Fermi-Dirac distribution function. The cold magnetized gas is analyzed as an example.
\end{itemize}

\chapter{The importance of spin}
\label{chap:intro}
\noindent All fundamental particles known in physics have a non-zero quantized spin angular momentum with the exception of the Higgs boson which is a scalar with spin-0. All other confirmed elementary particles (such as electrons, quarks, photons, etc...) have values of either spin-1/2 or spin-1. Particles with even values of spin are known as bosons while half-integer particles with spin are called fermions. Composite particles (such as atomic nuclei) can exhibit more exotic spin values and fundamental particles with higher spins such as spin-3/2 or spin-2 graviton are commonly predicted in beyond-standard-model (BSM) physics.

In the realm of the Poincar{\'e} group of spacetime symmetry (rotations, boosts and translations) transformations, each particle can be uniquely labeled by two distinct Casimir invariants: mass and spin. These two operators commute with all generators of the Poincar{\'e} group and act as labels which represent a particle. Therefore in a relativistic context, mass and spin are of fundamental importance on equal footing.

If a particle is electrically charged, then by virtue of its spin it will have a magnetic dipole moment. Most neutral particles with spin, though not all, will also have magnetic dipoles though for more complex reasons. Therefore the magnetic behavior of particle is an important window into probing one of the most fundamental properties in physics. As quantum mechanics is not well described in terms of forces or accelerations (except in the context of Ehrenfest-style equations), there is no simple operator description of torque and spin-forces despite having played a key role in the development of quantum mechanics. For a short historical overview of spin and its relationship to angular momentum, see~\cite{Ohanian:1986wg}.

This introduction serves to motivate the fundamental concepts of spin, magnetic moment and electromagnetism which have played a crucial role in the history physics and will be explored in the subsequent research chapters. Magnetic (and electric) dipoles, anomalous magnetic moments (AMM), and the Dirac and Dirac-Pauli wave equations which describe spin-1/2 fermions are covered in \rsec{sec:mom}. The Klein-Gordon-Pauli equation is introduced in \rsec{sec:kgp}. Lastly, \rsec{sec:flrw} covers topics in $\Lambda\mathrm{CDM}$ cosmology which are particular relevance to \rchap{chap:cosmo}. This chapter will also serve to establish notation and mathematical conventions. SI units will be used unless otherwise stated.



\section{Quantum magnetic dipoles and wave equations}
\label{sec:mom}
In classical theory, when charges rotate or circulate in some manner, a magnetic field is produced characterized by the magnetic dipole moment of the system. An Amp{\`e}rian loop of wire with a current is the quintessential example. This concept can be transplanted into quantum theory for spinning particles where the natural size of the magnetic moment of a particle (in this context a charged lepton) is given by the magneton value
\begin{gather}
    \label{mag:1}
    \mu_{\ell}\equiv\frac{e\hbar}{2m_{\ell}}
\end{gather}
where the lepton (denoted by $\ell$) has charge $e$ and mass $m_{\ell}$. 

A quick word on notation: Euclidean three-vectors and matrices will be denoted by boldface font. If indices are specifically printed, they will be done so using Latin indices such as $s_{i}$. Inner products of three-vectors will be noted via $\bb{a}\cdot\bb{b}=a_{i}b_{i}$ using Einstein summation notation where repeated indices are summed over. For electrons, \req{mag:1} is referred to as the Bohr magneton $\mu_{B}$. The non-relativistic spin operator $\bb{S}$ for a spin-1/2 particle is defined as
\begin{gather}
    \label{qspin:1}
    \bb{S}=\frac{\hbar}{2}\bb{\sigma}=\frac{\hbar}{2}\left(\sigma_{1},\,\sigma_{2},\,\sigma_{3}\right)^\mathrm{T}\,,
\end{gather}
where $\bb{\sigma}$ is the three-vector comprised of the familiar $2\times2$ Pauli matrices which act upon two-component spinors $\chi=(\chi_{1},\chi_{2})^\mathrm{T}$. Spinor indices will be suppressed or noted with Latin indices. The algebra defined by the commutators of the Pauli matrices serves as a representation of $SU(2)$ group structure
\begin{gather}
    \label{pauli:1}
    \{\sigma_{i},\sigma_{j}\}=2\delta_{ij}\,,\qquad
    [\sigma_{i},\sigma_{j}] = 2i\varepsilon_{ijk}\sigma_{k}\,,
\end{gather}
where $\varepsilon_{ijk}$ is the totally antisymmetric Levi-Civita symbol and $\delta_{ij}$ is the Kronecker delta.

The relativistic theory of spin-1/2 fermions however necessitates a four-component spinor $\psi=(\psi_{1},\psi_{2},\psi_{3},\psi_{4})^\mathrm{T}$ which as Dirac famously noted accommodates the required degrees of freedom for particles and antiparticles of both spin up $(\uparrow)$ and spin down $(\downarrow)$ eigenstates. The Hamiltonian density (in the Dirac representation) for the magnetic dipole moment interaction is given by
\begin{gather}
	\label{pauli:2}
    \mathcal{H}_\mathrm{int} = \frac{e\hbar}{2m_{\ell}}\psi^{\dag}
    \begin{pmatrix}
        -\bb{\sigma}\cdot\bb{B} & i\bb{\sigma}\cdot\bb{E}/c\\
        -i\bb{\sigma}\cdot\bb{E}/c & \bb{\sigma}\cdot\bb{B}
    \end{pmatrix}
    \psi\,,
\end{gather}
where $\psi^{\dag}$ is the complex conjugate transpose of the $\psi$ spinor. The electric $\bb{E}$ and magnetic $\bb{B}$ fields are defined in terms of the scalar potential $V$ and vector potential $\bb{A}$ in the usual way.
\begin{align}
    \label{eb:1}
    \bb{E}=-\bb{\nabla}V-\frac{\partial\bb{A}}{\partial t}\,,\qquad
    \bb{B}=\bb{\nabla}\times\bb{A}\,.
\end{align}

In the non-relativistic limit for particle states, the lower (antiparticle) components of $\psi$ are suppressed by $|\bb{p}|/mc$. We can approximate the particle states in terms of two-component spinors $\chi$ to first order as
\begin{align}
    \label{approx:1}
    \psi\approx\left(\chi,\ \frac{\bb{\sigma}\cdot\bb{\pi}}{2m_{\ell}c}\chi\right)^\mathrm{T}\,,\qquad \bb{\pi}=\bb{p}-e\bb{A}\,.
\end{align}
A more rigorous method of obtaining non-relativistic Hamiltonian can be found in~\cite{Foldy:1949wa}. The operator $\bb{\pi}$ is the kinetic momentum operator written in terms of canonical momentum $\bb{p}$ and vector potential $\bb{A}$. Making use of the identity
\begin{align}
    \sigma_{i}\sigma_{j} = \delta_{ij} + i\varepsilon_{ijk}\sigma_{k}\,,
\end{align}
we insert \req{approx:1} into \req{pauli:2} yielding to order $\mathcal{O}(1/m^{3})$
\begin{gather}
    \label{ham:1}
    \mathcal{H}_\mathrm{int} \approx -\chi^{\dag}\left(\frac{e\hbar}{2m_{\ell}}\bb{\sigma}\cdot\bb{B}
    +\frac{ie\hbar}{4m_{\ell}^{2}c^{2}}\Big[(\bb{\sigma}\cdot\bb{E}),(\bb{\sigma}\cdot\bb{\pi})\Big]\right)\chi\\
    \label{ham:2}
    \mathcal{H}_\mathrm{int} \approx -\chi^{\dag}\left(\frac{e\hbar}{2m_{\ell}}\bb{\sigma}\cdot\bb{B}
    +\frac{e\hbar^{2}}{4m_{\ell}^{2}c^{2}}\bb{\nabla}\cdot\bb{E}
    +\frac{e\hbar}{4m_{\ell}^{2}c^{2}}\bb{\sigma}\cdot\left(\bb{E}\times\bb{\pi}-\bb{\pi}\times\bb{E}\right)\right)\chi\,.
\end{gather}
Keeping only up to first order, the dipole interaction \req{pauli:2} reduces to 
\begin{gather}
	\label{pauli:3}
    \mathcal{H}_\mathrm{int} \approx -\frac{e\hbar}{2m_{\ell}}\chi^{\dag}\bb{\sigma}\cdot\bb{B}\chi\,,
\end{gather}
which is the expected non-relativistic quantum dipole term. The second and third terms in \req{ham:2} can be interpreted as a Darwin term $\sim\bb{\nabla}\cdot\bb{E}$ sensitive to charge density and spin orbit coupling $\sim\bb{\sigma}\cdot(\bb{E}\times\bb{p})$. We will return to relativistic notation and concepts in \rsec{sec:dp}.

The magnetic moment operator $\bb{\mu}$, as suggested by \req{pauli:3} is defined in terms of the Pauli matrices as
\begin{gather}
    \label{mag:3}
    \bb{\mu}=g\left(\frac{e\hbar}{2m_{\ell}}\right)\frac{\bb{\sigma}}{2}=g\mu_{\ell}\frac{\bb{\sigma}}{2}\,,\qquad\mu\equiv\frac{g}{2}\mu_{\ell}\,,
\end{gather}
where $\mu$ is the `total magneton' value representing the full magnetic moment. The parameter $g$ in \req{mag:3} is the gyromagnetic ratio (or $g$-factor) of the particle. The `natural' value is $g\!=\!2$. While this prediction is normally attributed to the Dirac equation, it justified from the construction of the kinetic energy operator in the Schr{\"o}dinger-Pauli equation; see \rsec{sec:unique} and~\cite{sakurai1967advanced}.

In non-relativistic quantum mechanics, the time-dependant Schr{\"o}dinger-Pauli (SP) equation (with Hamiltonian $H_\mathrm{SP}$) for a charged particle is given by
\begin{gather}
	\label{sp:1}
    {H}_{\mathrm{SP}}\chi=\left(\frac{1}{2m_{\ell}}\bb{\pi}^{2}-\bb{\mu}\cdot\bb{B}+e{ V}\right)\chi=i\hbar\frac{\partial}{\partial t}\chi\,,\qquad
    \bb{\pi}=\bb{ p}-e{\bb{A}}\,,
\end{gather}
where $\chi$ is again a two-component spinor. It is well known that \req{sp:1} is obtainable from the Dirac equation (see \rsec{sec:dp}) in the non-relativistic limit.

Before moving on, we will verify that the SP \req{sp:1} contains within it an expression of the Stern-Gerlach force which was used to first provide evidence of the quantization of angular momentum~\citep{Gerlach:1922zz}. To accomplish this, we will work in the Heisenberg representation where operators obey the following equation of motion
\begin{align}
    \label{h:1}
    i\hbar\frac{d\bb{O}}{dt}=[\bb{O},H]+
    i\hbar\frac{\partial\bb{O}}{\partial t}\,,
\end{align}
To obtain a `force' in quantum mechanics we need to find the time derivative of the kinematic momentum operator $\bb{\pi}$ which is given by
\begin{gather}
    \label{h:2}
    \frac{d\bb{\pi}}{dt}=-\frac{i}{\hbar}[\bb{\pi},H_\mathrm{SP}]+\frac{\partial\bb{\pi}}{\partial t}=-\frac{i}{\hbar}\left[\bb{\pi},\frac{(\bb{\sigma}\cdot\bb{\pi})^{2}}{2m}+eV\right]+\frac{\partial\bb{\pi}}{\partial t}\,,\\
    \label{h:3}
    \frac{\partial\bb{\pi}}{\partial t} = -\frac{\partial e\bb{A}}{\partial t}\,,\qquad
    [\pi_{i},\pi_{j}]=ie\hbar\varepsilon_{ijk}B_{k}\,,\qquad
    [\pi_{i},B_{j}]=-i\hbar\nabla_{i}B_{j}\,.
\end{gather}
After some derivation and making use of the identities in \req{h:3}, we arrive at the quantum analog of the Lorentz force for particles with spin
\begin{gather}
    \label{ehren:1}
    \boxed{\frac{d\bb{\pi}}{dt}=e\bb{E}+\frac{e}{2m_{\ell}}(\bb{\pi}\times\bb{B}-\bb{B}\times\bb{\pi})+\frac{e\hbar}{2m_{\ell}}\sigma_{i}\bb{\nabla}B_{i}}\,.
\end{gather}
The last term in the expression is the Stern-Gerlach force which is sensitive to inhomogeneous magnetic fields. We also note this equation is suggestive of the `Amp{\'e}rian' dipole force which is in the direction of the gradient $\bb{\nabla}$ rather than the `Gilbertian' type of dipole force which is in the direction of the field $\bb{B}$; see \rsec{sec:cspin}. \req{ehren:1} can be connected to our classical understanding by taking the expectation value and casting it as an Ehrenfest-style theorem~\citep{Ehrenfest:1927swx}.

\subsection{Anomalous magnetic moment}
In nature there is no particle with exactly $g\!=\!2$. As seen in \rt{tab:gfactor}, composite particles often deviate from $g\!=\!2$ greatly as the $g$-factor of a composite particle is related to its internal composition. In the case of the neutron and proton, the internal quarks themselves are responsible in a nontrivial fashion~\citep{Chang:2015qxa}. The comparison between three listed isotopes of hydrogen also displays how magnetic moments can `cancel out' or add together: While the deuterium nucleus value of $g$ is suppressed by the extra neutron, the two neutrons in the tritium nucleus balance one another returning the ratio into one manifestly similar to the proton. This reasoning however only works as a heuristic and non-perturbative Lattice QCD computations~\citep{Detmold:2019ghl} are needed to obtain the magnetic moments of hadrons with great accuracy.

When $g\neq2$ (which is true for all physical particles with magnetic moment; composite of otherwise) the anomalous magnetic moment (AMM) can be defined via 
\begin{gather}
    \label{amm:1}
    a\equiv\frac{g}{2}-1\,,\qquad
    a\frac{e\hbar}{2m_{\ell}}\rightarrow\delta\mu\equiv\mu-\mu_{\ell}\,,
\end{gather}
where $a$ is the anomaly parameter. We also introduce $\delta\mu$ as the anomalous magneton which will be helpful in our proposal to connect mass and magnetic moment in \rsec{sec:ikgp} and \rsec{sec:numoment}.

\begin{table}
	\centering
\begin{tabular}{r|c|l}
    particle & category & $g$-factor\\
    \hline
	electron & elementary & -2.002\ 319\ 304\ 362\ 56(35)\\
	muon & elementary & -2.002\ 331\ 8418(13)\\
	tau & elementary & -2.036(34)\\
	neutron & composite & -3.826\ 085\ 45(90)\\
	proton & composite & \ 5.585\ 694\ 6893(16)\\
	deuteron & composite & \ 0.857\ 438\ 2338(22)\\
	triton & composite & \ 5.957\ 924\ 931(12)\\
\end{tabular}
	\caption{The $g$-factor of various particles found in~\cite{Tiesinga:2021myr,ParticleDataGroup:2022pth}.}
	\label{tab:gfactor}
\end{table}

The anomalous magnetic moment of a particle can arise from a variety of physical sources with the most famous being the one-loop vacuum polarization contribution to the electron first computed by~\cite{Schwinger:1951nm}. In that work, the first correction to $g$ is given by
\begin{gather}
    a_{e} = \frac{\alpha}{2\pi}\,,\qquad
    \alpha\equiv\frac{1}{4\pi\varepsilon_{0}}\frac{e^{2}}{\hbar c}\,,
\end{gather}
where $\alpha$ is the fine structure constant with an approximate value of $1/137$. The measurement of the electron's $g$-factor is among the most precise measurements in all of physics~\citep{Tiesinga:2021myr}. Precision measurements of the muon's anomalous magnetic moment are rapidly improving~\citep{Muong-2:2023cdq}. This makes the study of magnetic moment, and spin, an exciting area of physical research as new developments continue today.

\subsection{Dirac and Dirac-Pauli equations}
\label{sec:dp}
\noindent While it is always beneficial to be well-appraised of non-relativistic mechanics, nature is intrinsically relativistic and therefore this dissertation must be as well. The relativistic generalization of \req{sp:1} is the Dirac equation given by
\begin{gather}
    \label{dirac:1a}
    \left(\gamma_{\alpha}\left(i\hbar\partial^{\alpha} - eA^{\alpha}\right)-m_{\ell}c\right)\psi=0\,,\\
    \label{dirac:1b}
    \pi^{\alpha}=i\hbar{\widetilde\nabla}^{\alpha}=i\hbar\partial^{\alpha}-eA^{\alpha}\,.
\end{gather}
The wave function $\psi$ in \req{dirac:1a} is understood to be a four-component spinor and $\widetilde\nabla^{\alpha}$ in \req{dirac:1b} is the covariant derivative. $\pi^{\alpha}$ is the four-vector version of the kinetic momentum versus the four-momentum $p^{\alpha}=i\hbar\partial^{\alpha}$. Four-vectors and tensors in this work will be denoted by Greek indices. Inner products of four-vectors will be noted by $a\cdot b=a^{\alpha}\eta_{\alpha\beta}b^{\beta}=a^{\alpha}b_{\alpha}$ again following Einstein notation. The four-derivative $\partial^{\alpha}$ and four-potential $A^{\alpha}$ are defined as
\begin{gather}
    \label{dirac:2}
    \partial^{\alpha}=\left(\frac{1}{c}\frac{\partial}{\partial t},\,-\bb{\nabla}\right)\,,\qquad A^{\alpha}=\left(\frac{V}{c},\,\bb{A}\right)\,.
\end{gather}
We have written the Dirac equation here in the covariant form where $\gamma^{\alpha}$ are the gamma matrices which obey the anticommuting Clifford algebra
\begin{gather}
    \label{gamma:1}
    \{\gamma_{\alpha},\gamma_{\beta}\}=\gamma_{\alpha}\gamma_{\beta} + \gamma_{\beta}\gamma_{\alpha} = 2\eta_{\alpha\beta}\,,\\
    \eta_{\alpha\beta}=\mathrm{diag}(+1,-1,-1,-1)\,,
\end{gather}
where $\eta_{\alpha\beta}$ is the flat spacetime Minkowski metric tensor defined with a positive time metric signature. The metric tensor is also responsible for raising and lowering covariant and contravariant indices e.g. $a_{\alpha}=\eta_{\alpha\beta}a^{\beta}$. As $\gamma^{\alpha}$ are also spinor matrices, the commutator in \req{gamma:1} carries implicit spinor indices which here computes to the $4\times4$ identity matrix $\mathbbm{1}_{4}$ (which is suppressed). We also introduce the `fifth' gamma matrix $\gamma^{5}$ which anticommutes with $\gamma^{\alpha}$ and the following standard conventions following~\cite{Itzykson:1980rh}
\begin{alignat}{1}
	\label{conventions:1} \bb{\alpha}=\gamma^{0}\bb{\gamma}\,,\indent \bb{\Sigma}=\gamma^{5}\bb{\alpha}\,,\indent \gamma^{5}=i\gamma^{0}\gamma^{1}\gamma^{2}\gamma^{3}\,,\indent \gamma^{2}_{5}=1\,.
\end{alignat}

As mentioned before, \req{dirac:1a} predicts $g\!=\!2$ which is a standard calculation in many textbooks. The most straight-forward manner to generalize the Dirac equation allowing for an anomalous magnetic moment is to add a Pauli term proportional to the anomalous parameter $a$. While in most texts, the anomaly is given in terms of $g-2$ or $a$, we wish to keep our equations generalized to fermions of any given charge $e$ and magnetic moment $\mu$. 

Therefore we make use of the substitution in \req{amm:1} and write the Dirac-Pauli~(DP) equation as
\begin{gather}
	\label{dp:1}
    \left(\gamma_{\alpha}\left(i\hbar\partial^{\alpha} - eA^{\alpha}\right) - m_{\ell}c - \delta\mu\frac{1}{2c}\sigma_{\alpha\beta}F^{\alpha\beta}\right)\psi=0\,,
\end{gather}
where the antisymmetric spin tensor $\sigma_{\alpha\beta}$ is defined in terms of the commutator of the gamma matrices
\begin{alignat}{1}
	\label{sigma:1}
    \sigma_{\alpha\beta}=\frac{i}{2}\left[\gamma_{\alpha},\gamma_{\beta}\right]=\frac{i}{2}\left(\gamma_{\alpha}\gamma_{\beta}-\gamma_{\beta}\gamma_{\alpha}\right)\,.
\end{alignat}
In the above we printed only the magnetic moment term; some remarks about electric dipole moments (EDM) and CP violation can be found in \rsec{sec:edm}.

Exact solutions to the DP equation are relatively scarce due to the complicating nature of the anomalous term. The most extensively studied solutions are those with high symmetries or constant external fields \citep{Thaller:1992ji}. When the anomalous part $\delta\mu$ is zero, the Dirac equation is recovered. $F^{\alpha\beta}$ is the standard antisymmetric electromagnetic field tensor defined by
\begin{gather}
    \label{em:1}
    F^{\alpha\beta} = \partial^{\alpha}A^{\beta} - \partial^{\beta}A^{\alpha} = 
    \begin{pmatrix}
        0        & -E_{1}/c  & -E_{2}/c  & -E_{3}/c\\
        E_{1}/c  & 0         & -B_{3}    & B_{2}\\
        E_{2}/c  & B_{3}     & 0         & -B_{1}\\
        E_{3}/c  & -B_{2}    & B_{1}     & 0
    \end{pmatrix}\,.
\end{gather}
The electromagnetic field tensor can also be defined in terms of the commutators of the covariant derivative \req{dirac:1b} as
\begin{align}
    \label{curve:1}
    \left[\widetilde\nabla^{\alpha},\widetilde\nabla^{\beta}\right]=
    \frac{ie}{\hbar}F^{\alpha\beta}\,.
\end{align}
It is also useful to define the Hodge dual of the electromagnetic field tensor
\begin{gather}
    \label{em:2}
    F_{\alpha\beta}^{*} = \frac{1}{2}\varepsilon_{\alpha\beta\mu\nu}F^{\mu\nu} = 
    \begin{pmatrix}
        0        & -B_{1}  & -B_{2}  & -B_{3}\\
        B_{1}  & 0         & -E_{3}/c    & E_{2}/c\\
        B_{2}  & E_{3}/c     & 0         & -E_{1}/c\\
        B_{3}  & -E_{2}/c    & E_{1}/c     & 0
    \end{pmatrix}\,,
\end{gather}
where we use the four-dimensional fully antisymmetric Levi-Civita pseudo-tensor $\varepsilon_{\alpha\beta\mu\nu}$ with the $\varepsilon_{0123}=+1$ convention. The contracted portion $\sigma_{\alpha\beta}F^{\alpha\beta}$ in the Pauli term in \req{dp:1} can be further expressed as
\begin{alignat}{1}
	\label{dp:2} \frac{1}{2}\sigma_{\alpha\beta}F^{\alpha\beta} = i\bb{\alpha}\cdot\bb{E}/c-\bb{\Sigma}\cdot\bb{B} = i\gamma^{0}\bb{\gamma}\cdot\bb{E}/c-\gamma^{5}\gamma^{0}\bb{\gamma}\cdot\bb{B}\,,
\end{alignat}
which captures that relativistic magnetic moments should be sensitive to electric as well as magnetic fields as required by Lorentz transformations of the $\bb{E}$ and $\bb{B}$ fields. We note that \req{dp:2} is the matrix which appears in \req{pauli:2} specifically in the Dirac representation of $\bb{\alpha}$ and $\bb{\Sigma}$. This should be unsurprising if one considers how the non-relativistic dipole form must generalize under Lorentz boosts which mix electric and magnetic fields.

The DP equation can be obtained from perturbative QED as an effective field theory for leptons due to vacuum polarization; see standard texts \cite{Itzykson:1980rh,Schwartz:2014sze}. However, if a particle's anomalous magnetic moment is not sourced by perturbative QFT, then the Pauli term introduced in \req{dp:1} must be added by hand \emph{ad hoc} or obtained via non-perturbative means such as Lattice calculations~\citep{Aoyama:2020ynm}. This is the case for the hadronic contribution to anomalous magnetic moment of leptons as well as any composite particle such as the proton or neutron whose moment is determined by internal structure~\citep{Proceedings:2012ulb,Green:2015wqa}.

Therefore we can describe the AMM as an added Lagrangian interaction term
\begin{gather}
    \label{lamm:1}
    \mathcal{L}_\mathrm{DP,AMM} = -{\bar\psi}\left(\delta\mu\frac{1}{2}\sigma_{\alpha\beta}F^{\alpha\beta}\right)\psi\,,
\end{gather}
where ${\bar\psi}=\psi^{\dagger}\gamma^{0}$ is the Dirac adjoint. While the focus of this dissertation is not on quantum field theory (QFT), it is valuable to note that the Pauli Lagrangian term in \req{lamm:1} is considered 5-dimensional as the $\psi$ fields have natural units of $[\mathrm{length}]^{-3/2}$ as determined from the Dirac Lagrangian
\begin{gather}
    \label{ld:1}
    \mathcal{L}_\mathrm{D}/c=\bar\psi\left(i\hbar\gamma_{\alpha}\widetilde\nabla^{\alpha}-m_{\ell}c\right)\psi\,,\qquad \mathcal{L}_\mathrm{DP} = \mathcal{L}_\mathrm{D} + \mathcal{L}_\mathrm{DP,AMM}\,.
\end{gather}

To demonstrate, we note that the electromagnetic field tensor has natural units of $F^{\alpha\beta}\sim[\mathrm{length}]^{-2}$. Therefore the product $\psi\sigma_{\alpha\beta}F^{\alpha\beta}\psi$ has natural units of $[\mathrm{length}]^{-5}$ and the coefficient of \req{lamm:1} (given by $\delta\mu$) has to compensate with $\delta\mu\sim[\mathrm{length}]^{1}$. This makes the DP Lagrangian unsuitable for renormalization which is an essential feature required for well-behaved QFTs. While this does not stop us using DP as an effective QFT with some natural cutoff scale responsible for the anomalous moment, it does reduce the usefulness of the equation as a general description of quantum dipole moments.

As such, there is no reason to expect non-perturbative sources of magnetic moment to strictly adhere to the DP form. Additionally, the DP equation has the physically inelegant consequence of splitting the spin dynamics of fermions into (a) natural $g\!=\!2$ behavior (see \rsec{sec:unique}) encompassed by the spinor structure of the Dirac equation and (b) the anomalous behavior contained in the Pauli term.

\section{Klein-Gordon-Pauli equation}
\label{sec:kgp}
\noindent While the DP equation is more commonly studied, there exists an alternative wave equation which describes the magnetic behavior of fermions called the Klein-Gordon-Pauli (KGP) equation. This first introduced by~\cite{Fock:1937dy} and found usefulness in the quantum electrodynamics~\citep{Feynman:1951gn} and in studying weak interactions~\citep{Feynman:1958ty} due to the ease of describing chiral states.

The KGP equation is generally considered to be the `square' of the Dirac equation as unlike the Dirac or DP equations, it is a second order equation wave equation for the four-component spinor $\Psi$
\begin{alignat}{1}
	\label{kgp:1} \left((i\hbar\partial^{\alpha}-eA^{\alpha})^{2}-m_{\ell}^{2}c^{2}-g\mu_{\ell}m_{\ell}\frac{1}{2}\sigma_{\alpha\beta}F^{\alpha\beta}\right)\Psi=0\,.
\end{alignat}
The initial benefit of the KGP formulation is that the wave equation fully commutes with $\gamma^{5}$ making eigen-functions explicitly good chiral states.

\req{kgp:1} is mathematically similar to the Klein-Gordon equation which describes charged scalar particles. In the same manner as scalar-QED, the squared covariant derivative contains a $e^{2}A^{2}$ term which in QFT results in the presence of a 4-vertex seagull interaction~\citep{Schwartz:2014sze} at tree-level.

It is important to emphasize that the KGP \req{kgp:1} and DP \req{dp:1} are distinct wave equations which do not share solutions except when $g\!=\!2$ whereas both reduce to the Dirac \req{dirac:1a}. We will clarify on the relationship between the KGP and Dirac equations here by rewriting the Dirac equation in \req{dirac:1a} as
\begin{alignat}{1}
	\label{do:1} \mathcal{D}_{\pm}=i\hbar\gamma_{\alpha}\widetilde\nabla^{\alpha}\pm m_{\ell}c\,,\qquad
    \mathcal{D}_{-}\psi=0\,,
\end{alignat}
with a `Dirac operator' $\mathcal{D}_{\pm}$ defined in terms of positive and negative mass. This operator has the following properties
\begin{gather}
    \label{do:2}
    \mathcal{D}_{-}=-\gamma^{5}\mathcal{D}_{+}\gamma^{5}\,,\qquad
    [\mathcal{D}_{+},\mathcal{D}_{-}]=0\,.
\end{gather}
Ignoring the proportionality factor of $\sqrt{\hbar/m_{\ell}c}$ which accommodate the units of $\psi$ versus $\Psi$, we can complete the square of the Dirac equation via the substitution
\begin{gather}
    \label{do:3a}
    \psi\rightarrow\mathcal{D}_{+}\Psi\,,\qquad
    \mathcal{D}_{-}\psi \rightarrow \mathcal{D}_{-}\mathcal{D}_{+}\Psi\,,\\
	\label{do:3}
    \mathcal{D}_{+}\mathcal{D}_{-}\Psi=\left(-\hbar^{2}\widetilde\nabla^{2}-m_{\ell}^{2}c^{2}-e\hbar\frac{1}{2}\sigma_{\alpha\beta}F^{\alpha\beta}\right)\Psi=0\,.
\end{gather}
This procedure yields the KGP equation for $g\!=\!2$. This algebraic `square root' will be elaborated on further in \rsec{sec:unique}.

For $g\!\neq\!2$ the relationship between the DP and KGP equation becomes more complicated. Instead of a clean algebraic separation, the substitution between $\psi$ and $\Psi$ requires an infinite series expansion resulting from the non-local inverse substitution 
\begin{alignat}{1}
	\label{nonlocal:1} \Psi\rightarrow\frac{1}{\mathcal{D}_{+}}\psi = \frac{1}{m_{\ell}c}\left(1 - \frac{\hbar}{m_{\ell}c}i\gamma_{\alpha}\widetilde\nabla^{\alpha} - \frac{\hbar^{2}}{m_{\ell}^{2}c^{2}}\left(\gamma_{\alpha}\widetilde\nabla^{\alpha}\right)^{2} + \ldots\right)\psi\,.
\end{alignat}
The expansion in \req{nonlocal:1} is considered non-local because it requires an infinite number of initial conditions to determine.

While this procedure `square roots' the KGP equation $(\mathrm{\sqrt{KGP}})$, the resulting AMM Pauli Lagrangian \req{lamm:1} picks up an infinite number of derivative and field terms which makes the theory rather unpalatable.
\begin{gather}
    \label{lamm:2}
    \mathcal{L}_{\sqrt{\mathrm{KGP}}} = \mathcal{L}_\mathrm{D}+\mathcal{L}_\mathrm{\sqrt{KGP},AMM}\,,\\
    \label{lamm:3}
    \mathcal{L}_\mathrm{\sqrt{KGP},AMM} = -{\bar\psi}\left(\delta\mu\frac{1}{2}\sigma_{\alpha\beta}F^{\alpha\beta}\left(1 - \frac{\hbar}{m_{\ell}c}i\gamma_{\alpha}\widetilde\nabla^{\alpha} - \frac{\hbar^{2}}{m_{\ell}^{2}c^{2}}\left(\gamma_{\alpha}\widetilde\nabla^{\alpha}\right)^{2} + \ldots\right)\right)\psi\,.
\end{gather}
We note each term in \req{lamm:3} is preceded by powers of the reduced Compton wavelength $\lambda_\mathrm{C}\!\equiv\!\hbar/m_{\ell}c$ therefore the $\sqrt{\mathrm{KGP}}$ model still might be of interest to study assuming the physical system that admits a reasonable cutoff.

While the first term present in \req{lamm:3} is indeed the correct $\mathcal{L}_\mathrm{DP,AMM}$ term, the resulting non-local behavior ultimately breaks the unitarity of the theory making it unsuitable as a fundamental particle theory~\citep{Veltman:1997am}. While the above is suggestive that there exists no unitary transform between the KGP and DP wave equations, we do not claim it as an absolute proof. If a generalized description of $g\!\neq\!2$ magnetic moment exists as a good quantum field theory, then likely non-minimal electromagnetic terms are required to maintain both renormalization and unitarity.

\subsection{Features of the KGP Lagrangian}
\label{sec:lagrangian}
\noindent Before continuing to specific physical problems, we consider how current conversation functions in the KGP formulation of fermions and how it might differ from the Dirac current $\mathcal{J}_\mathrm{D}^{\mu}\propto-i\bar\psi\gamma^{\mu}\psi$. The KGP equation can be obtained from a Lagrangian not dissimilar to the Klein-Gordon Lagrangian~\citep{Delgado-Acosta:2010ita,DelgadoAcosta:2015ikk,Espin:2015bja} and has the expression
\begin{gather}
\label{lagrangian:1} \mathcal{L}_\mathrm{KGP}/c^{2}=\left(i\hbar{\widetilde\nabla}^{\mu}\right)^{\dag}\bar{\Psi}h_{\mu\nu}\left(i\hbar{\widetilde\nabla}^{\mu}\right)\Psi-m^{2}c^{2}\bar{\Psi}\Psi\,,\qquad h_{\mu\nu}=\eta_{\mu\nu}-i\frac{g}{2}\sigma_{\mu\nu}\,.
\end{gather}
The matrix $h_{\mu\nu}$ acts an `effective' metric which has been modified to account for the presence of an AMM. We note that the field $\Psi$ must have units $[\mathrm{length}]^{-1}$ such that the Lagrangian density itself has natural units of $[\mathrm{length}]^{-4}$.

In comparison to the DP AMM Lagrangian \req{lamm:1}, the KGP magnetic moment Lagrangian obtained from \req{lagrangian:1} is
\begin{align}
    \label{lagrangian:2}
    \mathcal{L}_\mathrm{KGP,MM}/c^{2} = -{\bar\Psi}\left(\frac{g}{2}\mu_{\ell} m_{\ell}\sigma_{\alpha\beta}F^{\alpha\beta}\right)\Psi\,.
\end{align}
While they are mathematically similar both being `Pauli terms', there are some important differences. Here the combination of fields $\Psi$ and $F^{\alpha\beta}$ have natural units $[\mathrm{length}]^{-4}$ and the coupling coefficient $\mu m_{\ell}\!\propto\!g$ is manifestly dimensionless. This means the KGP Lagrangian is at first inspection renormalizable which is an improvement over the DP Lagrangian~\citep{Rafelski:2022bsv}. Literature however suggests that the KGP Lagrangian requires additional fermion self-interactions $\mathcal{L}_\mathrm{int}\sim\mathcal{O}(\bar\Psi\Psi)^{2}$ to be fully renormalizable (unless $g\!=\!0,\pm2$) which are not forbidden at tree-level~\citep{Angeles-Martinez:2011wpn,Vaquera-Araujo:2012jlk}.

The conserved current obtained from \req{lagrangian:1} can be expressed as
\begin{gather}
\label{norm:2}
\mathcal{J}^{\mu}=-\frac{1}{c^{2}}\frac{\partial\mathcal{L}}{\partial eA_{\mu}}\equiv 
 \mathcal{J}^\mu_{\mathrm{Conv}}+\mathcal{J}^\mu_{\mathrm{Mag}}\,, \\
\mathcal{J}^{\mu}=\bar{\Psi}\left(i\hbar{\widetilde\nabla}^{\mu}\right)\Psi + \left(i\hbar{\widetilde\nabla}^{\mu}\right)^{\dag}\bar{\Psi}\Psi + i\frac{g}{2}\bar{\Psi}\sigma^{\mu\nu}\left(i\hbar{\widetilde\nabla}_{\nu}\right)\Psi + i\frac{g}{2}\left(i\hbar{\widetilde\nabla}_{\nu}\right)^{\dag}\bar{\Psi}\sigma^{\nu\mu}\Psi\,.
\end{gather}
The conserved current \req{norm:2} can be interpreted as the sum of a convection current $\mathcal{J}_{\mathrm{Conv}}$ and magnetization current $\mathcal{J}_{\mathrm{Mag}}$.
\begin{align}
\label{norm:3a}\mathcal{J}^{\mu}_{\mathrm{Conv}}&=\bar{\Psi}\left(i\hbar{\widetilde\nabla}^{\mu}\right)\Psi + \left(i\hbar{\widetilde\nabla}^{\mu}\right)^{\dag}\bar{\Psi}\Psi\,,\\
\label{norm:3b} 
\mathcal{J}^{\mu}_{\mathrm{Mag}}&=\frac{g}{2}\hbar{\partial}_{\nu}\left(\bar{\Psi}\sigma^{\nu\mu}\Psi\right)\;.
\end{align}
 This is nearly identical to the Gordon Decomposition of the Dirac current $\mathcal{J}_\mathrm{D}^{\mu}$, with the exception that the magnetization current is proportional to $g$-factor.
 
The covariant derivative happens to simplify as $\widetilde\nabla\rightarrow\partial$ in \req{norm:3b} such that the current is a divergence of the spin density $\bar\Psi\sigma_{\mu\nu}\Psi$. Because of the antisymmetry of $\sigma_{\mu\nu}$ the spin tensor, the magnetization current is conserved independently of the the charge current. That both are independently conserved indicates conservation in both charge $(e)$ and magnetic moment $(\mu)$.

\subsection{The special case of g = 2}
\label{sec:unique}
There is a strong predilection in nature towards $g\!=\!2$ which can be explained by the requirements of kinetic operator in quantum mechanics. Rather than taking the non-relativistic limit of the Dirac equation, $g\!=\!2$ can also be derived as a consequence of replacing the definition of the inner product for vectors which accounts for spinor structure via an argument attributed to R. P. Feynman; see footnote in Chap. 3.2 of~\cite{sakurai1967advanced}.

The Schr{\"o}dinger equation can be converted into the SP \req{sp:1} via the following replacement
\begin{alignat}{1}
	\label{nat:0}
    \bb{\pi}^{2}\rightarrow(\bb{\sigma}\cdot\bb{\pi})^{2}=\pi_{i}\sigma_{i}\sigma_{j}\pi_{j}\,.
\end{alignat}
Because the $2\times2$ Pauli matrices $\sigma_{i}$ all anticommute, we can write down the relation
\begin{alignat}{1}
	\label{nat:1}
    (\bb{\sigma}\cdot\bb{a})(\bb{\sigma}\cdot\bb{b})=\bb{a}\cdot\bb{b}+i\bb{\sigma}\cdot(\bb{a}\times\bb{b})\,.
\end{alignat}
The non-relativistic kinetic energy (KE) Hamiltonian from \req{sp:1} then reads as
\begin{align}
	\label{nat:2}
    {H}_{\mathrm{SP,KE}}=\frac{1}{2m}\left(\bb{\sigma}\cdot\bb{\pi}\right)^{2}=\frac{1}{2m}\bb{\pi}^{2}+i\bb{\sigma}\cdot(\bb{\pi}\times\bb{\pi})=\frac{1}{2m}\bb{\pi}^{2}-\frac{e\hbar}{2m}\bb{\sigma}\cdot\bb{B}\,.
\end{align}
As the kinetic momentum operator $\bb{\pi}$ does not self-commute, its cross product is non-zero resulting in a magnetic moment term with magneton size $\mu_{\ell}=e\hbar/2m_{\ell}$ and $g\!=\!2$. Therefore, we see there is conceptual value in replacing the inner dot product with a more intricate algebraic structure; in this case: $\delta_{ij}\rightarrow\sigma_{i}\sigma_{j}$.

The natural gyromagnetic ratio then appears to arise from the $SU(2)$ Lie algebra representation that the Pauli matrices describe and electromagnetic minimal coupling. The natural scale of the magnetic moment can be interpreted as originating from group symmetry requirements on charged particles.

An almost identical argument that $g$-factor arises from spin-structure and electromagnetic coupling can be made for the relativistic case as well. First we consider the quantum analog to the energy-momentum relation
\begin{alignat}{1}
	\label{analog:1} \eta_{\alpha\beta}p^{\alpha}p^{\beta}\Phi=m^{2}c^{2}\Phi\,.
\end{alignat}
\req{analog:1} as written evaluates to the Klein-Gordon equation on scalar field $\Phi$ where the four-momentum is written in the position basis $p^{\alpha}\rightarrow i\hbar\partial^{\alpha}$. Much like the non-relativistic example, we can introduce spin by replacing the momentum inner product with one sensitive to a Clifford algebra~\citep{Weinberg:1995mt}. Rather than the Pauli matrices, the relativistic replacement utilizes the gamma matrices yielding
\begin{alignat}{1}
	\label{eq:spin:03}
    \eta_{\alpha\beta}\rightarrow\gamma_{\alpha}\gamma_{\beta}\,,\qquad
    \gamma_{\alpha}\gamma_{\beta}p^{\alpha}p^{\beta}\Psi&=m^{2}c^{2}\Psi\,.
\end{alignat}
Here $\Psi$ is understood to be a four-component spinor unlike in \req{analog:1}. The corresponding $4\times4$ matrix contraction identity analog to \req{nat:1} is then
\begin{alignat}{1}
	\label{eq:spin:04} \gamma_{\alpha}\gamma_{\beta}a^{\alpha}b^{\beta}=\eta_{\alpha\beta}a^{\alpha}b^{\beta}-i\sigma_{\alpha\beta}a^{\alpha}b^{\beta}\,.
\end{alignat}

In both the relativistic and non-relativistic cases, the distinction between spin-1/2 and spinless particles is only made apparent in the kinematics in the presence of electromagnetic fields. For minimal coupling $\pi^{\alpha}=p^{\alpha}-eA^{\alpha}$ we take advantage of the fact that any tensor product of vectors can be decomposed as a sum of commuting (symmetric) and anticommuting (antisymmetric) parts
\begin{align}
	\label{eq:spin:06} \pi^{\alpha}\pi^{\beta}=\frac{1}{2}\left\{\pi^{\alpha},\pi^{\beta}\right\}+
    \frac{1}{2}\left[\pi^{\alpha},\pi^{\beta}\right]\,,\qquad
    \gamma^{\alpha}\gamma^{\beta}=\frac{1}{2}\left\{\gamma^{\alpha},\gamma^{\beta}\right\}+
    \frac{1}{2}\left[\gamma^{\alpha},\gamma^{\beta}\right]\,.
\end{align}
From the above and \req{eq:spin:03} and \req{curve:1} we obtain
\begin{align}
	\label{eq:spin:07b} \gamma_{\alpha}\gamma_{\beta}\pi^{\alpha}\pi^{\beta}\Psi=
    \left(\eta_{\alpha\beta}\pi^{\alpha}\pi^{\beta}-\frac{e\hbar}{2}\sigma_{\alpha\beta}F^{\alpha\beta}\right)\Psi=m^{2}c^{2}\Psi\,.
\end{align}
\req{eq:spin:07b} is the square of the Dirac equation with precisely $g\!=\!2$ but in a different sense than the argument established in \req{do:3}. Rather than squaring the Dirac equation, from this perspective, we are enlarging the structure of the energy-momentum relation. The spin-1 Proca equations and spin-3/2 Rarita-Schwinger equations can also be justified via this line of reasoning with different replacements for the field and inner-product definition.

How $g\!\neq\!2$ AMM `breaks' the inner product substitution is seen more explicitly when we write the effective metric tensor $h_{\mu\nu}$ from \req{lagrangian:1} as
\begin{gather}
    \label{spinstruc}
    h_{\mu\nu}=\frac{1}{2}\{\gamma_{\mu},\gamma_{\nu}\}+\frac{1}{2}(1+a)[\gamma_{\mu},\gamma_{\nu}]=\gamma_{\mu}\gamma_{\nu}+\frac{a}{2}[\gamma_{\mu},\gamma_{\nu}]\,.
\end{gather}
The anomalous part in \req{spinstruc} is inconveniently unable to be packaged as the elementary tensor product of two four-vectors like in \req{eq:spin:03}. The same issue occurs with any anomalous EDM. We suggest that more elaborate algebraic structures might accommodate such terms more naturally though we leave that to future work.

Furthermore, compelling arguments can be made that all elementary particles of any spin must have a natural gyromagnetic factor of $g\!=\!2$; though we mention a competing idea is Belinfante's conjecture of $g\!=\!1/s$. To paraphrase the arguments by~\citet*{Ferrara:1992yc}, $g\!=\!2$ is likely the natural scale for particles of any spin because:
\begin{enumerate}[nosep]
	\item The W boson, as the only known higher spin charged elementary particle, has at tree level $g\!=\!2$ via a Proca-like equation.
	\item The relativistic TBMT torque equation is the same for any classical spin value and is most simple when the anomalous moment is zero.
	\item For arbitrary spin, $g\!=\!2$ facilitates finite Compton scattering cross sections without additional physical requirements.
	\item For charged interacting particles with arbitrary spin, open bosonic and super-symmetric string theory predicts $g\!=\!2$.
\end{enumerate}
We would like to add the additional argument that rotating charged black holes described by the Kerr-Newman metric also have a magnetic dipole moment with fixed $g\!=\!2$ character~\citep{Carter:1968rr}. This illustrates that in some sense the spin of a black hole is `particle-like' and dissimilar to the orbital Amp{\`e}rian motion of matter which has an orbital $g$-factor of $g_\mathrm{L}\!=\!1$; see \rsec{sec:homogeneous}.

While the above provide a nice justification for why particles should tend to this specific $g$-factor, the reality is no particle has exactly $g\!=\!2$ with all of them displaying some form of anomaly. The charged leptons come the closest to the natural value, but famously have vacuum polarization contributions~\citep{Schwinger:1951nm} from QED, non-perturbative hadronic contributions~\citep{Jegerlehner:2017gek}, and potentially BSM interactions~\citep{Knecht:2003kc} contributing to their anomalous magnetic dipole moment.

While the perturbative approach has proven to be exceedingly successful for the charged leptons, it is not appropriate for particles whose moments are dramatically different from $g\!=\!2$ or if the origin of the anomaly comes from internal structure such as the hadrons whose moments are determined by non-perturbative QCD~\citep{Pacetti:2014jai} and not weakly coupled $\alpha\sim1/137\ll1$ EM vacuum structure.

\section{A few words on cosmology}
\label{sec:flrw}
\noindent This section introduces some necessary concepts which will be useful in describing the magnetization of the electron-positron primordial plasma in \rchap{chap:cosmo}. We operate under the $\Lambda$ Cold Dark Matter $(\Lambda\mathrm{CDM})$ model of cosmology where the contemporary universe is approximately 69\% dark energy, 26\% dark matter, 5\% baryons, and $<1$\% photons and neutrinos in energy density~\citep{Davis:2003ad,Planck:2018vyg}. The standard picture of the universe's evolution is outlined in \rf{fig:cosmo}.

The Friedmann-Lema{\^i}tre-Robertson-Walker (FLRW) line element and metric~\citep{weinberg1972gravitation} in spherical coordinates is
\begin{gather}
    \label{FLRW} ds^2=dt^2-a^2(t)\left[\frac{dr^2}{1-kr^{2}}+r^{2}d\theta^2+r^{2}\sin\theta^{2}d\phi^2\right]\,,\\
    g_{\alpha\beta}=
    \begin{pmatrix}
        1&0&0&0\\
        0&-\frac{a^{2}(t)}{1-kr^{2}}&0&0\\
        0&0&-a^{2}(t)r^{2}&0\\
        0&0&0&-a^{2}(t)r^{2}\sin\theta^{2}
    \end{pmatrix}\,.
\end{gather}
The Gaussian curvature $k$ informs the spatial shape of the universe with the following possibilities: infinite flat Euclidean $(k=0)$, finite spherical but unbounded $(k=+1)$, or infinite hyperbolic saddle-shaped $(k=-1)$. Observation indicates our universe is flat or nearly so.

\begin{figure}[ht]
 \centering
 \includegraphics[width=0.95\linewidth]{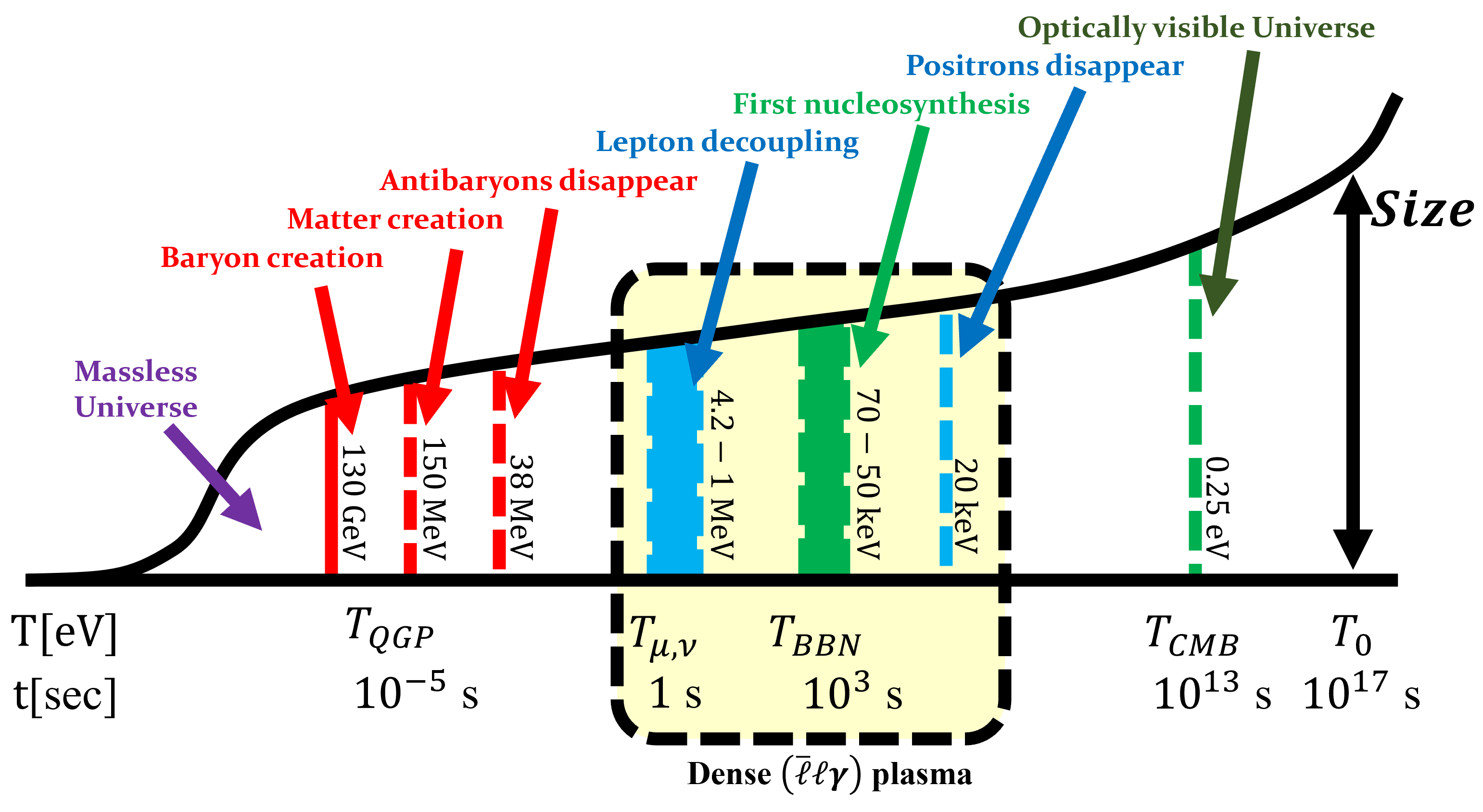}
 \caption{A schematic of the universe's evolution since the Big Bang. The region of interest studied in this dissertation is emphasized (in the highlighted box) to contain a dense nearly charge neutral matter-antimatter plasma.}
 \label{fig:cosmo} 
\end{figure}

The scale factor $a(t)$ denotes the change of proper distances $L(t)$ over time as
\begin{gather}
    L(t)=L_{0}\frac{a_{0}}{a(t)}\rightarrow L(z)=L_{0}(1+z)\,,
\end{gather}
where $z$ is the redshift and $L_{0}$ the comoving length. In an expanding (or contracting) universe which is both homogeneous and isotropic. This implies volumes change with $V(t)=V_{0}/a^{3}(t)$ where $V_{0}=L_{0}^{3}$ is the comoving Cartesian volume. The evolutionary expansion of the universe is then traditionally defined in terms of the Hubble parameter $H(t)$ following the conventions in~\cite{weinberg1972gravitation}
\begin{gather}
  \label{Friedmann:1} H(t)^{2}\equiv\left(\frac{\dot a}{a}\right)^2=\frac{8\pi G_{N}}{3}\rho_\mathrm{total},\qquad \rho_\mathrm{total}(t)=\rho_{\Lambda}+\rho_\mathrm{DM}(t)+\rho_\mathrm{Baryons}(t)+\ldots\\
  \label{Friedmann:2}
  \frac{\ddot a}{a}=-qH^2,\qquad 
q\equiv -\frac{a\ddot a}{\dot a^2},\qquad \dot H=-H^2(1+q).
\end{gather}
where $G_N$ is the Newtonian constant of gravitation. \req{Friedmann:1} and \req{Friedmann:2} are also known as the Friedmann equations. The total density $\rho_\mathrm{total}$ is the sum of all contributions from any form of matter, radiation or field. This includes but is not limited to: dark energy $(\Lambda)$, dark matter (DM), baryons (B), leptons $(\ell,\nu)$ and photons $(\gamma)$. Depending on the age of the universe, the relative importance of each group changes as each dilutes different under expansion with dark energy infamously remaining constant in density and accelerating the universe today.

The parameter $q$ is the cosmic deceleration parameter, which for historical reasons is positive for $q>0$. This sign convention was chosen before the discovery of dark energy under the tacit assumption that the universe would be decelerating. The value of $q$ depends on the energy content of the universe: The early universe was radiation dominated $(q = 1)$, subsequently matter dominated $(q = 1/2)$, and lastly the contemporary universe is undergoing a transition from matter to dark energy dominated approaching the asymptotic value of $q = -1$; see~\cite{Rafelski:2013yka}.

We can consider the expansion to be an adiabatic process~\citep{Abdalla:2022yfr} which results in a smooth shifting of the relevant dynamical quantities. As the universe undergoes isotropic expansion, the temperature decreases as 
\begin{gather}
 \label{tscale}
 T(t)=T_{0}\frac{a_{0}}{a(t)}\rightarrow T(z)=T_{0}(1+z)\,,
\end{gather}
where $z$ is the redshift. The entropy within a comoving volume is kept constant until gravitational collapse effects become relevant. The comoving temperature $T_{0}$ is given by the the present CMB temperature $T_{0}=2.7{\rm\ K}\simeq2.3\times10^{-4}\eV$~\citep{Planck:2018vyg}, with contemporary scale factor $a_{0}=1$.

As the universe expands, redshift reduces the momenta of particles lowering their contribution to the energy content of the universe. This cosmic redshift is written as
\begin{alignat}{1}
  \label{Redshift} p_{i}(t) = p_{i,0}\frac{a_{0}}{a(t)}\,.
\end{alignat}
Momentum (and the energy of massless particles $E=pc$) scales with the same factor as temperature. The energy of massive free particles in the universe however scales differently based on their momentum (and thus temperature).

When hot and relativistic, particle energy decreases inversely with scale factor like radiation. As the particles transition to non-relativistic (NR) energies, they decrease with the inverse square of the scale factor
\begin{alignat}{1}
    \label{EScale} E(t) = E_{0}\frac{a_{0}}{a(t)}\xrightarrow{\mathrm{NR}}\  E_{0}\frac{a_{0}^{2}}{a(t)^{2}}\,.
\end{alignat}
This occurs because of the functional dependence of energy on momentum in the relativistic $E\sim p$ versus non-relativistic $E\sim p^{2}$ cases.

\chapter{Dynamics of charged particles with arbitrary magnetic moment}
\label{chap:moment}
\noindent In \rsec{sec:mom}, we addressed two different models of introducing anomalous magnetic moment in QM: 
\begin{enumerate}
\item[(a)] the Dirac-Pauli (DP) first order equation which is the Dirac equation where $g$-factor is precisely fixed to the $g\!=\!2$, with the addition of an incremental Pauli term; and
\item[(b)] the Klein-Gordon-Pauli (KGP) second order equation which ``squares'' the Dirac equation and thereafter allows the magnetic moment $\bb{\mu}$ to vary independently of charge and mass, unlike Dirac theory.
\end{enumerate} 
These two approaches coincide when the anomaly $a$ vanishes. However, all particles that have magnetic moments differ from the Dirac value $g\!=\!2$, either due to their composite nature or due to the quantum vacuum fluctuation effect.

We find that even a small magnetic anomaly has a large effect in the limit of strong fields generated by massive magnetar stars~\citep{Kaspi:2017fwg}. Therefore it is not clear that the tacit assumption of $g\!=\!2$ in the case of strong fields~\citep{Rafelski:1976ts,Greiner:1985ce,Rafelski:2016ixr} is prudent~\citep{Evans:2018kor}. This argument is especially applicable to tightly bound composite particles such as protons and neutrons where the large anomalous magnetic moment can be taken as an external prescribed parameter unrelated to the elementary quantum vacuum fluctuations. It is then of particular interest to study the dynamical behavior of these particles in fields of magnetar strength. This interest carries over to the environment of strong fields created in focus of ultra-intense laser pulses and the associated particle production processes~\citep{Dunne:2014qda,Hegelich:2014tda}. We consider also precision spectroscopic experiments and recognize consequences even in the weak coupling limit.

This chapter reviews our work done in exploring relativistic dynamics with arbitrary magnetic dipoles in both a quantum mechanical and classical context. \rsec{sec:homogeneous} and \rsec{sec:coulomb} and covers analytic solutions for the Klein-Gordon-Pauli (KGP) equation in the presence of homogeneous magnetic fields and the Coulomb problem for hydrogen-like atoms. Comparisons with the Dirac-Pauli (DP) and Dirac solutions are made and novel consequences for strong fields are discussed in \rsec{sec:sb}. \rsec{sec:ikgp} explores extensions to KGP which combine mass and magnetic moment into a dynamical mass which is sensitive to electromagnetic fields. This work is primarily based on~\cite{Steinmetz:2018ryf}.

Relativistic classical spin dynamics is discussed in \rsec{sec:cspin} and is based on our work in~\cite{Rafelski:2017hce}. We propose in \rsec{sec:magpotential} a covariant form of magnetic dipole potential which modifies the Lorentz force, extends the Thomas-Bargmann-Michel-Telegdi (TMBT) equation, and reproduces the Stern-Gerlach force in the non-relativistic limit. \rsec{sec:ampgil} demonstrates that this magnetic potential also serves to unify both the Amp{\`e}rian and Gilbertian pictures of dipole moments.

\section{Homogeneous magnetic fields}
\label{sec:homogeneous}
\noindent The case of the homogeneous magnetic field, sometimes referred to as the Landau problem, provides a stepping stone in which to examine the consequences of quantum spin dynamics in a concrete analytical fashion. We present here an abbreviated analysis and the full treatment of this solution in terms of Ladder operators can be found in~\cite{Steinmetz:2018ryf} while alternative approaches are shown in texts such as~\cite{Itzykson:1980rh}. We assume a constant magnetic field in the $z$-direction
\begin{alignat}{1}
	\label{homogeneous:1} \bb{B}=(0,0,B)\,.
\end{alignat}
For our choice of gauge, there are two common options: (a) the Landau $\bb{A}_\mathrm{L}$ gauge and (b) the symmetric $\bb{A}_\mathrm{S}$ gauge
\begin{alignat}{1}
	\label{homogeneous:2} \bb{A}_\mathrm{L}=B(0,x,0)\,,\indent \bb{A}_\mathrm{S}=\frac{B}{2}(-y,x,0)\,.
\end{alignat}
As the system has a manifest rotational symmetry perpendicular to the direction of the homogeneous field, we will choose the symmetric gauge $\bb{A}_\mathrm{S}$ which preserves this symmetry explicitly.

Before we examine relativistic wave equations, it will be helpful to first consider the non-relativistic Schr{\"o}dinger-Pauli case as the KGP-Landau problem can be written as equivalent to the Schr{\"o}dinger-Pauli Hamiltonian. We consider energy eigenstates of and electron with $m_{e}$ described by \req{sp:1} under \req{homogeneous:1} as
\begin{alignat}{1}
	\label{homogeneous:3} \chi\rightarrow\chi_\mathrm{E}\exp\left(-\frac{iEt}{\hbar}\right)\,,\qquad\left(\frac{1}{2m_{e}}\bb{\pi}^{2}-\bb{\mu}\cdot\bb{B}\right)\chi_\mathrm{E}=E\chi_\mathrm{E}\,,
\end{alignat}
where $\mu$ is the magnitude of the magnetic moment as defined in \req{mag:3}. \req{homogeneous:3} can be further rewritten using angular momentum $\bb{L}$ and the symmetric gauge \req{homogeneous:2} as
\begin{gather}
	\label{homogeneous:4} \left(\frac{1}{2m_{e}}\bb{p}^{2}+\frac{e^{2}B^{2}}{8m_{e}}(x^{2}+y^{2})-\frac{eB}{2m_{e}}L_{3}-\mu B\sigma_{3}\right)\chi_\mathrm{E}=E\chi_\mathrm{E}\,,\\
    \bb{L}=\bb{r}\times\bb{p}\,,\qquad L_{i}=\varepsilon_{ijk}x_{j}p_{k}\,.
\end{gather}

The above can be broken into a set of three mutually commuting Hamiltonian operators:
\begin{itemize}[nosep]
    \item[(a)] Free particle Hamiltonian (Free)
    \item[(b)] Quantum harmonic oscillator (HO)
    \item[(c)] Zeeman interaction (ZI)
\end{itemize}
given by
\begin{align}
	\label{homogeneous:7a}
    \mathrm{(a)}\quad&H_{\mathrm{Free}}=\frac{p_{3}^{2}}{2m}\,,\\
	\label{homogeneous:7b}
    \mathrm{(b)}\quad&H_\mathrm{HO}\,=\frac{1}{2m_{e}}\left(p_{1}^{2}+p_{2}^{2}\right) + \frac{1}{2}m_{e}\omega^{2}(x^{2}+y^{2})\,,\\
	\label{homogeneous:7c}
    \mathrm{(c)}\quad&H_{\mathrm{ZI}}\ \,=-\mu_{B}B\frac{L_{3}}{\hbar}-\mu B\sigma_{3}\,,
\end{align}
which forms the total Hamiltonian
\begin{align}
    \label{homogeneous:7}
    H_\mathrm{total} = H_\mathrm{Free} + H_\mathrm{HO} + H_\mathrm{Mag.}\,.
\end{align}
The cyclotron frequency appears in \req{homogeneous:7b} as $2\omega=\omega_\mathrm{C}=eB/m_{e}$. We note that the Zeeman \req{homogeneous:7c} is usually expressed as
\begin{alignat}{1}
	\label{homogeneous:7d}
    H_{\mathrm{ZI}}=-\frac{e}{2m}\left(g_\mathrm{L}\bb{L}+g\bb{S}\right)\cdot\bb{B}\,,\qquad g_\mathrm{L}=1\,,
\end{alignat}
with $\bb{S}$ defined in \req{qspin:1}. We see explicitly that the orbital gyromagnetic ratio $g_\mathrm{L}$ which is a coefficient to the angular momentum operator $\bb{L}$ is unity unlike for spin. We refer back to our comment about black hole rotation in \rsec{sec:unique} as spin-like rather than orbital-like in terms of its magnetic behavior. As all the above Hamiltonian operators are mutually commuting, the energy eigenvalue of the total Hamiltonian is the sum of the individual energy eigenvalues. Our remaining goal will be to convert the KGP eigenvalue equation into the above three non-relativistic Hamiltonian operators. 

We now return to the KGP equation and write expand \req{kgp:1} for the Landau problem with energy eigenstates $\Psi_\mathrm{E}$ yielding
\begin{alignat}{1}
    \label{lan09}
    \left(\frac{E^{2}}{c^{2}}-m_{e}^{2}c^{2}-\bb{p}^{2}-\frac{1}{4}e^{2}B^{2}\left(x^{2}+y^{2}\right)+eBL_{3}+2\mu m_{e} B\sigma_{3}\right)\Psi_\mathrm{E}=0\;.
\end{alignat} 
We introduce the substitutions 
\begin{alignat}{1}
    \label{lan10}
    E\to m^\prime c^{2}\,,\quad \frac{E^{2}-m_{e}^{2}c^{4}}{2E}\to E^\prime\;,
\end{alignat} 
and recast KGP \req{lan09} into a Schr{\"o}dinger-style Hamiltonian equation
\begin{gather}
	\label{lan11} \left(\frac{1}{2m'}\bb{p}^{2}+\frac{e^{2}B^{2}}{8m'}(x^{2}+y^{2})-\frac{eB}{2m'}L_{3}-\mu\left(\frac{m_{e}}{m'}\right)B\sigma_{3}\right)\Psi_\mathrm{E}=E'\Psi_\mathrm{E}\,,
\end{gather}
which matches the non-relativistic Hamiltonian presented in \req{homogeneous:7}.

The energy eigenvalues of \req{homogeneous:7} are given by
\begin{alignat}{1}
    \label{lan23}
    E^\prime_{n,s}(p_{3},B)&=\frac{p_{3}^{2}}{2m^\prime }+\frac{e\hbar B}{m^\prime}\left(n+\frac{1}{2}\right)-\mu B\left(\frac{m_{e}}{m'}\right)s\,,
\end{alignat}
where $n\in1,2,3\ldots$ is the Landau orbital quantum number and $s\in\pm1$ is the spin quantum number. The physical relativistic energies can be obtained by undoing the substitutions in \req{lan10} yielding from \req{lan23}
\begin{gather}
\label{lan24}
E^{2}_{n,s}(p_{3},B)=m_{e}^{2}c^{4}+p_{3}^{2}c^{2}+2e\hbar c^{2}B\left(n+\frac{1}{2}\right)-2\mu B m_{e}c^{2}s\,,\\
\label{lan24b}
E_{n,s}=\pm\sqrt{m_{e}^{2}c^{4}+p_{3}^{2}c^{2}+2e\hbar c^{2}B\left(n+\frac{1}{2}\right)-2\mu B m_{e}c^{2}s}\;.
\end{gather}
This expression for the relativistic Landau levels is the same as found by~\cite{Weisskopf:1936hya} for the Dirac equation setting $g\!=\!2$ in \req{lan24b}. The Landau orbital part and spin portions can be combined when the magnetic moment is expressed in terms of $e/m$, but the form in \req{lan24b} keeps it generalized for the case of neutral particles.

Restricting ourselves to the positive energy spectrum, the non-relativistic reduction of \req{lan24b} can be carried out in powers of $1/m$ in the large mass limit yielding
\begin{alignat}{1}
\label{lan27} E_{n,s}|_\mathrm{NR}&=m_{e}c^{2}+\frac{p_{3}^{2}}{2m_{e}}+\mu_{B}B\left(2n+1-\frac{g}{2}s\right)-\frac{p_{3}^{4}}{8m_{e}^{3}c^{2}}\\ \notag &\!-\!\frac{p_{3}^{2}}{2m_{e}}\frac{\mu_{B}B}{m_{e}c^{2}}\left(2n+1-\frac{g}{2}s\right)\!-\!\frac{\mu_{B}^{2}B^{2}}{2m_{e}c^{2}}\left(2n+1-\frac{g}{2}s\right)^{2}\!+\!\mathcal{O}(1/m_{e}^{5})\;,\end{alignat}
which contains the expected terms such as the non-relativistic kinetic energy in the z-direction, the first relativistic correction to kinetic energy, the Landau energies, and cross terms that behave like modifications to the mass of the particle.

The KGP-Landau levels above the ground state lose their (accidental) degeneracy for $g\neq 2$. This is shown schematically in \rf{f04}. The anomaly also causes the ground state to be pushed downward, such that $E^{2}<m^{2}$; if the anomaly and the magnetic field are large enough, states above the ground state are also pushed below the rest mass energy of the particle.

\begin{figure}
 \centering
 \includegraphics[clip, trim=0.0cm 0.0cm 9.0cm 7.0cm,width=0.6\linewidth]{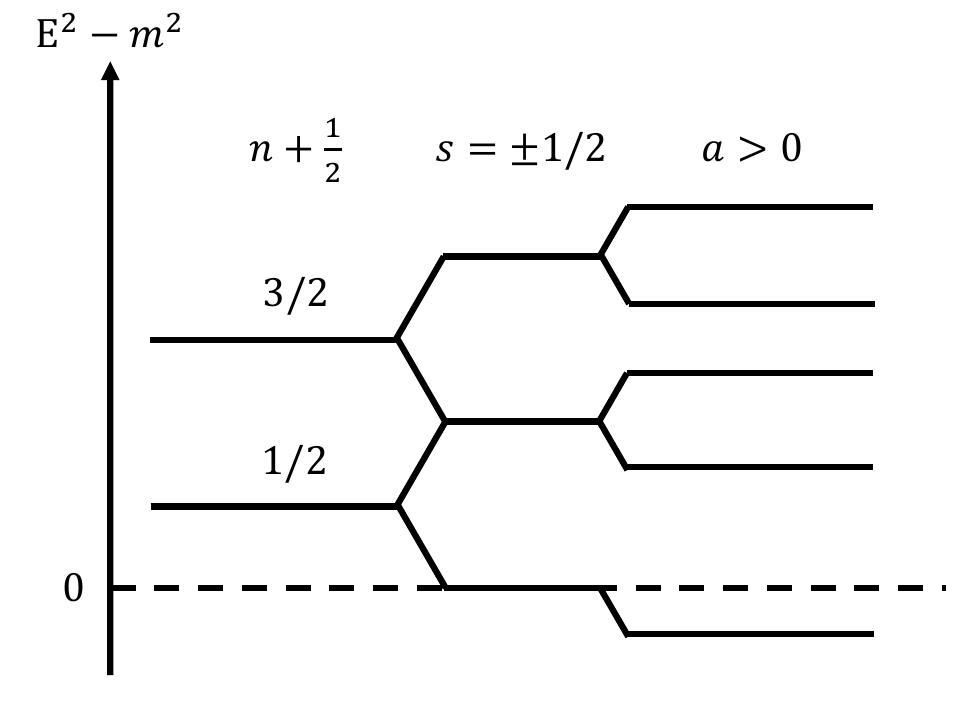}
 \caption[]{Diagram organizing the KGP-Landau levels for particles with zero z-component momentum. The Landau levels $n$ and spin $s$ serve to split levels while the anomaly $a$ controls the degeneracy among levels.}
 \label{f04}
\end{figure}

However, we recognize a periodicity considering the energy as a function of $g$. We recall that in Eq.~\eqref{lan24b} $n=0, 1, 2\ldots$. As $g$ varies, each time $gs/2$ crosses an integer value, for a different value of $n$ the energy eigenvalue $E$ repeat as a function of changing $g$. All possible values of energy $E$ are reached (at fixed $m$ and $p^2_3$) for $-2\le g\le 2$. Moreover, while for almost all $g\ne 2$ the degeneracy is completely broken, this periodicity implies that energy degeneracy is restored for values~\cite{Evans:2022ygl,Evans:2022fsu} 
\begin{alignat}{1}
\label{lan26}
g_{k}/2=1+k\,,\qquad
\lambda_\mathrm{L}'=\lambda_\mathrm{L}-ks\;,\qquad
\lambda_\mathrm{L}=n+\frac{1}{2}-s\;,
\end{alignat}
where $k=0,\pm1,\pm2,\ldots$ The Landau levels Eq.\,\eqref{lan24} contain an infinite number of degenerate levels bounded from below. Certain states change the sign of the magnetic energy and their total energies become unphysical in the limit that $g_{k}B$ becomes large; for even $k$ there are $k/2$ such states and for odd $k$ there are $(k+1)/2$.

It is useful to compare the KGP solution to the the Landau levels for the DP equation which we obtained by~\cite{Tsai:1971zma}. The DP-Landua energy eigenstates are given by
\begin{subequations}
\begin{alignat}{1}
\label{lan25} 
E^{2}_{n,s}(p_{3},B)|_\mathrm{DP} =&\left(\!\!\sqrt{\displaystyle m_{e}^{2}c^{4}\!+\!2e\hbar c^{2}B\left(n+\frac{1}{2}-s\right)}-\frac{eB\hbar}{2m_{e}}(g-2)s\!\right)^{2}\!\!\!+p_{3}^{\!\!2}c^{2},\\[0.4cm]
\label{lan25b}
E_{n,s}|_\mathrm{DP} =\pm &\sqrt{\!\left(\!\!\sqrt{\displaystyle m_{e}^{2}c^{4}+2e\hbar c^{2}B\left(n+\frac{1}{2}-s\right)}\!-\!\frac{eB\hbar}{2m_{e}}(g-2)s\!\right)^{\!\!2}\!\!\!+p_{3}^{2}c^{2}}\;,
\end{alignat}
\end{subequations}
which in our opinion fails Dirac's principle of mathematical beauty when compared to the KGP result Eq.~\eqref{lan24b}. Both Eqs.~\eqref{lan24b} and \eqref{lan25b} have the correct non-relativistic reduction at the lowest order though, the latter obscures the physical interpretation.

The most egregious issue with the DP-Landau levels is that, in a perturbative expansion, it includes cross terms between the $g\!=\!2$ magnetic moment and anomalous terms in $a=(g-2)/2$; thus the result does not depend on the particle magnetic moment alone; there is a functional dependence on the magnetic anomaly $a$. The presence of these cross terms implies that above first order the results cannot be given in terms of the full magnetic moment alone. In comparison, for the KGP-Landau levels Eq.\,\eqref{lan25b}, the entire effect of magnetic moment is contained in a single term.

\section{Hydrogen-like atoms}
\label{sec:coulomb}
The Coulomb problem, or sometimes referred to as the Kepler problem, provides us an important application of any quantum theory to explore. As the hydrogen-like atoms are the the most well understood atomic system in physics, any non-minimal behavior especially for high-$Z$ systems can lead to consequences for the resulting spectral lines. We take the Coulomb potential to be
\begin{equation}
	\label{eq:coulomb:01} eV_\mathrm{C}\equiv e A^{0}=\frac{Z \alpha\hbar c}{r}\;,\qquad \bb{A}=0\;.
\end{equation}
The KGP-Coulomb problem with arbitrary magnetic moment can be solved analytically and we will briefly sketch out the solution and its consequences.

For energy states $\Psi=e^{-iEt/\hbar}\Psi_\mathrm{E}$ the KGP equation yields the following differential equation
\begin{alignat}{1}
	\label{cou02} \left(\frac{E^{2}-m^{2}c^{4}}{\hbar^{2}c^{2}}+\frac{Z^{2}\alpha^{2}}{r^{2}}+\frac{2E}{\hbar c}\frac{Z\alpha}{r}+\frac{1}{r}\frac{\partial\bb{L}^{2}}{\partial r^{2}}r-\frac{\bb{L}^{2}/\hbar^{2}}{r^{2}}-\frac{g}{2}Z\alpha\frac{i\bb{\alpha}\cdot\bb{\hat{r}}}{r^{2}}\right)\Psi_\mathrm{E}=0\;.
\end{alignat}
We recast the squared angular momentum operator $\bb{L}^{2}$ with the Dirac spin-alignment operator
\begin{alignat}{1}
\label{cou04} &\mathcal{K}=\gamma^{0}\left(1+\bb{\Sigma}\cdot\frac{\bb{L}}{\hbar}\right)\;,\indent \bb{L}^{2}/\hbar^{2}=\mathcal{K}\left(\mathcal{K}-\gamma^{0}\right)\;.
\end{alignat}
The operator $\mathcal{K}$ commutes with $\bb{\alpha}\cdot\bb{\hat{r}}$ and its eigenvalues are given as either positive or negative integers $\kappa=\pm(j+1/2)$ where $j$ is the total angular momentum quantum number. By grouping all terms proportional to $1/r^{2}$, we see the effective angular momentum eigenvalues take on non-integer values which in the limit of classical mechanics corresponds to orbits which do not close. The non-integer eigenvalues depends explicitly on $g$-factor.

The difficulty of this equation is that the effective angular momentum operator is non-diagonal in spinor space due to the presence of $\bb{\alpha}\cdot\bb{\hat{r}}$ which mixes upper and lower components. The effective radial potential within \req{cou02} is then
\begin{alignat}{1}
	\label{temp} \boxed{V_\mathrm{eff}=-\frac{2E}{\hbar c}\frac{Z\alpha}{r}-\frac{Z^{2}\alpha^{2}}{r^{2}}+\frac{g}{2}Z\alpha\frac{i\bb{\alpha}\cdot\bb{\hat{r}}}{r^{2}}}\,.
\end{alignat}
We emphasize that the distinguishing characteristic which separates the KGP solutions (and Dirac for $g\!=\!2$) from the Klein-Gordon solutions is the last term in the effective potential \req{temp}. This is the dipole-charge interaction term which exists only because of the relativistic expression for the magnetic dipole and is entirely absent non-relativistically. 

Following the procedure of~\cite{Martin:1958zz}, we introduce the operator
\begin{alignat}{1}
\label{cou07} &\mathfrak{L}=-\gamma^{0}\mathcal{K}-\frac{g}{2}Z\alpha(i\bb{\alpha}\cdot\bb{\hat{r}})\;,
\end{alignat}
but with the novel modification that $g$-factor directly appears in the second term. This operator is also sometimes referred to as the Temple operator, therefore we will refer to it as the $g$-Temple operator. This operator then commutes with the spin-alignment operator $\mathcal{K}$ and has eigenvalues
\begin{alignat}{1}
\label{cou08} &\Lambda=\pm\sqrt{\kappa^{2}-\displaystyle\frac{\displaystyle g^{2}}{4}Z^{2}\alpha^{2}}\;,\end{alignat}
where the absolute values are denoted as $\lambda=|\Lambda|$. The angular momentum contributions to \req{cou02} can then be replaced by
\begin{alignat}{1}
\label{cou09} &\mathcal{K}(\mathcal{K}-\gamma^{0})-Z^{2}\alpha^{2}-\frac{g}{2}Z\alpha(i\bb{\alpha}\cdot\bb{\hat{r}})=\mathfrak{L}(\mathfrak{L}+1)+\left(\frac{g^{2}}{4}-1\right)Z^{2}\alpha^{2}.
\end{alignat}
If the $g$-factor is taken to be $g\!=\!2$, then the differential \req{cou02} reverts to the one discussed in~\cite{Martin:1958zz}. The coefficient $g^{2}/4-1$ will be commonly seen to precede new more complicated terms, which conveniently vanish for $|g|=2$ demonstrating that as function of $g$ there is a ``cusp''~\citep{Rafelski:2022bsv} for $|g|=2$. This will become especially evident when we discuss strongly bound systems which behave very differently for $|g|<2$ versus $|g|>2$. 

We omit further derivation which can be found in~\cite{Steinmetz:2018ryf}. We find the resulting energy levels of the KGP-Coulomb equation to be 
\begin{gather}
\label{cou17} E_{\pm\lambda}^{n_{r},j}=mc^{2}\left[1+\displaystyle\frac{Z^{2}\alpha^{2}}{\left(n_{r}+1/2+\nu\right)^{2}}\right]^{-1/2}\,,\\
\label{cou17b} \nu=\sqrt{(\lambda\pm1/2)^{2}+\left(\frac{g^{2}}{4}-1\right)Z^{2}\alpha^{2}}\,,\qquad
\lambda=\sqrt{\displaystyle(j+1/2)^{2}-\frac{\displaystyle g^{2}}{4}Z^{2}\alpha^{2}}\;.
\end{gather}
where $n_{r}$ is the node quantum number which takes on the values $n_{r}=0,1,2,\ldots$. \req{cou17} is the same ``Sommerfeld-style'' expression for energy that we can obtain from the Dirac or KG equations. The difference between them arises from the expression of the relativistic angular momentum which depends on $g$-factor for the KGP equation. The KGP eigenvalues \req{cou17} for arbitrary spin were obtained by~\cite{Niederle:2004bx} using a tensor approach.

In the limit that $g\rightarrow2$ for the Dirac case the expressions for $\lambda$ and $\nu$ reduce to
\begin{subequations}
\begin{alignat}{1}
\label{glimit01} &\lim_{g\rightarrow2}\lambda=\sqrt{\displaystyle(j+1/2)^{2}-Z^{2}\alpha^{2}},\\
&\lim_{g\rightarrow2}\nu_{\pm\lambda}=\lambda\pm1/2\;.
\end{alignat}
\end{subequations}
This procedure requires taking the root of perfect squares; therefore, the sign information is lost in \req{glimit01}. As long as $Z^{2}\alpha^{2}<3/4$ we can drop the absolute value notation as $\nu$ is always positive. The energy is then given by
\begin{alignat}{1}
\label{glimit02}
E_{\pm\lambda}^{n_{r},j}=mc^{2}\left[1+\displaystyle\frac{Z^{2}\alpha^{2}}{\left(n_{r}\begin{smallmatrix} +1 \\ +0 \end{smallmatrix}+\sqrt{\displaystyle(j+1/2)^{2}-Z^{2}\alpha^{2}}\right)^{2}}\right]^{-1/2}\;.
\end{alignat}
The $\begin{smallmatrix} +1 \\ +0 \end{smallmatrix}$ notation is read as the upper value corresponding to the $+\lambda$ states and the lower value corresponding to the $-\lambda$ states. 

The ground state energy (with: $n_{r}=0,\ \Lambda<0,\ j=1/2$) is therefore
\begin{alignat}{1}
\label{glimit07} &E^{0,1/2}_{-\lambda(j=1/2)}=mc^{2}\sqrt{1-Z^{2}\alpha^{2}}\;,\end{alignat}
as expected for the Dirac-Coulomb ground state. \req{glimit02} reproduces the Dirac-Coulomb energies and also contains a degeneracy between states of opposite $\lambda$ sign, same $j$ quantum number and node quantum numbers offset by one
\begin{alignat}{1}
\label{glimit03} &E^{n_{r}+1,j}_{-\lambda}=E^{n_{r},j}_{+\lambda}\;,\end{alignat}
which corresponds to the degeneracy between $2S_{1/2}$ and $2P_{1/2}$ states. There is no degeneracy for the $E^{0,j}_{-\lambda}$ states. 

In the limit that $g\rightarrow 0$, which is the KG case, the expressions are given by
\begin{subequations}
\begin{alignat}{1}
\label{glimit04} &\lim_{g\rightarrow0}\lambda=j+1/2,\\
&\lim_{g\rightarrow0}\nu_{\pm\lambda}=\sqrt{\left(j\begin{smallmatrix} +1 \\ +0 \end{smallmatrix}\right)^{2}-Z^{2}\alpha^{2}}\;,
\end{alignat}
\end{subequations}
which reproduces the correct expressions for the energy levels for the Klein-Gordon case 
\begin{alignat}{1}
\label{glimit05} E_{\pm\lambda}^{n_{r},j}=mc^{2}\left[1+\displaystyle\frac{Z^{2}\alpha^{2}}{\left(n_{r}+1/2+\displaystyle\sqrt{\left(j\begin{smallmatrix} +1 \\ +0 \end{smallmatrix}\right)^{2}-Z^{2}\alpha^{2}}\right)^{2}}\right]^{-1/2}\;,
\end{alignat}
except that in this limit we are still considering the total angular moment quantum number $j$ rather than orbital momentum quantum number $\ell$. It is interesting to note that the KG-Coulomb problem's energy formula contains $\ell+1/2$, which matches identically to our half-integer $j$ values; therefore, this artifact of spin, untethered and invisible by the lack of magnetic moment, does not alter the energies of the states. The degeneracy in energy levels are given by 
\begin{alignat}{1}
\label{glimit06} &E^{n_{r},j+1}_{-\lambda}=E^{n_{r},j}_{+\lambda}\;,\end{alignat}
with levels of opposite $\lambda$ sign, same node quantum number and shifted $j$ values by one. In a similar fashion to the Dirac case, here we have no degeneracy for $E^{n_{r},1/2}_{-\lambda}$ states.

\subsection{Non-relativistic Coulomb problem energies} \label{nonrel}
The first regime of interest to understand the effect of variable $g$ in the KGP-Coulomb problem is the non-relativistic limit characterized by the weak binding of low-Z atoms. We now will convert from $n_{r}$, $j$ and $\pm\lambda$ to the familiar quantum numbers of $n$, $j$ and $\ell$ allowing for easy comparison with the hydrogen spectrum in standard notation. We start by expanding \req{cou17} in powers of $Z\alpha$ to compare to the known hydrogen spectrum. 

To order $\mathcal{O}(Z^{4}\alpha^{4})$ the energy levels are given by
\begin{alignat}{1}
\label{nonrel01} \frac{E^{n_{r},j}_{\pm\lambda}}{mc^{2}}=1&-\frac{1}{2}\frac{Z^{2}\alpha^{2}}{(n_{r}+1/2+(\nu_{\pm\lambda})|_{Z=0})^{2}}+\frac{(\nu_{\pm\lambda})|_{Z=0}'Z^{3}\alpha^{3}}{(n_{r}+1/2+(\nu_{\pm\lambda})|_{Z=0})^{3}}\\ 
\notag&+\frac{1}{2}\frac{(3/4-3(\nu_{\pm\lambda})|_{Z=0}'^{2})Z^{4}\alpha^{4}}{(n_{r}+1/2+(\nu_{\pm\lambda})|_{Z=0})^{4}}+\frac{1}{2}\frac{(\nu_{\pm\lambda})|_{Z=0}^{\prime\prime} Z^{4}\alpha^{4}}{(n_{r}+1/2+(\nu_{\pm\lambda})|_{Z=0})^{3}}+\mathcal{O}(Z^{6}\alpha^{6})\;,\end{alignat}
where primed $\nu_{\pm\lambda}$ indicate derivatives with respect to $Z\alpha$. These derivatives evaluate to
\begin{alignat}{1}
\label{nonrel02} &(\nu_{\pm\lambda})|_{Z=0}=j+1/2\pm1/2,\\ \notag &(\nu_{\pm\lambda})|_{Z=0}'=0,\\ \notag &(\nu_{\pm\lambda})|_{Z=0}^{\prime\prime}=\frac{(g^{2}/4-1)}{j+1/2\pm1/2}-\frac{g^{2}/4}{j+1/2}\;.
\end{alignat}
\req{nonrel01} then simplifies to
\begin{alignat}{1}
\label{nonrel03} \frac{E^{n_{r},j}_{\pm\lambda}}{mc^{2}}=1&-\frac{1}{2}\frac{Z^{2}\alpha^{2}}{\left(n_{r}+j\begin{smallmatrix}+3/2 \\ +1/2\end{smallmatrix}\right)^{2}}+\frac{3}{8}\frac{Z^{4}\alpha^{4}}{\left(n_{r}+j\begin{smallmatrix}+3/2 \\ +1/2\end{smallmatrix}\right)^{4}}\\ 
\notag &+\frac{1}{2}\left(\frac{(g^{2}/4-1)}{j\begin{smallmatrix}+1 \\ +0\end{smallmatrix}}-\frac{g^{2}/4}{j+1/2}\right)\frac{Z^{4}\alpha^{4}}{\left(n_{r}+j\begin{smallmatrix}+3/2 \\ +1/2\end{smallmatrix}\right)^{3}}+\mathcal{O}(Z^{6}\alpha^{6})\;.
\end{alignat}
In the non relativistic limit, the node quantum number corresponds to the principle quantum number via $n_{r}=n'-j-1/2$ with $n'=1,2,3\ldots$ Using \req{nonrel03} and \req{cou07} we see that in the non relativistic limit $+\lambda$ corresponds to $\kappa>0$ or anti-aligned spin-angular momentum with $j=\ell-1/2$ and $\ell\geq1$. Conversely $-\lambda$ corresponds to $\kappa<0$ or aligned spin-angular momentum with $j=\ell+1/2$. 

With all this input we arrive at
\begin{alignat}{1}
\label{nonrel04} \frac{E^{n,j}_{\begin{smallmatrix}
\kappa>0 \\ \kappa<0
\end{smallmatrix}}}{mc^{2}}=1&-\frac{1}{2}\frac{Z^{2}\alpha^{2}}{\left(n'\begin{smallmatrix}
+1 \\ +0
\end{smallmatrix}\right)^{2}}+\frac{3}{8}\frac{Z^{4}\alpha^{4}}{\left(n'\begin{smallmatrix}
+1 \\ +0
\end{smallmatrix}\right)^{4}}
\\ \notag &+\frac{1}{2}
\frac{(g^{2}/4-1)}{j\begin{smallmatrix}
+1 \\ +0
\end{smallmatrix}}\frac{Z^{4}\alpha^{4}}{\left(n'\begin{smallmatrix}
+1 \\ +0
\end{smallmatrix}\right)^{3}}-\frac{1}{2}\frac{g^{2}/4}{j+1/2}\frac{Z^{4}\alpha^{4}}{\left(n'\begin{smallmatrix}
+1 \\ +0
\end{smallmatrix}\right)^{3}}+\mathcal{O}(Z^{6}\alpha^{6})\;.
\end{alignat}
Lastly we recast, for the $\kappa>0$ states, the principle quantum number as $n'+1\rightarrow n$ with $n\geq2$ and we simply relabel $n'\rightarrow n$ for $\kappa<0$ states. This allows \req{nonrel04} to be completely written in terms of $n$, $j$, and $\ell$ as
\begin{alignat}{1}
\label{nonrel09} \frac{E^{n,j}_{\ell}}{mc^{2}}=1&-\frac{1}{2}\frac{Z^{2}\alpha^{2}}{n^{2}}+\frac{3}{8}\frac{Z^{4}\alpha^{4}}{n^{4}}\\
\notag &+\frac{1}{2}\frac{(g^{2}/4-1)}{\ell+1/2}\frac{Z^{4}\alpha^{4}}{n^{3}}-\frac{1}{2}\frac{g^{2}/4}{j+1/2}\frac{Z^{4}\alpha^{4}}{n^{3}}+\mathcal{O}(Z^{6}\alpha^{6})\;,\end{alignat}
where it is understood that $n-\ell\geq1$, this condition allows us to write what was previously described in \req{nonrel04} as two distinct spectra now as a single energy spectra. In the limit $g\rightarrow2$ or $g\rightarrow0$ the correct expansion to order $Z^{4}\alpha^4$ of the Dirac or KG energies are obtained. In the following we explore some consequences of our principal non-relativistic result, \req{nonrel09}.

\subsection{g-Lamb Shift between 2S and 2P orbitals} \label{lamb}
The breaking of degeneracy in \req{nonrel09} between states of differing $\ell$ orbital quantum number, but the same total angular momentum $j$ and principle quantum number $n$ is responsible for the Lamb shift due to anomalous magnetic moment. The only term in \req{nonrel09} (up to order $Z^{4}\alpha^{4}$) that breaks the degeneracy between the $E^{n,j}_{\ell=j+1/2}$ and $E^{n,j}_{\ell=j-1/2}$ states for $n\geq2$ is the fourth term. This is unsurprising as it depends exclusively on quantum number $\ell$ and $n$. The lowest order Lamb shift due to anomalous magnetic moment is then
\begin{alignat}{1}
\label{lamb01} \frac{\Delta E_{\mathrm{gLamb}}^{n,j}}{mc^{2}}&=E^{n,j}_{\ell=j-1/2}-E^{n,j}_{\ell=j+1/2}=\left(g^{2}/8-1/2\right)\left(\frac{1}{j}-\frac{1}{j+1}\right)\frac{Z^{4}\alpha^{4}}{n^{3}}\;.
\end{alignat}
For the $2S_{1/2}$ and $2P_{1/2}$ states \req{lamb01} reduces to
\begin{alignat}{1}
\label{lamb02} \frac{\Delta E_{\mathrm{gLamb}}^{2S_{1/2}-2P_{1/2}}}{mc^{2}}&=\left(g^{2}/8-1/2\right)\frac{Z^{4}\alpha^{4}}{6}=\left(a+a^{2}/2\right)\frac{Z^{4}\alpha^{4}}{6}\;.
\end{alignat}
Our result in \req{lamb01} and \req{lamb02} is sensitive to $g^{2}/8-1/2=a+a^{2}/2$. Traditionally the Lamb shift due to an anomalous lepton magnetic moment is obtained perturbatively~\citep{Itzykson:1980rh} by considering the DP equation which is sensitive to $g/2-1=a$ the shift takes on the expression at lowest order
\begin{alignat}{1}
\label{lamb03} \frac{\Delta E_{\mathrm{gLamb,DP}}^{2S_{1/2}-2P_{1/2}}}{mc^{2}}&=\left(\frac{g-2}{2}\right)\frac{Z^{4}\alpha^{4}}{6}=a\frac{Z^{4}\alpha^{4}}{6}\;.
\end{alignat}
It is of experimental interest to resolve this discrepancy between the first order DP equation and the second order fermion formulation KGP. We recall the present day values
\begin{subequations}
\begin{alignat}{1}
\label{aeFULL} a_e&=1159.65218091(26)\times 10^{-6}\simeq \frac{\alpha}{2\pi}\;,\\
a_\mu-a_e&=6.2687(6)\times 10^{-6}\;.
\end{alignat}
\end{subequations}
The largest contribution to the anomalous moment for charged leptons is, as indicated the lowest order QED Schwinger result $a=\alpha/2\pi$. For the KGP approach, the anomalous $g$-factor mixes contributions of different powers of fine structure $\alpha$. Precision values for the fundamental constants are taken from~\cite{Tiesinga:2021myr}. For the $2S_{1/2}-2P_{1/2}$ states, the shift is
\begin{alignat}{1}
\label{lamb04} &\frac{\Delta E_{\mathrm{gLamb}}^{2S_{1/2}-2P_{1/2}}}{mc^{2}}=\frac{Z^{4}\alpha^{5}}{12\pi}+\frac{Z^{4}\alpha^{6}}{48\pi^{2}}\;.
\end{alignat}
The scale of the discrepancy between KGP and DP for the hydrogen atom is then
\begin{alignat}{1}
\label{lamb05} \Delta &E_{\mathrm{gLamb,KGP}}^{2S_{1/2}-2P_{1/2}}\!-\!\Delta E_{\mathrm{gLamb,DP}}^{2S_{1/2}-2P_{1/2}}\!=\!\frac{\alpha^{6}mc^{2}}{48\pi^{2}}\\ \notag&\;\;=\!1.62881214\times10^{-10}\;\mathrm{eV} =39.3845030\;\mathrm{kHz}\;,\end{alignat}
without taking into account the standard corrections such as reduced mass, recoil, radiative, or finite nuclear size; for more information on those corrections please refer to~\cite{Jentschura:1996zz,Eides:2000xc,Tiesinga:2021myr}. It is to be understood that the corrections presented here are illustrative of the effect magnetic moment has on the spectroscopic levels, but that further work is required to compare these to experiment: for example we look here on behavior of point particles only.

While the discrepancy is small for the hydrogen system, it is $\approx 40$ kHz and will be visible in this or next generation's spectroscopic experiments. The discrepancy is also non-negligible for hydrogen-like exotics such as proton-antiproton because the proton $g$-factor is much larger
\begin{alignat}{1}\label{gpaFULL}
&g_p=5.585694702(17)\;,\qquad a_p=1.792847351(9)\;. 
\end{alignat} 
The discrepancy for the proton-antiproton system is
\begin{alignat}{1}
\label{lamb06} \Delta E_{\mathrm{gLamb,KGP}}^{2S_{1/2}-2P_{1/2}}-\Delta E_{\mathrm{gLamb,DP}}^{2S_{1/2}-2P_{1/2}}&=0.71268151\;\mathrm{eV}\;.
\end{alignat} 

\subsection{g-Fine structure effects within P orbitals} \label{fs}
The fifth term in \req{nonrel09}, which depends on $j$ and $n$, will shift the levels due to an anomalous moment, but does not contribute to the Lamb shift. Rather this expression, which contains the spin-orbit $\vec{L}\cdot\vec{S}$ coupling, is responsible for the fine structure splittings. From \req{nonrel09} the fine structure splitting is given by
\begin{alignat}{1}
\label{fs00} \frac{\Delta E_{\mathrm{gFS}}^{n,\ell}}{mc^{2}}&=E^{n,j=\ell+1/2}_{\ell}-E^{n,j=\ell-1/2}_{\ell}=\left(g^{2}/8\right)\left(\frac{1}{\ell}-\frac{1}{\ell+1}\right)\frac{Z^{4}\alpha^{4}}{n^{3}}\;.
\end{alignat}
The splitting between the $2P_{3/2}$ and $2P_{1/2}$ states is therefore
\begin{alignat}{1}
\label{fs01} \frac{\Delta E_{\mathrm{gFS}}^{2P_{3/2}-2P_{1/2}}}{mc^{2}}&=\left(g^{2}/8\right)\frac{Z^{4}\alpha^{4}}{16}=\left(1/2+a+a^{2}/2\right)\frac{Z^{4}\alpha^{4}}{16}\;.
\end{alignat}
In comparison the fine structure dependence on $g$-factor in the DP equation is given as
\begin{alignat}{1}
\label{fs02} \frac{\Delta E_{\mathrm{gFS,DP}}^{2P_{3/2}-2P_{1/2}}}{mc^{2}}&=\left(\frac{g-1}{2}\right)\frac{Z^{4}\alpha^{4}}{16}=\left(1/2+a\right)\frac{Z^{4}\alpha^{4}}{16}\;.
\end{alignat}
Just as in the case of the Lamb shift, we find that the KGP and DP equations disagree for fine structure splitting. For the hydrogen atom this discrepancy is 
\begin{alignat}{1}
\label{fs03} \Delta &E_{\mathrm{gFS,KGP}}^{2P_{3/2}-2P_{1/2}}\!-\!\Delta E_{\mathrm{gFS,DP}}^{2P_{3/2}-2P_{1/2}}\!=\!\frac{\alpha^{6}mc^{2}}{128\pi^{2}}\\ \notag&\;\;=\!6.10804553\times10^{-11}\;\mathrm{eV}=14.7691885\;\mathrm{kHz}\;,\end{alignat}
and for proton-antiproton, the fine structure splitting discrepancy is
\begin{alignat}{1}
\label{fs04} \Delta E_{\mathrm{gFS,KGP}}^{2P_{3/2}-2P_{1/2}}-\Delta E_{\mathrm{gFS,DP}}^{2P_{3/2}-2P_{1/2}}&=0.26725557\;\mathrm{eV}\;.
\end{alignat}
For fine structure of the muonic-hydrogen system, the KGP-DP discrepancy is
\begin{alignat}{1}
\label{fs05} \Delta E_{\mathrm{gFS,KGP}}^{2S_{1/2}-2P_{1/2}}-\Delta E_{\mathrm{gFS,DP}}^{2S_{1/2}-2P_{1/2}}&=\!1.272774\times10^{-8}\;\mathrm{eV}\;
\end{alignat} 
We can make a general observation that non minimal magnetic coupling, such as we have studied in the DP and KGP cases, enlarge energy level splittings. The above shows that these discrepancies will remain when calculating within more realistic finite nuclear size context.

\section{Particles in strong electromagnetic fields}
\label{sec:sb}
Care must be taken when interpreting the results presented in strong electromagnetic fields; see~\cite{Gonoskov:2021hwf,Fedotov:2022ely,Sainte-Marie:2023aqn}. Strong EM fields are produced in heavy-ion collisions forming quark-gluon-plasma~\citep{Grayson:2022asf}. For physical electrons the AMM interaction is the result of vacuum fluctuations whose strength also depends on the strength of the field. For example in the large magnetic field limit a QED computation shows that the ground state is instead of \req{lan24b} given by~\cite{Jancovici:1969exc}
\begin{alignat}{1} \label{vacfl01}
E_{0}\approx mc^{2}+\frac{\alpha}{4\pi}mc^{2}\ \mathrm{ln}^{2}\left(\frac{2e\hbar B}{m^{2}c^{3}}\right)\,,
\end{alignat}
which even for enormous magnetic fields does not deviate significantly from the rest mass-energy of the electron. Further the AMM radiative corrections approach zero for higher Landau levels~\citep{Ferrer:2015wca,Hackebill:2022uxv}. Therefore the AMM in the case of electrons does not have a significant effect in highly magnetized environments such as those found in astrophysics (magnetars).

The situation is different for composite particles such as the proton, neutron and light nuclei whose anomalous magnetic moments are dominated by their internal structure and not by vacuum fluctuations. In this situation we expect that the AMM interaction in high magnetic fields remains significant. Therefore, asking whether the DP or KGP equations better describes the dynamics of composite hadrons and atomic nuclei in presence of magnetar strength fields is a relevant question despite the standard choice in literature being the DP equation~\citep{Broderick:2000pe}. The same question can be asked for certain exotic hydrogen-like atoms where the particles have anomalous moments which can be characterized as an external parameter. 

\subsection{Strong homogeneous magnetic fields}
\label{sec:sbl}
The magnetic moment anomaly can flip the sign of the magnetic energy for the least excited states causing the gap between particle and antiparticle states to decrease with magnetic field strength. Setting $p_z=0$ in \req{lan24}, we show in~\rf{f01} that the energy of the lowest KGP Landau eigenstate $n=0, s=1/2$ reaches zero where the gap between particle and antiparticle states vanishes for the field
\begin{subequations}
\begin{alignat}{1}\label{Bcrit}
&B_\mathrm{crit}^{e}=\frac{{B}_{S}^{e}}{a_{e}}\simeq861{B}_{S}^{e} =3.8006\times10^{16}\;\mathrm{G}\;,\\
&B_\mathrm{crit}^{p}=\frac{{B}_{S}^{p}}{a_{p}}\simeq\frac{1}{1.79}{B}_{S}^{p}=8.3138\times10^{19}\;\mathrm{G}\;,
\end{alignat} 
\end{subequations}
where ${B}_{S}$ is the so-called Schwinger critical field~\citep{Schwinger:1951nm}.
\begin{subequations}
\begin{alignat}{1}\label{Bsch}
{B}_{S}^{e}\equiv\frac{{m_{e}^2}c^2}{e\hbar}=\frac{m_{e}c^2}{2\mu_B}=4.4141\times 10^{13}\;\mathrm{G}\;,\\
{B}_{S}^{p}\equiv\frac{{m_{p}^2}c^2}{e\hbar}=\frac{m_{p}c^2}{2\mu_N}=1.4882\times 10^{20}\;\mathrm{G}\;.
\end{alignat}
\end{subequations}
The numerical results are evaluated for the anomalous moment of the electron and proton, given by \req{aeFULL} and \req{gpaFULL}. At the critical field strength $B_\mathrm{crit}$ the Hamiltonian loses self-adjointness and the KGP loses its predictive properties. The Schwinger critical field \req{Bsch} denotes the boundary when electrodynamics is expected to behave in an intrinsically nonlinear fashion, and the equivalent electric field configurations become unstable~\citep{Labun:2008re}. However, it is possible that the vacuum is stabilized by strong magnetic fields~\citep{Evans:2018kor}.
 
\begin{figure}[ht]
     \centering
     \includegraphics[clip, trim=0.0cm 0.0cm 0.0cm 0.5cm,width=0.95\textwidth]{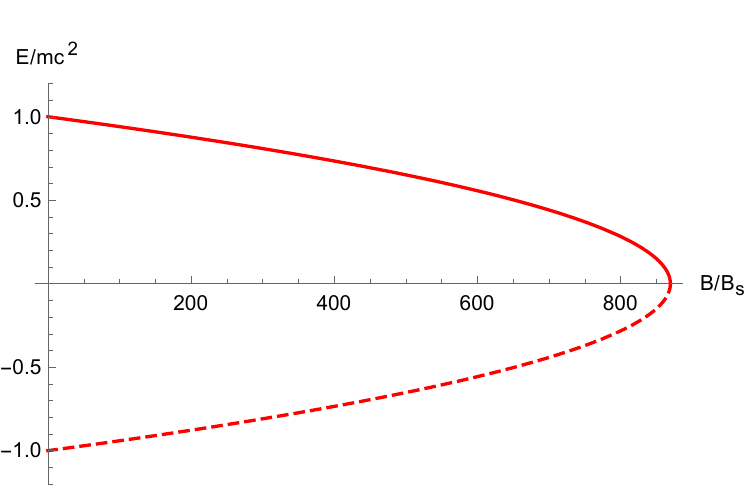}
     \caption{The $n=0$, $s=1/2$ ground state for a KGP electron given by \req{lan24b} with $g/2-1=\alpha/2\pi$ in a homogeneous magnetic field. We consider the particle with no $z$-direction momentum. The particle state (solid red) and antiparticle (dashed red) are presented.}
     \label{f01}
\end{figure}

The critical magnetic fields as shown in \req{Bcrit} appear in discussion of magnetars~\citep{Kaspi:2017fwg}. The magnetar field is expected to be more than 100-fold that of the Schwinger critical magnetic field which is on the same order of magnitude as $B_\textrm{crit}$ for an electron. While the critical field for a proton exceeds that of a magnetar, the dynamics of protons (and neutrons) in such fields is nevertheless significantly modified. A correct description of magnetic moment therefore has relevant consequences to astrophysics. 

\rf{f02} shows analogous reduction in particle/anti\-particle energy gap for the DP equation. In this case the vanishing point happens at a larger magnetic field strength. This time the solutions continue past this point, but require allowing the states to cross into the opposite continua which we consider nonphysical. We are not satisfied with either model's behavior though the KGP description is preferable. However, it is undesirable that both KGP and DP solutions loose physical meaning and vacuum stability in strong magnetic fields.

\begin{figure}[ht]
     \centering
     \includegraphics[clip, trim=0.0cm 0.0cm 0.0cm 0.5cm,width=0.95\textwidth]{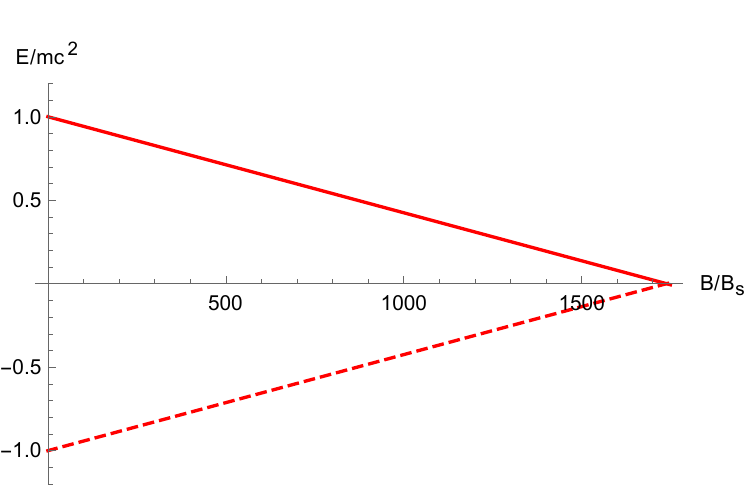}
     \caption{The $n=0$, $s=1/2$ ground state for a DP electron given by \req{lan25b} with $g/2-1=\alpha/2\pi$ in a homogeneous magnetic field. We consider the particle with no $z$-direction momentum. The particle state (solid red) and antiparticle (dashed red) are presented.}
     \label{f02}
\end{figure}

\subsection{High-Z hydrogen-like atoms}
\label{sec:sbc}
\noindent For the case of $g\!=\!2$ hydrogen-like systems with large $Z$ nuclei, there is extensive background related to the long study of the solutions of the Dirac equation~\citep{Rafelski:1976ts,Greiner:1985ce,Rafelski:2016ixr}. For $g\ne 2$ and $1/r$ singular potential we refer back to the exact expression for the KGP energy levels in \req{cou17}. In the situation of critical electric fields, states lose self-adjointness for large $Z$ in both the Dirac $g\!=\!2$ case~\citep{Gesztesy:1984hd} and for KGP $g\!\neq\!2$. \cite{Thaller:1992ji} notes that the DP-Coulomb solutions retain self-adjointness via `diving' states. For KGP $|g|<2$, states vanish similar to Dirac energy levels for the $1/r$ singular potential, but if $|g|\!>\!2$ there is merging of particle to particle states (and antiparticle to antiparticle) for states of the same total angular momentum quantum number $j$, but opposite spin orientations.

This merging state behavior can be seen in~\rf{f03}, which shows the meeting of the $1S_{1/2}$ and $2P_{1/2}$ states when $|g|\!>\!2$. For $|g|\!<\!2$ there is no state merging, but the solution is discontinuous in the sense that even for $1S_{1/2}$ we see a maximum allowed value of $Z$ at a finite energy. This is reminiscent of the Dirac $g\!=\!2$ behavior we are familiar with for the $2P_{1/2}$ state (see upper dashed blue line in~\rf{f03}) and many other $g\!=\!2$ eigenstates. We know from study of numerical solutions of the Dirac equation that the regularization of the Coulomb potential by a finite nuclear size removes this singular behavior. It remains to be seen how this exactly works in the context of the KGP equation allowing for the magnetic anomaly.

\begin{figure}[ht]
    \centering
    \includegraphics[width=\linewidth]{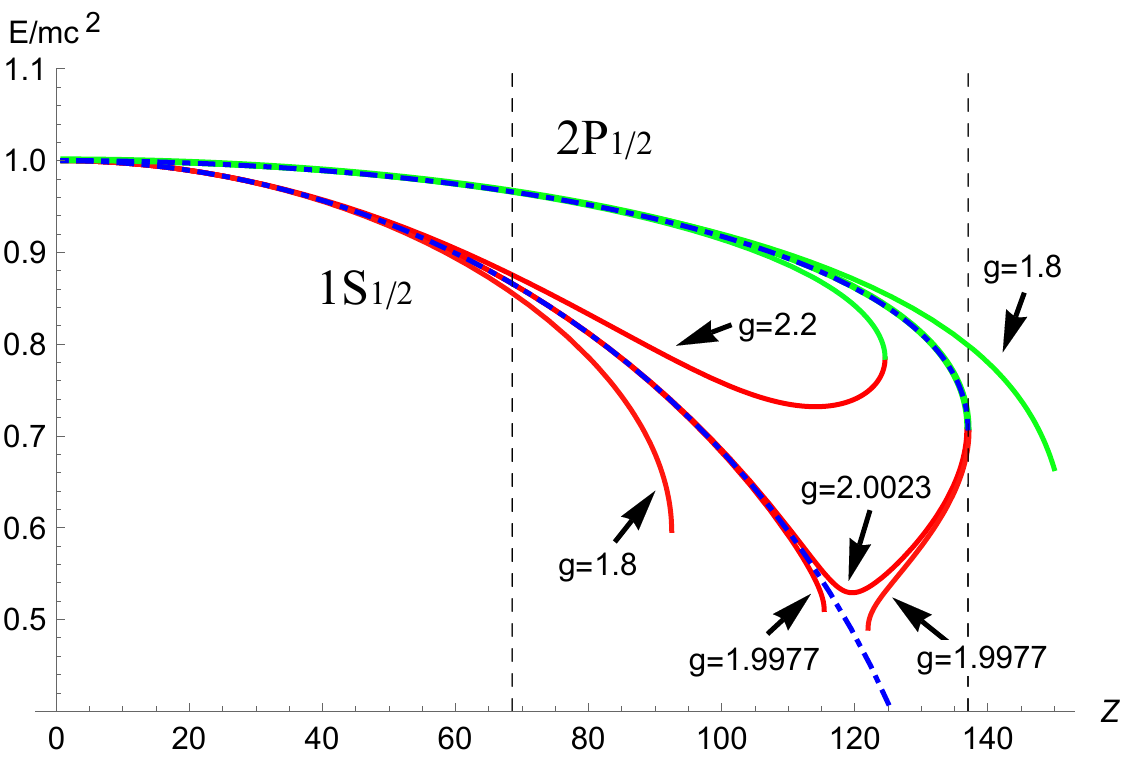}
     \caption{The KGP $1S_{1/2}$ (lower red curves) and $2P_{1/2}$ (upper green curves) energy levels for $g$-factor values $g\!=\!\{1.8,\ 1.9977,\ 2.0023,\ 2.2\}$ are shown for large $Z$ hydrogen-like atoms. The curves for the Dirac $g\!=\!2$ case for (lower dashed blue) $1S_{1/2}$ and (upper dashed blue) $2P_{1/2}$ are also presented.}
    \label{f03}
\end{figure}

\cite{Thaller:1992ji} presented numerically computed DP equation energy levels for large $Z$ hydrogen like atoms. These numerical solutions involve crossings in energy levels between states with the same total angular quantum number $j$, but differing spin orientations such as $1S_{1/2}$ and $2P_{1/2}$; these states also have the behavior of diving into the antiparticle lower continuum even for $1/r$-potential. These features are not present for the KGP-Coulomb solution. However, there is a similarity between the numerical solutions of the DP equation and our analytical KGP solutions, because for $|g|>2$ the merging states as described above correspond to the crossing states in the DP solution.

The DP equation also allows for the so-called `super-positronium' states as described by~\cite{Barut:1975hz,Barut:1976hs}. Such states represent resonances due to the magnetic interaction that reside incredibly close to the center of the atom i.e $\sqrt{\langle r^{2}\rangle}\approx a\alpha\hbar/mc$, but this feature is absent from the KGP formation of the Coulomb problem as all KGP-Coulomb wave functions which can be normalized can be successfully matched to their Dirac ($g\!=\!2$) companions.

\begin{figure}[ht]
    \centering
    \includegraphics[clip, trim=0.0cm 0.0cm 16.0cm 0.0cm,width=0.90\linewidth]{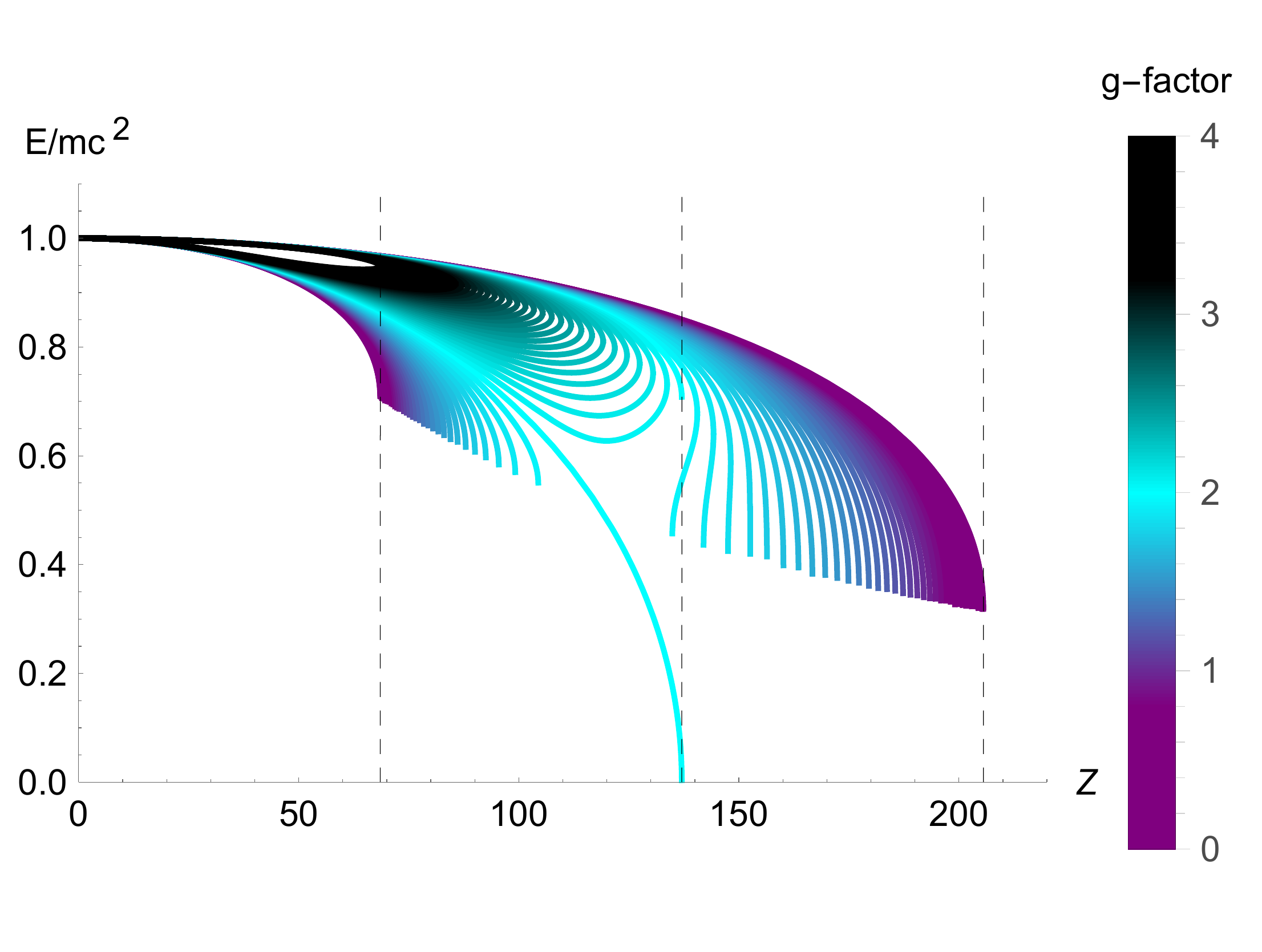}
     \caption{The KGP $1S_{1/2}$ and $2P_{1/2}$ states are plotted for hydrogen-like energies for $220>Z>0$. The spread of lines corresponds to a spread of $g$-factor values: $4>g>0$. Integer multiples of $Z=137/2$ are marked with vertical dashed lines. The separation between any two adjacent curves is $|g_{i}-g_{i+1}|=0.05$. The unique curve which dives towards and stops at the boundary $E=0$ is the Dirac $g=2$ ground state.}
    \label{fig:gspec1}
\end{figure}

\begin{figure}[ht]
    \centering
    \includegraphics[clip, trim=0.0cm 0.0cm 8.0cm 0.0cm,width=0.90\linewidth]{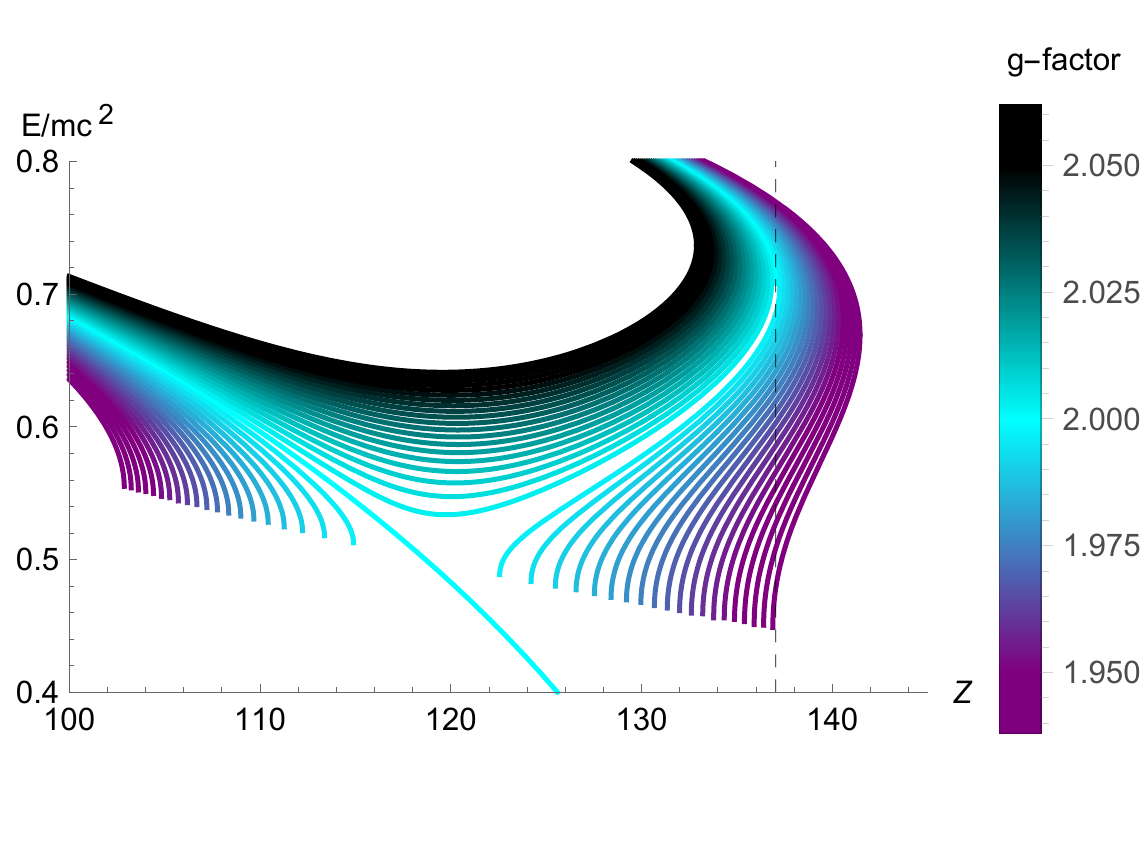}
     \caption{A close up of the KGP $1S_{1/2}$ and $2P_{1/2}$ states between $2.0625>g>1.9375$. The separation between any two adjacent curves is $|g_{i}-g_{i+1}|=0.003125$. The value $Z=137$ is marked with a vertical dashed line.}
    \label{fig:gspec2}
\end{figure}

Because analytical solutions of the DP-Coulomb problem are not available, unlike our results for KGP, it is hard to pinpoint precisely the origin of the diving and crossing state behavior. However, we can hypothesize that the problems arise due to the pathological structure of DP equation where the magnetic anomaly rather than full magnetic moment appears. On the other hand KGP framework for large $Z$ shows interesting and well-behaved analytical behavior.

\begin{figure}[ht]
    \centering
    \includegraphics[clip, trim=0.0cm 0.0cm 8.0cm 0.0cm,width=0.90\linewidth]{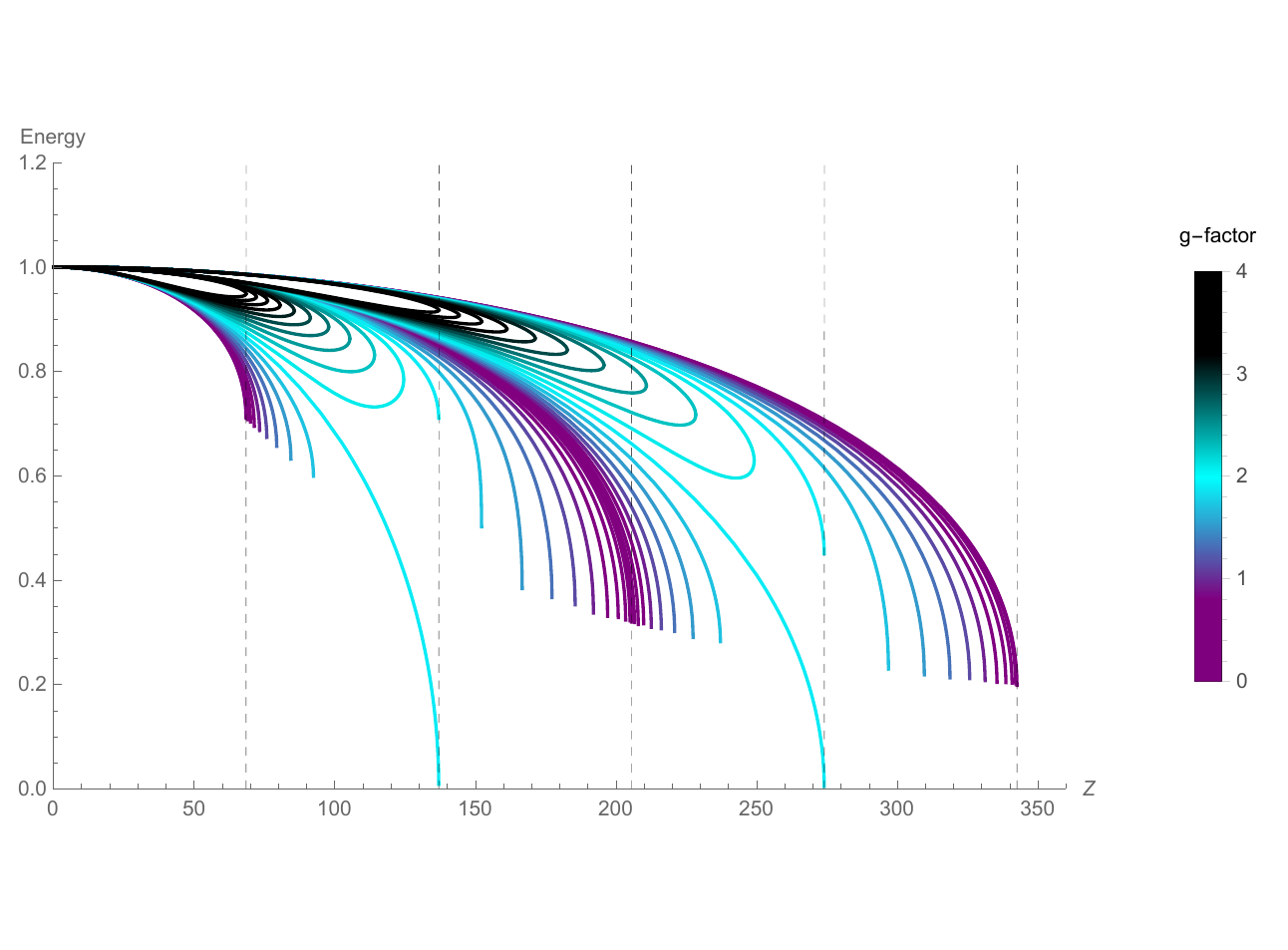}
     \caption{Energy level curves are plotted for total angular momentum quantum number $j=1/2,\,3/2$ with the relativistic principle quantum number $n_{r}=0$. The separation between any two adjacent curves is $|g_{i}-g_{i+1}|=0.2$. Dashed vertical lines indicate integer multiples of $Z=137/2$.}
    \label{fig:gspec3}
\end{figure}

We further can explore the $g$-dependency on the $1S_{1/2}$ and $2P_{1/2}$ states by plotting a spectrum of $g$-factor values as is done in \rf{fig:gspec1}. Purple regions are where $g\rightarrow0$ and the energies resembles the familiar Klein-Gordon case. As $g\to0$, small changes in $g$-factor only lead to modest changes in the energies for large $Z$ systems. The black curves represent where $g\rightarrow4$. A unique feature of $g>2$ fermions is that after a certain point, certain states become \emph{less} bound with increasing $Z$. These rising curves represent spin anti-aligned $-\lambda$ levels which become nonphysical (e.g the slope becomes vertical) and merge with their spin aligned $+\lambda$ counterparts precisely where those states also become nonphysical.

The cyan curves in \rf{fig:gspec1} are where $g\approx2$ and the states resemble the Dirac case. At exactly $g\!=\!2$ there is a unique behavior where the $1S_{1/2}$ state dashes downward and terminates at $Z\!=\!1/\alpha$. This path is unique and does not occur for any $g\neq2$. Additionally, the $g\!=\!2$ state does not smoothly connect with the $g\approx2$ solutions for large $Z$ hydrogen-like atoms. This is more visible in \rf{fig:gspec2} (note the same purple-cyan-black color scheme is used but for different $g$-factor values) which plots a variety of $g$-factor values near $g\approx2$. More specifically, small changes in $g$-factor lead to large deviations in the energies which is represented by the lack of dense lines near $g\!=\!2$ indicating the ``cusp-like'' nature of the Dirac case for very small anomalies.

The pattern of merging states with the same angular momentum first discussed in \rf{f03} repeats itself for higher values of $j$ total angular momentum. We show this explicitly in \rf{fig:gspec3} where mergers between $j=1/2$ states occur in the left lobe while mergers between $j=3/2$ states occur in the right lobe for $g>2$.

\section{Combination of mass and magnetic moment}
\label{sec:ikgp}
While we have thus far focused on the DP and KGP models for magnetic dipole moments, the non-uniqueness of spin dynamics allows us to invent further non-linear EM models which all in the non-relativistic limit yield the non-relativistic QM magnetic dipole Hamiltonian. One such extension to quantum spin dynamics is to note the close relationship mass and magnetic moment share in the KGP formalism which we describe below.

Neutral particle electromagnetic interactions are of interest~\citep{Bruce:2020xer,Bruce:2021cva,Bruce:2021fph} allowing for a variety of model-building. The `Landau' energies for neutral particles (\emph{i.e.} neutron with mass $m_\mathrm{N}$) 
in \req{lan24b} simplifies to
\begin{gather}
\label{neutral:1}
E_{n,s}|_{e=0}\rightarrow E_{s}(\bb{p},B)=\sqrt{m_\mathrm{N}^{2}c^{4}+\bb{p}^{2}c^{2}-2\mu B m_\mathrm{N}c^{2}s}\;,
\end{gather}
which is just the free particle motion with a magnetic dipole energy. We note the correspondence between the quantized Landau orbitals and continuous transverse momentum $\bb{p}^{2}=p_{3}^{2}+p_\mathrm{T}^{2}$. We can define an `effective' polarization mass given by
\begin{gather}
\label{neutral:2}
\tilde{m}^{2}_{s}(B)=m^{2}c^{4}-2\mu B mc^{2}s\,.
\end{gather}
This effective polarization mass (which also is easily defined for charged particles) will find use when we consider cosmic thermodynamics and plasmas in \rchap{chap:cosmo}.

Inspired by \req{neutral:2}, we can write a unified dipole-mass as
\begin{alignat}{1}
    \label{eq:ext:01} \boxed{\widetilde{m}(\bb{E},\bb{B}) = m +\mu\frac{\sigma_{\alpha\beta}F^{\alpha\beta}}{2c^{2}}}\,,
\end{alignat}
which satisfies the wave equation
\begin{gather}
\label{eq:ext:02} \left((i\hbar\partial_{\mu}-eA_{\mu})^{2}-\widetilde{m}^{2}(\bb{E},\bb{B})c^{2}\right)\Psi=0\,,\\
\label{IKGP01} \left(\left(i\hbar\partial_{\mu}-eA_{\mu}\right)^{2}-\left(mc+\mu\frac{\sigma^{\alpha\beta}F_{\alpha\beta}}{2c}\right)^{2}\right)\Psi=0\;.
\end{gather}
This modified KGP formulation then requires spin sensitive and explicit electromagnetic component to the charged lepton mass. In terms of QFT, \req{IKGP01} results in higher order vertex diagrams coupling fermions to photons. This ansatz is inspired by classical theory where charged particles should be understood to derive at least some of their mass from their electromagnetic fields. Dynamical mass is the driving motivation behind our work in \rchap{chap:neutrino}~\citep{Rafelski:2023zgp} in the context of dynamic neutrino flavor mixing. It is also a useful mathematical tool in \rchap{chap:cosmo}~\citep{Steinmetz:2023nsc} in the context of cosmic magnetism.

We note that \req{eq:ext:01} suggests that there may by some perturbative connection between particle mass and magnetic moment. This relationship would only manifest in strong fields where non-minimal coupled electromagnetism may be large. To this end we propose the following as a possible ansatz
\begin{alignat}{1}
	\label{mass:eq:10} \widetilde{m}(\bb{E},\bb{B}) = m\,\exp\left(\frac{\mu}{m}\frac{\sigma_{\alpha\beta}F^{\alpha\beta}}{2c^{2}}\right)\,,
\end{alignat}
which for weak fields $F\rightarrow0$ reduces to the prior form \req{eq:ext:01}. We will not explore \req{mass:eq:10} further in this work.

The approach in \req{eq:ext:01} is superficially similar to the model proposed by~\cite{Frenkel:1926zz} in classical mechanics by giving the particle a spin dependent mass of the form $m\sim\Sigma_{\mu\nu}F^{\mu\nu}$ where $\Sigma_{\mu\nu}$ is the covariant generalization of the classical magnetic and electric dipole; a more detailed exploration can be found in~\cite{Formanek:2020ojr}. \req{IKGP01} is however distinct in that the mass is allowed off-diagonal components in spinor space (a subspace which doesn't exist classically). The dipole-mass \req{eq:ext:01} is off-diagonal in spinor space in the Dirac representation and no longer commutes likes a scalar.

\req{eq:ext:01} also differs from the regular KGP equation by the presence of an additional quadratic interaction which we can evaluate using \req{dp:2} in the Dirac representation as
\begin{gather}
    \label{quad:1} \delta V=-\frac{\mu^{2}}{4}\left(\gamma_{\alpha}\gamma_{\beta}F^{\alpha\beta}\right)^{2}=\mu^{2}
    \begin{pmatrix}
        i\bb{\sigma}\cdot\bb{E}/c & -\bb{\sigma}\cdot\bb{B} \\
        -\bb{\sigma}\cdot\bb{B} & i\bb{\sigma}\cdot\bb{E}/c
    \end{pmatrix}
    \begin{pmatrix}
        i\bb{\sigma}\cdot\bb{E}/c & -\bb{\sigma}\cdot\bb{B} \\
        -\bb{\sigma}\cdot\bb{B} & i\bb{\sigma}\cdot\bb{E}/c
    \end{pmatrix}\,,\\
    \label{quad:2}
    \delta V = \mu^{2}
    \begin{pmatrix}
        \bb{B}^{2}-\bb{E}/c^{2} & -i\bb{E}\cdot\bb{B} \\
        -i\bb{E}\cdot\bb{B} & \bb{B}^{2}-\bb{E}/c^{2}
    \end{pmatrix}=2\mu^{2}
    \begin{pmatrix}
        \mathcal{S} & -i\mathcal{P} \\
        -i\mathcal{P} & \mathcal{S}
    \end{pmatrix}\,.
\end{gather}
We define the invariants of the electromagnetic field tensor $F^{\alpha\beta}$ in \req{quad:2}  letting us write the above more compactly as
\begin{alignat}{1}
    \label{inv:1} \mathcal{S}=\frac{1}{2}(\bb{B}^{2}-\bb{E}^{2}/c^{2})\,,\qquad
    \mathcal{P}=\bb{E}\cdot\bb{B}/c\,,\qquad
    \delta V=2\mu^{2}\left(\mathcal{S}-i\gamma^{5}\mathcal{P}\right)\,.
\end{alignat}
We note that $\sigma_{\alpha\beta}F^{\alpha\beta}/2$ can also be written in terms of its four eigenvalues
\begin{align}
    \label{inv:2}
    \lambda_{1}=+\mathcal{S}+i\mathcal{P}\,,\qquad
    \lambda_{2}=+\mathcal{S}-i\mathcal{P}\,,\qquad
    \lambda_{3}=-\mathcal{S}+i\mathcal{P}\,,\qquad
    \lambda_{4}=-\mathcal{S}-i\mathcal{P}\,.
\end{align}
This represents only one possible non-linear extension to electromagnetism in relativistic quantum mechanics of which there are a family of extensions~\citep{Foldy:1952a}.

For the homogeneous magnetic field \req{IKGP01} can be solved in much the same way as the KGP equation in \rsec{sec:homogeneous}. One obtains energy eigenvalues by noting the simple shift that occurs in the mass of $m^2\rightarrow m^{2}+\mu^{2}{B}^{2}$ quadratic in the magnetic field. Quadratic (spin independent) contributions are often referred to as scalar polarization~\citep{Holstein:2013kia}. The resulting energy levels are
\begin{alignat}{1}
\label{IKGP05} E_{n,s}(\bb{B})=\sqrt{m_{e}^{2}c^{4}+\mu^{2}{B}^{2}+p_{3}^{2}c^{2}+2e\hbar c^{2}B\left(n+\frac{1}{2}\right)-2\mu B m_{e}c^{2}s}
\end{alignat}
Interestingly in ultra-high magnetic fields $(B\gg{B}_{S})$, \req{IKGP05} approximates
\begin{alignat}{1}
\label{IKGP07} E\approx\mu B\;.
\end{alignat}
This is not dissimilar to the non-relativistic case where the magnetic energy is simply proportional to the magnetic field.

\begin{figure}[ht]
 \centering
 \includegraphics[clip, trim=0.0cm 0.0cm 0.0cm 0.5cm,width=\linewidth]{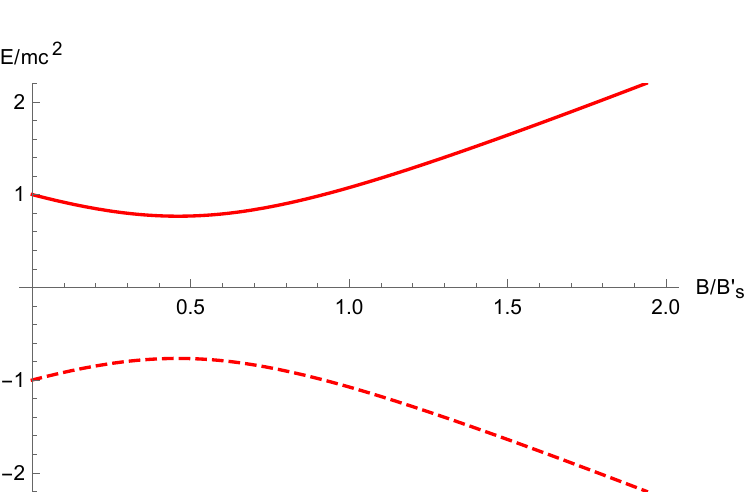}
 \caption{The $n=0$, $s=1/2$ ground state for a IKGP proton given by Eq.~\eqref{IKGP05} with $g=5.58$ in a homogeneous magnetic field. The magnetic minimum is well visible for particles with larger anomalous moment such as proton. We consider the particle with no $z$-direction momentum. The particle state (solid red) and antiparticle (dashed red) are presented. The magnetic field scale is ${B}'_{S}=(m^2_p/m^2_e)B_{S}$.}
\label{f05}
\end{figure}

The most striking feature is that the ground state remains physical for all values of magnetic field when an anomalous moment is included. The self-adjointness of the system is not lost for some critical magnetic field strength. It can be then thought that the magnetic field provides a stabilizing influence on the system. Rather, there exists a ``magnetic minimum'' located for $n=0, s=1/2$ at
\begin{alignat}{1}
\label{IKGP06} B_{\mathrm{min}}&=\frac{4mc^{2}}{g^{2}\mu_{B}}a\;,
\end{alignat}
which for the electron is
\begin{alignat}{1}
B_{\mathrm{min}}^{e}&=\frac{8a_e}{g_e^2}B_{S}
=1.02126\times10^{11}\;\mathrm{G}\;.
\end{alignat}
The minimum for a proton is in comparison
\begin{alignat}{1}
B_{\mathrm{min}}^{p}&=\frac{8a_p}{g_p^2} \frac{m_p^2}{m_e^2}B_{S}
=6.841\times10^{19}\;\mathrm{G}\;,
\end{alignat}
which can be seen in~\rf{f05}. Here it is understood that for the calculation of the proton's magnetic minimum, the nuclear mass and magneton was used rather than the electron Bohr magneton. For large enough g-factor, excited states also contain a minimum, but for any nonzero anomalous moment the ground state always does.

\section{Classical relativistic spin dynamics}
\label{sec:cspin}
\noindent We turn away from quantum mechanical approaches to inspect the classical analogue of spin dynamics. Considering the Poincar{\'e} group of space-time symmetry
transformations~\citep{Weinberg:1995mt,greiner2012quantum}, it was established by Wigner that elementary particles are group representations and can be characterized by eigenvalues of the Poincar{\'e} group's two Casimir operators:
\begin{gather}
    C_{1}\equiv p_{\mu}p^{\mu} = p^{2} = m^{2}c^{2}\,,\\
    C_{2}\equiv w_{\mu}w^{\mu} = w^{2}\,,\qquad w_{\alpha}=\frac{1}{2}\varepsilon_{\alpha\beta\mu\nu}M^{\beta\mu}p^{\nu}\,,
\end{gather}
where $w^{\alpha}$ is the Pauli-Lubanski pseudo-vector. Here $M^{\alpha\beta}$ is the relativistic tensor expression for the angular momentum defined via
\begin{gather}
    M^{\alpha\beta} = x^{\alpha}p^{\beta}-x^{\beta}p^{\alpha} + S^{\alpha\beta}
\end{gather}
where $S^{\alpha\beta}$ is the spin angular momentum tensor. The spin tensor $S^{\alpha\beta}$ can be understood via the classical (Cl.) four-spin in the rest frame (RF) as
\begin{gather}
    \label{fourspin}
    s^{\alpha}\vert_\mathrm{RF} = (0,\bb{s}_\mathrm{Cl.})\,,\qquad s_{\alpha}s^{\alpha} = -\bb{s}_\mathrm{Cl.}^{2}\,,\qquad s_{\alpha} = \frac{1}{2mc}\varepsilon_{\alpha\beta\mu\nu}S^{\beta\mu}p^{\nu}\,,
\end{gather}
where $\bb{s}_\mathrm{Cl.}$ is the classical Euclidean three-spin (not to be confused with the quantum operator $\bb{s}$). We also note that to make the units correct, the Pauli-Lubanski pseudo-vector and the four-spin are proportional by a factor of $\sqrt{C_{1}}=mc$. Quantum mechanically~\citep{Ohlsson:2011zz} $S^{\alpha\beta}$ appears as $\sim\sigma^{\alpha\beta}$ which we've already identified as the spin tensor in Dirac theory defined via the $\gamma^{\alpha}$ matrices.

\subsection{Covariant magnetic potential and modified Lorentz force}
\label{sec:magpotential}
We are interested in elementary particles with electric charge $e$, and elementary magnetic dipole charge $d=\mu/|\bb{s}_\mathrm{Cl.}|$. Therefore the covariant dynamics must be extended beyond the Lorentz force to incorporate the Stern–Gerlach (SG) force. To achieve a generalization we introduce~\citep{Rafelski:2017hce} the covariant magnetic potential
\begin{gather}
    \label{bpot:1}
    B_{\alpha}(x,s)\equiv F_{\alpha\beta}^{*}s^{\beta}
\end{gather}
As $s_{\alpha}$ is a pseudo-vector; the product in \req{bpot:1} results in a polar 4-vector $B_{\alpha}$. We note that the magnetic dipole potential $B_{\alpha}$ by construction in terms of the antisymmetric field pseudo-vector $F_{\alpha\beta}^{*}$ satisfies
\begin{gather}
    \label{bpot:2}
    \partial_{\alpha}B^{\alpha}=0\,,\qquad s\cdot B=0 \rightarrow B\cdot\frac{ds}{d\tau}+s\cdot\frac{dB}{d\tau}\,,
\end{gather}
where $\tau$ is the proper time.

The zeroth component of the covariant potential in the rest frame \req{bpot:1} reproduces the classical magnetic dipole energy given by
\begin{gather}
    U_\mathrm{Mag.}=dB^{0}=dF^{0\beta}s_{\beta}=-\bb{\mu}_\mathrm{Cl.}\cdot\bb{B}\,,\qquad \bb{\mu}_\mathrm{Cl.}=\mu\frac{\bb{s}_\mathrm{Cl.}}{|\bb{s}_\mathrm{Cl.}|}\,.
\end{gather}
We can then define a covariant magnetic field tensor from the potential \req{bpot:1} which generalizes the covariant Lorentz force $F_\mathrm{L}^{\alpha}$ as
\begin{gather}
    \label{LSG02}
    G^{\alpha\beta}=\partial^{\alpha}B^{\beta}-\partial^{\beta}B^{\alpha}= \partial^{\alpha}F^{*\beta\gamma}s_{\gamma}-\partial^{\beta}F^{*\alpha\gamma}s_{\gamma}\,,\\
    \label{LSG01}
    F_\mathrm{L}^{\alpha}\equiv\frac{dp^{\alpha}}{d\tau}=eF^{\alpha\beta}u_{\beta}+dG^{\alpha\beta}u_{\beta}\,,
\end{gather}
where $u_{\alpha}$ is the four-velocity and as previously stated, $e$ and $d$ are the electric and dipole charges. While the first term in \req{LSG01} is the standard Lorentz force, the second term is a covariant formulation of the SG force.

Because the spin precession is sensitive to the force on a particle, the presence of a SG force will induce precession terms which are second order in spin. The torque on the magnetic moment of the particle can be determined via the properties of the four-spin. Namely $s^{\alpha}$ is orthogonal~\citep{schwinger1974spin} to the four-velocity yielding
\begin{gather}
    \label{bpot:4}
    u\cdot\frac{ds}{d\tau}+\frac{du}{d\tau}\cdot s
\end{gather}
The spin torque equations can be obtained~\citep{Thomas:1926dy,Bargmann:1959gz} by inserting the Lorentz force (in our case the modified Lorentz force) that corresponds to \req{LSG01} yielding
\begin{alignat}{1}
  \notag\frac{\mathrm{d}s^{\mu}}{\mathrm{d}\tau}&=(1+\tilde{a})\frac{e}{m}F^{\mu\nu}s_{\nu}-\tilde{a}\frac{e}{m}u^{\mu}\left(u_{\alpha}F^{\alpha\beta}s_{\beta}\right)/c^{2}\\
  \label{LSG03}&+(1+\tilde{b})\frac{d}{m}G^{\mu\nu}s_{\nu}-\tilde{b}\frac{d}{m}u^{\mu}\left(u_{\alpha}G^{\alpha\beta}s_{\beta}\right)/c^{2}\,.
\end{alignat}
The constants $\tilde{a}$ and $\tilde{b}$ are arbitrary allowing for extra terms not forbidden by special relativity. With $d=0$, \req{LSG03} are known as the Thomas-Bargmann-Michel-Telegdi (TMBT) equations. In the standard derivation of relativistic spin precession, in the TBMT equation, the $\tilde{a}$ constant is associated with the anomalous magnetic moment.

In allowing for spin precession sourced by a Stern-Gerlach dipole force, an additional constant $\tilde{b}$ must be introduced. The terms in \req{LSG02} involving the $G$ tensor are spin precession directly originating from dipole forces. In homogeneous electromagnetic fields, \req{LSG02} reduces to the standard TBMT equation. These dynamical torque equations have found use in describing neutral and charged systems classically~\citep{Formanek:2019cga,Formanek:2021mcp} and inspired further efforts to improve covariant dynamics~\citep{Formanek:2020ojr}. The incorporation of electric dipole moments (EDM)~\citep{Asenjo:2023mon} into the above formalism would be a novel extension of our work here to further generalize the TBMT equations.

\subsection{Equivalency of Gilbertian and Amp{\`e}rian Stern-Gerlach forces}
\label{sec:ampgil}
We can identify the covariant formulation described in \req{LSG01} with the SG force in two different ways which define different interpretations of the magnetic dipole. Explicitly magnetic forces in the non-relativistic limit manifest in two variants:
\begin{alignat}{1}
\label{gil:0}
\bb{F}_\mathrm{G} = (\bb{\mu}\cdot\bb{\nabla})\bb{B}\ &:\ \mathrm{Gilbert\ dipole}\,,\\
\label{amp:0}
\bb{F}_\mathrm{A} =\bb{\nabla}(\bb{\mu}\cdot\bb{B})\ &:\ \mathrm{Amp\grave{e}re\ dipole}\,.
\end{alignat}
The Gilbert model describes the dipole moment of two magnetic monopole charges; and the Amp{\`e}re dipole is generated by a loop of electric current. They differ in the directionality of the Euclidean three-force: The Gilbertian dipole force is in the direction of magnetic field while the Amp{\`e}rian force is in the direction of gradient. The two forces are related via
\begin{alignat}{1}
\label{grad:1}
\bb{\nabla}(\bb{\mu}\cdot\bb{B}) = (\bb{\mu}\cdot\bb{\nabla})\bb{B}+\bb{\mu}\times(\bb{\nabla}\times\bb{B})\,,
\end{alignat}
with the assumption that the dipole is spatially independent. The second term in the above equation can be rewritten using Amp{\`e}re's law
\begin{alignat}{1}
\label{grad:2}
\bb{\mu}\times(\bb{\nabla}\times\bb{B}) = \bb{\mu}\times\Big(\mu_{0}\bb{J}+\frac{1}{c^{2}}\frac{\partial\bb{E}}{\partial t}\Big)
\end{alignat}
where $\mu_{0}$ is understood to be the vacuum permeability. The important takeaway is that the Amp{\`e}rian (gradient direction) and Gilbertian (field direction) forces are related via a term ultimately sensitive to the external current $\bb{J}$ of the system.

Before we show how this manifests in the covariant formulation, we recall the orthogonality of four-velocity and four-spin given in \req{bpot:4}. This property allows us to write the covariant Lorentz force in \req{LSG01} in two equivalent forms which we will call the Amp{\`e}rian $F_\mathrm{L,A}^{\alpha}$ and Gilbertian $F_\mathrm{L,G}^{\alpha}$ covariant expressions for reasons that will become obvious soon
\begin{align}
\label{amp}
F_\mathrm{L,A}^{\alpha} &= eF^{\alpha\beta}u_{\beta} + ds_{\gamma}u_{\beta}(\partial^{\alpha}F^{*\beta\gamma}-\partial^{\beta}F^{*\alpha\gamma})\,,\\
\label{gil}
F_\mathrm{L,G}^{\alpha} &= eF^{\alpha\beta}u_{\beta} -d(s_{\gamma}\partial^{\gamma})F^{*\alpha\beta}u_{\beta}+ds_{\gamma}T^{\gamma\alpha\beta}u_{\beta}\,.
\end{align}
The tensor $T^{\alpha\mu\nu}$ is the completely antisymmetric dual tensor to electric four-current
\begin{align}
    T^{\alpha\mu\nu} = \partial^{\alpha}F^{*\mu\nu}+\partial^{\mu}F^{*\nu\alpha}+\partial^{\nu}F^{*\alpha\mu}\,.
\end{align}

We will now show that the expression $F_\mathrm{L,A}^{\alpha}$ depends on an Amp{\`e}rian dipole and the proper time derivative of the fields. Recognizing that $d/d\tau=u\cdot\partial$, \req{amp} can be rewritten as
\begin{align}
F_\mathrm{L,A}^{\alpha} = eF^{\alpha\beta}u_{\beta} + ds_{\gamma}\left(\partial^{\alpha}F^{*\beta\gamma}u_{\beta}-\frac{d}{d\tau}F^{*\alpha\gamma}\right)\,.
\end{align}
We evaluate the four-force in the instantaneous rest frame (RF) of the particle such that $u^{\alpha}=(c,0)$, $d/d\tau=\partial/\partial t$ and $s^{\alpha}=(0,\bb{s}_\mathrm{Cl.})$. Noting that $\bb{\mu}=dc\bb{s}_\mathrm{Cl.}$ this yields
\begin{alignat}{1}
\label{amp:1}
F_\mathrm{L,A}^{0}\vert_\mathrm{RF} &= -\frac{1}{c}\bb{\mu}\cdot\left(\frac{\partial\bb{B}}{\partial t}-\frac{\partial\bb{B}}{\partial t}\right)=0\,,\\
\label{amp:2}
\bb{F}_\mathrm{L,A}\vert_\mathrm{RF} &= e\bb{E} + \bb{\nabla}(\bb{\mu}\cdot\bb{B})-\frac{1}{c^{2}}\bb{\mu}\times\frac{\partial\bb{E}}{\partial t}\,.
\end{alignat}
As we see in \req{amp:2}, the second term corresponds to the Amp{\`e}rian SG force written in \req{amp:0}.

We can accomplish the same procedure for $F_\mathrm{L,G}^{\alpha}$ which explicitly depends on a covariant Gilbertian dipole and spin coupling to current. The four-force yields
\begin{alignat}{1}
\label{gil:1}
F_\mathrm{L,G}^{0}\vert_\mathrm{RF} &=0\\
\label{gil:2}
\bb{F}_\mathrm{L,G}\vert_\mathrm{RF} &= e\bb{E} +(\bb{\mu}\cdot\bb{\nabla})\bb{B}+\left(\bb{\nabla}(\bb{\mu}\cdot\bb{B})-(\bb{\mu}\cdot\bb{\nabla})\bb{B}-\frac{1}{c^{2}}\bb{\mu}\times\frac{\partial\bb{E}}{\partial t}\right)
\end{alignat}
If we cancel out the second and forth terms in \req{gil:2}, we re-obtain the Amp{\`e}rian three-force written in \req{amp:2}.

However, making use of the expressions in \req{grad:1} and Amp{\`e}re's circuital law \req{grad:2}, we can rewrite the above terms yielding
\begin{alignat}{1}
\bb{F}_\mathrm{L,G}\vert_\mathrm{RF} &= e\bb{E} +(\bb{\mu}\cdot\bb{\nabla})\bb{B}+\bb{\mu}\times\mu_{0}\bb{J}
\end{alignat}
which is the Gilbertian expression for the SG force as given in \req{gil:0}. Our covariant formulation then requires that
\begin{align}
    \boxed{\bb{F}_\mathrm{L,A}\vert_\mathrm{RF} \equiv \bb{F}_\mathrm{L,G}\vert_\mathrm{RF}}
\end{align}
Thus we find an equivalence between the Gilbertian and Amp{\`e}rian dipoles from the covariant introduction of intrinsic magnetic dipole in our covariant dynamics.


\chapter{Dynamic neutrino flavor mixing through transition moments}
\label{chap:neutrino}
We proposed in~\cite{Rafelski:2023zgp} that neutrino flavors are remixed when exposed to strong EM fields travelling as a superposition distinct from the vacuum propagation of free neutrinos. Neutrino mixing is an important topic for studying BSM physics as flavor mixing only occurs in the presence of massive neutrinos allowing for the misalignment between the flavor basis which participates in left-chiral $SU(2)_{L}$ weak interactions and the mass basis which are the propagating neutrino states.

We discuss the neutrino anomalous magnetic moment (AMM) in \rsec{sec:numoment}. We narrow our analysis for Majorana neutrinos which are allowed only transition magnetic moments which couple different flavors electromagnetically, but do not violate CPT symmetry. Transition moments however notably break lepton number conservation. We discuss the standard flavor mixing program in \rsec{sec:nuflavor} and the effective Lagrangian density in \rsec{sec:nuaction} containing both Majorana mass and transition moments. In \rsec{sec:nuem} we discuss the chirality of the relativistic Pauli dipole.

The two-flavor neutrino model is evaluated explicitly in \rsec{sec:nutoy} and the remixed electromagnetic-mass eigenstates are obtained in \rsec{sec:zmixing}. We obtain in \rsec{sec:emmass} an EM-mass basis, distinct from flavor and free-particle mass basis, which mixes flavors as a function of EM fields. The case of two nearly degenerate neutrinos is studied in detail in \rsec{sec:nulimits}. Moreover we show solutions relating to full EM field tensor which result in covariant expressions allowing for both magnetic and electrical fields.

\section{Electromagnetic characteristics of neutrinos}
\label{sec:numoment}
We study the connection between Majorana neutrino transition magnetic dipole moments~\citep{Fujikawa:1980yx,Shrock:1980vy,Shrock:1982sc} and neutrino flavor oscillation. Neutrino electromagnetic (EM) properties have been considered before~\citep{Schechter:1981hw,Giunti:2014ixa,Popov:2019nkr,Dvornikov:2019sfo,Chukhnova:2019oum} including the effect of oscillation in magnetic fields~\citep{Lim:1987tk,Akhmedov:1988uk,Pal:1991pm,Elizalde:2004mw,Akhmedov:2022txm}. The influence of transition magnetic moments on solar neutrinos is expected~\citep{Martinez-Mirave:2023fyb}, but difficult to measure due to the lack of knowledge of solar magnetism near the core.

The case of transition moments has the mathematical characteristics of an off-diagonal mass which is distinct from normal direct dipole moment behavior. EM field effects are also distinct from weak interaction remixing within matter, {\it i.e.\/} the Mikheyev-Smirnov-Wolfenstein effect~\citep{Wolfenstein:1977ue,Mikheyev:1985zog,Mikheev:1986wj,Smirnov:2003da}.



The size of the neutrino magnetic dipole moment can be constrained as follows: The lower bound is found by higher order standard model interactions with the minimal extension of neutrino mass $m_{\nu}$ included~\citep{Fujikawa:1980yx,Shrock:1980vy,Shrock:1982sc}. The upper bound is derived from reactor, solar and astrophysical experimental observations~\citep{Giunti:2015gga,Canas:2015yoa,Studenikin:2016ykv,AristizabalSierra:2021fuc}. The bounds are expressed in terms of the electron Bohr magneton $\mu_{B}$ as
\begin{align}
    \label{bound:1}
    \frac{e\hbar G_{F}m_{\nu}c^{2}}{8\pi^{2}\sqrt{2}} \sim 10^{-20}\mu_{B}<\mu_{\nu}^\mathrm{eff}<10^{-10}\mu_{B}\,,\qquad\mu_{B}=\frac{e\hbar}{2m_{e}}
\end{align}
where $G_{F}$ is the Fermi constant and $\mu_{\nu}^\mathrm{eff}$ is the effective and characteristic size of the neutrino magnetic moment. In \req{bound:1}, the lower bound was estimated using a characteristic mass of $m_{\nu}\sim0.1~\mathrm{eV}$. From cosmological studies, the sum of neutrino masses is estimated~\citep{Planck:2018vyg} to be $\sum_{i}m_{i}<0.12$~eV; the effective electron (anti)neutrino mass is bounded~\citep{KATRIN:2021uub} by $m_{e}^{\nu}<0.8$~eV.

\section{Neutrino flavor mixing and electromagnetic fields}
\label{sec:nuflavor}
Oscillation of neutrino flavors observed in experiment is in general interpreted as being due to a difference in neutrino mass and flavor eigenstates. This misalignment between the two representations is described as rotation of the neutrino flavor $N$-vector where $N=3$ is the observed number of generations. The unitary mixing matrix $V_{\ell k}$ allows for the change of basis between mass $(k)$ and flavor $(\ell)$ eigenstates via the transform 
\begin{alignat}{1}
    \label{basis:1} \nu_{\ell}=V_{\ell k}\nu_{k}\,\rightarrow
    \begin{pmatrix}
        \nu_{e}\\
        \nu_{\mu}\\
        \nu_{\tau}
    \end{pmatrix}=
    \begin{pmatrix}
        V_{e1} & V_{e2} & V_{e3}\\
        V_{\mu1} & V_{\mu2} & V_{\mu3}\\
        V_{\tau1} & V_{\tau2} & V_{\tau3}
    \end{pmatrix}
    \begin{pmatrix}
        \nu_{1}\\
        \nu_{2}\\
        \nu_{3}
    \end{pmatrix}\,,
\end{alignat}
where $\nu_{\ell}$ is the neutrino four-spinor written in the flavor basis while in the mass basis we use $\nu_{k}$ with $k\in1,2,3$.

The parameterization of the components of the mixing matrix depends on the Dirac or Majorana-nature of the neutrinos. First we recall the Dirac neutrino mixing matrix $U_{\ell k}$ in the standard parameterization~\citep{Xing:2014wwa,Schwartz:2014sze} 
\begin{alignat}{1}
    \label{rotation:1} U_{\ell k} =
    \begin{pmatrix}
         c_{12}c_{13} & s_{12}c_{13} & s_{13}e^{-i\delta}\\
         -s_{12}c_{23} - c_{12}s_{13}s_{23}e^{i\delta} & c_{12}c_{23} - s_{12}s_{13}s_{23}e^{i\delta} & c_{13}s_{23}\\
         s_{12}s_{23} - c_{12}s_{13}c_{23}e^{i\delta}& -c_{12}s_{23} - s_{12}s_{13}c_{23}e^{i\delta} & c_{13}c_{23}
    \end{pmatrix}\,,
\end{alignat}
where $c_{ij} = \mathrm{cos}(\theta_{ij})$ and $s_{ij} = \mathrm{sin}(\theta_{ij})$. In this convention, the three mixing angles $(\theta_{12}, \theta_{13}, \theta_{23})$ are understood to be the Euler angles for generalized rotations and $\delta$ is the CP-violating complex phase. 

For the Majorana case we must allow a greater number of complex phases: Majorana neutrinos allow up to two additional complex phases $\rho$ and $\sigma$ which along with $\delta$ participate in CP-violation. A parameterization is achieved by introducing an additional phase matrix $P_{kk'}$
\begin{alignat}{1}
    \label{phases:1} &V_{\ell k} = U_{\ell k'}P_{k'k}\,,\\
    \label{phases:3} &P_{kk'} = \mathrm{diag}(e^{i\rho},e^{i\sigma},1)\,.
\end{alignat}
The mixing matrix $V_{\ell k}$ defined in \req{phases:1} can then be used to transform the symmetric mass matrix $M_{\ell\ell'}$ from the flavor basis into the diagonal mass basis 
\begin{align}
    \label{diag:0}
    M_{\ell\ell'}=
    \begin{pmatrix}
        m_{ee}^{\nu} & m_{e\mu}^{\nu} & m_{e\tau}^{\nu}\\
        m_{e\mu}^{\nu} & m_{\mu\mu}^{\nu} & m_{\mu\tau}^{\nu}\\
        m_{e\tau}^{\nu} & m_{\mu\tau}^{\nu} & m_{\tau\tau}^{\nu}
    \end{pmatrix}\,,\qquad
    M_{\ell\ell'}^\mathrm{T}=M_{\ell\ell'}\,,\\
    \label{diag:1}
    V_{\ell k}^\mathrm{T}M_{\ell\ell'}V_{\ell'k'} = M_{kk'} = m_{k}\delta_{kk'} = \mathrm{diag}(m_{1},m_{2},m_{3})\,.
\end{align}
We note the Majorana mass matrix is symmetric due to the anticommuting nature of the neutrino fields $\bar\nu\nu=-\nu^\mathrm{T}\bar\nu^\mathrm{T}$ and is in general complex~\citep{Adhikary:2013bma,giunti2007fundamentals} though it will be taken to be fully real in this work. There are many interesting models for mass matrices which were pioneered by~\cite{Fritzsch:1995dj,Fritzsch:1998xs,Fritzsch:1999ee,Xing:2000ik} in the leptonic sector. The masses $m_{k}$ are taken to be real and positive labelling the free propagating states of the three neutrinos.

\subsection{Effective Majorana neutrino Lagrangian}
\label{sec:nuaction}
Given the mass matrix defined in \req{diag:0}, the Majorana mass term in the Lagrangian can be written in the flavor basis as
\begin{alignat}{1}
    \label{mass:1} -\mathcal{L}_{\mathrm{mass}}^{\mathrm{Maj.}}=\frac{1}{2}\bar\nu_{\ell}M_{\ell\ell'}\nu_{\ell'}=
    -\frac{1}{2}\nu_{L,\ell}^\mathrm{T}C^{\dag}M_{\ell\ell'}\nu_{L,\ell'}+\mathrm{h.c.}\,,
\end{alignat}
where the Majorana fields are written as $\nu=\nu_{L}+C(\bar\nu_{L})^\mathrm{T}$. The field $\nu_{L}$ refers to left-handed Weyl four-component spinors. Charged conjugated fields are written as $\nu^{c}=C(\bar\nu)^\mathrm{T}$. The charge conjugation operator $C$ is defined in the usual way in~\cite{Itzykson:1980rh}; p.692.

Given these conventions, we can extend our consideration to include the electromagnetic interaction of neutrinos which is possible if neutrinos are equipped with a magnetic moment matrix $\mu_{\ell\ell'}$. We allow for a fixed \emph{external} electromagnetic field tensor $F^{\alpha\beta}_\mathrm{ext}(x^{\mu})$ which imparts a force on the neutrino fields. We emphasize that $F^{\alpha\beta}_\mathrm{ext}$ is not dynamical in our formulation and consists of real functions over four-position and not field operators.

We generalize the AMM Pauli Lagrangian in \req{lamm:1} to account for the Majorana fields in the flavor basis as
\begin{align}
\label{moment:1}
    -\mathcal{L}_{\mathrm{AMM}}^\mathrm{Maj.}=\frac{1}{2}\bar\nu_{\ell}\left(\mu_{\ell\ell'}\frac{1}{2}\sigma_{\alpha\beta}F^{\alpha\beta}_\mathrm{ext}\right)\nu_{\ell'}=
    -\frac{1}{2}\nu_{L,\ell}^\mathrm{T}C^{\dag}\left(\mu_{\ell\ell'}\frac{1}{2}\sigma_{\alpha\beta}F^{\alpha\beta}_\mathrm{ext}\right)\nu_{L,\ell'}+\mathrm{h.c.}
\end{align}
The operator $\sigma_{\alpha\beta}$ is the $4\times 4$ spin tensor defined in \req{sigma:1}. We would like to point out some interesting features of the Pauli term most notably that the spin tensor itself is not Hermitian with
\begin{align}
\label{notherm:1}
\sigma_{\alpha\beta}^{\dag} = \gamma_{0}\sigma_{\alpha\beta}\gamma_{0}\,.
\end{align}
However, the conjugate of the Lagrangian term in \req{moment:1}
\begin{align}
\left(\nu^{\dag}\gamma_{0}\sigma_{\alpha\beta}F^{\alpha\beta}_\mathrm{ext}\nu\right)^{\dag} = \nu^{\dag}\sigma_{\alpha\beta}^{\dag}F^{\alpha\beta}_\mathrm{ext}\gamma_{0}\nu = \nu^{\dag}\gamma_{0}\sigma_{\alpha\beta}F^{\alpha\beta}_\mathrm{ext}\nu\,,
\end{align}
is Hermitian. More about the spin tensor's properties will be elaborated on in \rsec{sec:numoment}.

The Majorana magnetic moment matrix acts in flavor space. It satisfies the following constraints~\citep{Giunti:2014ixa} for CPT symmetry reasons and the anticommuting nature of fermions
\begin{alignat}{1}
\label{props:1}
\mu_{\ell\ell'}^{\dag}=\mu_{\ell\ell'}\,,\qquad
\mu_{\ell\ell'}^\mathrm{T}=-\mu_{\ell\ell'}\,,
\end{alignat}
{\it i.e.\/} the AMM matrix $\mu_{\ell\ell'}$ is Hermitian and fully anti-symmetric. This requires that the transition magnetic moment elements are purely imaginary while all diagonal AMM matrix elements vanish
\begin{align}
\label{mu:1}
\mu_{\ell\ell'}=
\begin{pmatrix}
\mu_{ee} & \mu_{e\mu} & \mu_{e\tau} \\
\mu_{\mu e} & \mu_{\mu\mu} & \mu_{\mu\tau} \\
\mu_{\tau e} & \mu_{\tau\mu} & \mu_{\tau\tau}
\end{pmatrix}\xrightarrow{\mathrm{Majorana}}
\mu_{\ell\ell'}=
\begin{pmatrix}
0 & i\mu_{e\mu} & -i\mu_{e\tau} \\
-i\mu_{e\mu} & 0 & i\mu_{\mu\tau} \\
i\mu_{e\tau} & -i\mu_{\mu\tau} & 0
\end{pmatrix}\,.
\end{align}

We can combine the mass term in~\req{mass:1} and AMM contribution in~\req{moment:1} into a single effective Lagrangian
\begin{align}
\label{massmom:1}
\mathcal{L}_\mathrm{eff}^\mathrm{Maj.} &= \mathcal{L}_\mathrm{kinetic}^\mathrm{Maj.} + \mathcal{L}_\mathrm{mass}^\mathrm{Maj.} + \mathcal{L}_\mathrm{AMM}^\mathrm{Maj.}\,,\\
\label{massmom:2}
\mathcal{L}_\mathrm{eff}^\mathrm{Maj.} &= \mathcal{L}_\mathrm{kinetic}^\mathrm{Maj.} - \frac{1}{2}\bar\nu_{\ell}\left(M_{\ell\ell'}+\mu_{\ell\ell'}\frac{1}{2}\sigma_{\alpha\beta}F^{\alpha\beta}_\mathrm{ext}\right)\nu_{\ell'}\;.
\end{align}
\req{massmom:2} is our working Lagrangian. For later convenience we define the generalized mass-dipole matrix $\mathcal{M}_{\ell\ell'}$ present in \req{massmom:2} as
\begin{align}
\label{massmom:3}
\mathcal{M}_{\ell\ell'}(\bb{E},\bb{B})\equiv M_{\ell\ell'}+\mu_{\ell\ell'}\frac{1}{2}\sigma_{\alpha\beta}F^{\alpha\beta}_\mathrm{ext}\,,\qquad \mathcal{M}_{\ell\ell'}^{\dag}=\gamma_{0}\mathcal{M}_{\ell\ell'}\gamma_{0}\,.
\end{align}
As neutrinos must propagate as energy eigenstates, our objective is to find the eigenvalues of \req{massmom:2} rather than \req{diag:1}. As the mass eigenvalues are modified by the presence the EM interactions $m\rightarrow\widetilde m(\bb{E},\bb{B})$ so will the mixing matrix, leading to modifications of \req{phases:1}. These electromagnetic components then facilitate time-dependant oscillation among the free-particle mass eigenstates~\citep{Giunti:2014ixa}.

Additionally we may consider matter effects via the weak interaction. Electron (anti)neutrinos passing through matter preferentially interact via weak charge-current (via the $W^{\pm}$ boson) with electrons which make up the bulk of charged leptons in most matter. The neutral-current (via the $Z_{0}$ boson) however affects all flavors and couples to the neutrons within the medium as the electron and proton contributions cancel in charge neutral matter. This can be represented, MSW effect aside, as the weak charge-current $V_{CC}$ and neutral-current $V_{NC}$ effective potentials~\citep{Pal:1991pm,greiner2009gauge} which contribute to the action as
\begin{align}
\label{matter:1}
\mathcal{L}_\mathrm{matter}^\mathrm{Maj.} &= \bar\nu_{\ell}(\gamma_{0}V_{\ell\ell'})\nu_{\ell'}\,,\qquad
V_{\ell\ell'} = 
\begin{pmatrix}
V_{CC}+V_{NC} & 0 & 0\\
0 & V_{NC} & 0\\
0 & 0 & V_{NC}
\end{pmatrix}\,,\\
V_{CC} &= \sqrt{2}G_{F}\hbar^{2}c^{2}n_{e}\,,\qquad V_{NC} = -\frac{1}{2}\sqrt{2}G_{F}\hbar^{2}c^{2}n_{n}\,.
\end{align}
The coefficient $G_{F}$ is the Fermi constant, $n_{e}$ is the number density of electron matter and $n_{n}$ is the number density of neutrons within the medium. We note that $V_{\ell\ell'}\gamma_{0}$ behaves like the zeroth component of a vector-potential. As written, \req{matter:1} is approximately true for non-relativistic matter.

\subsection{Chiral properties of the relativistic Pauli dipole}
\label{sec:nuem}
While the Pauli dipole was introduced and discussed in \rsec{sec:dp}, we will further elaborate on details directly relevant to neutrinos. The electromagnetic dipole behavior of the neutrino depends on mathematical properties of the tensor product $\sigma_{\alpha\beta}F^{\alpha\beta}_\mathrm{ext}$. We prefer to work in the Weyl (chiral) spinor representation where the EM contribution is diagonal in spin space. Therefore we evaluate the product $\sigma_{\alpha\beta}F^{\alpha\beta}_\mathrm{ext}$ in the Weyl representation following~\cite{Feynman:1958ty} yielding
\begin{align}
\label{chiral:1}
-\frac{1}{2}\sigma_{\alpha\beta}F^{\alpha\beta}_\mathrm{ext}=
\begin{pmatrix}
\bb{\sigma}\cdot(\bb{B}+i\bb{E}/c) & 0\\
0 & \bb{\sigma}\cdot(\bb{B}-i\bb{E}/c)
\end{pmatrix}\equiv
\begin{pmatrix}
\bb{\sigma}\cdot\bb{f}_{+} & 0 \\
0 & \bb{\sigma}\cdot\bb{f}_{-}
\end{pmatrix}\,,
\end{align}
where we introduced the complex electromagnetic field form $\bb{f}_{\pm}=\bb{B}\pm i\bb{E}/c$ showing sensitivity to both magnetic and electric fields. The eigenvalues of \req{chiral:1} were also discussed in \rsec{sec:ikgp}. As this expression is diagonal in the Weyl representation, it does not exchange handedness when acting upon a state. This is explicitly understood by the fact that \req{chiral:1} commutes with $\gamma^{5}$. Since left and right-handed neutrinos are not remixed by magnetic moments, sterile right-handed neutrinos do not need to be introduced. We can also see explicitly in \req{chiral:1} its non-Hermitian character, see \req{massmom:2}, of the EM spin-field coupling. Specifically this is mirrored in the complex field's $\bb{f}_{\pm}$ relation to its complex conjugate $(\bb{f}_{\pm})^{*}=\bb{f}_{\mp}$. The complex EM fields have a Hermitian $(\bb{B})$ and anti-Hermitian $(i\bb{E})$ part.

Taking the product of $\bb{f}_{\pm}$ with its complex conjugate we find
\begin{align}
\label{cross:1}
\frac{1}{2}\left(\bb{\sigma}\cdot\bb{f}_{\pm}\right)\left(\bb{\sigma}\cdot\bb{f}_{\mp}\right)=T_\mathrm{ext}^{00}\mp \sigma_{i}T_\mathrm{ext}^{0i}\,,
\end{align}
where we recognize the stress-energy tensor $T_\mathrm{ext}^{\alpha\beta}$ component $T_\mathrm{ext}^{00}$ for field energy density and $T_\mathrm{ext}^{0i}$ momentum density respectively
\begin{align}
T_\mathrm{ext}^{00}=\frac{1}{2}\left(B^{2}+E^{2}/c^{2}\right)\,,\qquad
T_\mathrm{ext}^{0i}=\frac{1}{c}\varepsilon_{ijk}E_{j}B_{k}\,.
\end{align}
As we will see in \rsec{sec:nutoy}, \req{cross:1} will appear in the EM-mass eigenvalues of our effective Lagrangian \req{massmom:1}. Using the identity in \req{chiral:1} and \req{cross:1} we also find the interesting relationship
\begin{align}
\label{cross:2}
\frac{1}{2}\left(\frac{1}{2}\sigma_{\alpha\beta}F^{\alpha\beta}_\mathrm{ext}\right)\left(\frac{1}{2}\sigma_{\alpha\beta}F^{\alpha\beta}_\mathrm{ext}\right)^{\dag}=
\gamma_{0}\left(T_\mathrm{ext}^{00}\gamma_{0}+T_\mathrm{ext}^{0i}\gamma_{i}\right)\,.
\end{align}
Now that we have elaborated on the relevant EM field identities, we turn back to the magnetic dipole and flavor rotation problem.

\section{Electromagnetic-flavor mixing for two generations}
\label{sec:nutoy}
Considering experimental data on neutrino oscillations, it is understood that either the two lighter (normal hierarchy) or the two heavier (inverted hierarchy) neutrino states are close together in mass. If the electromagnetic properties of the neutrino do indeed lead to flavor mixing effects, then it is likely the closer pair of neutrino mass states that are most sensitive to the phenomenon we explore. In the spirit of~\cite{Bethe:1986ej}, we therefore explore the $N=2$ two generation $(\nu_{e},\nu_{\mu})$ toy model.

Following the properties established in \req{props:1} and \req{massmom:3} we write down the two generation mass and dipole matrices as
\begin{alignat}{1}
\label{mix:1} M_{\ell\ell'}= 
\begin{pmatrix}
m_{e}^{\nu} & {\delta m}\\
{\delta m} & m_{\mu}^{\nu}
\end{pmatrix}\,,\qquad
\mu_{\ell\ell'} = 
\begin{pmatrix}
0 & i\delta\mu\\
-i\delta\mu & 0
\end{pmatrix}\,.
\end{alignat}
The AMM coupling $\delta\mu$ is taken to be real with a pure imaginary coefficient. While the mass elements $(m_{e}^{\nu},m_{\mu}^{\nu},{\delta m})$ are generally complex, we choose in our toy model for them to be fully real
\begin{align}
\label{choice:1}
m_{e}^{\nu}=(m_{e}^{\nu})^{*}\,,\qquad
m_{\mu}^{\nu}=(m_{\mu}^{\nu})^{*}\,,\qquad
\delta m=\delta m^{*}\,,
\end{align}
making the mass matrix $M_{\ell\ell'}$ Hermitian. This allows us to more easily evaluate and emphasize the EM contributions to mixing rather than complications arising from the mass matrix.

Using \req{mix:1} and \req{choice:1}, we write the mass-dipole matrix in \req{massmom:3} in terms of $2\times2$ flavor components as
\begin{align}
\label{mix:2}
\mathcal{M}_{\ell\ell'} = 
\begin{pmatrix}
m_{e}^{\nu} & {\delta m}+i\delta\mu\sigma_{\alpha\beta}F^{\alpha\beta}_\mathrm{ext}/2\\
{\delta m}-i\delta\mu\sigma_{\alpha\beta}F^{\alpha\beta}_\mathrm{ext}/2 & m_{\mu}^{\nu}
\end{pmatrix}\,,\qquad
\mathcal{M}_{\ell\ell'}^{\dag}=\gamma_{0}\mathcal{M}_{\ell\ell'}\gamma_{0}\,.
\end{align}
As noted before, this matrix is not Hermitian due to the inclusion of the spin tensor, therefore it is not guaranteed to satisfy an algebraic eigenvalue equation in its present form which is a requirement for well behaved masses.

This can be remedied by recalling that any arbitrary complex matrix can be diagonalized into its real eigenvalues $\lambda_{j}$ by the biunitary transform
\begin{align}
\label{biunitary:1}
W_{\ell j}^{\dag}\mathcal{M}_{\ell\ell'}Y_{\ell'j'}=\lambda_{j}\delta_{jj'}\,,
\end{align}
where $Y_{\ell j}$ and $W_{\ell j}$ are both unitary matrices. Taking the complex conjugate of \req{biunitary:1}, we arrive at
\begin{align}
\label{biunitary:2}
(W_{\ell j}^{\dag}\mathcal{M}_{\ell\ell'}Y_{\ell'j'})^{\dag} = 
Y_{\ell j'}^{\dag}\gamma_{0}\mathcal{M}_{\ell\ell'}\gamma_{0}W_{\ell' j}=\lambda_{j}\delta_{jj'}\,,\\
Y_{\ell j}=\gamma_{0}W_{\ell j}\rightarrow
W_{\ell j}^{\dag}\mathcal{M}_{\ell\ell'}\gamma_{0}W_{\ell'j'}=\lambda_{j}\delta_{jj'}\,. 
\end{align}
As $Y_{\ell j}$ and $W_{\ell j}$ are related by a factor of $\gamma_{0}$ based on the conjugation properties of \req{mix:2}, this lets us eliminate $Y_{\ell j}$ and diagonalize using a single unitary matrix $W_{\ell j}$. The related matrix $\mathcal{M}_{\ell\ell'}\gamma_{0}$ is Hermitian
\begin{align}
\label{herm:1}
(\mathcal{M}_{\ell\ell'}\gamma_{0})^{\dag} = \mathcal{M}_{\ell\ell'}\gamma_{0}\,,
\end{align}
and also equivalent to the root of the Hermitian product of \req{mix:2}
\begin{align}
(\mathcal{M}\mathcal{M}^{\dag})_{\ell\ell'} = \left((\mathcal{M}\gamma_{0})(\mathcal{M}\gamma_{0})\right)_{\ell\ell'}\,.
\end{align}
Therefore a suitable unitary transformation $W_{\ell j}$ rotates flavor $\ell$-states into magnetized mass $j$-states. The eigenvalues $\lambda_{j}^{2}$ of $(\mathcal{M}\mathcal{M}^{\dag})_{\ell\ell'}$ are the squares of both signs of the eigenvalues of $\mathcal{M}_{\ell\ell'}\gamma_{0}$. We write this property (with flavor indices suppressed) as
\begin{align}
W^{\dag}(\mathcal{M}\mathcal{M}^{\dag})W &= W^{\dag}(\mathcal{M}\gamma_{0})WW^{\dag}(\mathcal{M}\gamma_{0})W = \mathrm{diag}(\lambda_{1}^{2},\lambda_{2}^{2})\,.
\end{align}
We associate $\lambda_{j}=\widetilde m_{j}(\bb{E},\bb{B})$ with $j\in1,2$ as the effective EM-mass states which are field dependant in this basis. 

\subsection{Separating electromagnetic-mass mixing into two rotations}
\label{sec:zmixing}
The matrix $W_{\ell j}$ mixes flavor states into a new basis distinct from the free-particle case however this rotation must smoothly connect with the free-particle case in the limit that the electromagnetic fields go to zero. We proceed to evaluate $W_{\ell j}$ breaking the rotation into two separate unitary transformations:
\begin{itemize}[nosep]
    \item [(a)] Rotation matrix $V_{\ell k}^{\dag}(\ell\rightarrow k)$ converting from flavor to free-particle mass
    \item [(b)] Rotation matrix $Z_{kj}^{\mathrm{ext}\dag}(k\rightarrow j)$ converting from free-particle mass to EM-mass
\end{itemize}
Guided by \req{basis:1} we write
\begin{align}
\label{zrot:1}
\nu_{j} = W^{\dag}_{\ell j}\nu_{\ell} = Z_{kj}^{\mathrm{ext}\dag}V_{\ell k}^{\dag}\nu_{\ell}\,.
\end{align}
In the limit that the EM fields go to zero, the electromagnetic rotation becomes unity $Z_{kj}^\mathrm{ext}\rightarrow\delta_{kj}$ thereby ensuring the EM-mass basis and free-particle mass basis become equivalent. The rotation $Z_{kj}^\mathrm{ext}$ can then be interpreted as the external field forced rotation. While our argument above is done explicitly for the two generation case, it can be generalized to accommodate three generations of neutrinos as well.

According to \req{diag:1}, the mass matrix in \req{mix:1} can be diagonalized in the two generation case by a one parameter unitary mixing matrix $V_{\ell k}$ given by
\begin{align}
\label{rot:1}
V_{\ell k}(\theta)=
\begin{pmatrix}
\cos\theta & \sin\theta\\
-\sin\theta & \cos\theta
\end{pmatrix}\,.
\end{align}
For a real Hermitian $2\times 2$ mass matrix, the rotation matrix $V_{\ell k}$ is real and only depends on the angle $\theta$. The explicit form of the EM-field related rotation $Z_{kj}^\mathrm{ext}$ introduced in \req{zrot:1} is
\begin{align}
\label{zrot:2}
Z_{kj}^\mathrm{ext}(\omega,\phi)=
\begin{pmatrix}
\cos\omega & e^{i\phi}\sin\omega\\
-e^{-i\phi}\sin\omega & \cos\omega
\end{pmatrix}\,,\qquad
W_{\ell j}(\theta,\omega,\phi)=V_{\ell k}(\theta)Z_{kj}^\mathrm{ext}(\omega,\phi)\,,
\end{align}
where $Z_{kj}^\mathrm{ext}$ depends on the real angle $\omega$ and complex phase $\phi$. The full rotation $W_{\ell j}$ therefore depends on three parameters when broken into free-particle rotation and EM rotation.

The eigenvalues of the original Hermitian mass matrix in \req{mix:1} are given by
\begin{align}
\label{massroot:1}
m_{1,2}=\frac{1}{2}\left(m_{e}^{\nu}+m_{\mu}^{\nu}\mp\sqrt{|\Delta m_{0}|^{2}+4\delta m^{2}}\right)\,,\qquad
|\Delta m_{0}|=|m_{\mu}^{\nu}-m_{\mu}^{e}|\,.
\end{align}
We assign $m_{1}$ to the lower mass $(-)$ root and $m_{2}$ with the larger mass $(+)$ additive root. The rotation angle $\theta$ in \req{rot:1} is then given by
\begin{align}
\label{massroot:2}
\sin2\theta=\sqrt{\frac{4\delta m^{2}}{|\Delta m_{0}|^{2}+4\delta m^{2}}}\,,\qquad
\cos2\theta=\sqrt{\frac{|\Delta m_{0}|^{2}}{|\Delta m_{0}|^{2}+4\delta m^{2}}}\,.
\end{align}

In our toy model, the off-diagonal imaginary transition magnetic moment  $\mu_{\ell\ell'}$ commutes with the real valued mixing matrix $V_{\ell k}$ and the following relations hold
\begin{align}
\label{commuting:1}
V_{\ell k}^{\dag}\mu_{\ell\ell'}V_{\ell' k'}=(V^{\dag}V)_{k\ell'}\mu_{\ell'k'}=\mu_{kk'}=
\begin{pmatrix}
0 & i\delta\mu\\
-i\delta\mu & 0
\end{pmatrix}\,.
\end{align}
We see that the Majorana transition dipoles in our model are off-diagonal in both flavor and mass basis. Therefore the real parameter unitary matrix in \req{commuting:1} cannot rotate a pure imaginary matrix at least in the two generation case. We apply the rotation in \req{rot:1} to \req{herm:1} yielding
\begin{align}
\label{herm:2}
V_{\ell k}^{\dag}(\mathcal{M}_{\ell\ell'}\gamma_{0})V_{\ell' k'} &= 
V_{\ell k}^{\dag}M_{\ell\ell'}\gamma_{0}V_{\ell' k'} +
V_{\ell k}^{\dag}(\mu_{\ell\ell'}\sigma_{\alpha\beta}\gamma_{0}F^{\alpha\beta}_\mathrm{ext}/2)V_{\ell' k'}\,,\\
\label{herm:3}
V_{\ell k}^{\dag}(\mathcal{M}_{\ell\ell'}\gamma_{0})V_{\ell' k'} &= 
\begin{pmatrix}
m_{1}\gamma_{0} & i\delta\mu\sigma_{\alpha\beta}\gamma_{0}F^{\alpha\beta}_\mathrm{ext}/2\\
-i\delta\mu\sigma_{\alpha\beta}\gamma_{0}F^{\alpha\beta}_\mathrm{ext}/2 & m_{2}\gamma_{0}
\end{pmatrix}\equiv
\begin{pmatrix}
\mathcal{A} & i\mathcal{C}\\
-i\mathcal{C} & \mathcal{B}
\end{pmatrix}\,,
\end{align}
where we have defined implicitly the Hermitian elements $(\mathcal{A},\mathcal{B},\mathcal{C})$.  Applying now both rotations 
to \req{herm:1} yields
\begin{align}
    \label{herm:4}
    W_{\ell j}^{\dag}(\mathcal{M}_{\ell\ell'}\gamma_{0})W_{\ell' j'} &= 
    Z^{\mathrm{ext}\dag}
    \begin{pmatrix}
        \mathcal{A} & i\mathcal{C}\\
        -i\mathcal{C} & \mathcal{B}
    \end{pmatrix}Z^\mathrm{ext}=\lambda_{j}\delta_{jj'}\,.
\end{align}
\req{herm:4} is therefore the working matrix equation which needs to be solved to identify the EM rotation parameters. As discussed before, this means that the rotation angle $\omega$ and phase $\phi$ are in general functions of electromagnetic fields.

\subsection{Effective electromagnetic-mass eigenvalues}
\label{sec:emmass}
We will now solve for the rotation parameters necessary to define the EM-mass basis which acts as a distinct propagating basis for neutrinos in external fields. Considering that the $j$-columns vectors $\bb{v}^{(j)}$ of $Z_{kj}^\mathrm{ext}$ as eigenvectors for each eigenvalue $\lambda_{j}$
\begin{align}
    \label{herm:5}
    Z_{kj}^\mathrm{ext}=v_{k}^{(j)}=
    \begin{pmatrix}
        \bb{v}^{1} & \bb{v}^{2}
    \end{pmatrix}\,,
\end{align}
\req{herm:4} has the meaning of an eigenvalue equation
\begin{align}
    \label{herm:6}
    \begin{pmatrix}
        \mathcal{A} & i\mathcal{C}\\
        -i\mathcal{C} & \mathcal{B}
    \end{pmatrix}Z^\mathrm{ext}=
    Z^\mathrm{ext}\begin{pmatrix}
        \lambda_{1} & 0\\
        0 & \lambda_{2}
    \end{pmatrix}\rightarrow
    \begin{pmatrix}
        \mathcal{A} & i\mathcal{C}\\
        -i\mathcal{C} & \mathcal{B}
    \end{pmatrix}\bb{v}^{(j)}=\lambda_{j}\bb{v}^{(j)}\,.
\end{align}
Given the eigenvalue equation defined in \req{herm:6}, the effective EM-masses are then solutions to the characteristic polynomial
\begin{align}
\label{poly:1}
(\mathcal{A}-\lambda_{j}\gamma_{0})(\mathcal{B}-\lambda_{j}\gamma_{0})-\mathcal{C}^{2}=0\,,
\end{align}
which we obtained by taking the determinant of \req{herm:6} over flavor but not spin space. It is useful to define the following identities for the off-diagonal element
\begin{align}
\label{poly:1a}
\mathcal{C}^{2} = 
\delta\mu^{2}\left(\frac{1}{2}\sigma_{\alpha\beta}F^{\alpha\beta}_\mathrm{ext}\right)\left(\frac{1}{2}\sigma_{\alpha\beta}F^{\alpha\beta}_\mathrm{ext}\right)^{\dag}=
2\delta\mu^{2}\gamma_{0}\left(T_\mathrm{ext}^{00}\gamma_{0}+T_\mathrm{ext}^{0i}\gamma_{i}\right)\,,
\end{align}
and for the diagonal elements
\begin{align}
(\mathcal{B}-\mathcal{A})^{2} = |m_{2}-m_{1}|^{2} = |\Delta m|^{2}\,,\qquad (\mathcal{A}+\mathcal{B})\gamma_{0} = m_{1} + m_{2}\,.
\end{align}
\req{poly:1a} was obtained using the expression in \req{cross:2}. Because of the spinor behavior of each element, the eigenvalues are obtained with $\gamma_{0}$ coefficients. \req{poly:1} therefore has the roots $\lambda_{1,2} = \widetilde m_{1,2}(\bb{E},\bb{B})$
\begin{align}
\label{poly:2}
\widetilde m_{1,2}(\bb{E},\bb{B})\! =\! \frac{1}{2}\left(m_{1}\!+\!m_{2}\!\mp\!\sqrt{|\Delta m|^{2}\!+\!8\delta\mu^{2}\gamma_{0}\left(T_\mathrm{ext}^{00}\gamma_{0}+T_\mathrm{ext}^{0i}\gamma_{i}\right)}\right)\!,\\
\label{poly:3}
\boxed{\widetilde m_{1,2}(\bb{E},\bb{B})\! =\! \frac{m_{1}\!+\!m_{2}}{2}\!\mp\!\frac{1}{2}\sqrt{|\Delta m|^{2}\!+\!8\delta\mu^{2}\gamma_{0}\left(\gamma_{0}\frac{1}{2}\left(B^{2}\!+\!\frac{E^{2}}{c^{2}}\right)\!+\!\bb{\gamma}\!\cdot\!(\frac{\bb{E}}{c}\times\bb{B})\right)}}
\end{align}
The EM-mass eigenstates $\widetilde m(\bb{E},\bb{B})$ depends on the energy density $T_\mathrm{ext}^{00}$ of the EM field and the spin projection along the EM momentum density $T_\mathrm{ext}^{0i}$. However the coefficient $\delta\mu^{2}$ is presumed to be very small, therefore the EM contribution only manifests in strong EM fields or where the free-particle case has very nearly or exactly degenerate masses, $\Delta m\to 0$. When the the electromagnetic fields go to zero, the EM-masses in \req{poly:3} reduce as expected to the free-particle result.

The complex phase in \req{zrot:2} has the value $\phi=\pi(n-1/2)$ with $n\in0,\pm1,\pm2...$ making the complex exponential in \req{zrot:2} pure imaginary. Curiously, the phase is not field dependant, but tied to the fact that the Majorana moments are pure imaginary quantities. Complex phases in mixing matrices are generally associated with CP violation such as the Dirac phase $\delta$ in \req{rotation:1} which suggests that CP violation in the neutrino sector can be induced in the presence of external EM fields. Some implications of CP violation from transition moments are discussed in~\cite{Nieves:1981zt}. Analysis of the three generation case is required to show this explicitly, but we postulate that the constant valued complex phases would be replaced with field dependant quantities $\delta\to\delta(\bb{E},\bb{B})$.

We note that the solution in \req{poly:3} actually contain four distinct EM-mass eigenstates $\widetilde m_{j}^{s}(\bb{E},\bb{B})$ with the lower $(j=1)$ and upper $(j=2)$ masses and the additional spin splitting from the alignment $(s=+1)$ or anti-alignment $(s=-1)$ of the neutrino spin with the momentum density of the external EM field. Spin splitting vanishes for the pure electric or magnetic field cases. For good spin eigenstates $s\in\pm1$, we can rewrite \req{poly:1a} with EM fields explicitly as
\begin{align}
\label{spinsplit:1}
\mathcal{C}^{2}_{s}(\bb{E},\bb{B})=2\delta\mu^{2}\left(\frac{1}{2}(B^{2}+E^{2}/c^{2})+s|\bb{E}/c\times\bb{B}|\right)\,.
\end{align}
The above expression within the square is positive definite, therefore \req{spinsplit:1} is always real. Spin splitting requires that we consider separate rotations for each spin state as the rotation angle $\omega_{s}$ depends on the spin quantum number
\begin{align}
\label{zrot:3}
\sin2\omega_{s}=\sqrt{\frac{4\mathcal{C}_{s}^{2}}{|\Delta m|^{2}+4\mathcal{C}_{s}^{2}}}\,,\qquad
\cos2\omega_{s}=\sqrt{\frac{|\Delta m|^{2}}{|\Delta m|^{2}+4\mathcal{C}_{s}^{2}}}\,.
\end{align}
The expressions in \req{zrot:3} are mathematically similar to that of the free-particle case written in \req{massroot:2} in the two flavor generation model with the off-diagonal mass being replaced with the EM dependant quantity $\mathcal{C}_{s}$.

\section{Strong field (degenerate mass) and weak field limits}
\label{sec:nulimits}
The rotation angles in \req{zrot:3} reveal two distinct limits where EM-masses are dominated by either:
\begin{itemize}[nosep]
    \item[(a)] Intrinsic mass splitting $\mathcal{C}_{s}\ll|\Delta m|^{2}$ with $\omega_{s}\rightarrow0$
    \item[(b)] EM mass splitting $\mathcal{C}_{s}\gg|\Delta m|^{2}$ with $\omega_{s}\rightarrow\pi/4$
\end{itemize}
For the first case where the masses are not degenerate or the fields are weak, we obtain the expansion
\begin{align}
\label{series:1}
\lim_{\mathcal{C}_{s}\ll|\Delta m|^{2}}\widetilde m_{1,2}^{s}(E,B)=\frac{1}{2}\left(m_{1}+m_{2}\mp|\Delta m|\left(1+\frac{2\mathcal{C}_{s}^{2}}{|\Delta m|^{2}}+\ldots\right)\right)\,,
\end{align}
which as stated before reduces to the free-particle case at lowest order.

In the opposite limit, where the masses are very nearly degenerate or fields are strong, the EM-mass eigenvalues in \req{poly:3} can be approximated by the series
\begin{align}
\label{series:2}
\lim_{\mathcal{C}_{s}\gg|\Delta m|^{2}}\widetilde m_{1,2}^{s}(E,B)=\frac{1}{2}\left(m_{1}+m_{2}\mp2\mathcal{C}_{s}\left(1+\frac{|\Delta m|^{2}}{8\mathcal{C}_{s}^{2}}+\ldots\right)\right)
\end{align}
For fully degenerate free-particle masses $m_{1}=m_{2}$, this reduces to
\begin{align}
\label{series:2a}
\lim_{|\Delta m|^{2}\to0}\widetilde m_{1,2}^{s}(E,B)=m_{1}\mp\mathcal{C}_{s}\,.
\end{align}
\req{series:2a} indicates that for degenerate free-particle masses, the effective splitting $|\Delta m_\mathrm{EM}|\equiv\mathcal{C}_{s}$ between masses arises purely from the electromagnetic interaction of the neutrinos. We return to this interesting insight in our final comments.

Because of the bounds in \req{bound:1} on the effective neutrino magnetic moment, we can estimate the field strength required for an external magnetic field to generate an electromagnetic mass splitting of $|\Delta m_\mathrm{EM}|=10^{-3}$~eV which is a reasonable comparison to intrinsic splitting based on the experimental limits on neutrino masses. Using the upper limit for the neutrino effective moment of $\mu_{\nu}^\mathrm{eff}\sim10^{-10}\mu_{B}$ we obtain
\begin{align}
\label{estimate:1}
\left.\frac{\mathcal{C}_{s}}{\mu_{\nu}^\mathrm{eff}}\right\rvert_{\vec{E}=0}=\frac{10^{-3}\,\mathrm{eV}}{10^{-10}\mu_{B}}\approx1.7\times10^{11}\,\mathrm{T}\,.
\end{align}
This is near the upper bound of the magnetic field strength of magnetars~\citep{Kaspi:2017fwg} which are of the order $10^{11}$~Tesla. In this situation, the EM contribution to the mass splitting rivals the estimated inherent splitting~\citep{ParticleDataGroup:2022pth} of the two closer in mass neutrinos. Primordial magnetic fields~\citep{Grasso:2000wj} in the Early Universe may also present an environment for significant EM neutrino flavor mixing as both the external field strength and density of neutrinos would be very large~\citep{Rafelski:2023emw}. The magnetic properties of neutrinos may also have contributed alongside the charged leptons in magnetization in the Early Universe~\citep{Steinmetz:2023nsc} prior to recombination. 

While the above estimate was done with astrophysical systems in mind, we note that strong electrical fields should also produce EM-mass splitting. Therefore environments near to high $Z$-nuclei is also of interest~\citep{Bouchiat:1974kt,Bouchiat:1997mj,Safronova:2017xyt} as weak interactions violate parity. Should neutrinos have abnormally large transition magnetic dipole moments, then they should exhibit mass splitting from the neutrino's electromagnetic dipole interaction which may compete with the intrinsic mass differences of the free-particles.

\chapter{Matter-antimatter origin of cosmic magnetism}
\label{chap:cosmo}
\noindent We investigate the hypothesis that the observed intergalactic magnetic fields (IGMF) are primordial in nature, predating the recombination epoch. Specifically, we explore the role of the extremely large electron-positron $(e^{+}e^{-})$ pair abundance in the temperature range of $2000\keV>T>20\keV$ which only disappeared after Big Bang nucleosynthesis (BBN). We review the status of cosmic magnetism in \rsec{sec:universe} which motivates our study. \rsec{sec:abundance} discusses the extreme electron-positron abundance during this epoch. The statistical and thermodynamic theory of the electron-positron gas is described in \rsec{sec:theory}. \rsec{sec:magnetization} describes the relativistic paramagnetism of the electron-positron gas. We propose in \rsec{sec:ferro} a model of self-magnetization caused by spin polarization within the individual species in the gas.

This chapter serves primarily as a review of our work in~\cite{Steinmetz:2023nsc} and portions of~\cite{Rafelski:2023emw} where we propose that the early universe electron-positron plasma was a highly magnetized environment. We will use natural units $(c=\hbar=k_{B}=1)$ unless otherwise noted.

\begin{center}
    \textbf{NOTE: The letter $\mu$ within this chapter will refer exclusively to charged chemical potential and \emph{not} magnetic moment as before.}
\end{center}

\section{Short survey of magnetism in the universe}
\label{sec:universe}
\noindent Macroscopic domains of magnetic fields have been found in all astrophysical environments from compact objects (stars, planets, etc.); interstellar and intergalactic space; and surprisingly in deep extra-galactic void spaces. Considering the ubiquity of magnetic fields in the universe~\citep{Giovannini:2017rbc,Giovannini:2003yn,Kronberg:1993vk}, we search for a common primordial mechanism initiate the diversity of magnetism observed today. In this chapter, IGMF will refer to experimentally observed intergalactic fields of any origin while primordial magnetic fields (PMF) refers to fields generated via early universe processes possibly as far back as inflation. The conventional elaboration of the origins for cosmic PMFs are detailed in~\citep{Gaensler:2004gk,Durrer:2013pga,AlvesBatista:2021sln}.

IGMF are notably difficult to measure and difficult to explain. The bounds for IGMF at a length scale of $1{\rm\ Mpc}$ are today~\citep{Neronov:2010gir,Taylor:2011bn,Pshirkov:2015tua,Jedamzik:2018itu,Vernstrom:2021hru}
\begin{gather}
 \label{igmf}
 10^{-8}{\rm\ G}>B_\mathrm{IGMF}>10^{-16}{\rm\ G}\,.
\end{gather}
We note that generating PMFs with such large coherent length scales is nontrivial~\citep{Giovannini:2022rrl} though currently the length scale for PMFs are not well constrained~\citep{AlvesBatista:2021sln}. Faraday rotation from distant radio active galaxy nuclei (AGN)~\citep{Pomakov:2022cem} suggest that neither dynamo nor astrophysical processes would sufficiently account for the presence of magnetic fields in the universe today if the IGMF strength was around the upper bound of $B_\mathrm{IGMF}\simeq30-60{\rm\ nG}$ as found in~\cite{Vernstrom:2021hru}. Such strong magnetic fields would then require that at least some portion of the IGMF arise from primordial sources that predate the formation of stars.

Magnetized baryon inhomogeneities which in turn would produce anisotropies in the cosmic microwave background (CMB)~\citep{Jedamzik:2013gua,Abdalla:2022yfr}. \cite{Jedamzik:2020krr} propose further that the presence of a magnetic field of $B_\mathrm{PMF}\simeq0.1{\rm\ nG}$ could be sufficient to explain the Hubble tension.

\begin{figure}[ht]
    \centering
    \includegraphics[width=0.95\textwidth]{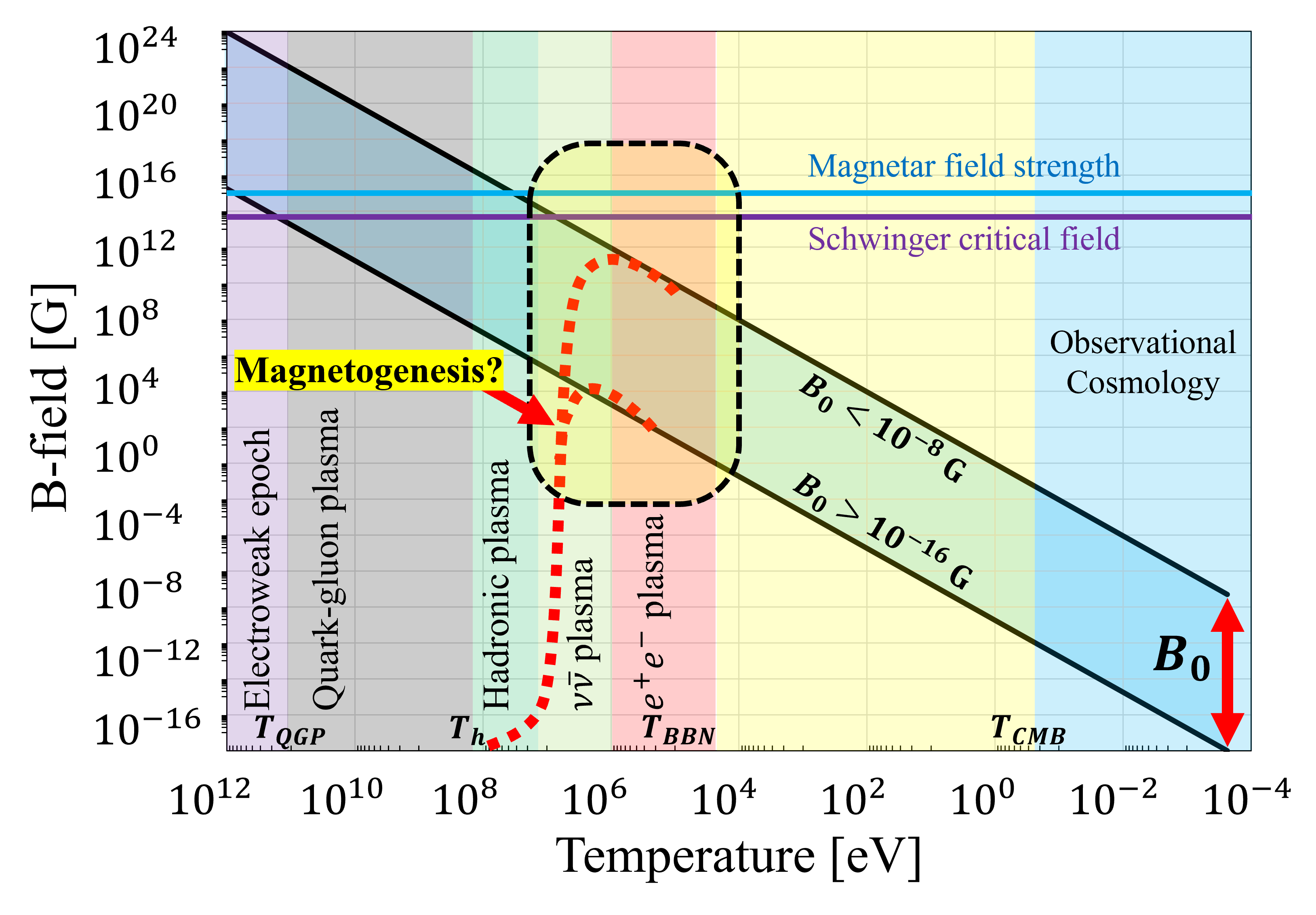}
    \caption{Qualitative plot of the primordial magnetic field strength over cosmic time. All figures are printed in temporal sequence in the expanding universe beginning with high temperatures (and early times) on the left and lower temperatures (and later times) on the right.}
    \label{fig:pmf}
\end{figure}

Our motivating hypothesis is outlined qualitatively in \rf{fig:pmf} where PMF evolution is plotted over the temperature history of the universe. The descending blue band indicates the range of possible PMF strengths. The different epochs of the universe according to $\Lambda\mathrm{CDM}$ are delineated by temperature. The horizontal lines mark two important scales: (a) the Schwinger critical field strength given by
\begin{align}
    \label{crit:1}
    B_\mathrm{C} = \frac{m_{e}^{2}}{e}\simeq4.41\times10^{13}\,\mathrm{G}\,.
\end{align}
where electrodynamics is expected to display nonlinear characteristics and (b) the upper field strength seen in magnetars of $\sim10^{15}\,\mathrm{G}$. A schematic of magnetogenesis is drawn with the dashed red lines indicating spontaneous formation of the PMF within the early universe plasma itself. The $e^{+}e^{-}$ era is notably the final epoch where antimatter exists in large quantities in the cosmos~\citep{Rafelski:2023emw}.

\section{Electron-positron abundance}
\label{sec:abundance}
\noindent As the universe cooled below temperature $T\!=\!m_{e}$ (the electron mass), the thermal electron and positron comoving density depleted by over eight orders of magnitude. At $T_\mathrm{split}=20.3\keV$, the charged lepton asymmetry (mirrored by baryon asymmetry and enforced by charge neutrality) became evident as the surviving excess electrons persisted while positrons vanished entirely from the particle inventory of the universe due to annihilation.

\begin{figure}[ht]
 \centering
\includegraphics[width=0.95\textwidth]{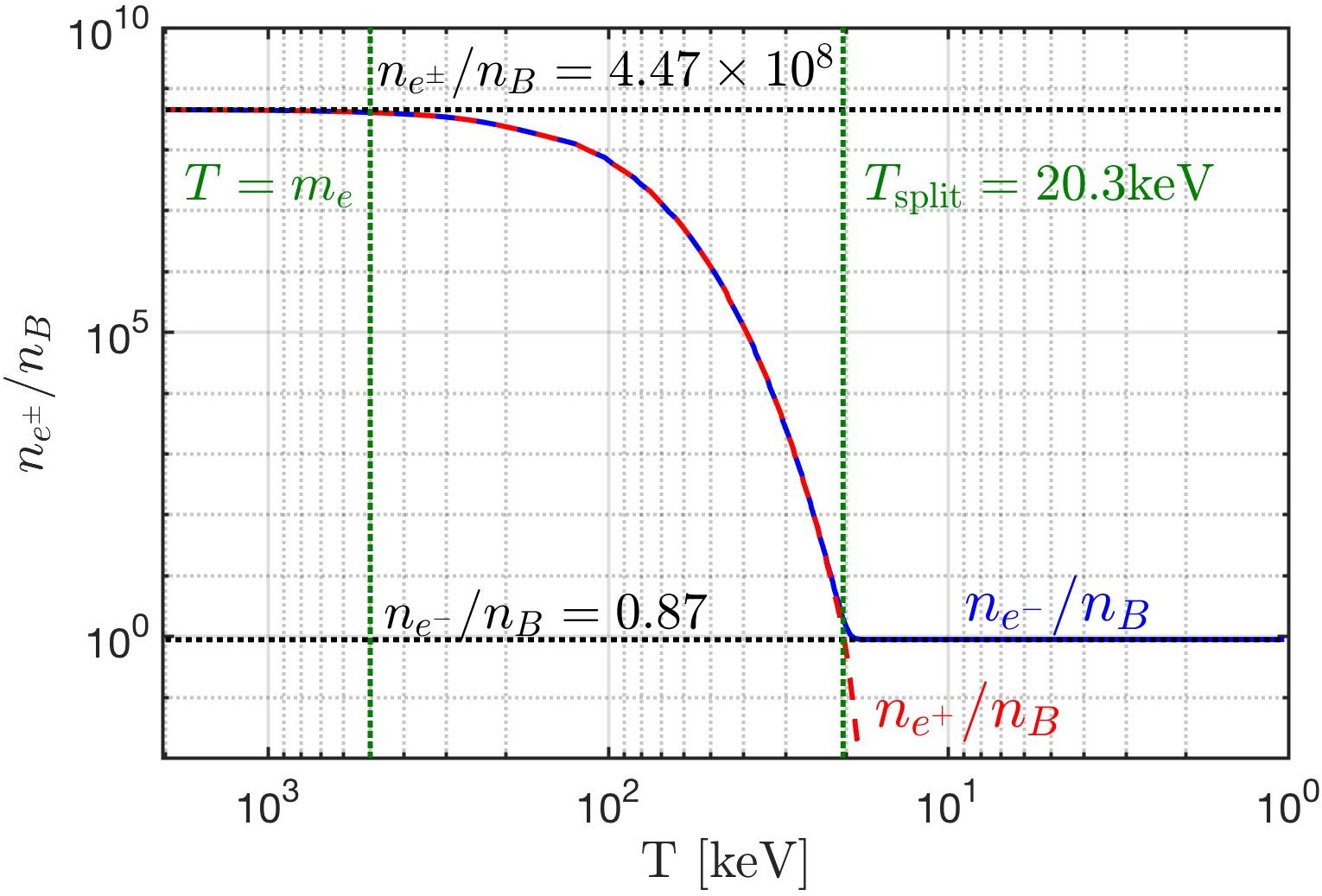}
 \caption{Number density of electron $e^{-}$ and positron $e^{+}$ to baryon ratio $n_{e^{\pm}}/n_{B}$ as a function of photon temperature in the universe. See text for further details. In this work we measure temperature in units of energy (keV) thus we set the Boltzmann constant to $k_{B}=1$. Figure courtesy of Cheng Tao Yang.}
 \label{fig:densityratio} 
\end{figure}

The electron-to-baryon density ratio $n_{e^{-}}/n_{B}$ is shown in \rf{fig:densityratio} as the solid blue line while the positron-to-baryon ratio $n_{e^{+}}/n_{B}$ is represented by the dashed red line. These two lines overlap until the temperature drops below $T_\mathrm{split}=20.3\keV$ as positrons vanish from the universe marking the end of the $e^{+}e^{-}$ plasma and the dominance of the electron-proton $(e^{-}p)$ plasma. The two vertical dashed green lines denote temperatures $T\!=\!m_{e}\simeq511\keV$ and $T_\mathrm{split}=20.3\keV$. These results were obtained using charge neutrality and the baryon-to-photon content (entropy) of the universe; see details in~\cite{Rafelski:2023emw}. The two horizontal black dashed lines denote the relativistic $T\gg m_e$ abundance of $n_{e^{\pm}}/n_{B}=4.47\times10^{8}$ and post-annihilation abundance of $n_{e^{-}}/n_{B}=0.87$. Above temperature $T\simeq85\keV$, the $e^{+}e^{-}$ primordial plasma density exceeded that of the Sun's core density $n_{e}\simeq6\times10^{26}{\rm\ cm}^{-3}$~\citep{Bahcall:2000nu}. 

Conversion of the dense $e^{+}e^{-}$ pair plasma into photons reheated the photon background~\citep{Birrell:2014uka} separating the photon and neutrino temperatures. The $e^{+}e^{-}$ annihilation and photon reheating period lasted no longer than an afternoon lunch break. Because of charge neutrality, the post-annihilation comoving ratio $n_{e^{-}}/n_{B}=0.87$~\citep{Rafelski:2023emw} is slightly offset from unity in~\rf{fig:densityratio} by the presence of bound neutrons in $\alpha$ particles and other neutron containing light elements produced during BBN epoch.

The abundance of baryons is itself fixed by the known abundance relative to photons~\citep{ParticleDataGroup:2022pth} and we employed the contemporary recommended value $n_B/n_\gamma=6.09\times 10^{-10}$. The resulting chemical potential needs to be evaluated carefully to obtain the behavior near to $T_\mathrm{split}=20.3\keV$ where the relatively small value of chemical potential $\mu$ rises rapidly so that positrons vanish from the particle inventory of the universe while nearly one electron per baryon remains. The detailed solution of this problem is found in \cite{Fromerth:2012fe,Rafelski:2023emw} leading to the results shown in \rf{fig:densityratio}.

\section{Theory of thermal matter-antimatter plasmas}
\label{sec:theory}
\noindent To evaluate magnetic properties of the thermal $e^{+}e^{-}$ pair plasma we take inspiration from Ch. 9 of Melrose's treatise on magnetized plasmas~\citep{melrose2008quantum}. We focus on the bulk properties of thermalized plasmas in (near) equilibrium.

We consider a homogeneous magnetic field domain defined along the $z$-axis as
\begin{gather}
    \label{homoB:1}
    \bb{B}=(0,\,0,\,B)\,,
\end{gather}
with magnetic field magnitude $|\bb{B}|=B$. Following \rchap{chap:moment}, we reprint the microscopic energy (\req{lan24b} in different notation) of the charged relativistic fermion within a homogeneous magnetic field given by
\begin{align}
 \label{cosmokgp}
 E^{n}_{\sigma,s}(p_{z},{B})=\sqrt{m_{e}^{2}+p_{z}^{2}+e{B}\left(2n+1+\frac{g}{2}\sigma s\right)}\,,
\end{align}
where $n\in0,1,2,\ldots$ is the Landau orbital quantum number, $p_{z}$ is the momentum parallel to the field axis and the electric charge is $e\equiv q_{e^{+}}=-q_{e^{-}}$. The index $\sigma$ in \req{cosmokgp} differentiates electron $(e^{-};\ \sigma=+1)$ and positron $(e^{+};\ \sigma=-1)$ states. The index $s$ refers to the spin along the field axis: parallel $(\uparrow;\ s=+1)$ or anti-parallel $(\downarrow;\ s=-1)$ for both particle and antiparticle species.

\begin{figure}[ht]
 \centering
 \includegraphics[width=0.95\linewidth]{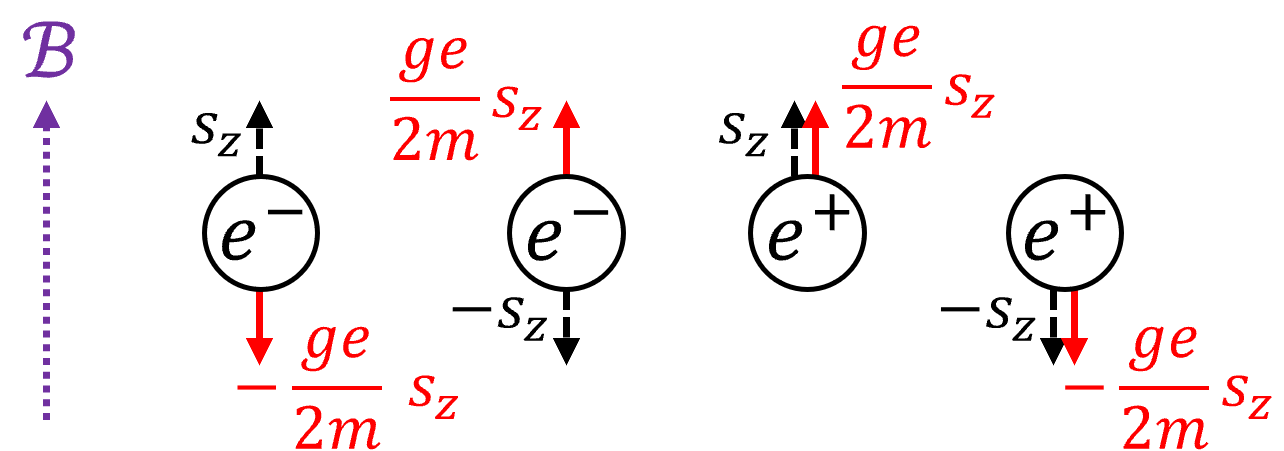}\Bstrut\\
 \begin{tabular}{ r|c|c| }
 \multicolumn{1}{r}{}
 & \multicolumn{1}{c}{aligned: $s=+1$}
 & \multicolumn{1}{c}{anti-aligned: $s=-1$} \\
 \cline{2-3}
 electron: $\sigma=+1$ & $U_{\rm Mag}>0$ & $U_{\rm Mag}<0$ \TBstrut\\
 \cline{2-3}
 positron: $\sigma=-1$ & $U_{\rm Mag}<0$ & $U_{\rm Mag}>0$ \TBstrut\\
 \cline{2-3}
 \end{tabular}\\
 \caption{Organizational schematic of matter-antimatter $(\sigma)$ and polarization $(s)$ states with respect to the sign of the non-relativistic magnetic dipole energy $U_{\rm Mag}$ obtainable from~\req{cosmokgp}.}
 \label{fig:schematic}
\end{figure}

The reason \req{cosmokgp} distinguishes between electrons and positrons is to ensure the correct non-relativistic limit for the magnetic dipole energy is reached. Following the conventions found in \cite{Tiesinga:2021myr}, we set the gyro-magnetic factor $g\equiv g_{e^{+}}=-g_{e^{-}}>0$ such that electrons and positrons have opposite $g$-factors and opposite magnetic moments relative to their spin; see \rf{fig:schematic}.

We recall the conventions established in \rsec{sec:flrw}. Conservation of magnetic flux requires that the magnetic field through a comoving surface $L_{0}^{2}$ remain unchanged. The magnetic field strength under expansion~\citep{Durrer:2013pga} starting at some initial time $t_{0}$ is then given by
\begin{gather}
 \label{bscale}
 B(t)=B_{0}\frac{a^{2}_{0}}{a^{2}(t)}\rightarrow B(z)=B_{0}\left(1+z\right)^{2}\,,
\end{gather}
where $B_{0}$ is the comoving value obtained from the contemporary value of the magnetic field today. Magnetic fields in the cosmos generated through mechanisms such as dynamo or astrophysical sources do not follow this scaling~\citep{Pomakov:2022cem}. It is only in deep intergalactic space where matter density is low are magnetic fields preserved (and thus uncontaminated) over cosmic time.

From \req{tscale} and \req{bscale} there emerges a natural ratio of interest which is conserved over cosmic expansion 
\begin{gather}
 \label{tbscale}
 \boxed{b\equiv\frac{e{B}(t)}{T^{2}(t)}=\frac{e{B}_{0}}{T_{0}^{2}}\equiv b_0={\rm\ const.}}\\
 10^{-3}>b_{0}>10^{-11}\,,
\end{gather}
given in natural units ($c=\hbar=k_{B}=1$). We computed the bounds for this cosmic magnetic scale ratio by using the present day IGMF observations given by \req{igmf} and the present CMB temperature $T_{0}=2.7{\rm\ K}\simeq2.3\times10^{-4}\eV$~\citep{Planck:2018vyg}.

\subsection{Eigenstatess of magnetic moment in cosmology}
\label{sec:protection}

As statistical properties depend on the characteristic Boltzmann factor $E/T$, another interpretation of \req{tbscale} in the context of energy eigenvalues (such as those given in \req{cosmokgp}) is the preservation of magnetic moment energy relative to momentum under adiabatic cosmic expansion. The Boltzmann statistical factor is given by
\begin{alignat}{1}
    \label{Boltz} x\equiv\frac{E}{T}\,.
\end{alignat}
We can explore this relationship for the magnetized system explicitly by writing out \req{Boltz} using the KGP energy eigenvalues written in \req{cosmokgp} as
\begin{alignat}{1}
    \label{XExplicit} x_{\sigma,s}^{n} = \frac{E_{\sigma,s}^{n}}{T} = \sqrt{\frac{m_{e}^{2}}{T^{2}}+\frac{p_{z}^{2}}{T^{2}}+\frac{eB}{T^{2}}\left(2n+1+\frac{g}{2}\sigma s\right)}\,.
\end{alignat}

Introducing the expansion scale factor $a(t)$ via \req{tscale}, \req{bscale} and \req{tbscale}. The Boltzmann factor can then be written as
\begin{alignat}{1}
    \label{xscale:1} x_{\sigma,s}^{n}(a(t)) = \sqrt{\frac{m_{e}^{2}}{T^{2}(t_{0})}\frac{a(t)^{2}}{a_{0}^{2}}+\frac{p_{z,0}^{2}}{T_{0}^{2}}+\frac{eB_{0}}{T_{0}^{2}}\left(2n+1+\frac{g}{2}\sigma s\right)}\,.
\end{alignat}
This reveals that only the mass contribution is dynamic over cosmological time. The constant of motion $b_{0}$ defined in \req{tbscale} is seen as the coefficient to the Landau and spin portion of the energy. For any given eigenstate, the mass term drives the state into the non-relativistic limit while the momenta and magnetic contributions are frozen by initial conditions. 

In comparison, the Boltzmann factor for the DP energy eigenvalues are given by
\begin{alignat}{1}
    \label{xscaledp:1} x_{\sigma,s}^{n}\vert_\mathrm{DP} = \sqrt{\left(\sqrt{\frac{m_{e}^{2}}{T^{2}}+\frac{eB}{T^{2}}\left(2n+1+\sigma s\right)}+\frac{eB}{2m_{e}T}\left(\frac{g}{2}-1\right)\sigma s\right)^{2}+\frac{p_{z}^{2}}{T^{2}}}\,,
\end{alignat}
which scales during FLRW expansion as
\begin{multline}
    \label{xscaledp:2} x_{\sigma,s}^{n}(a(t))\vert_\mathrm{DP} =\\ \sqrt{\left(\sqrt{\frac{m_{e}^{2}}{T_{0}^{2}}\frac{a(t)^{2}}{a_{0}^{2}}+\frac{eB_{0}}{T_{0}^{2}}\left(2n+1+\sigma s\right)}+\frac{eB_{0}}{2m_{e}T_{0}}\frac{a_{0}}{a(t)}\left(\frac{g}{2}-1\right)\sigma s\right)^{2}+\frac{p_{z,0}^{2}}{T_{0}^{2}}}\,.
\end{multline}
While the above expression is rather complicated, we note that the KGP~\req{xscale:1} and DP~\req{xscaledp:1} Boltzmann factors both reduce to the Sch{\"o}dinger-Pauli limit as $a(t)\rightarrow\infty$ thereby demonstrating that the total magnetic moment is protected under the adiabatic expansion of the universe.

As noted in in \rsec{sec:ikgp} and \rsec{sec:emmass}, higher order non-minimal magnetic contributions can be introduced to the Boltzmann factor such as $\sim(e/m)^{2}B^{2}/T^{2}$. The reasoning above suggests that these terms are suppressed over cosmological time driving the system into minimal electromagnetic coupling with the exception of the anomalous magnetic moment. It is interesting to note that cosmological expansion then serves to `smooth out' the characteristics of more complex electrodynamics erasing them from a statistical perspective in favor of minimal-like dynamics.

\subsection{Magnetized fermion partition function}
\label{sec:partition}
\noindent To obtain a quantitative description of the above evolution, we study the bulk properties of the relativistic charged/magnetic gasses in a nearly homogeneous and isotropic primordial universe via the thermal Fermi-Dirac or Bose distributions.

The grand partition function for the relativistic Fermi-Dirac ensemble is given by the standard definition
\begin{alignat}{1}
    \label{part:1} \ln\mathcal{Z}_\mathrm{total}=\sum_{\alpha}\ln\left(1+\Upsilon_{\alpha_{1}\ldots\alpha_{m}}\exp\left(-\frac{E_{\alpha}}{T}\right)\right)\,,\qquad\Upsilon_{\alpha_{1}\ldots\alpha_{m}}=\lambda_{\alpha_{1}}\lambda_{\alpha_{2}}\ldots\lambda_{\alpha_{m}}
\end{alignat}
where we are summing over the set all relevant quantum numbers $\alpha=(\alpha_{1},\alpha_{2},\ldots,\alpha_{m})$. We note here the generalized the fugacity $\Upsilon_{\alpha_{1}\ldots\alpha_{m}}$ allowing for any possible deformation caused by pressures effecting the distribution of any quantum numbers.

In the case of the Landau problem, there is an additional summation over $\widetilde{G}$ which represents the occupancy of Landau states~\citep{greiner2012thermodynamics} which are matched to the available phase space within $\Delta p_{x}\Delta p_{y}$. If we consider the orbital Landau quantum number $n$ to represent the transverse momentum $p_{T}^{2}=p_{x}^{2}+p_{y}^{2}$ of the system, then the relationship that defines $\widetilde{G}$ is given by
\begin{alignat}{1}
    \label{phase:1} \frac{L^{2}}{(2\pi)^{2}}\Delta p_{x}\Delta p_{y}=\frac{eBL^{2}}{2\pi}\Delta n\,,\qquad\widetilde{G}=\frac{eBL^{2}}{2\pi}\,.
\end{alignat}
The summation over the continuous $p_{z}$ is replaced with an integration and the double summation over $p_{x}$ and $p_{y}$ is replaced by a single sum over Landau orbits
\begin{alignat}{1}
    \label{phase:2}
    \sum_{p_{z}}\rightarrow\frac{L}{2\pi}\int^{+\infty}_{-\infty}dp_{z}\,,\qquad\sum_{p_{x}}\sum_{p_{y}}\rightarrow\frac{eBL^{2}}{2\pi}\sum_{n}\,,
\end{alignat}
where $L$ defines the boundary length of our considered volume $V=L^{3}$.

The partition function of the $e^{+}e^{-}$ plasma can be understood as the sum of four gaseous species
\begin{align}
    \label{partition:0}    
    \ln\mathcal{Z}_{e^{+}e^{-}}=\ln\mathcal{Z}_{e^{+}}^{\uparrow}+\ln\mathcal{Z}_{e^{+}}^{\downarrow}+\ln\mathcal{Z}_{e^{-}}^{\uparrow}+\ln\mathcal{Z}_{e^{-}}^{\downarrow}\,,
\end{align}
of electrons and positrons of both polarizations $(\uparrow\downarrow)$. The change in phase space written in \req{phase:2} modify the magnetized $e^{+}e^{-}$ plasma partition function from \req{part:1} into
\begin{gather}
     \label{partition:1}
     \ln\mathcal{Z}_{e^{+}e^{-}}=\frac{e{B}V}{(2\pi)^{2}}\sum_{\sigma}^{\pm1}\sum_{s}^{\pm1}\sum_{n=0}^{\infty}\int_{-\infty}^{\infty}\mathrm{d}p_{z}\left[\ln\left(1+\lambda_{\sigma}\xi_{\sigma,s}\exp\left(-\frac{E_{\sigma,s}^{n}}{T}\right)\right)\right]\,\\
    \label{partition:2}    
    \Upsilon_{\sigma,s} =\lambda_{\sigma}\xi_{\sigma,s} = \exp{\frac{\mu_{\sigma}+\eta_{\sigma,s}}{T}}\,,
\end{gather}
where the energy eigenvalues $E_{\sigma,s}^{n}$ are given in \req{cosmokgp}. The index $\sigma$ in \req{partition:1} is a sum over electron and positron states while $s$ is a sum over polarizations. The index $s$ refers to the spin along the field axis: parallel $(\uparrow;\ s=+1)$ or anti-parallel $(\downarrow;\ s=-1)$ for both particle and antiparticle species.

We are explicitly interested in small asymmetries such as baryon excess over antibaryons, or one polarization over another. These are described by \req{partition:2} as the following two fugacities:
\begin{itemize}[nosep]
 \item[(a)] Chemical fugacity $\lambda_{\sigma}$
 \item[(b)] Polarization fugacity $\xi_{\sigma,s}$
\end{itemize}
For matter $(e^{-};\ \sigma=+1)$ and antimatter $(e^{+};\ \sigma=-1)$ particles, a nonzero relativistic chemical potential $\mu_{\sigma}=\sigma\mu$ is caused by an imbalance of matter and antimatter. While the primordial electron-positron plasma era was overall charge neutral, there was a small asymmetry in the charged leptons (namely electrons) from baryon asymmetry~\citep{Fromerth:2012fe,Canetti:2012zc} in the universe. Reactions such as $e^{+}e^{-}\leftrightarrow\gamma\gamma$ constrains the chemical potential of electrons and positrons~\citep{Elze:1980er} as 
\begin{align}
 \label{cpotential}
 \mu\equiv\mu_{e^{-}}=-\mu_{e^{+}}\,,\qquad
 \lambda\equiv\lambda_{e^{-}}=\lambda_{e^{+}}^{-1}=\exp\frac{\mu}{T}\,,
\end{align}
where $\lambda$ is the chemical fugacity of the system.

We can then parameterize the chemical potential of the $e^{+}e^{-}$ plasma as a function of temperature $\mu\rightarrow\mu(T)$ via the charge neutrality of the universe which implies
\begin{align}
 \label{chargeneutrality}
 n_{p}=n_{e^{-}}-n_{e^{+}}=\frac{1}{V}\lambda\frac{\partial}{\partial\lambda}\ln\mathcal{Z}_{e^{+}e^{-}}\,.
\end{align}
In \req{chargeneutrality}, $n_{p}$ is the observed total number density of protons in all baryon species. The chemical potential defined in \req{cpotential} is obtained from the requirement that the positive charge of baryons (protons, $\alpha$ particles, light nuclei produced after BBN) is exactly and locally compensated by a tiny net excess of electrons over positrons.

We then introduce a novel polarization fugacity $\xi_{\sigma,s}$ and polarization potential $\eta_{\sigma,s}=\sigma s\eta$. We propose the polarization potential follows analogous expressions as seen in \req{cpotential} obeying
\begin{align}
 \label{spotential}
 \eta\equiv\eta_{+,+}=\eta_{-,-}\,,\quad\eta=-\eta_{\pm,\mp}\,,\quad\xi_{\sigma,s}\equiv\exp{\frac{\eta_{\sigma,s}}{T}}\,.
\end{align}
An imbalance in polarization within a region of volume $V$ results in a nonzero polarization potential $\eta\neq0$. Conveniently since antiparticles have opposite signs of charge and magnetic moment, the same magnetic moment is associated with opposite spin orientations. A completely particle-antiparticle symmetric magnetized plasma will have therefore zero total angular momentum.

\subsubsection{Euler-Maclaurin integration}
\label{sec:eulermac}
\noindent Before we proceed with the Boltzmann distribution approximation which makes up the bulk of our analysis, we will comment on the full Fermi-Dirac distribution analysis. The Euler-Maclaurin formula~\citep{abramowitz1988handbook} is used to convert the summation over Landau levels $n$ into an integration given by
\begin{multline}
    \label{eulermaclaurin}\sum^{b}_{n=a}f(n)-\int^{b}_{a}f(x)dx = \frac{1}{2}\left(f(b)+f(a)\right)\\
    +\sum_{i=1}^{j}\frac{b_{2i}}{(2i)!}\left(f^{(2i-1)}(b)-f^{(2i-1)}(a)\right)+R(j)\,,
\end{multline}
where $b_{2i}$ are the Bernoulli numbers and $R(j)$ is the error remainder defined by integrals over Bernoulli polynomials. The integer $j$ is chosen for the level of approximation that is desired. Euler-Maclaurin integration is rarely convergent, and in this case serves only as an approximation within the domain where the error remainder is small and bounded; see~\cite{greiner2012thermodynamics} for the non-relativistic case. In this analysis, we keep the zeroth and first order terms in the Euler-Maclaurin formula. We note that regularization of the excess terms in \req{eulermaclaurin} is done in the context of strong field QED~\citep{greiner2008quantum} though that is outside our scope.

Using \req{eulermaclaurin} allows us to convert the sum over $n$ quantum numbers in \req{partition:1} into an integral. Defining
\begin{alignat}{1}
    \label{Func} f_{\sigma,s}^{n}=\ln\left(1+\Upsilon_{\sigma,s}\exp\left(-\frac{E_{\sigma,s}^{n}}{T}\right)\right)\,,
\end{alignat}
\req{partition:1} for $j=1$ becomes
\begin{multline}
    \label{PartFuncTwo} \ln\mathcal{Z}_{e^{+}e^{-}} = \frac{e{B}V}{(2\pi)^{2}}\sum_{\sigma,s}^{\pm1}\int_{-\infty}^{+\infty}dp_{z}\\
    \left(\int_{0}^{+\infty}dn f_{\sigma,s}^{n} + \frac{1}{2}f_{\sigma,s}^{0} + \frac{1}{12}\frac{\partial f_{\sigma,s}^{n}}{\partial n}\bigg\rvert_{n=0} + R(1)\right)
\end{multline}
It will be useful to rearrange \req{cosmokgp} by pulling the spin dependency and the ground state Landau orbital into the mass writing
\begin{gather}
 \label{effmass:1}
 E^{n}_{\sigma,s}={\tilde m}_{\sigma,s}\sqrt{1+\frac{p_{z}^{2}}{{\tilde m}_{\sigma,s}^{2}}+\frac{2e{B}n}{{\tilde m}_{\sigma,s}^{2}}}\,,\\
 \label{effmass:2}
 \varepsilon_{\sigma,s}^{n}(p_{z},{B})=\frac{E_{\sigma,s}^{n}}{{\tilde m}_{\sigma,s}}\,,\qquad{\tilde m}_{\sigma,s}^{2}=m_{e}^{2}+e{B}\left(1+\frac{g}{2}\sigma s\right)\,,
\end{gather}
where we introduced the dimensionless energy $\varepsilon^{n}_{\sigma,s}$ and effective polarized mass ${\tilde m}_{\sigma,s}$ which is distinct for each spin alignment and is a function of magnetic field strength ${B}$. The effective polarized mass ${\tilde m}_{\sigma,s}$ allows us to describe the $e^{+}e^{-}$ plasma with the spin effects almost wholly separated from the Landau characteristics of the gas when considering the plasma's thermodynamic properties.

With the energies written in this fashion, we recognize the first term in \req{PartFuncTwo} as mathematically equivalent to the free particle fermion partition function with a re-scaled mass $m_{\sigma,s}$. The phase-space relationship between transverse momentum and Landau orbits in \req{phase:1} and \req{phase:2} can be succinctly described by
\begin{gather}
    p_{T}^{2} \sim 2eBn\,,\qquad2p_{T}dp_{T} \sim 2eBdn\,,\qquad d\bb{p}^{3}=2\pi p_{T}dp_{T}dp_{z}\\
    \frac{eBV}{(2\pi)^{2}}\int_{-\infty}^{+\infty}dp_{z}\int_{0}^{+\infty}dn \rightarrow \frac{V}{(2\pi)^{3}}\int d\bb{p}^{3}
\end{gather}
which recasts the first term in \req{PartFuncTwo} as
\begin{align}
    \label{FreePart}
    \ln\mathcal{Z}_{e^{+}e^{-}} = \frac{V}{(2\pi)^{3}}\sum_{\sigma,s}^{\pm1}\int d\bb{p}^{3}\ln\left(1+\Upsilon_{\sigma,s}\exp{\left(-\frac{m_{\sigma,s}\sqrt{1+p^{2}/m_{\sigma,s}^{2}}}{T}\right)}\right)+\ldots
\end{align}
As we will see in the proceeding section, this separation of the `free-like' partition function can be reproduced in the Boltzmann distribution limit as well. This marks the end of the analytic analysis without approximations.

\subsection{Boltzmann approach to electron-positron plasma}
\label{sec:boltzmann}
\noindent Since we address the temperature interval $200\keV>T>20\keV$ where the effects of quantum Fermi statistics on the $e^{+}e^{-}$ pair plasma are relatively small, but the gas is still considered relativistic, we will employ the Boltzmann approximation to the partition function in \req{partition:1}. However, we extrapolate our results for presentation completeness up to $T\simeq 4m_{e}$.

\begin{table}[ht]
 \centering
 \begin{tabular}{ r|c|c| }
 \multicolumn{1}{r}{}
 & \multicolumn{1}{c}{aligned: $s=+1$}
 & \multicolumn{1}{c}{anti-aligned: $s=-1$} \\
 \cline{2-3}
 electron: $\sigma=+1$ & $+\mu+\eta$ & $+\mu-\eta$ \TBstrut\\
 \cline{2-3}
 positron: $\sigma=-1$ & $-\mu-\eta$ & $-\mu+\eta$ \TBstrut\\
 \cline{2-3}
 \end{tabular}\\\,\Bstrut\\
 \caption{Organizational schematic of matter-antimatter $(\sigma)$ and polarization $(s)$ states with respect to the chemical $\mu$ and polarization $\eta$ potentials as seen in~\req{partitionpower:2}. Companion to \rt{fig:schematic}.}
 \label{fig:org}
\end{table}

The partition function shown in equation \req{partition:1} can be rewritten removing the logarithm as
\begin{gather}
\label{partitionpower:1}
\ln{\mathcal{Z}_{e^{+}e^{-}}}=\frac{e{B}V}{(2\pi)^{2}}\sum_{\sigma,s}^{\pm1}\sum_{n=0}^{\infty}\sum_{k=1}^{\infty}\int_{-\infty}^{+\infty}\mathrm{d}p_{z}
\frac{(-1)^{k+1}}{k}\exp\left({k\frac{\sigma\mu+\sigma s\eta-{\tilde m}_{\sigma,s}\varepsilon^{n}_{\sigma,s}}{T}}\right)\,,\\
\label{bapprox} 
\sigma\mu+\sigma s\eta-{\tilde m}_{\sigma,s}\varepsilon_{\sigma,s}^{n}<0\,,
\end{gather}
which is well behaved as long as the factor in \req{bapprox} remains negative. We evaluate the sums over $\sigma$ and $s$ as
\begin{multline}
    \label{partitionpower:2}
    \ln{\mathcal{Z}_{e^{+}e^{-}}}=\frac{e{B}V}{(2\pi)^{2}}\sum_{n=0}^{\infty}\sum_{k=1}^{\infty}\int_{-\infty}^{+\infty}\mathrm{d}p_{z}\frac{(-1)^{k+1}}{k}\times\\
    \left(\ \exp\left(k\frac{+\mu+\eta}{T}\right)\exp\left(-k\frac{{\tilde m}_{+,+}\varepsilon_{+,+}^{n}}{T}\right)\right.
    +\exp\left(k\frac{+\mu-\eta}{T}\right)\exp\left(-k\frac{{\tilde m}_{+,-}\varepsilon_{+,-}^{n}}{T}\right)\qquad\\
    +\exp\left(k\frac{-\mu-\eta}{T}\right)\exp\left(-k\frac{{\tilde m}_{-,+}\varepsilon_{-,+}^{n}}{T}\right)
    +\left.\exp\left(k\frac{-\mu+\eta}{T}\right)\exp\left(-k\frac{{\tilde m}_{-,-}\varepsilon_{-,-}^{n}}{T}\right)\right)
\end{multline}
We note from \rf{fig:schematic} that the first and forth terms and the second and third terms share the same energies via
\begin{align}
    \label{partitionpower:3}
    \varepsilon_{+,+}^{n}=\varepsilon_{-,-}^{n}\,,\qquad
    \varepsilon_{+,-}^{n}=\varepsilon_{-,+}^{n}\,.\qquad
    \varepsilon_{+,-}^{n}<\varepsilon_{+,+}^{n}\,,
\end{align}

\req{partitionpower:3} allows us to reorganize the partition function with a new magnetization quantum number $s'$ which characterizes paramagnetic flux increasing states $(s'=+1)$ and diamagnetic flux decreasing states $(s'=-1)$. This recasts \req{partitionpower:2} as
\begin{multline}
    \label{partitionpower:4}
    \ln{\mathcal{Z}_{e^{+}e^{-}}}=\frac{e{B}V}{(2\pi)^{2}}\sum_{s'}^{\pm1}\sum_{n=0}^{\infty}\sum_{k=1}^{\infty}\int_{-\infty}^{+\infty}\mathrm{d}p_{z}\frac{(-1)^{k+1}}{k}\\
    \left[2\xi_{s'}\cosh\frac{k\mu}{T}\right]\exp\left(-k\frac{{\tilde m}_{s'}\varepsilon_{s'}^{n}}{T}\right)
\end{multline}
with dimensionless energy $\varepsilon_{s'}^{n}$, polarization mass $\tilde{m}_{s'}$, and polarization $\eta_{s'}$ redefined in terms of the moment orientation quantum number $s'$
\begin{gather}
    {\tilde m}_{s'}^{2}=m_{e}^{2}+e{B}\left(1-\frac{g}{2}s'\right)\,,\\
    \eta\equiv\eta_{+}=-\eta_{-}\qquad\xi\equiv\xi_{+}=\xi_{-}^{-1}\,,\qquad\xi_{s'}=\xi^{\pm1}=\exp\left(\pm\frac{\eta}{T}\right)\,.
\end{gather}

We introduce the modified Bessel function $K_{\nu}$ (see Ch. 10 of~\cite{Letessier:2002ony}) of the second kind
\begin{gather}
\label{besselk}
K_{\nu}\left(\frac{m}{T}\right)=\frac{\sqrt{\pi}}{\Gamma(\nu-1/2)}\frac{1}{m}\left(\frac{1}{2mT}\right)^{\nu-1}
\int_{0}^{\infty}\mathrm{d}p\,p^{2\nu-2}\exp\left({-\frac{m\varepsilon}{T}}\right)\,,\\
\nu>1/2\,,\qquad\varepsilon=\sqrt{1+p^{2}/m^{2}}\,,
\end{gather}
allowing us to rewrite the integral over momentum in \req{partitionpower:4} as
\begin{align}
 \label{besselkint}
 \frac{1}{T}\int_{0}^{\infty}\!\!\mathrm{d}p_{z}\exp\!\left(\!{-\frac{k{\tilde m}_{s'}\varepsilon_{s'}^{n}}{T}}\!\right)\!=\!W_{1}\!\!\left(\frac{k{\tilde m}_{s'}\varepsilon_{s'}^{n}(0,{B})}{T}\right)\,.
\end{align}
The function $W_{\nu}$ serves as an auxiliary function of the form $W_{\nu}(x)=xK_{\nu}(x)$. The notation $\varepsilon(0,{B})$ in \req{besselkint} refers to the definition of dimensionless energy found in \req{effmass:2} with $p_{z}=0$. The standard Boltzmann distribution is obtained by summing only $k=1$ and neglecting the higher order terms.

We take advantage again of Euler-Maclaurin integration \req{eulermaclaurin} and integrate the partition function. After truncation of the series and error remainder, the partition function \req{partitionpower:1} can then be written in terms of modified Bessel $K_{\nu}$ functions of the second kind and cosmic magnetic scale $b_{0}$, yielding
\begin{gather}
    \label{boltzmann}
    \boxed{\ln\mathcal{Z}_{e^{+}e^{-}}\simeq\frac{T^{3}V}{\pi^{2}}\sum_{s'}^{\pm1}\left[\xi_{s'}\cosh{\frac{\mu}{T}}\right]
    \left(x_{s'}^{2}K_{2}(x_{s'})+\frac{b_{0}}{2}x_{s'}K_{1}(x_{s'})+\frac{b_{0}^{2}}{12}K_{0}(x_{s'})\right)}\,,\\
    \label{xfunc}
    x_{s'}=\frac{{\tilde m}_{s'}}{T}=\sqrt{\frac{m_{e}^{2}}{T^{2}}+b_{0}\left(1-\frac{g}{2}s'\right)}\,.
\end{gather}
The latter two terms in \req{boltzmann} proportional to $b_{0}K_{1}$ and $b_{0}^{2}K_{0}$ are the uniquely magnetic terms present in powers of magnetic scale \req{tbscale} containing both spin and Landau orbital influences in the partition function. The $K_{2}$ term is analogous to the free Fermi gas~\citep{greiner2012thermodynamics} being modified only by spin effects.

This `separation of concerns' can be rewritten as
\begin{gather}
    \label{spin}
    \ln\mathcal{Z}_\mathrm{S}=\frac{T^{3}V}{\pi^{2}}\sum_{s'}^{\pm1}\left[\xi_{s'}\cosh{\frac{\mu}{T}}\right]\left(x_{s'}^{2}K_{2}(x_{s'})\right)\,,\\
    \label{spinorbit}
    \ln\mathcal{Z}_\mathrm{SO}=\frac{T^{3}V}{\pi^{2}}\sum_{s'}^{\pm}\left[\xi_{s'}\cosh{\frac{\mu}{T}}\right]
    \left(\frac{b_{0}}{2}x_{s'}K_{1}(x_{s'})+\frac{b_{0}^{2}}{12}K_{0}(x_{s'})\right)\,,        
\end{gather}

where the spin (S) and spin-orbit (SO) partition functions can be considered independently. When the magnetic scale $b_{0}$ is small, the spin-orbit term \req{spinorbit} becomes negligible leaving only paramagnetic effects in \req{spin} due to spin. In the non-relativistic limit, \req{spin} reproduces a quantum gas whose Hamiltonian is defined as the free particle (FP) Hamiltonian plus the magnetic dipole (MD) Hamiltonian which span two independent Hilbert spaces $\mathcal{H}_\mathrm{FP}\otimes\mathcal{H}_\mathrm{MD}$. The non-relativistic limit is further discussed in \rsec{sec:nrboltz}.

Writing the partition function as \req{boltzmann} instead of \req{partitionpower:1} has the additional benefit that the partition function remains finite in the free gas $({B}\rightarrow0)$ limit. This is because the free Fermi gas and \req{spin} are mathematically analogous to one another. As the Bessel $K_{\nu}$ functions are evaluated as functions of $x_{\pm}$ in \req{xfunc}, the `free' part of the partition $K_{2}$ is still subject to spin magnetization effects. In the limit where ${B}\rightarrow0$, the free Fermi gas is recovered in both the Boltzmann approximation $k=1$ and the general case $\sum_{k=1}^{\infty}$.

\subsection{Non-relativistic limit of the magnetized partition function}
\label{sec:nrboltz}
While we label the first term in \req{FreePart} as the `free' partition function, this is not strictly true as the partition function dependant on the magnetic-mass we defined in \req{effmass:2}. When determining the magnetization of the quantum Fermi gas, derivatives of the magnetic field $B$ will not fully vanish on this first term which will resulting in an intrinsic magnetization which is distinct from the Landau levels.

This represents magnetization that arises from the spin magnetic energy rather than orbital contributions. To demonstrate this, we will briefly consider the weak field limit for $g=2$. The effective polarized mass for electrons is then
\begin{align}
  \label{MagMassPlus}
  \tilde{m}_{+}^{2}&=m_{e}^{2}\,,\\
  \label{MagMassMinus}
  \tilde{m}_{-}^{2}&=m_{e}^{2}+2eB\,,
\end{align}
with energy eigenvalues
\begin{align}
  \label{EPlus}
  E_{n}^{+}&=\sqrt{p_{z}^{2}+m_{e}^{2}+2eBn}\,,\\
  \label{EMinus}
  E_{n}^{-}&=\sqrt{\left(E_{n}^{+}\right)^{2}+2eB}\,.
\end{align}
The spin anti-aligned states in the non-relativistic (NR) limit reduce to
\begin{align}
  \label{EMinusNR} E_{n}^{-}\vert_\mathrm{NR}\approx E_{n}^{+}\vert_\mathrm{NR}+\frac{eB}{m_{e}}\,.
\end{align}
This shift in energies is otherwise not influenced by summation over Landau quantum number $n$, therefore we can interpret this energy shift as a shift in the polarization potential from \req{spotential}. The polarization potential is then
\begin{align}
  \label{SpinChem} \eta_{e}^{\pm}=\eta_{e}\pm\frac{eB}{2m_{e}}\,,
\end{align}
allowing us to rewrite the partition function in \req{partitionpower:1} as
\begin{gather}
  \label{PartTotalNR} \ln\mathcal{Z}_{e^{-}}\vert_{NR}=\frac{eBV}{(2\pi)^{2}}\sum_{s'}^{\pm}\sum_{n=0}^{\infty}\sum_{k=1}^{\infty}\int_{-\infty}^{+\infty}dp_{z}\frac{(-1)^{k+1}}{k}2\cosh(k\beta\eta_{e}^{s'})\lambda^{k}\exp(-k\epsilon_{n}/T)\,,\\
  \label{NREnergy} \epsilon_{n}=m_{e}+\frac{p_{z}^{2}}{2m_{e}}+\frac{eB}{2m_{e}}\left(n+1\right)\,.
\end{gather}

\req{PartTotalNR} is then the traditional NR quantum harmonic oscillator partition function with a spin dependant potential shift differentiating the aligned and anti-aligned states. We note that in this formulation, the spin contribution is entirely excised from the orbital contribution. Under Euler-Maclaurin integration, the now spin-independant Boltzmann factor can be further separated into `free' and Landau quantum parts as was done in \req{FreePart} for the relativistic case. We note however that the inclusion of anomalous magnetic moment spoils this clean separation.

\subsection{Electron-positron chemical potential}
\label{sec:chem}
\noindent In presence of a magnetic field in the Boltzmann approximation, the charge neutrality condition \req{chargeneutrality} becomes
\begin{gather}
 \label{chem}
 \sinh\frac{\mu}{T}=n_{p}\frac{\pi^{2}}{T^{3}}
 \left[\sum_{s'}^{\pm1}\xi_{s'}\!\left(\!x_{s'}^{2}K_{2}(x_{s'})\!+\!\frac{b_{0}}{2}x_{s'}K_{1}(x_{s'})\!+\!\frac{b_{0}^{2}}{12}K_{0}(x_{s'}\!)\!\right)\!\right]^{-1}\!.
\end{gather}
\req{chem} is fully determined by the right-hand-side expression if the spin fugacity is set to unity $\eta=0$ implying no external bias to the number of polarizations except as a consequence of the difference in energy eigenvalues. In practice, the latter two terms in \req{chem} are negligible to chemical potential in the bounds of the primordial $e^{+}e^{-}$ plasma considered and only becomes relevant for extreme (see \rf{fig:chemicalpotential}) magnetic field strengths well outside our scope.

\begin{figure}[ht]
 \centering
 \includegraphics[clip, trim=0.0cm 0.0cm 0.0cm 0.0cm,width=0.95\linewidth]{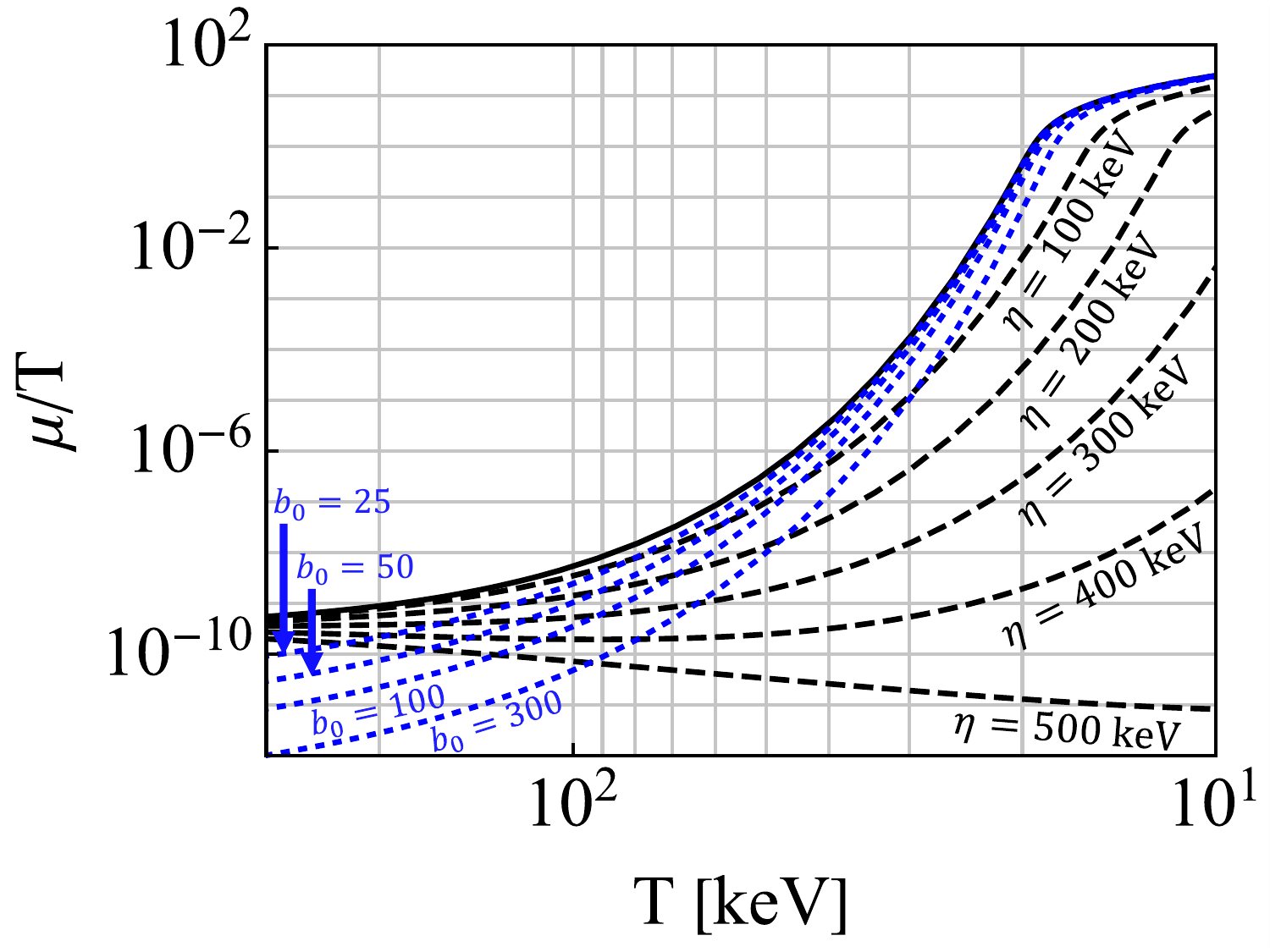}
 \caption{The chemical potential over temperature $\mu/T$ is plotted as a function of temperature with differing values of spin potential $\eta$ and magnetic scale $b_{0}$.}
 \label{fig:chemicalpotential}
\end{figure}

\req{chem} simplifies if there is no external magnetic field $b_{0}=0$ into
\begin{align}
    \label{simpchem:1}
    \sinh\frac{\mu}{T}=n_{p}\frac{\pi^{2}}{T^{3}}\left[2\cosh\frac{\eta}{T}\left(\frac{m_{e}}{T}\right)^{2}K_{2}\left(\frac{m_{e}}{T}\right)\right]^{-1}\,.
\end{align}

In \rf{fig:chemicalpotential} we plot the chemical potential $\mu/T$ in \req{chem} and \req{simpchem:1} which characterizes the importance of the charged lepton asymmetry as a function of temperature. Since the baryon (and thus charged lepton) asymmetry remains fixed, the suppression of $\mu/T$ at high temperatures indicates a large pair density which is seen explicitly in \rf{fig:densityratio}. The black line corresponds to the $b_{0}=0$ and $\eta=0$ case. 

The para-diamagnetic contribution from \req{spinorbit} does not appreciably influence $\mu/T$ until the magnetic scales involved become incredibly large well outside the observational bounds defined in \req{igmf} and \req{tbscale} as seen by the dotted blue curves of various large values $b_{0}=\{25,\ 50,\ 100,\ 300\}$. The chemical potential is also insensitive to forcing by the spin potential until $\eta$ reaches a significant fraction of the electron mass $m_{e}$ in size. The chemical potential for large values of spin potential $\eta=\{100,\ 200,\ 300,\ 400,\ 500\}\,\keV$ are also plotted as dashed black lines with $b_{0}=0$.

It is interesting to note that there are crossing points where a given chemical potential can be described as either an imbalance in spin-polarization or presence of external magnetic field. While spin potential suppresses the chemical potential at low temperatures, external magnetic fields only suppress the chemical potential at high temperatures.

The profound insensitivity of the chemical potential to these parameters justifies the use of the free particle chemical potential (black line) in the ranges of magnetic field strength considered for cosmology. Mathematically this can be understood as $\xi$ and $b_{0}$ act as small corrections in the denominator of \req{chem} if expanded in powers of these two parameters.

\section{Relativistic paramagnetism of electron-positron gas}
\label{sec:magnetization}
\noindent The total magnetic flux within a region of space can be written as the sum of external fields and the magnetization of the medium via
\begin{align}
 \label{totalmag}
 {B}_\mathrm{total} = {B} + \mathcal{M}\,.
\end{align}
For the simplest mediums without ferromagnetic or hysteresis considerations, the relationship can be parameterized by the susceptibility $\chi$ of the medium as
\begin{align}
 \label{susceptibility}
 {B}_\mathrm{total} = (1+\chi){B}\,,\qquad \mathcal{M} = \chi{B}\,,\qquad \chi\equiv\frac{\partial\mathcal{M}}{\partial{B}}\,,
\end{align}
with the possibility of both paramagnetic materials $(\chi>1)$ and diamagnetic materials $(\chi<1)$. The $e^{+}e^{-}$ plasma however does not so neatly fit in either category as given by \req{spin} and \req{spinorbit}. In general, the susceptibility of the gas will itself be a field dependant quantity.

In our analysis, the external magnetic field always appears within the context of the magnetic scale $b_{0}$, therefore we can introduce the change of variables
\begin{align}
 \frac{\partial b_{0}}{\partial{B}}=\frac{e}{T^{2}}\,.
\end{align}
The magnetization of the $e^{+}e^{-}$ plasma described by the partition function in \req{boltzmann} can then be written as
\begin{align}
 \label{defmagetization}
 \mathcal{M}\equiv\frac{T}{V}\frac{\partial}{\partial{B}}\ln{\mathcal{Z}_{e^{+}e^{-}}} = \frac{T}{V}\left(\frac{\partial b_{0}}{\partial{B}}\right)\frac{\partial}{\partial b_{0}}\ln{\mathcal{Z}_{e^{+}e^{-}}}\,,
\end{align}
Magnetization arising from other components in the cosmic gas (protons, neutrinos, etc.) could in principle also be included. Localized inhomogeneities of matter evolution are often non-trivial and generally be solved numerically using magneto-hydrodynamics (MHD)~\citep{melrose2008quantum,Vazza:2017qge,Vachaspati:2020blt} or with a suitable Boltzmann-Vlasov transport equation. An extension of our work would be to embed magnetization into transport theory~\citep{Formanek:2021blc}. In the context of MHD, primordial magnetogenesis from fluid flows in the electron-positron epoch was considered in~\cite{Gopal:2004ut,Perrone:2021srr}.

We introduce dimensionless units for magnetization ${\mathfrak M}$ by defining the critical field strength
\begin{align}
 {B}_{C}\equiv\frac{m_{e}^{2}}{e}\,,\qquad{\mathfrak M}\equiv\frac{\mathcal{M}}{{B}_{C}}\,.
\end{align}
The scale ${B}_{C}$ is where electromagnetism is expected to become subject to non-linear effects, though luckily in our regime of interest, electrodynamics should be linear. We note however that the upper bounds of IGMFs in \req{igmf} (with $b_{0}=10^{-3}$; see \req{tbscale}) brings us to within $1\%$ of that limit for the external field strength in the temperature range considered.

The total magnetization ${\mathfrak M}$ can be broken into the sum of magnetic moment parallel ${\mathfrak M}_{+}$ and magnetic moment anti-parallel ${\mathfrak M}_{-}$ contributions
\begin{align}
\label{g2mag}
{\mathfrak M}={\mathfrak M}_{+}+{\mathfrak M}_{-}\,.
\end{align}
We note that the expression for the magnetization simplifies significantly for $g\!=\!2$ which is the `natural' gyro-magnetic factor~\citep{Evans:2022fsu,Rafelski:2022bsv} for Dirac particles. For illustration, the $g\!=\!2$ magnetization from \req{defmagetization} is then
\begin{align}
 \label{g2magplus}
 {\mathfrak M}_{+}&=\frac{e^{2}}{\pi^{2}}\frac{T^{2}}{m_{e}^{2}}\xi\cosh{\frac{\mu}{T}}\left[\frac{1}{2}x_{+}K_{1}(x_{+})+\frac{b_{0}}{6}K_{0}(x_{+})\right]\,,\\
 \label{g2magminus}
 -{\mathfrak M}_{-}&=\frac{e^{2}}{\pi^{2}}\frac{T^{2}}{m_{e}^{2}}\xi^{-1}\cosh{\frac{\mu}{T}}
 \left[\left(\frac{1}{2}+\frac{b_{0}^{2}}{12x_{-}^{2}}\right)x_{-}K_{1}(x_{-})+\frac{b_{0}}{3}K_{0}(x_{-})\right]\,,\\
 x_{+}&=\frac{m_{e}}{T}\,,\qquad
 x_{-}=\sqrt{\frac{m_{e}^{2}}{T^{2}}+2b_{0}}\,.
\end{align}
As the $g$-factor of the electron is only slightly above two at $g\simeq2.00232$~\citep{Tiesinga:2021myr}, the above two expressions for ${\mathfrak M}_{+}$ and ${\mathfrak M}_{-}$ are only modified by a small amount because of anomalous magnetic moment (AMM) and would be otherwise invisible on our figures.

\subsection{Evolution of electron-positron magnetization}
\label{sec:paramagnetism}
\noindent In \rf{fig:magnet}, we plot the magnetization as given by \req{g2magplus} and \req{g2magminus} with the spin potential set to unity $\xi=1$. The lower (solid red) and upper (solid blue) bounds for cosmic magnetic scale $b_{0}$ are included. The external magnetic field strength ${B}/{B}_{C}$ is also plotted for lower (dotted red) and upper (dotted blue) bounds. Since the derivative of the partition function governing magnetization may manifest differences between Fermi-Dirac and the here used Boltzmann limit more acutely, out of abundance of caution, we indicate extrapolation outside the domain of validity of the Boltzmann limit with dashes.

\begin{figure}[ht]
 \centering
 \includegraphics[width=0.95\linewidth]{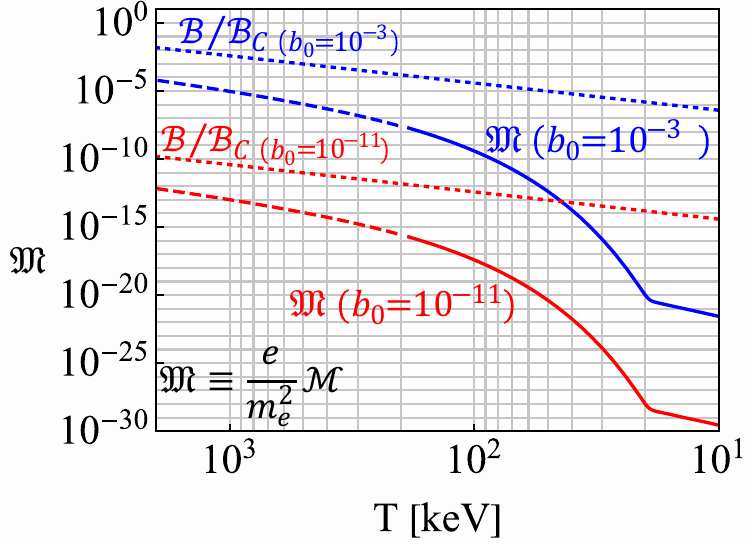}
 \caption{The magnetization ${\mathfrak M}$, with $g\!=\!2$, of the primordial $e^{+}e^{-}$ plasma is plotted as a function of temperature. Figure made in collaboration with Cheng Tao Yang.}
 \label{fig:magnet} 
\end{figure}

We see in \rf{fig:magnet} that the $e^{+}e^{-}$ plasma is overall paramagnetic and yields a positive overall magnetization which is contrary to the traditional assumption that matter-antimatter plasma lack significant magnetic responses of their own in the bulk. With that said, the magnetization never exceeds the external field under the parameters considered which shows a lack of ferromagnetic behavior. 

The large abundance of pairs causes the smallness of the chemical potential seen in~\rf{fig:chemicalpotential} at high temperatures. As the universe expands and temperature decreases, there is a rapid decrease of the density $n_{e^{\pm}}$ of $e^{+}e^{-}$ pairs. This is the primary the cause of the rapid paramagnetic decrease seen in \rf{fig:magnet} above $T\!=\!21\keV$. At lower temperatures $T<21\keV$ there remains a small electron excess (see~\rf{fig:densityratio}) needed to neutralize proton charge. These excess electrons then govern the residual magnetization and dilutes with cosmic expansion.

An interesting feature of \rf{fig:magnet} is that the magnetization in the full temperature range increases as a function of temperature. This is contrary to Curie's law~\citep{greiner2012thermodynamics} which stipulates that paramagnetic susceptibility of a laboratory material is inversely proportional to temperature. However, Curie's law applies to systems with fixed number of particles which is not true in our situation; see \rsec{sec:perlepton}.

A further consideration is possible hysteresis as the $e^{+}e^{-}$ density drops with temperature. It is not immediately obvious the gas's magnetization should simply `degauss' so rapidly without further consequence. If the very large paramagnetic susceptibility present for $T\simeq m_{e}$ is the origin of an overall magnetization of the plasma, the conservation of magnetic flux through the comoving surface ensures that the initial residual magnetization is preserved at a lower temperature by Faraday induced kinetic flow processes however our model presented here cannot account for such effects.

Early universe conditions may also apply to some extreme stellar objects with rapid change in $n_{e^{\pm}}$ with temperatures above $T\!=\!21\keV$. Production and annihilation of $e^{+}e^{-}$ plasmas is also predicted around compact stellar objects~\citep{Ruffini:2009hg,Ruffini:2012it} potentially as a source of gamma-ray bursts.

\subsection{Dependency on g-factor}
\label{sec:gfac}

\noindent As discussed at the end of \rsec{sec:magnetization}, the AMM of $e^{+}e^{-}$ is not relevant in the present model. However out of academic interest, it is valuable to consider how magnetization is effected by changing the $g$-factor significantly.

\begin{figure}[ht]
 \centering
 \includegraphics[width=0.95\textwidth]{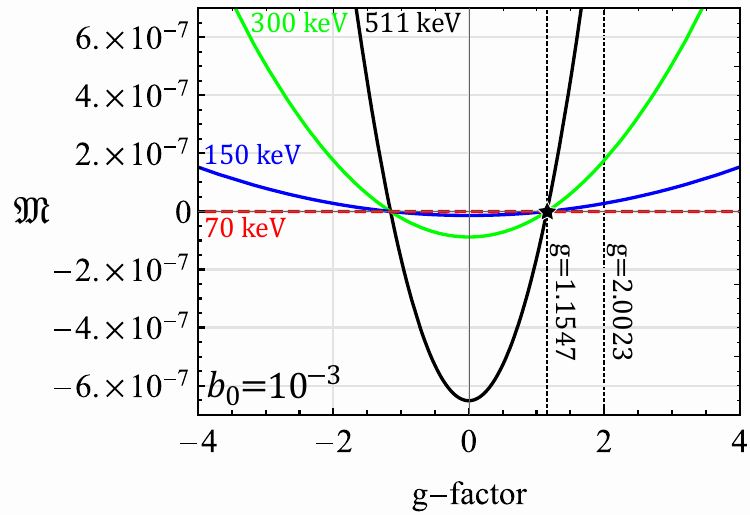}
 \caption{The magnetization $\mathfrak M$ as a function of $g$-factor plotted for several temperatures with magnetic scale $b_{0}=10^{-3}$ and polarization fugacity $\xi=1$.}
 \label{fig:gfac} 
\end{figure}

The influence of AMM would be more relevant for the magnetization of baryon gasses since the $g$-factor for protons $(g\approx5.6)$ and neutrons $(g\approx3.8)$ are substantially different from $g\!=\!2$. The influence of AMM on the magnetization of thermal systems with large baryon content (neutron stars, magnetars, hypothetical bose stars, etc.) is therefore also of interest~\citep{Ferrer:2019xlr,Ferrer:2023pgq}.

\req{g2magplus} and \req{g2magminus} with arbitrary $g$ reintroduced is given by
\begin{gather}
\label{arbg:1}
{\mathfrak M}=\frac{e^{2}}{\pi^{2}}\frac{T^{2}}{m_{e}^{2}}\sum_{s'}^{\pm1}\xi_{s'}\cosh{\frac{\mu}{T}}
\left[C^{1}_{s'}(x_{s'})K_{1}(x_{s'})+C^{0}_{s'}K_{0}(x_{s'})\right]\,,\\
\label{arbg:2}
C^{1}_{s'}(x_{\pm}) = \left[\frac{1}{2}-\left(\frac{1}{2}-\frac{g}{4}s'\right)\left(1+\frac{b^2_0}{12x^{2}_{s'}}\right)\right]x_{s'}\,,\qquad
C^{0}_{s'} = \left[\frac{1}{6}-\left(\frac{1}{4}-\frac{g}{8}s'\right)\right]b_0\,,
\end{gather}
where $x_{s'}$ was previously defined in \req{xfunc}.

In \rf{fig:gfac}, we plot the magnetization as a function of $g$-factor between $4>g>-4$ for temperatures $T\!=\!\{511,\ 300,\ 150,\ 70\}\keV$. We find that the magnetization is sensitive to the value of AMM revealing a transition point between paramagnetic $({\mathfrak M}>0)$ and diamagnetic gasses $({\mathfrak M}<0)$. Curiously, the transition point was numerically determined to be around $g\simeq1.1547$ in the limit $b_{0}\rightarrow0$. The exact position of this transition point however was found to be both temperature and $b_{0}$ sensitive, though it moved little in the ranges considered.

It is not surprising for there to be a transition between diamagnetism and paramagnetism given that the partition function (see \req{spin} and \req{spinorbit}) contained elements of both. With that said, the transition point presented at $g\approx1.15$ should not be taken as exact because of the approximations used to obtain the above results. 

It is likely that the exact transition point has been altered by our taking of the Boltzmann approximation and Euler-Maclaurin integration steps. It is known that the Klein-Gordon-Pauli solutions to the Landau problem in \req{cosmokgp} have periodic behavior~\citep{Steinmetz:2018ryf,Evans:2022fsu,Rafelski:2022bsv} for $|g|=k/2$ (where $k\in1,2,3\ldots$).

These integer and half-integer points represent when the two Landau towers of orbital levels match up exactly. Therefore, we propose a more natural transition between the spinless diamagnetic gas of $g=0$ and a paramagnetic gas is $g=1$. A more careful analysis is required to confirm this, but that our numerical value is close to unity is suggestive.

\subsection{Magnetization per lepton}
\label{sec:perlepton}
\noindent Despite the relatively large magnetization seen in \rf{fig:magnet}, the average contribution per lepton is only a small fraction of its overall magnetic moment indicating the magnetization is only loosely organized. Specifically, the magnetization regime we are in is described by
\begin{align}
 \label{fractionalmagnetization}
 \mathcal{M}\ll\mu_{B}\frac{N_{e^{+}}+N_{e^{-}}}{V}\,,\qquad\mu_{B}\equiv\frac{e}{2m_{e}}\,,
\end{align}
where $\mu_{B}$ is the Bohr magneton and $N=nV$ is the total particle number in the proper volume V. To better demonstrate that the plasma is only weakly magnetized, we define the average magnetic moment per lepton given by along the field ($z$-direction) axis as
\begin{align}
 \label{momentperlepton}
 \vert\vec{m}\vert_{z}\equiv\frac{\mathcal{M}}{n_{e^{-}}+n_{e^{+}}}\,,\qquad\vert\vec{m}\vert_{x}=\vert\vec{m}\vert_{y}=0\,.
\end{align}
Statistically, we expect the transverse expectation values to be zero. We emphasize here that despite $|\vec{m}|_{z}$ being nonzero, this doesn't indicate a nonzero spin angular momentum as our plasma is nearly matter-antimatter symmetric. The quantity defined in \req{momentperlepton} gives us an insight into the microscopic response of the plasma.

\begin{figure}[ht]
 \centering
 \includegraphics[clip, trim=0.0cm 0.0cm 0.0cm 0.0cm,width=0.95\textwidth]{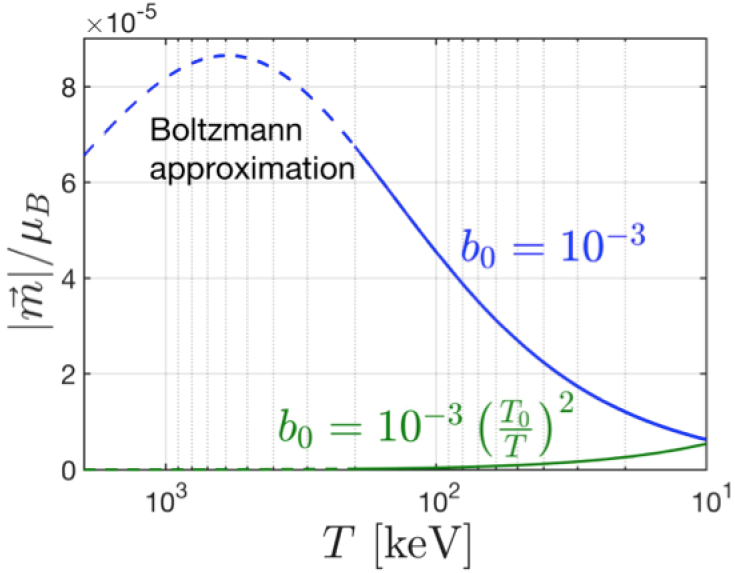}
 \caption{The magnetic moment per lepton $\vert\vec{m}\vert_{z}$ along the field axis as a function of temperature. Figure made in collaboration with Cheng Tao Yang.}
 \label{fig:momentperlepton}
\end{figure}

The average magnetic moment $\vert\vec{m}\vert_{z}$ defined in \req{momentperlepton} is plotted in \rf{fig:momentperlepton} which displays how essential the external field is on the `per lepton' magnetization. The $b_{0}=10^{-3}$ case (blue curve) is plotted in the Boltzmann approximation. The dashed lines indicate where this approximation is only qualitatively correct. For illustration, a constant magnetic field case (solid green line) with a comoving reference value chosen at temperature $T_{0}=10\keV$ is also plotted.

If the field strength is held constant, then the average magnetic moment per lepton is suppressed at higher temperatures as expected for magnetization satisfying Curie's law. The difference in \rf{fig:momentperlepton} between the non-constant (blue solid curve) case and the constant field (solid green curve) case demonstrates the importance of the conservation of primordial magnetic flux in the plasma, required by \req{bscale}. While not shown, if \rf{fig:momentperlepton} was extended to lower temperatures, the magnetization per lepton of the constant field case would be greater than the non-constant case which agrees with our intuition that magnetization is easier to achieve at lower temperatures. This feature again highlights the importance of flux conservation in the system and the uniqueness of the primordial cosmic environment.

\section{Polarization potential and ferromagnetism}
\label{sec:ferro}
\noindent Up to this point, we have neglected the impact that a nonzero spin potential $\eta\neq0$ (and thus $\xi\neq1$) would have on the primordial $e^{+}e^{-}$ plasma magnetization. In the limit that $(m_{e}/T)^2\gg b_0$ the magnetization given in \req{arbg:1} and \req{arbg:2} is entirely controlled by the spin fugacity $\xi$ asymmetry generated by the spin potential $\eta$ yielding up to first order $\mathcal{O}(b_{0})$ in magnetic scale
\begin{multline}
 \label{ferro}
 \lim_{m_{e}^{2}/T^{2}\gg b_0}{\mathfrak M}=\frac{g}{2}\frac{e^{2}}{\pi^{2}}\frac{T^{2}}{m_{e}^{2}}\sinh{\frac{\eta}{T}}\cosh{\frac{\mu}{T}}\left[\frac{m_{e}}{T}K_{1}\left(\frac{m_{e}}{T}\right)\right]\\
 +b_{0}\left(g^{2}-\frac{4}{3}\right)\frac{e^{2}}{8\pi^{2}}\frac{T^{2}}{m_{e}^{2}}\cosh{\frac{\eta}{T}}\cosh{\frac{\mu}{T}}K_{0}\left(\frac{m_{e}}{T}\right)
 +\mathcal{O}\left(b_{0}^{2}\right)
\end{multline}

Given \req{ferro}, we can understand the spin potential as a kind of `ferromagnetic' influence on the primordial gas which allows for magnetization even in the absence of external magnetic fields. This interpretation is reinforced by the fact the leading coefficient is $g/2$.

We suggest that a variety of physics could produce a small nonzero $\eta$ within a domain of the gas. Such asymmetries could also originate statistically as while the expectation value of free gas polarization is zero, the variance is likely not.

As $\sinh{\eta/T}$ is an odd function, the sign of $\eta$ also controls the alignment of the magnetization. In the high temperature limit \req{ferro} with strictly $b_{0}=0$ assumes a form of to lowest order for brevity
\begin{align}
 \label{hiTferro}
 \lim_{m_{e}/T\rightarrow0}{\mathfrak M}\vert_{b_{0}=0}=\frac{g}{2}\frac{e^{2}}{\pi^{2}}\frac{T^{2}}{m_{e}^{2}}\frac{\eta}{T}\,,
\end{align}

While the limit in \req{hiTferro} was calculated in only the Boltzmann limit, it is noteworthy that the high temperature (and $m\rightarrow0$) limit of Fermi-Dirac distributions only differs from the Boltzmann result by a proportionality factor. 

The natural scale of the $e^{+}e^{-}$ magnetization with only a small spin fugacity ($\eta<1\eV$) fits easily within the bounds of the predicted magnetization during this era if the IGMF measured today was of primordial origin. The reason for this is that the magnetization seen in \req{g2magplus}, \req{g2magminus} and \req{ferro} are scaled by $\alpha{B}_{C}$ where $\alpha$ is the fine structure constant.

\subsection{Hypothesis of ferromagnetic self-magnetization}
\label{sec:self}
\noindent One exploratory model we propose is to fix the spin polarization asymmetry, described in \req{spotential}, to generate a homogeneous magnetic field which dissipates as the universe cools down. In this model, there is no external primordial magnetic field $({B}_\mathrm{PMF}=0)$ generated by some unrelated physics, but rather the $e^{+}e^{-}$ plasma itself is responsible for the field by virtue of spin polarization.

\begin{figure}[ht]
 \centering
 \includegraphics[width=0.5\textwidth]{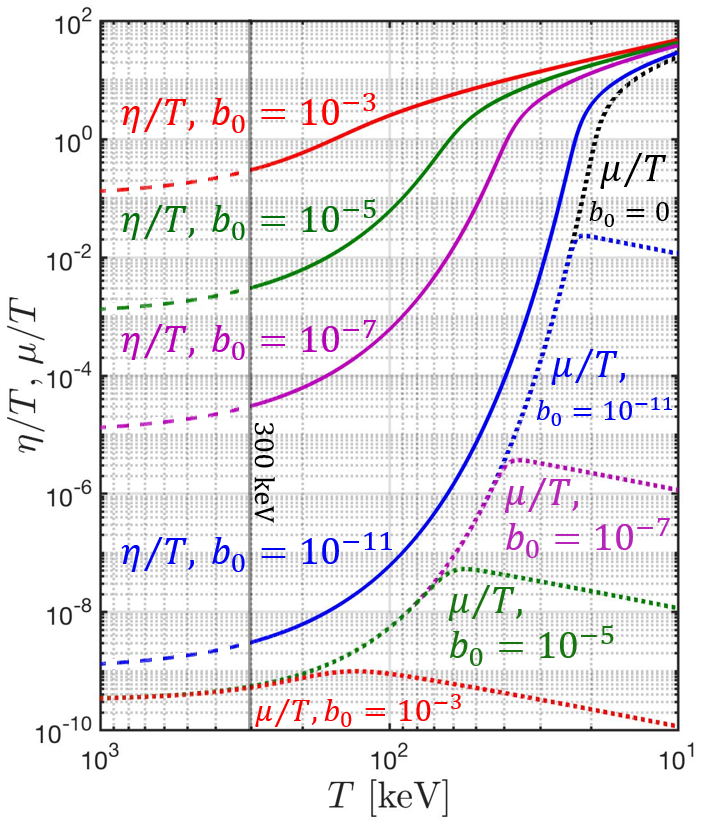}
 \caption{The spin potential $\eta$ and chemical potential $\mu$ are plotted under the assumption of self-magnetization through a nonzero spin polarization in bulk of the plasma. Figure made in collaboration with Cheng Tao Yang.}
 \label{fig:self} 
\end{figure}

This would obey the following assumption of
\begin{align}
 \label{selfmag}
 {\mathfrak M}(b_{0})=\frac{\mathcal{M}(b_0)}{{B}_{C}}\longleftrightarrow\frac{B}{{B}_{C}}=b_{0}\frac{T^{2}}{m_{e}^{2}}\,,
\end{align}
which sets the total magnetization as a function of itself. The spin polarization described by $\eta\rightarrow\eta(b_{0},T)$ then becomes a fixed function of the temperature and magnetic scale. The underlying assumption would be the preservation of the homogeneous field would be maintained by scattering within the gas (as it is still in thermal equilibrium) modulating the polarization to conserve total magnetic flux.

The result of the self-magnetization assumption in \req{selfmag} for the potentials is plotted in \rf{fig:self}. The solid lines indicate the curves for $\eta/T$ for differing values of $b_{0}=\{10^{-11},\ 10^{-7},\ 10^{-5},\ 10^{-3}\}$ which become dashed above $T\!=\!300\keV$ to indicate that the Boltzmann approximation is no longer appropriate though the general trend should remain unchanged.

\begin{figure}[ht]
 \centering
 \includegraphics[width=0.95\textwidth]{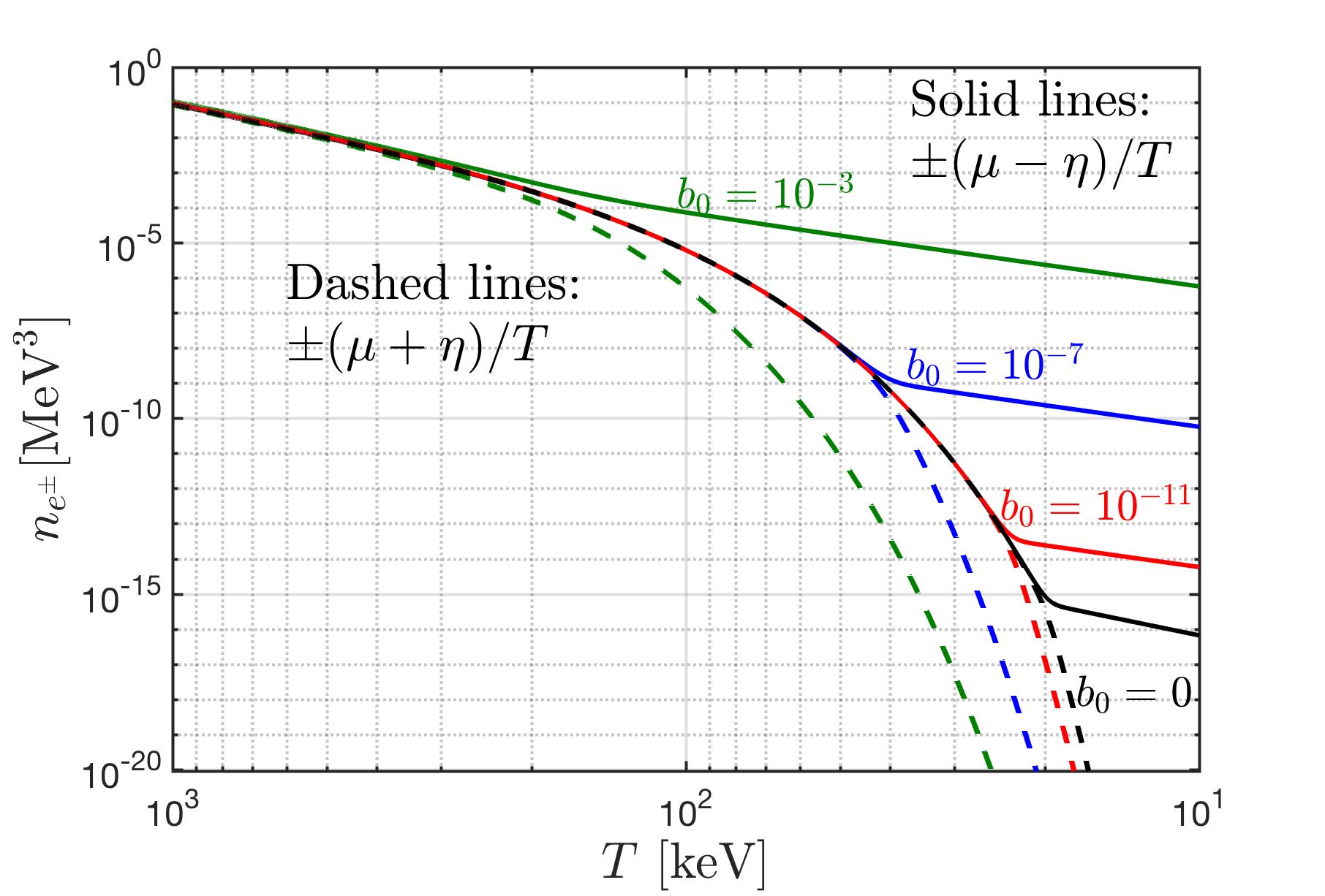}
 \caption{The number density $n_{e^{\pm}}$ of polarized electrons and positrons under the self-magnetization model for differing values of $b_{0}$. Figure courtesy of Cheng Tao Yang.}
 \label{fig:polarswap} 
\end{figure}

The dotted lines are the curves for the chemical potential $\mu/T$. At high temperatures we see that a relatively small $\eta/T$ is needed to produce magnetization owing to the large densities present. \rf{fig:self} also shows that the chemical potential does not deviate from the free particle case until the spin polarization becomes sufficiently high which indicates that this form of self-magnetization would require the annihilation of positrons to be incomplete even at lower temperatures.

This is seen explicitly in~\rf{fig:polarswap} where we plot the numerical density of particles as a function of temperature for spin aligned $(+\eta)$ and spin anti-aligned $(-\eta)$ species for both positrons $(-\mu)$ and electrons $(+\mu)$. Various self-magnetization strengths are also plotted to match those seen in~\rf{fig:self}. The nature of $T_{\rm split}$ changes under this model since antimatter and polarization states can be extinguished separately. Positrons persist where there is insufficient electron density to maintain the magnetic flux. Polarization asymmetry therefore appears physical only in the domain where there is a large number of matter-antimatter pairs.

\subsection{Matter inhomogeneities in the cosmic plasma}
\label{sec:inhomogeneous}
\noindent In general, an additional physical constraint is required to fully determine $\mu$ and $\eta$ simultaneously as both potentials have mutual dependency (see \rsec{sec:ferro}). We note that spin polarizations are not required to be in balanced within a single species to preserve angular momentum.

The CMB~\citep{Planck:2018vyg} indicates that the early universe was home to domains of slightly higher and lower baryon densities which resulted in the presence of galactic super-clusters, cosmic filaments, and great voids seen today. However, the CMB, as measured today, is blind to the localized inhomogeneities required for gravity to begin galaxy and supermassive black hole formation.

Such acute inhomogeneities distributed like a dust~\citep{Grayson:2023flr} in the plasma would make the proton density sharply and spatially dependant $n_{p}\rightarrow n_{p}(x)$ which would directly affect the potentials $\mu(x)$ and $\eta(x)$ and thus the density of electrons and positrons locally. This suggests that $e^{+}e^{-}$ may play a role in the initial seeding of gravitational collapse. If the plasma were home to such localized magnetic domains, the nonzero local angular momentum within these domains would provide a natural mechanism for the formation of rotating galaxies today.

Recent measurements by the James Webb Space Telescope (JWST)~\citep{Yan:2022sxd,adams2023discovery,arrabal2023spectroscopic} indicate that galaxy formation began surprisingly early at large redshift values of $z\gtrsim10$ within the first 500 million years of the universe requiring gravitational collapse to begin in a hotter environment than expected. The observation of supermassive black holes already present~\citep{CEERSTeam:2023qgy} in this same high redshift period (with millions of solar masses) indicates the need for local high density regions in the early universe whose generation is not yet explained and likely need to exist long before the recombination epoch.


\chapter{Outlook and key results}
\label{chap:outlook}

\subsection*{Chapter 2: Dynamics of charged particles with arbitrary magnetic moment}
\label{sec:chap2}
We highlighted the comparison of magnetic moment dynamics of DP and KGP formulation of relativistic quantum mechanics. The DP equation breaks up the magnetic moment into an underlying spinor structure part inherent to the Dirac equation, and a dedicated anomalous part. In contrast, for the KGP, the entire effect of magnetic moment is contained in a single Pauli term irrespective of the magnetic moment\rq s size. We find that the two models disagree in their predicted energy levels for the homogeneous magnetic field and the Coulomb field.

For the KGP-Landau levels, Eq.\,\eqref{lan24b}, we have a simple dependence on the full magnetic moment $g$-factor and have the correct non-relativistic reduction at lowest order. This simplicity allows, for the KGP equation, the straightforward analysis of physical systems and elegant expressions for their solutions. Thus Dirac's beauty principle favors heavily the KGP considering the Landau levels.

In the case of the Coulomb field there are weak fields differences in both the Lamb shift and fine structure; the contribution to the Lamb shift and fine structure splitting are proportional in KGP to: $g^{2}/8-1/2$, Eq.\,\eqref{lamb01} and $g^{2}/8$, Eq.\,\eqref{fs00} respectively, rather than: DP $g/2-1$ Eq.\,\eqref{lamb03}) and $g/2-1/2$, Eq.\,\eqref{fs02}) respectively.

For strong fields, both DP and KGP share the behavior of a shrinking particle/antiparticle gap for the ground state when $|g|>2$, though the expressions differ from each other. The models are extremely different in both their predicted energy values and the ultimate fate of the states as $B$ increases. For the KGP equation, the gap vanishes in very strong magnetic fields Eq.\,\eqref{Bcrit}. KGP also shows state-merging behavior for states of the same total angular momentum.

\subsection*{Chapter 3: Dynamic neutrino flavor mixing through transition moments}
\label{sec:chap3}
We have incorporated electromagnetic effects in the Majorana neutrino mixing matrix by introducing an anomalous transition magnetic dipole moment. We have described the formalism for three generations of neutrinos and explicitly explored the two generation case as a toy model. 

In the two generation case, we determined the effect of electric and magnetic fields on flavor rotation in~\req{zrot:2} by introducing an electromagnetic flavor unitary rotation $Z_{kj}^\mathrm{ext}$. We presented remixed mass eigenstates $\widetilde m(E,B)$ in~\req{poly:3} which are the propagating mass-states in a background electromagnetic field $F^{\alpha\beta}_\mathrm{ext}(x^{\mu})$. These EM-mass eigenstates were also further split by spin aligned and anti-aligned states relative to the external field momentum density. There is much left to do to explore further the nascent connection between the spin and the flavor via transition magnetic moments. 

The transition dipole moments are the origin of dynamical flavor mixing. While our focus was on Majorana neutrinos, Dirac-type fermions (neutrinos included) may also have non-zero transition  dipole moments. These could  remix flavor in the presence of strong external background fields. Here  quarks  are of special interest because they are not only electrically charged, but have color charge as well. This means quarks could in principle possess one, or both, EM and color-charge transition dipole moments, leading to dynamical effects in the CKM mixing matrix within hadrons as well as in quark-gluon plasma.

More speculatively, as transition dipoles act as a mechanism to generate mass by virtue of EM energy density $T_\mathrm{ext}^{00}$ as seen in~\req{poly:3}, an analogous consequence of our work could arise in the presence of a dark vector field in the Universe coupled to neutrinos, resulting in off-diagonal masses in flavor. Massless neutrinos could then obtain dynamical non-zero masses in the Universe by virtue of their interactions originating in dark transition moments.

\subsection*{Chapter 4: Matter-antimatter origin of cosmic magnetism}
\label{sec:chap4}
We characterized the primordial magnetic properties of the early universe before recombination. We studied the temperature range of $2000\keV$ to $20\keV$ where all of space was filled with a hot dense electron-positron plasma (to the tune of 450 million pairs per baryon) which occurred within the first few minutes after the Big Bang. We note that our chosen period also includes the era of Big Bang Nucleosynthesis.

We found that subject to a primordial magnetic field, the early universe electron-positron plasma has a significant paramagnetic response due to magnetic moment polarization. We considered the interplay of charge chemical potential, baryon asymmetry, anomalous magnetic moment, and magnetic dipole polarization on the nearly homogeneous medium.

This novel approach to high temperature magnetization shows that the $e^{+}e^{-}$-plasma paramagnetic response (see~\req{g2magplus} and~\req{g2magminus}) is dominated by the varying abundance of electron-positron pairs, decreasing with decreasing $T$ for $T\!<\!m_{e}c^2$. This is unlike conventional laboratory cases where the number of magnetic particles is constant. 

We find that electron-positron magnetization rapidly vanishes as the number of pairs depletes as the universe cools. This therefore presents an opportunity for induced currents to facilitate inhomogeneities in the early universe. We also presented a simple model of self-magnetization of the primordial electron-positron plasma which indicates that only a small polarization asymmetry is required to generate significant magnetic flux when the universe was very hot and dense.

\chapter{Future research efforts}
\label{chap:future}

This chapter contains unpublished work and is exploratory; though we hope to complete this research in the near future. \rsec{sec:quarks} extends the KGP wave equation to also include chromomagnetic dipole moments necessary for quarks in quantum chromodynamics (QCD). We show explicitly that the dipoles are linear in field tensors and do not mix spin degrees of freedom.

In \rsec{sec:nucp}, we explore CP violation in the neutrino sector in terms of the Jarlskog invariant and modifications to the mass matrix. The CP symmetry of the electric dipole is discussed. This section follows the conventions found in \rchap{chap:neutrino}.

The final \rsec{sec:kk} is a short description of a personal passion project to integrate spin dynamics into the five dimensional theory of Kaluza-Klein which unifies electromagnetism and gravitation classically. While I have written much on the topic, it remains unfinished and will only occupy a small section of this dissertation.

\section{Klein-Gordon-Pauli extensions to non-Abelian fields}
\label{sec:quarks}

The KGP approach to wave equations is robust and is useful not only for electromagnetic interactions: This is a first look at adding the strong interaction into KGP inspired in part by~\cite{Labun:2012ra}. Quarks participate in the strong color interaction of quantum chromodynamics (QCD); therefore they should present a $g$-factor for both their electromagnetic dipole $g_\mathrm{EM}$ as well as their QCD chromomagnetic dipole $g_\mathrm{QCD}$.

Since color charges follows a more complex $SU(3)$ group structure unlike the more straight forward $U(1)$ of electromagnetism, the ``color magnetism'' of QCD requires more than just the analogous Pauli term to describe color dipole moments owing due to the fact that QCD has non-Abelian (non-commuting) gluon gauge fields $\mathcal{A}^{\alpha}$.

The quarks (like all particles) should obey the quantum mechanical analogue of the energy-momentum relation seen in \req{eq:spin:03} with the only theoretical difference being a modified covariant derivative. The covariant derivative, written in terms of kinetic momentum, should appear as
\begin{alignat}{1}
    \label{eq:spin:08}
    \mathrm{EM+QCD}:\qquad i\hbar\widetilde\nabla=\pi^{\alpha}=p^{\alpha}-q_\mathrm{EM}A^{\alpha}-q_\mathrm{QCD}\mathcal{A}^{\alpha}\,,
\end{alignat}
where $q_\mathrm{EM}$ is the electric charge of the quarks $q_\mathrm{EM}/e\in\pm1/3,\pm2/3$ and $q_\mathrm{QCD}$ is the color charge coupling strength. In many texts the symbol $g_{s}$ is used for the color coupling strength, but we circumvent that notation using $q_\mathrm{QCD}$ to avoid confusion with $g$-factor.

We follow the conventions in~\cite{greiner2006qcd}. The eight $3\times3$ Gell-Mann matrices $\lambda^{a}$ are embedded into each independent field as
\begin{alignat}{1}
	\label{eq:spin:09} \mathcal{A}^{\alpha}\equiv\frac{1}{2}\lambda^{a}\mathcal{A}^{\alpha}_{a}\,,\qquad a\in1\ldots8\,,\qquad
    [\lambda^{a},\lambda^{b}]=\frac{i}{2}f^{abc}\lambda^{c}\,,
\end{alignat}
where $\mathcal{A}_{a}^{\alpha}$ are the individual fields for each gluon species in a given representation and $f^{abc}$ is the $SU(3)$ antisymmetric structure function. The non-commuting behavior of the Gell-Mann matrices captures the non-Abelian structure of the gauge fields. The gluon field strength tensor $\mathcal{G}^{\alpha\beta}$ is then
\begin{gather}
	\label{eq:spin:10a}
    \mathcal{G}^{\alpha\beta} = \partial^{\alpha}\mathcal{A}^{\beta} -\partial^{\beta}\mathcal{A}^{\alpha} + \frac{i}{\hbar}q_\mathrm{QCD}\left[\mathcal{A}^{\alpha},\mathcal{A}^{\beta}\right]\,,\\
	\label{eq:spin:10b} \left[\mathcal{A}^{\alpha},\mathcal{A}^{\beta}\right] =
    \frac{1}{4}\mathcal{A}^{\alpha}_{a}\mathcal{A}^{\beta}_{b}\left[\lambda^{a},\lambda^{b}\right] =
    \frac{i}{8}\mathcal{A}^{\alpha}_{a}\mathcal{A}^{\beta}_{b}f^{abc}\lambda_{c}\,.
\end{gather}

Following the same procedure in \rsec{sec:unique}, we can generalize the energy-momentum relation and obtain the EM+QCD variant of the KGP equation for quarks. We find that the resulting quark-KGP equation is
\begin{align}
    \label{eq:spin:11a}
    \gamma_{\alpha}\gamma_{\beta}\pi^{\alpha}\pi^{\beta} = \eta_{\alpha\beta}\pi^{\alpha}\pi^{\beta} - 
    \frac{q_\mathrm{EM}\hbar}{2}\sigma_{\alpha\beta}F^{\alpha\beta} - 
    \frac{q_\mathrm{QCD}\hbar}{2}\sigma_{\alpha\beta}\mathcal{G}^{\alpha\beta}\,,\\
	\label{eq:spin:11b} \left(\eta_{\alpha\beta}\pi^{\alpha}\pi^{\beta} - 
    \frac{q_\mathrm{EM}\hbar}{2}\sigma_{\alpha\beta}F^{\alpha\beta} - 
    \frac{q_\mathrm{QCD}\hbar}{2}\sigma_{\alpha\beta}\mathcal{G}^{\alpha\beta}\right)\Psi=m_{q}^{2}c^{2}\Psi_{q}\,,
\end{align}
which mirrors the electromagnetic case except for the extension of a chromomagnetic Pauli term~\citep{Morgan:1995te}. We note that $m_{q}$ is the quark mass and that the field $\Psi_{q}$ is a color triplet of spinors for: red, green, blue:
\begin{align}
    \Psi_{q}=
    \begin{pmatrix}
        \Psi_{r}\\
        \Psi_{g}\\
        \Psi_{b}
    \end{pmatrix}\,.
\end{align}
As only the non-commuting portion of \req{eq:spin:11b} (written explicitly in \req{eq:spin:10b}) is off-diagonal in color, this means that the additional non-commuting chromomagnetic term acts as a transition matrix between different quark colors.

This method (based on the commutator of the kinetic momentum) suggests that the color and electromagnetic $g$-factors both have a natural value of $g_\mathrm{EM}\!=\!g_\mathrm{QCD}\!=\!2$. We can generalize \req{eq:spin:11a} to allow for arbitrary EM and QCD dipole moments $g_\mathrm{EM}$ and $g_\mathrm{QCD}$ respectively as
\begin{align}
	\label{eq:spin:12}
    \boxed{\left(\eta_{\alpha\beta}\pi^{\alpha}\pi^{\beta}-\frac{g_\mathrm{EM}}{2}\frac{q_\mathrm{EM}\hbar}{2}\sigma_{\alpha\beta}F^{\alpha\beta}-\frac{g_\mathrm{QCD}}{2}\frac{q_\mathrm{QCD}\hbar}{2}\sigma_{\alpha\beta}\mathcal{G}^{\alpha\beta}\right)\Psi=m_{q}^{2}c^{2}\Psi}\,.
\end{align}

While in electromagnetism, DP and KGP approaches only differ in the presence of strong EM fields and are otherwise identical in the weak field limit, this cannot be equally said in QCD. The perturbative limit which justifies the DP approach for leptons in QED is possible due to the small value of the fine structure constant which is not true in QCD. Only for high momentum interactions (such as those present in quark-gluon-plasma (QGP) or in energetic collisions) is the perturbative approach applicable~\citep{Choudhury:2014lna}. We emphasize however that the DP approach is only valid where the dipole moment is obtainable via perturbative expansion which may not hold if the $g$-factor results from non-perturbative physics.

The dipole characteristics (both electromagnetic and chromomagnetic) of the top-quark is of particular interest~\citep{Labun:2012fg,Vryonidou:2018eyv} because of top-quark's strong coupling to the Higgs and potential BSM physics. We note that current studies focus on a DP approach to chromomagnetism~\citep{Zhang:2010dr,Zhang:2012muc,BuarqueFranzosi:2015jrv}. There is also the added complexity of both the chromomagnetic and magnetic $g$-factors differing from the natural value independently of one another $g_\mathrm{EM}\neq g_\mathrm{QCD}\neq2$. Further study of the KGP approach to chromomagnetism should be pursued. To our knowledge, there is no formulation of \req{eq:spin:12} as an effective field theory in the manner of~\cite{Fleming:2000ib,Bauer:2000yr} for quarks or for electromagnetically charged fermions as was discussed in \rsec{sec:lagrangian}.

\section{Electromagnetic forced neutrino CP violation}
\label{sec:nucp}
Electromagnetic processes in the neutrino sector may yield measurable effects in two aspects of neutrino physics: 
\begin{itemize}
    \item[(a)] Neutrino oscillation which is evidence of the non-zero mass eigenstates
    \item[(b)] Charge-parity (CP) violation in the neutrino sector which occurs due to the presence of at least three generations of neutrinos or additional CP violating interactions
\end{itemize}

Our motivation is to explore the effect of strong electromagnetic fields on neutrino CP violation by analysis of the electromagnetic dipole interaction and determining its influence on the Jarlskog invariant $J$~\citep{Jarlskog:1985ht,Jarlskog:1985cw,Jarlskog:2004be} which controls the size of CP violation.

One important aspect of neutrino physics is the size of the CP violation~\citep{Xing:2000ik,giunti2007fundamentals,Huber:2022lpm} which can yield insights not only in fundamental physics but cosmology as well. One of the goals of modern neutrino experiments such as DUNE~\cite{DUNE:2020jqi} is to better characterize such effects in astrophysical contexts such as supernova~\citep{DUNE:2020zfm,SajjadAthar:2021prg}, solar neutrinos~\citep{Akhmedov:2022txm} and magnetars~\citep{Lichkunov:2020zzx}. Additionally, there may be a connection~\citep{Pehlivan:2014zua,Balaji:2019fxd,Balaji:2020oig} between CP violation in the neutrino sector and the anomalous magnetic moment (AMM) of the neutrino.

Amplified CP violation is already expected in matter~\citep{Harrison:1999df} such as when travelling through the Earth's crust due to the abundance of electrons and the lack of muons and taus which preferentially affect the electron neutrinos via the weak interaction. This effect may also play a role in the primordial universe where neutrinos would propagate through incredibly matter dense gasses such as during the electron-positron epoch~\citep{Rafelski:2023emw}. Matter modifies CP violation because it changes the effective Hamiltonian of the propagating neutrinos, therefore we should look to other possible sources for modification of the Hamiltonian for CP violation amplification or suppression.

The source of direct CP violation in the neutrinos is ultimately attributable to the mismatch between the mass matrices of the charged flavors $(e,\mu,\tau)$ and the neutral flavors $(\nu_{e},\nu_{\mu},\nu_{\tau})$. The situation is analogous to the quark sector, where instead the relation is between the upper $(u,c,t)$ and lower quark $(d,s,b)$ flavors.

Therefore mass matrix for charged leptons does not commute with the mass matrix of the neutral leptons and cannot be simultaneously diagonalized except for special cases or degeneracy among the mass eigenstates. We can characterize CP violation by introducing the mixing matrices $V_{\ell k}$ (following the notation in \rchap{chap:neutrino}) which diagonalize the individual mass-matrices $M_{\ell\ell'}$ as follows
\begin{alignat}{1}
	\label{diagj:1}
    V_{\ell l}^{\dag}(M^{\nu}M^{\nu\dag})_{\ell\ell'}V_{\ell'k'} = D_{\ell\ell'}^{\nu} = \mathrm{diag}(m_{1}^{2},m_{2}^{2},m_{3}^{2})\,,\\
    \label{diagj:2}
    D_{\ell\ell'} = \mathrm{diag}(m_{e},m_{\mu},m_{\tau})\,,
\end{alignat}
We have specifically defined the charged leptons flavor states as being simultaneously mass eigenstates without a loss of generality. We will not consider oscillation among the charged leptons, though that may be an avenue of further study~\cite{Akhmedov:2007fk}.

\subsection{Jarlskog invariant}
\label{sec:jscalar}
The intrinsic CP violation inherent to these mass matrices can be described using the Jarlskog invariant $J$. We first define the commutator of the charged and neutral lepton mass matrices as 
\begin{alignat}{1}
	\label{comm:1} [M^{\nu}M^{\nu\dag},D^{2}]_{\ell\ell'} = C_{\ell\ell'}\,.
\end{alignat}
As the elements of the neutrino mixing matrix $V_{\ell k}$ can be experimentally determined, the size of the commutator can be expressed using~\req{diagj:1} and~\req{diagj:2} in terms of the mass eigenstates of the neutrino mass matrix
\begin{alignat}{1}
	\label{comm:2} [V(D^{\nu})^{2}V^{\dag},D^{2}]_{\ell\ell'} = C_{\ell\ell'}\,.
\end{alignat}
The matrix $C_{\ell\ell'}$ can be unwieldy, so following the procedure by~\citep{Jarlskog:1985ht,Jarlskog:1985cw,Jarlskog:2004be}, we take the determinant of $C_{\ell\ell'}$ in~\req{comm:2} which extracts the invariant quantity associated with the size of the CP violation present. Specifically we are interested in the imaginary portion given by
\begin{alignat}{1}
	\label{det:1} \mathrm{Im}\left[\mathrm{det}(C_{\ell\ell'})\right]=2\left(\Delta_{12}\Delta_{23}\Delta_{13}\right)\left(\Delta_{e\mu}\Delta_{\mu\tau}\Delta_{e\tau}\right)J\,,
\end{alignat}
where $J$ is the invariant quantity of interest. We define $\Delta_{ij}$ via the eigenstates of the mass matrices as
\begin{alignat}{1}
	\label{delta:1} \Delta_{ij}\equiv m^{2}_{i}-m^{2}_{j}\,.
\end{alignat}
We can also define the real portion of the determinant and define two quantities $R$ and $J$ together. These two scalars are written in terms of the components of the mixing matrix $V_{\ell k}$ as
\begin{alignat}{1}
	\label{j:1}
    {\cal J}_{ikjl} = \mathrm{Im}\left[V_{ij}V_{kl}V^{*}_{il}V^{*}_{kj}\right]=J\sum_{m,n}\epsilon_{ikm}\epsilon_{jln}\,,\\
    \label{j:2}
    {\cal R}_{ikjl} = \mathrm{Re}\left[V_{ij}V_{kl}V^{*}_{il}V^{*}_{kj}\right]=R\sum_{m,n}\epsilon_{ikm}\epsilon_{jln}\,.
\end{alignat}
The benefit of $J$ is it captures the `size' of CP violation in a single value and is identically zero for systems which preserve charge-parity symmetry. From \req{det:1} we can see that CP violation vanishes if the commutator of the mass matrices is purely real, and thus the mixing matrix is also purely real, or if there is degeneracy among the mass eigenstates which absorbs a degree of freedom.

While generally CP violation comes exclusively from the presence of three flavor generations in the Standard Model, we would like to expand the usage of the J-invariant to encompass a family of effects via modification of the mass matrix as was done in \rchap{chap:neutrino}. This would encompass CP violation from the number of flavor generations, physical systems (matter, strong fields, etc...) which effectively break CP, or in CP violating terms in the Lagrangian.

\subsection{Toy model: CP violation amplification}
\label{sec:amp}
\noindent While the specific change to neutrino mixing and CP violation depends on the model of the neutrino dipole moment, a demonstrative model would be to assume that the magnetic moment matrix was simply proportional to the natural mass matrix of the neutrino at low order. Therefore we substitute
\begin{alignat}{1}
	\label{simplek:1} M_{\ell\ell'}^{\nu}\rightarrow
    (M_{\ell\ell'}^{\nu})'=M_{\ell\ell'}^{\nu}(1+\kappa\sigma_{\alpha\beta}F^{\alpha\beta})\,.
\end{alignat}
If the same mechanism which produced neutrino masses also produced their dipoles through some BSM physics, this would not be an unreasonable assumption.

As the modified mass matrix commutes entirely with the original mass matrix, the mathematical structure of the commutator in \req{comm:1} is unchanged
\begin{alignat}{1}
	\label{commutes:1} \left[(M^{\nu})',M^{\nu}\right]=0\,\rightarrow
    \mathrm{Im}[\mathrm{det}(C_{\ell\ell'})] = \mathrm{Im}[\mathrm{det}(C_{\ell\ell'}')]\,.
\end{alignat}
The determinant calculation is then identical between the two mass matrices. While the overall determinant is fixed, the individual elements which make up the determinant are modified. This yields
\begin{alignat}{1}
	\label{det:0} \mathrm{Im}\left[\mathrm{det}(C_{\ell\ell'})\right] &= \left(\Delta_{12}\Delta_{23}\Delta_{13}\right)\left(\Delta_{e\mu}\Delta_{\mu\tau}\Delta_{e\tau}\right)J\,,\\
	\label{det:2} \mathrm{Im}\left[\mathrm{det}(C_{\ell\ell'}')\right] &= \left(\Delta_{12}'\Delta_{23}'\Delta_{13}'\right)\left(\Delta_{e\mu}\Delta_{\mu\tau}\Delta_{e\tau}\right)J'\,,
\end{alignat}
where we have added primes to denote new values due to remixing. A similar result occurs in matter in~\cite{Harrison:1999df}. The ratio of modified $J'$ to $J$ is the amplification (or suppression) of CP violation given by
\begin{alignat}{1}
	\label{ampj:1} \mathcal{R} = \frac{J'}{J} = \frac{\left(\Delta_{12}\Delta_{23}\Delta_{13}\right)}{\left(\Delta_{12}'\Delta_{23}'\Delta_{13}'\right)}\,.
\end{alignat}
In our simple proportionality model, this simplifies to
\begin{alignat}{1}
	\label{ampj:2} \mathcal{R} = \frac{1}{\left(1+\kappa\sigma_{\alpha\beta}F^{\alpha\beta}\right)^{6}}\,.
\end{alignat}
To first order, and evaluating $\sigma_{\alpha\beta}F^{\alpha\beta}$ for homogeneous magnetic fields, \req{ampj:2} can be expressed as
\begin{alignat}{1}
	\label{amp:3} \mathcal{R} = 1\pm12\kappa B+\mathcal{O}(B^{2})\,.
\end{alignat}
where we allow for aligned and anti-aligned spin states. This suggests the possibility that CP violation having spin dependence in strong EM fields.

\subsection{Electric dipole moments and CP symmetry}
\label{sec:edm}
Here we state the standard picture of CP violation through an electric dipole. The structure of the Pauli term in \req{lamm:1} informs us how to construct the relativistic electric dipole moment (EDM); see~\cite{Knecht:2003kc,Jegerlehner:2017gek}. The generalization to include the electric dipole is
\begin{alignat}{1}
	\label{edm:1} \delta\mu\rightarrow\delta\tilde{\mu}\equiv\delta\mu+i\epsilon\gamma^{5}\,,
\end{alignat}
where $\epsilon$ is the EDM of the particle. As the natural electric dipole within the Dirac equation is zero, the presence of $\epsilon$ is always considered anomalous. The EDM Pauli Lagrangian term is
\begin{gather}
    \label{ledm:1}
    \mathcal{L}_\mathrm{EDM} = -{\bar\psi}\left(i\epsilon\gamma^{5}\frac{1}{2}\sigma_{\alpha\beta}F^{\alpha\beta}\right)\psi\,,
\end{gather}
which is of interest because of the inclusion of $\gamma^{5}$. Taking advantage of the properties of $\gamma^{5}$, we can write the EDM in \req{ledm:1} as
\begin{gather}
    \label{ledm:3}
    \gamma^{5}\sigma_{\mu\nu}=\frac{i}{2}\varepsilon_{\mu\nu\alpha\beta}\sigma^{\alpha\beta}\,\rightarrow
    \mathcal{L}_\mathrm{EDM} = +{\bar\psi}\left(\epsilon\frac{1}{2}\sigma^{\alpha\beta}F_{\alpha\beta}^{*}\right)\psi\,.
\end{gather}
which is more closely analogous to the structure of the AMM in \req{lamm:1} making use of the dual form of the electromagnetic field tensor shown in \req{em:2}.

Following a procedure similar to the one found in \rsec{sec:mom}, \req{ledm:3} reduces in the non-relativistic limit to the Hamiltonian density EDM interaction
\begin{gather}
    \label{ledm:4}
    \mathcal{H}_\mathrm{EDM} \approx \epsilon\chi^{\dag}\bb{\sigma}\cdot\bb{E}\chi\,.
\end{gather}
The electric dipole is important because the presence of one would signify charge-parity (CP) violation in the theory which provides a method to distinguish between matter and antimatter. We discuss briefly using the parity (P) and time (T) symmetries of the relevant spin $\bb{s}$ and field vectors $(\bb{E},\bb{B})$ and their inner products. The even and odd symmetries of each are printed in \rt{fig:cp}.

\begin{table}[ht]
 \centering
 \begin{tabular}{ r|c|c|c|c|c| }
 \multicolumn{1}{r}{}
 & \multicolumn{1}{c}{$\bb{E}$}
 & \multicolumn{1}{c}{$\bb{B}$}
 & \multicolumn{1}{c}{$\bb{s}$}
 & \multicolumn{1}{c}{$\bb{s}\!\cdot\!\bb{E}$}
 & \multicolumn{1}{c}{$\bb{s}\!\cdot\!\bb{B}$} \\
 \cline{2-6}
 \begin{tabular}[x]{@{}c@{}}T symmetry\\ $(t\rightarrow-t)$\end{tabular} & \textbf{even} & odd & odd & odd & \textbf{even} \TBstrut\\
 \cline{2-6}
 \begin{tabular}[x]{@{}c@{}}P symmetry\\ $(\bb{x}\rightarrow-\bb{x})$\end{tabular} & odd & \textbf{even} & \textbf{even} & odd & \textbf{even} \TBstrut\\
 \cline{2-6}
 \begin{tabular}[x]{@{}c@{}}PT symmetry\\ $(x^{\alpha}\rightarrow-x^{\alpha})$\end{tabular} & odd & odd & odd & \textbf{even} & \textbf{even} \TBstrut\\
 \cline{2-6}
 \end{tabular}\\ \,\Bstrut\\
 \caption{Time (T), parity (P) and PT symmetries of electric $\bb{E}$, magnetic $\bb{B}$, spin $\bb{s}$ three-vectors and the inner products which describe the dipole Hamiltonian terms.}
 \label{fig:cp}
\end{table}

The EDM term \req{ledm:4} is overall T-odd and P-odd while the magnetic dipole is T-even and P-even. While EDM dipoles are common in molecular systems, no electric dipole has ever been measured for an elementary particle nor composite particles like the proton or neutron despite extensive searching. As a point of comparison, the EDM of the electron is excluded~\citep{ACME:2018yjb,Roussy:2022cmp} by a bound of $|\epsilon_{e}/c|<4.1\times10^{-30}\, e\,\mathrm{cm}$.

For CPT symmetry to hold, \rt{fig:cp} implies that both the electric and magnetic dipoles must be C-even. C-symmetry is a more complicated concept to discuss as it is only well-defined relativistically where particle and antiparticle states are simultaneously described by the theory. The Dirac spinor charge conjugates as
\begin{align}
    \label{c:1}
    C:\psi\rightarrow\psi_{c}=\eta_{c}C(\bar\psi)^\mathrm{T}= \eta_{c}C\gamma_{0}^\mathrm{T}\psi^{*}\,,
\end{align}
where $C$ is the charge conjugation matrix satisfying the conjugation relation
\begin{align}
    \label{c:2}
    -C\gamma_{\alpha}^\mathrm{T}C^{-1}=\gamma_{\alpha}\,,
\end{align}
and $\eta_{c}$ is an arbitrary complex phase. The exact matrix expression of $C$ depends on the representational basis used (Dirac, Weyl, Majorana, etc...). We're specifically interesting in the following conjugations
\begin{align}
    \label{c:3}
    C:\bar\psi\gamma_{\alpha}\psi\rightarrow-\bar\psi\gamma_{\alpha}\psi\,,\qquad
    C:\bar\psi\sigma_{\alpha\beta}\psi\rightarrow-\bar\psi\sigma_{\alpha\beta}\psi\,,\qquad
    C:A^{\alpha}\rightarrow-A^{\alpha}\,.
\end{align}
As the spin density $\bar\psi\sigma_{\alpha\beta}\psi$ and vector potential $A^{\alpha}$ (and thus $F^{\alpha\beta}$) are both odd under charge conjugation, the combination present in the AMM Lagrangian \req{lamm:1} is C-even under charge conjugation. The same is true of the EDM Lagrangian \req{ledm:3} which is more easily seen when cast in terms of the dual tensor.

We note that the fields $\psi$, $\bar\psi$ and $A^{\alpha}$ are considered dynamical \emph{operators} within a quantum field theory. C-symmetry is therefore broken when considering an externally fixed background field such as $A_\mathrm{ext}^{\alpha}(x)$ as $C:A_\mathrm{ext}^{\alpha}(x)\rightarrow A_\mathrm{ext}^{\alpha}(x)$. This distinction bears some importance when discussing the neutrino flavor rotation in \rchap{chap:neutrino}. There is also interest in EDMs in the background of curved spacetime~\citep{Filho:2023lqe}.

\section{Spin in 5D Kaluza-Klein theory}
\label{sec:kk}
The `miracle' of Kaluza-Klein~\citep{Kaluza:1921tu,Klein:1926tv} is that both 4D gravitation and electromagnetism emerge from a higher dimension 5D gravitational theory. While the specific importance of Kaluza's result to the physical world (if there is one) has yet to be revealed, the ideas of Kaluza-Klein have been extensively used to showcase the emergence of unified physics from higher dimensional geometries~\citep{Ortin:2015hya}.

In the modern context, we understand that Kaluza-Klein in its original form cannot strictly be correct as we are operating purely in the classical regime and we are missing the incorporation of the weak and strong interactions as well. With all that said, there is still value in exploring the implications of Kaluza-Klein in that Kaluza's miracle might very well not be an accident, but a natural result from a fuller more complete implementation of these ideas~\citep{Overduin:1997sri}.

The main goal of this research effort is to obtain the spin dynamics of a test particle in the context of a 5D Kaluza-Klein style theory in the same spirit as was accomplished for the general relativistic torque equations
\begin{alignat}{1}
  \label{STRESS06} \frac{Dp^{\mu}}{D\tau}+\frac{1}{2}u^{\nu}s^{\rho\sigma}R^{\mu}_{\ \nu\rho\sigma}=0\,,\qquad
  \frac{Ds^{\mu\nu}}{D\tau}+2u^{[\mu}p^{\nu]}=0\,,
\end{alignat}
which couples spin and precession to a curved spacetime. An additional term is also required to accommodate spin in the kinematic equation. \req{STRESS06} is known as the Mathisson-Papapetrou-Dixon (MPD) equations~\citep{Mathisson:1937zz,Papapetrou:1951pa,Dixon:1970zza,Mathisson:2010opl}. The rank-four tensor $R^{\mu}_{\ \nu\rho\sigma}$ is the Rienmann curvature defined by
\begin{alignat}{1}
  \label{STRESS08} R^{\mu}_{\ \nu\rho\sigma}V^{\nu}=2\nabla_{[\rho}\nabla_{\sigma]}V^{\mu}\,,
\end{alignat}
where $V^{\mu}$ is any arbitrary vector and the notation $[\rho,\sigma]$ indicates a commutator of indices. The significance of \req{STRESS06} is that particle motion will deviate away from traditional geodesic motion due to a coupling of the curvature to the spin.

To promote the above equations into the Kaluza-Klein framework, we consider Lorentz five-vectors $\hat{P}^{A}$ which are dressed with a hat and capital Latin letters indices $A\in(0-3,5)$. The position five-vector is then denoted by $\hat{x}^{A}=(x^{\mu},x_{5})$ where $x_{5}$ is the fifth coordinate position. The 5D Einstein-Hilbert action is
\begin{alignat}{1}
	\label{KALUZA01} \mathcal{S}[\hat{g}_{AB}]=\frac{c^{4}}{16\pi\hat{G}}\int\mathrm{d}\hat{x}^{5}\sqrt{-\hat{g}}\hat{R}\,.
\end{alignat}
The variable $\hat{G}$ is the 5D gravitational constant and $\hat{R}$ is the Ricci scalar in the 5D space-time. Kaluza-Klein theories are usually expressed as Ricci-flat theories with the scalar curvature $\hat{R}=0$, but in general nothing prevents us from including matter or other fields in the five-dimensional bulk by adding terms to the above action.

In a 5D spacetime under Klein's cylindrical condition, a particle under free-fall motion manifests as accelerated motion analogous to \req{STRESS06} caused by the electromagnetic force and an additional scalar force in the four-dimensional sector. Because particles with spin deviate from geodesics in free-fall, there should ultimately be spin precession generated by electromagnetism, the scalar field, and gravitation.

Therefore we propose a classical spin five-vector
\begin{alignat}{1}
	\label{KALUZA01a} \hat{s}^{A}=(s^{\alpha},s_{5})\,,
\end{alignat}
which relates the classical four-spin \req{fourspin} discussed in \rsec{sec:cspin} to a new fifth component of spin $s_{5}$. We note that analogously the fifth component of five-momentum is related to the mass of the particle; it is our suggestion that the fifth component of spin is related to the invariant spin magnitude. It is the study of this object that will be left to future publications.


\renewcommand{\baselinestretch}{1}
\small\normalsize

\bibliographystyle{uabibnat}
\bibliography{bibs/chap01intro-refs,bibs/chap02moment-refs,bibs/chap03neutrino-refs,bibs/chap04cosmo-refs,bibs/chap06future-refs}

\begin{thebibliography}{199}
\addcontentsline{toc}{chapter}{\bibname}
\providecommand{\natexlab}[1]{#1}
\expandafter\ifx\csname urlstyle\endcsname\relax
  \providecommand{\doi}[1]{doi:\discretionary{}{}{}#1}\else
  \providecommand{\doi}{doi:\discretionary{}{}{}\begingroup
  \urlstyle{rm}\Url}\fi

\bibitem[{Abdalla et~al.(2022)Abdalla, Abell{\'a}n, Aboubrahim, Agnello,
  Akarsu, Akrami, Alestas, Aloni, Amendola, Anchordoqui
  et~al.}]{Abdalla:2022yfr}
Abdalla, E., G.~F. Abell{\'a}n, A.~Aboubrahim, A.~Agnello, {\"O}.~Akarsu,
  Y.~Akrami, G.~Alestas, D.~Aloni, L.~Amendola, L.~A. Anchordoqui, et~al.
  (2022).
\newblock Cosmology intertwined: A review of the particle physics,
  astrophysics, and cosmology associated with the cosmological tensions and
  anomalies.
\newblock \emph{Journal of High Energy Astrophysics}, \textbf{34}, pp. 49--211.
\newblock \doi{10.1016/j.jheap.2022.04.002}.

\bibitem[{Abi et~al.(2020)}]{DUNE:2020jqi}
Abi, B. et~al. (2020).
\newblock {Long-baseline neutrino oscillation physics potential of the DUNE
  experiment}.
\newblock \emph{Eur. Phys. J. C}, \textbf{80}(10), p. 978.
\newblock \doi{10.1140/epjc/s10052-020-08456-z}.

\bibitem[{Abi et~al.(2021)}]{DUNE:2020zfm}
Abi, B. et~al. (2021).
\newblock {Supernova neutrino burst detection with the Deep Underground
  Neutrino Experiment}.
\newblock \emph{Eur. Phys. J. C}, \textbf{81}(5), p. 423.
\newblock \doi{10.1140/epjc/s10052-021-09166-w}.

\bibitem[{Abramowitz et~al.(1988)Abramowitz, Stegun, and
  Romer}]{abramowitz1988handbook}
Abramowitz, M., I.~A. Stegun, and R.~H. Romer (1988).
\newblock \emph{Handbook of mathematical functions with formulas, graphs, and
  mathematical tables}.
\newblock American Association of Physics Teachers.

\bibitem[{Adams et~al.(2023)Adams, Conselice, Ferreira
  et~al.}]{adams2023discovery}
Adams, N.~J., C.~J. Conselice, L.~Ferreira, et~al. (2023).
\newblock Discovery and properties of ultra-high redshift galaxies (9< z< 12)
  in the JWST ERO SMACS 0723 Field.
\newblock \emph{Monthly Notices of the Royal Astronomical Society},
  \textbf{518}(3), pp. 4755--4766.
\newblock \doi{10.1093/mnras/stac3347}.

\bibitem[{Adhikary et~al.(2013)Adhikary, Chakraborty, and
  Ghosal}]{Adhikary:2013bma}
Adhikary, B., M.~Chakraborty, and A.~Ghosal (2013).
\newblock {Masses, mixing angles and phases of general Majorana neutrino mass
  matrix}.
\newblock \emph{JHEP}, \textbf{10}, p. 043.
\newblock \doi{10.1007/JHEP10(2013)043}.
\newblock [Erratum: JHEP 09, 180 (2014)].

\bibitem[{Aghanim et~al.(2020)}]{Planck:2018vyg}
Aghanim, N. et~al. (2020).
\newblock {Planck 2018 results. VI. Cosmological parameters}.
\newblock \emph{Astron. Astrophys.}, \textbf{641}, p.~A6.
\newblock \doi{10.1051/0004-6361/201833910}.
\newblock [Erratum: Astron.Astrophys. 652, C4 (2021)].

\bibitem[{Aguillard et~al.(2023)}]{Muong-2:2023cdq}
Aguillard, D.~P. et~al. (2023).
\newblock {Measurement of the Positive Muon Anomalous Magnetic Moment to 0.20
  ppm}.
\newblock \emph{arXiv preprint}.
\newblock \doi{10.48550/arXiv.2308.06230}.

\bibitem[{Aker et~al.(2022)}]{KATRIN:2021uub}
Aker, M. et~al. (2022).
\newblock {Direct neutrino-mass measurement with sub-electronvolt sensitivity}.
\newblock \emph{Nature Phys.}, \textbf{18}(2), pp. 160--166.
\newblock \doi{10.1038/s41567-021-01463-1}.

\bibitem[{Akhmedov(1988)}]{Akhmedov:1988uk}
Akhmedov, E. (1988).
\newblock {Resonant Amplification of Neutrino Spin Rotation in Matter and the
  Solar Neutrino Problem}.
\newblock \emph{Phys. Lett. B}, \textbf{213}, pp. 64--68.
\newblock \doi{10.1016/0370-2693(88)91048-9}.

\bibitem[{Akhmedov and Mart\'\i{}nez-Mirav\'e(2022)}]{Akhmedov:2022txm}
Akhmedov, E. and P.~Mart\'\i{}nez-Mirav\'e (2022).
\newblock {Solar $ {\overline{\nu}}_e $ flux: revisiting bounds on neutrino
  magnetic moments and solar magnetic field}.
\newblock \emph{JHEP}, \textbf{10}, p. 144.
\newblock \doi{10.1007/JHEP10(2022)144}.

\bibitem[{Akhmedov(2007)}]{Akhmedov:2007fk}
Akhmedov, E.~K. (2007).
\newblock {Do charged leptons oscillate?}
\newblock \emph{JHEP}, \textbf{09}, p. 116.
\newblock \doi{10.1088/1126-6708/2007/09/116}.

\bibitem[{Andreev et~al.(2018)}]{ACME:2018yjb}
Andreev, V. et~al. (2018).
\newblock {Improved limit on the electric dipole moment of the electron}.
\newblock \emph{Nature}, \textbf{562}(7727), pp. 355--360.
\newblock \doi{10.1038/s41586-018-0599-8}.

\bibitem[{Angeles-Martinez and Napsuciale(2012)}]{Angeles-Martinez:2011wpn}
Angeles-Martinez, R. and M.~Napsuciale (2012).
\newblock {Renormalization of the QED of second order spin 1/2 fermions}.
\newblock \emph{Phys. Rev. D}, \textbf{85}, p. 076004.
\newblock \doi{10.1103/PhysRevD.85.076004}.

\bibitem[{Aoyama et~al.(2020)}]{Aoyama:2020ynm}
Aoyama, T. et~al. (2020).
\newblock {The anomalous magnetic moment of the muon in the Standard Model}.
\newblock \emph{Phys. Rept.}, \textbf{887}, pp. 1--166.
\newblock \doi{10.1016/j.physrep.2020.07.006}.

\bibitem[{Aristizabal~Sierra et~al.(2022)Aristizabal~Sierra, Miranda,
  Papoulias, and Garcia}]{AristizabalSierra:2021fuc}
Aristizabal~Sierra, D., O.~G. Miranda, D.~K. Papoulias, and G.~S. Garcia
  (2022).
\newblock {Neutrino magnetic and electric dipole moments: From measurements to
  parameter space}.
\newblock \emph{Phys. Rev. D}, \textbf{105}(3), p. 035027.
\newblock \doi{10.1103/PhysRevD.105.035027}.

\bibitem[{Asenjo and Koch(2023)}]{Asenjo:2023mon}
Asenjo, F.~A. and B.~Koch (2023).
\newblock {Extended BMT equations and the anomalous magnetic moment}.
\newblock \href{https://arxiv.org/abs/2309.16216}{arXiv:2309.16216}.

\bibitem[{Bahcall et~al.(2001)Bahcall, Pinsonneault, and Basu}]{Bahcall:2000nu}
Bahcall, J.~N., M.~H. Pinsonneault, and S.~Basu (2001).
\newblock Solar Models: Current Epoch and Time Dependences, Neutrinos, and
  Helioseismological Properties.
\newblock \emph{The Astrophysical Journal}, \textbf{555}(2), p. 990.
\newblock \doi{10.1086/321493}.

\bibitem[{Balaji et~al.(2020{\natexlab{a}})Balaji, Ramirez-Quezada, and
  Zhou}]{Balaji:2019fxd}
Balaji, S., M.~Ramirez-Quezada, and Y.-L. Zhou (2020{\natexlab{a}}).
\newblock {CP violation and circular polarisation in neutrino radiative decay}.
\newblock \emph{JHEP}, \textbf{04}, p. 178.
\newblock \doi{10.1007/JHEP04(2020)178}.

\bibitem[{Balaji et~al.(2020{\natexlab{b}})Balaji, Ramirez-Quezada, and
  Zhou}]{Balaji:2020oig}
Balaji, S., M.~Ramirez-Quezada, and Y.~L. Zhou (2020{\natexlab{b}}).
\newblock {CP violation in neutral lepton transition dipole moment}.
\newblock \emph{JHEP}, \textbf{12}, p. 090.
\newblock \doi{10.1007/JHEP12(2020)090}.

\bibitem[{Bargmann et~al.(1959)Bargmann, Michel, and Telegdi}]{Bargmann:1959gz}
Bargmann, V., L.~Michel, and V.~L. Telegdi (1959).
\newblock {Precession of the polarization of particles moving in a homogeneous
  electromagnetic field}.
\newblock \emph{Phys. Rev. Lett.}, \textbf{2}, pp. 435--436.
\newblock \doi{10.1103/PhysRevLett.2.435}.

\bibitem[{Barut and Kraus(1975)}]{Barut:1975hz}
Barut, A.~O. and J.~Kraus (1975).
\newblock {Resonances in e+ e- System Due to Anomalous Magnetic Moment
  Interactions}.
\newblock \emph{Phys. Lett. B}, \textbf{59}, pp. 175--178.
\newblock \doi{10.1016/0370-2693(75)90696-6}.

\bibitem[{Barut and Kraus(1976)}]{Barut:1976hs}
Barut, A.~O. and J.~Kraus (1976).
\newblock {Solution of the Dirac Equation with Coulomb and Magnetic Moment
  Interactions}.
\newblock \emph{J. Math. Phys.}, \textbf{17}, pp. 506--508.
\newblock \doi{10.1063/1.522932}.

\bibitem[{Batista and Saveliev(2021)}]{AlvesBatista:2021sln}
Batista, R.~A. and A.~Saveliev (2021).
\newblock The Gamma-ray Window to Intergalactic Magnetism.
\newblock \emph{Universe}, \textbf{7}(7).
\newblock ISSN 2218-1997.
\newblock \doi{10.3390/universe7070223}.

\bibitem[{Bauer et~al.(2001)Bauer, Fleming, Pirjol, and S.}]{Bauer:2000yr}
Bauer, C.~W., S.~Fleming, D.~Pirjol, and I.~W. S. (2001).
\newblock {An Effective field theory for collinear and soft gluons: Heavy to
  light decays}.
\newblock \emph{Phys. Rev. D}, \textbf{63}, p. 114020.
\newblock \doi{10.1103/PhysRevD.63.114020}.

\bibitem[{Bethe(1986)}]{Bethe:1986ej}
Bethe, H.~A. (1986).
\newblock {A Possible Explanation of the Solar Neutrino Puzzle}.
\newblock \emph{Phys. Rev. Lett.}, \textbf{56}, p. 1305.
\newblock \doi{10.1103/PhysRevLett.56.1305}.

\bibitem[{Birrell et~al.(2014)Birrell, Yang, and Rafelski}]{Birrell:2014uka}
Birrell, J., C.~T. Yang, and J.~Rafelski (2014).
\newblock {Relic Neutrino Freeze-out: Dependence on Natural Constants}.
\newblock \emph{Nucl. Phys. B}, \textbf{890}, pp. 481--517.
\newblock \doi{10.1016/j.nuclphysb.2014.11.020}.

\bibitem[{Bouchiat and Bouchiat(1974)}]{Bouchiat:1974kt}
Bouchiat, M.~A. and C.~C. Bouchiat (1974).
\newblock {Weak Neutral Currents in Atomic Physics}.
\newblock \emph{Phys. Lett. B}, \textbf{48}, pp. 111--114.
\newblock \doi{10.1016/0370-2693(74)90656-X}.

\bibitem[{Bouchiat and Bouchiat(1997)}]{Bouchiat:1997mj}
Bouchiat, M.~A. and C.~C. Bouchiat (1997).
\newblock {Parity violation in atoms}.
\newblock \emph{Rept. Prog. Phys.}, \textbf{60}, pp. 1351--1396.
\newblock \doi{10.1088/0034-4885/60/11/004}.

\bibitem[{Broderick et~al.(2000)Broderick, Prakash, and
  Lattimer}]{Broderick:2000pe}
Broderick, A., M.~Prakash, and J.~M. Lattimer (2000).
\newblock {The Equation of state of neutron star matter in strong magnetic
  fields}.
\newblock \emph{Astrophys. J.}, \textbf{537}, p. 351.
\newblock \doi{10.1086/309010}.

\bibitem[{Bruce(2021{\natexlab{a}})}]{Bruce:2021fph}
Bruce, S.~A. (2021{\natexlab{a}}).
\newblock {Model of electromagnetic interactions with spin-1/2 neutral
  particles}.
\newblock \emph{Mod. Phys. Lett. A}, \textbf{36}(33), p. 2150236.
\newblock \doi{10.1142/S0217732321502369}.

\bibitem[{Bruce(2021{\natexlab{b}})}]{Bruce:2021cva}
Bruce, S.~A. (2021{\natexlab{b}}).
\newblock {Neutron interactions with electromagnetic fields and single-neutron
  solitons}.
\newblock \emph{Int. J. Mod. Phys. A}, \textbf{36}(26), p. 2150185.
\newblock \doi{10.1142/S0217751X21501852}.

\bibitem[{Bruce and Diaz-Valdes(2020)}]{Bruce:2020xer}
Bruce, S.~A. and J.~F. Diaz-Valdes (2020).
\newblock {Relativistic neutron interaction with electric fields revisited}.
\newblock \emph{Eur. Phys. J. A}, \textbf{56}(7), p. 191.
\newblock \doi{10.1140/epja/s10050-020-00196-8}.

\bibitem[{Buarque~Franzosi and Zhang(2015)}]{BuarqueFranzosi:2015jrv}
Buarque~Franzosi, D. and C.~Zhang (2015).
\newblock {Probing the top-quark chromomagnetic dipole moment at
  next-to-leading order in QCD}.
\newblock \emph{Phys. Rev. D}, \textbf{91}(11), p. 114010.
\newblock \doi{10.1103/PhysRevD.91.114010}.

\bibitem[{Canas et~al.(2016)Canas, Miranda, Parada, Tortola, and
  Valle}]{Canas:2015yoa}
Canas, B.~C., O.~G. Miranda, A.~Parada, M.~Tortola, and J.~W.~F. Valle (2016).
\newblock {Updating neutrino magnetic moment constraints}.
\newblock \emph{Phys. Lett. B}, \textbf{753}, pp. 191--198.
\newblock \doi{10.1016/j.physletb.2015.12.011}.
\newblock [Addendum: Phys.Lett.B 757, 568--568 (2016)].

\bibitem[{Canetti et~al.(2012)Canetti, Drewes, and
  Shaposhnikov}]{Canetti:2012zc}
Canetti, L., M.~Drewes, and M.~Shaposhnikov (2012).
\newblock {Matter and Antimatter in the Universe}.
\newblock \emph{New J. Phys.}, \textbf{14}, p. 095012.
\newblock \doi{10.1088/1367-2630/14/9/095012}.

\bibitem[{Carter(1968)}]{Carter:1968rr}
Carter, B. (1968).
\newblock {Global structure of the Kerr family of gravitational fields}.
\newblock \emph{Phys. Rev.}, \textbf{174}, pp. 1559--1571.
\newblock \doi{10.1103/PhysRev.174.1559}.

\bibitem[{Chang et~al.(2015)Chang, Detmold, Orginos, Parreno, Savage, Tiburzi,
  and Beane}]{Chang:2015qxa}
Chang, E., W.~Detmold, K.~Orginos, A.~Parreno, M.~J. Savage, B.~C. Tiburzi, and
  S.~R. Beane (2015).
\newblock {Magnetic structure of light nuclei from lattice QCD}.
\newblock \emph{Phys. Rev. D}, \textbf{92}(11), p. 114502.
\newblock \doi{10.1103/PhysRevD.92.114502}.

\bibitem[{Choudhury and Lahiri(2015)}]{Choudhury:2014lna}
Choudhury, I.~D. and A.~Lahiri (2015).
\newblock {Anomalous chromomagnetic moment of quarks}.
\newblock \emph{Mod. Phys. Lett. A}, \textbf{30}(23), p. 1550113.
\newblock \doi{10.1142/S0217732315501138}.

\bibitem[{Chukhnova and Lobanov(2020)}]{Chukhnova:2019oum}
Chukhnova, A.~V. and A.~E. Lobanov (2020).
\newblock {Neutrino flavor oscillations and spin rotation in matter and
  electromagnetic field}.
\newblock \emph{Phys. Rev. D}, \textbf{101}(1), p. 013003.
\newblock \doi{10.1103/PhysRevD.101.013003}.

\bibitem[{Davis and Lineweaver(2004)}]{Davis:2003ad}
Davis, T.~M. and C.~H. Lineweaver (2004).
\newblock {Expanding confusion: common misconceptions of cosmological horizons
  and the superluminal expansion of the universe}.
\newblock \emph{Publ. Astron. Soc. Austral.}, \textbf{21}, p.~97.
\newblock \doi{10.1071/AS03040}.

\bibitem[{Delgado-Acosta et~al.(2015)Delgado-Acosta, Banda~Guzm\'an, and
  Kirchbach}]{DelgadoAcosta:2015ikk}
Delgado-Acosta, E.~G., V.~M. Banda~Guzm\'an, and M.~Kirchbach (2015).
\newblock {Gyromagnetic $g_s$ factors of the spin-1/2 particles in the
  (1/2$^+$-1/2$^-$-3/2$^-$) triad of the four-vector spinor, $\psi_\mu$,
  irreducibility and linearity}.
\newblock \emph{Int. J. Mod. Phys. E}, \textbf{24}(07), p. 1550060.
\newblock \doi{10.1142/S0218301315500603}.

\bibitem[{Delgado-Acosta et~al.(2011)Delgado-Acosta, Napsuciale, and
  Rodriguez}]{Delgado-Acosta:2010ita}
Delgado-Acosta, E.~G., M.~Napsuciale, and S.~Rodriguez (2011).
\newblock {Second order formalism for spin 1/2 fermions and Compton
  scattering}.
\newblock \emph{Phys. Rev. D}, \textbf{83}, p. 073001.
\newblock \doi{10.1103/PhysRevD.83.073001}.

\bibitem[{Detmold et~al.(2019)Detmold, Edwards, Dudek, Engelhardt, Lin, Meinel,
  Orginos, and Shanahan}]{Detmold:2019ghl}
Detmold, W., R.~G. Edwards, J.~J. Dudek, M.~Engelhardt, H.~W. Lin, S.~Meinel,
  K.~Orginos, and P.~Shanahan (2019).
\newblock {Hadrons and Nuclei}.
\newblock \emph{Eur. Phys. J. A}, \textbf{55}(11), p. 193.
\newblock \doi{10.1140/epja/i2019-12902-4}.

\bibitem[{Dixon(1970)}]{Dixon:1970zza}
Dixon, W.~G. (1970).
\newblock {Dynamics of extended bodies in general relativity. I. Momentum and
  angular momentum}.
\newblock \emph{Proc. Roy. Soc. Lond. A}, \textbf{314}, pp. 499--527.
\newblock \doi{10.1098/rspa.1970.0020}.

\bibitem[{Dunne(2014)}]{Dunne:2014qda}
Dunne, G.~V. (2014).
\newblock {Extreme quantum field theory and particle physics with IZEST}.
\newblock \emph{Eur. Phys. J. ST}, \textbf{223}(6), pp. 1055--1061.
\newblock \doi{10.1140/epjst/e2014-02156-4}.

\bibitem[{Durrer and Neronov(2013)}]{Durrer:2013pga}
Durrer, R. and A.~Neronov (2013).
\newblock Cosmological magnetic fields: their generation, evolution and
  observation.
\newblock \emph{The Astronomy and Astrophysics Review}, \textbf{21}, pp.
  1--109.
\newblock \doi{10.1007/s00159-013-0062-7}.

\bibitem[{Dvornikov(2019)}]{Dvornikov:2019sfo}
Dvornikov, M. (2019).
\newblock {Neutrino spin oscillations in external fields in curved spacetime}.
\newblock \emph{Phys. Rev. D}, \textbf{99}(11), p. 116021.
\newblock \doi{10.1103/PhysRevD.99.116021}.

\bibitem[{Ehrenfest(1927)}]{Ehrenfest:1927swx}
Ehrenfest, P. (1927).
\newblock {Bemerkung \"uber die angen\"aherte G\"ultigkeit der klassischen
  Mechanik innerhalb der Quantenmechanik}.
\newblock \emph{Z. Phys.}, \textbf{45}(7-8), pp. 455--457.
\newblock \doi{10.1007/BF01329203}.

\bibitem[{Eides et~al.(2001)Eides, Grotch, and Shelyuto}]{Eides:2000xc}
Eides, M.~I., H.~Grotch, and V.~A. Shelyuto (2001).
\newblock {Theory of light hydrogen - like atoms}.
\newblock \emph{Phys. Rept.}, \textbf{342}, pp. 63--261.
\newblock \doi{10.1016/S0370-1573(00)00077-6}.

\bibitem[{Elizalde et~al.(2004)Elizalde, Ferrer, and de~la
  Incera}]{Elizalde:2004mw}
Elizalde, E., E.~J. Ferrer, and V.~de~la Incera (2004).
\newblock {Neutrino propagation in a strongly magnetized medium}.
\newblock \emph{Phys. Rev. D}, \textbf{70}, p. 043012.
\newblock \doi{10.1103/PhysRevD.70.043012}.

\bibitem[{Elze et~al.(1980)Elze, Greiner, and Rafelski}]{Elze:1980er}
Elze, H.~T., W.~Greiner, and J.~Rafelski (1980).
\newblock {The relativistic Fermi gas revisited}.
\newblock \emph{J. Phys. G}, \textbf{6}, pp. L149--L153.
\newblock \doi{10.1088/0305-4616/6/9/003}.

\bibitem[{Espin(2015)}]{Espin:2015bja}
Espin, J. (2015).
\newblock \emph{{Second-order fermions}}.
\newblock Ph.D. thesis, Nottingham U.
\newblock \href{https://arxiv.org/abs/1509.05914}{arXiv:1509.05914}.

\bibitem[{Evans(2022)}]{Evans:2022ygl}
Evans, S. (2022).
\newblock \emph{{Nonperturbative Aspects in the QED Vacuum Related to Anomalous
  Magnetic Moment}}.
\newblock Ph.D. thesis, Arizona U.
\newblock URN/HDL: \href{https://hdl.handle.net/10150/665648}{10150/665648}.

\bibitem[{Evans and Rafelski(2018)}]{Evans:2018kor}
Evans, S. and J.~Rafelski (2018).
\newblock {Vacuum stabilized by anomalous magnetic moment}.
\newblock \emph{Phys. Rev. D}, \textbf{98}(1), p. 016006.
\newblock \doi{10.1103/PhysRevD.98.016006}.

\bibitem[{Evans and Rafelski(2022)}]{Evans:2022fsu}
Evans, S. and J.~Rafelski (2022).
\newblock {Emergence of periodic in magnetic moment effective QED action}.
\newblock \emph{Phys. Lett. B}, \textbf{831}, p. 137190.
\newblock \doi{10.1016/j.physletb.2022.137190}.

\bibitem[{Fedotov et~al.(2023)Fedotov, Ilderton, Karbstein, King, Seipt, Taya,
  and Torgrimsson}]{Fedotov:2022ely}
Fedotov, A., A.~Ilderton, F.~Karbstein, B.~King, D.~Seipt, H.~Taya, and
  G.~Torgrimsson (2023).
\newblock {Advances in QED with intense background fields}.
\newblock \emph{Phys. Rept.}, \textbf{1010}, pp. 1--138.
\newblock \doi{10.1016/j.physrep.2023.01.003}.

\bibitem[{Ferrara et~al.(1992)Ferrara, Porrati, and Telegdi}]{Ferrara:1992yc}
Ferrara, S., M.~Porrati, and V.~L. Telegdi (1992).
\newblock {$g=2$ as the natural value of the tree-level gyromagnetic ratio of
  elementary particles}.
\newblock \emph{Phys. Rev. D}, \textbf{46}, pp. 3529--3537.
\newblock \doi{10.1103/PhysRevD.46.3529}.

\bibitem[{Ferrer et~al.(2015)Ferrer, de~la Incera, Manreza~Paret,
  P\'erez~Mart\'\i{}nez, and Sanchez}]{Ferrer:2015wca}
Ferrer, E.~J., V.~de~la Incera, D.~Manreza~Paret, A.~P\'erez~Mart\'\i{}nez, and
  A.~Sanchez (2015).
\newblock {Insignificance of the anomalous magnetic moment of charged fermions
  for the equation of state of a magnetized and dense medium}.
\newblock \emph{Phys. Rev. D}, \textbf{91}(8), p. 085041.
\newblock \doi{10.1103/PhysRevD.91.085041}.

\bibitem[{Ferrer and Hackebill(2019)}]{Ferrer:2019xlr}
Ferrer, E.~J. and A.~Hackebill (2019).
\newblock {Thermodynamics of Neutrons in a Magnetic Field and its Implications
  for Neutron Stars}.
\newblock \emph{Phys. Rev. C}, \textbf{99}(6), p. 065803.
\newblock \doi{10.1103/PhysRevC.99.065803}.

\bibitem[{Ferrer and Hackebill(2023)}]{Ferrer:2023pgq}
Ferrer, E.~J. and A.~Hackebill (2023).
\newblock {The Importance of the Pressure Anisotropy Induced by Strong Magnetic
  Fields on Neutron Star Physics}.
\newblock \emph{J. Phys. Conf. Ser.}, \textbf{2536}(1), p. 012007.
\newblock \doi{10.1088/1742-6596/2536/1/012007}.

\bibitem[{Feynman(1951)}]{Feynman:1951gn}
Feynman, R.~P. (1951).
\newblock {An Operator calculus having applications in quantum
  electrodynamics}.
\newblock \emph{Phys. Rev.}, \textbf{84}, pp. 108--128.
\newblock \doi{10.1103/PhysRev.84.108}.

\bibitem[{Feynman and Gell-Mann(1958)}]{Feynman:1958ty}
Feynman, R.~P. and M.~Gell-Mann (1958).
\newblock {Theory of Fermi interaction}.
\newblock \emph{Phys. Rev.}, \textbf{109}, pp. 193--198.
\newblock \doi{10.1103/PhysRev.109.193}.

\bibitem[{Filho et~al.(2023)Filho, Hassanabadi, Reis, and
  Lisboa-Santos}]{Filho:2023lqe}
Filho, A. A.~A., H.~Hassanabadi, J.~A. A.~S. Reis, and L.~Lisboa-Santos (2023).
\newblock {Fermions with Electric Dipole Moment in curved spacetime}.
\newblock \emph{arXiv preprint}.
\newblock \href{https://arxiv.org/abs/2306.10897}{arXiv:2306.10897}.

\bibitem[{Fleming et~al.(2001)Fleming, Rothstein, and
  Leibovich}]{Fleming:2000ib}
Fleming, S., I.~Z. Rothstein, and A.~K. Leibovich (2001).
\newblock {Power counting and effective field theory for charmonium}.
\newblock \emph{Phys. Rev. D}, \textbf{64}, p. 036002.
\newblock \doi{10.1103/PhysRevD.64.036002}.

\bibitem[{Fock(1937)}]{Fock:1937dy}
Fock, V. (1937).
\newblock {Proper time in classical and quantum mechanics}.
\newblock \emph{Phys. Z. Sowjetunion}, \textbf{12}, pp. 404--425.
\newblock URL:
  \href{http://www.neo-classical-physics.info/uploads/3/4/3/6/34363841/fock_-_wkb_and_dirac.pdf}{http://www.neo-classical-physics.info/uploads/3/4/3/6/34363841/fock\_-\_wkb\_and\_dirac.pdf}.

\bibitem[{Foldy(1952)}]{Foldy:1952a}
Foldy, L.~L. (1952).
\newblock The Electromagnetic Properties of Dirac Particles.
\newblock \emph{Phys. Rev.}, \textbf{87}, pp. 688--693.
\newblock \doi{10.1103/PhysRev.87.688}.

\bibitem[{Foldy and Wouthuysen(1950)}]{Foldy:1949wa}
Foldy, L.~L. and S.~A. Wouthuysen (1950).
\newblock {On the Dirac theory of spin 1/2 particle and its nonrelativistic
  limit}.
\newblock \emph{Phys. Rev.}, \textbf{78}, pp. 29--36.
\newblock \doi{10.1103/PhysRev.78.29}.

\bibitem[{Formanek(2020)}]{Formanek:2020ojr}
Formanek, M. (2020).
\newblock \emph{{Revisiting Covariant Particle Dynamics}}.
\newblock Ph.D. thesis, Arizona U.
\newblock URN/HDL: \href{https://hdl.handle.net/10150/650804}{10150/650804}.

\bibitem[{Formanek et~al.(2018)Formanek, Evans, Rafelski, Steinmetz, and
  Yang}]{Formanek:2017mbv}
Formanek, M., S.~Evans, J.~Rafelski, A.~Steinmetz, and C.~T. Yang (2018).
\newblock {Strong fields and neutral particle magnetic moment dynamics}.
\newblock \emph{Plasma Phys. Control. Fusion}, \textbf{60}, p. 074006.
\newblock \doi{10.1088/1361-6587/aac06a}.

\bibitem[{Formanek et~al.(2021{\natexlab{a}})Formanek, Grayson, Rafelski, and
  M\"uller}]{Formanek:2021blc}
Formanek, M., C.~Grayson, J.~Rafelski, and B.~M\"uller (2021{\natexlab{a}}).
\newblock {Current-conserving relativistic linear response for collisional
  plasmas}.
\newblock \emph{Annals Phys.}, \textbf{434}, p. 168605.
\newblock \doi{10.1016/j.aop.2021.168605}.

\bibitem[{Formanek et~al.(2019)Formanek, Steinmetz, and
  Rafelski}]{Formanek:2019cga}
Formanek, M., A.~Steinmetz, and J.~Rafelski (2019).
\newblock {Classical neutral point particle in linearly polarized EM plane wave
  field}.
\newblock \emph{Plasma Phys. Control. Fusion}, \textbf{61}(8), p. 084006.
\newblock \doi{10.1088/1361-6587/ab242e}.

\bibitem[{Formanek et~al.(2020)Formanek, Steinmetz, and
  Rafelski}]{Formanek:2020zwc}
Formanek, M., A.~Steinmetz, and J.~Rafelski (2020).
\newblock {Radiation reaction friction: Resistive material medium}.
\newblock \emph{Phys. Rev. D}, \textbf{102}(5), p. 056015.
\newblock \doi{10.1103/PhysRevD.102.056015}.

\bibitem[{Formanek et~al.(2021{\natexlab{b}})Formanek, Steinmetz, and
  Rafelski}]{Formanek:2021mcp}
Formanek, M., A.~Steinmetz, and J.~Rafelski (2021{\natexlab{b}}).
\newblock {Motion of classical charged particles with magnetic moment in
  external plane-wave electromagnetic fields}.
\newblock \emph{Phys. Rev. A}, \textbf{103}(5), p. 052218.
\newblock \doi{10.1103/PhysRevA.103.052218}.

\bibitem[{Frenkel(1926)}]{Frenkel:1926zz}
Frenkel, J. (1926).
\newblock {Die Elektrodynamik des rotierenden Elektrons}.
\newblock \emph{Z. Phys.}, \textbf{37}, pp. 243--262.
\newblock \doi{10.1007/BF01397099}.

\bibitem[{Fritzsch and Xing(1996)}]{Fritzsch:1995dj}
Fritzsch, H. and Z.~Z. Xing (1996).
\newblock {Lepton mass hierarchy and neutrino oscillations}.
\newblock \emph{Phys. Lett. B}, \textbf{372}, pp. 265--270.
\newblock \doi{10.1016/0370-2693(96)00107-4}.

\bibitem[{Fritzsch and Xing(1998)}]{Fritzsch:1998xs}
Fritzsch, H. and Z.~Z. Xing (1998).
\newblock {Large leptonic flavor mixing and the mass spectrum of leptons}.
\newblock \emph{Phys. Lett. B}, \textbf{440}, pp. 313--318.
\newblock \doi{10.1016/S0370-2693(98)01106-X}.

\bibitem[{Fritzsch and Xing(2000)}]{Fritzsch:1999ee}
Fritzsch, H. and Z.~Z. Xing (2000).
\newblock {Mass and flavor mixing schemes of quarks and leptons}.
\newblock \emph{Prog. Part. Nucl. Phys.}, \textbf{45}, pp. 1--81.
\newblock \doi{10.1016/S0146-6410(00)00102-2}.

\bibitem[{Fromerth et~al.(2012)Fromerth, Kuznetsova, Labun, Letessier, and
  Rafelski}]{Fromerth:2012fe}
Fromerth, M.~J., I.~Kuznetsova, L.~Labun, J.~Letessier, and J.~Rafelski (2012).
\newblock {From Quark-Gluon Universe to Neutrino Decoupling: 200 \ensuremath{<}
  T \ensuremath{<} 2MeV}.
\newblock \emph{Acta Phys. Polon. B}, \textbf{43}(12), pp. 2261--2284.
\newblock \doi{10.5506/APhysPolB.43.2261}.

\bibitem[{Fujikawa and Shrock(1980)}]{Fujikawa:1980yx}
Fujikawa, K. and R.~E. Shrock (1980).
\newblock {The Magnetic Moment of a Massive Neutrino and Neutrino Spin
  Rotation}.
\newblock \emph{Phys. Rev. Lett.}, \textbf{45}, p. 963.
\newblock \doi{10.1103/PhysRevLett.45.963}.

\bibitem[{Gaensler et~al.(2004)Gaensler, Beck, and Feretti}]{Gaensler:2004gk}
Gaensler, B.~M., R.~Beck, and L.~Feretti (2004).
\newblock The origin and evolution of cosmic magnetism.
\newblock \emph{New Astronomy Reviews}, \textbf{48}(11-12), pp. 1003--1012.
\newblock \doi{10.1016/j.newar.2004.09.003}.

\bibitem[{Gerlach and Stern(1922)}]{Gerlach:1922zz}
Gerlach, W. and O.~Stern (1922).
\newblock {Das magnetische Moment des Silberatoms}.
\newblock \emph{Z. Phys.}, \textbf{9}, pp. 353--355.
\newblock \doi{10.1007/BF01326984}.

\bibitem[{Gesztesy et~al.(1985)Gesztesy, Simon, and Thaller}]{Gesztesy:1984hd}
Gesztesy, F., B.~Simon, and B.~Thaller (1985).
\newblock On the selfadjointness of Dirac operators with anomalous magnetic
  moment.
\newblock \emph{Proceedings of the American Mathematical Society},
  \textbf{94}(1), pp. 115--118.
\newblock \doi{10.2307/2044962}.

\bibitem[{Giovannini(2004)}]{Giovannini:2003yn}
Giovannini, M. (2004).
\newblock {The Magnetized universe}.
\newblock \emph{Int. J. Mod. Phys. D}, \textbf{13}, pp. 391--502.
\newblock \doi{10.1142/S0218271804004530}.

\bibitem[{Giovannini(2018)}]{Giovannini:2017rbc}
Giovannini, M. (2018).
\newblock {Probing large-scale magnetism with the Cosmic Microwave Background}.
\newblock \emph{Class. Quant. Grav.}, \textbf{35}(8), p. 084003.
\newblock \doi{10.1088/1361-6382/aab17d}.

\bibitem[{Giovannini(2023)}]{Giovannini:2022rrl}
Giovannini, M. (2023).
\newblock The scaling of primordial gauge fields.
\newblock \emph{Physics Letters B}, \textbf{842}, p. 137967.
\newblock ISSN 0370-2693.
\newblock \doi{10.1016/j.physletb.2023.137967}.

\bibitem[{Giunti and Kim(2007)}]{giunti2007fundamentals}
Giunti, C. and C.~W. Kim (2007).
\newblock \emph{Fundamentals of neutrino physics and astrophysics}.
\newblock Oxford university press.
\newblock \doi{10.1093/acprof:oso/9780198508717.001.0001}.

\bibitem[{Giunti et~al.(2016)Giunti, Kouzakov, Li, Lokhov, Studenikin, and
  Zhou}]{Giunti:2015gga}
Giunti, C., K.~A. Kouzakov, Y.~F. Li, A.~V. Lokhov, A.~Studenikin, and S.~Zhou
  (2016).
\newblock {Electromagnetic neutrinos in laboratory experiments and
  astrophysics}.
\newblock \emph{Annalen Phys.}, \textbf{528}, pp. 198--215.
\newblock \doi{10.1002/andp.201500211}.

\bibitem[{Giunti and Studenikin(2015)}]{Giunti:2014ixa}
Giunti, C. and A.~Studenikin (2015).
\newblock {Neutrino electromagnetic interactions: a window to new physics}.
\newblock \emph{Rev. Mod. Phys.}, \textbf{87}, p. 531.
\newblock \doi{10.1103/RevModPhys.87.531}.

\bibitem[{Gonoskov et~al.(2022)Gonoskov, Blackburn, Marklund, and
  Bulanov}]{Gonoskov:2021hwf}
Gonoskov, A., T.~G. Blackburn, M.~Marklund, and S.~S. Bulanov (2022).
\newblock {Charged particle motion and radiation in strong electromagnetic
  fields}.
\newblock \emph{Rev. Mod. Phys.}, \textbf{94}(4), p. 045001.
\newblock \doi{10.1103/RevModPhys.94.045001}.

\bibitem[{Gopal and Sethi(2005)}]{Gopal:2004ut}
Gopal, R. and S.~Sethi (2005).
\newblock {Generation of magnetic field in the pre-recombination era}.
\newblock \emph{Mon. Not. Roy. Astron. Soc.}, \textbf{363}, pp. 521--528.
\newblock \doi{10.1111/j.1365-2966.2005.09442.x}.

\bibitem[{Grasso and Rubinstein(2001)}]{Grasso:2000wj}
Grasso, D. and H.~R. Rubinstein (2001).
\newblock {Magnetic fields in the early universe}.
\newblock \emph{Phys. Rept.}, \textbf{348}, pp. 163--266.
\newblock \doi{10.1016/S0370-1573(00)00110-1}.

\bibitem[{Grayson et~al.(2022)Grayson, Formanek, Rafelski, and
  Mueller}]{Grayson:2022asf}
Grayson, C., M.~Formanek, J.~Rafelski, and B.~Mueller (2022).
\newblock {Dynamic magnetic response of the quark-gluon plasma to
  electromagnetic fields}.
\newblock \emph{Phys. Rev. D}, \textbf{106}(1), p. 014011.
\newblock \doi{10.1103/PhysRevD.106.014011}.

\bibitem[{Grayson et~al.(2023)Grayson, Yang, Formanek, and
  Rafelski}]{Grayson:2023flr}
Grayson, C., C.~T. Yang, M.~Formanek, and J.~Rafelski (2023).
\newblock {Electron\textendash{}positron plasma in BBN: Damped-dynamic
  screening}.
\newblock \emph{Annals Phys.}, \textbf{458}, p. 169453.
\newblock \doi{10.1016/j.aop.2023.169453}.

\bibitem[{Green et~al.(2015)Green, Meinel, Engelhardt, Krieg, Laeuchli, Negele,
  Orginos, Pochinsky, and Syritsyn}]{Green:2015wqa}
Green, J., S.~Meinel, M.~Engelhardt, S.~Krieg, J.~Laeuchli, J.~Negele,
  K.~Orginos, A.~Pochinsky, and S.~Syritsyn (2015).
\newblock {High-precision calculation of the strange nucleon electromagnetic
  form factors}.
\newblock \emph{Phys. Rev. D}, \textbf{92}(3), p. 031501.
\newblock \doi{10.1103/PhysRevD.92.031501}.

\bibitem[{Greiner and M{\"u}ller(2009)}]{greiner2009gauge}
Greiner, W. and B.~M{\"u}ller (2009).
\newblock \emph{Gauge theory of weak interactions}.
\newblock Springer.
\newblock \doi{10.1007/978-3-540-87843-8}.

\bibitem[{Greiner and M{\"u}ller(2012)}]{greiner2012quantum}
Greiner, W. and B.~M{\"u}ller (2012).
\newblock \emph{Quantum mechanics: symmetries}.
\newblock Springer Science \& Business Media.
\newblock \doi{10.1007/978-3-642-57976-9}.
\newblock [orig. pub. 1984].

\bibitem[{Greiner et~al.(2012{\natexlab{a}})Greiner, Muller, and
  Rafelski}]{Greiner:1985ce}
Greiner, W., B.~Muller, and J.~Rafelski (2012{\natexlab{a}}).
\newblock \emph{{Quantum electrodynamics of strong fields}}.
\newblock Springer Berlin, Heidelberg.
\newblock \doi{10.1007/978-3-642-82272-8}.
\newblock [orig. pub. 1985].

\bibitem[{Greiner et~al.(2012{\natexlab{b}})Greiner, Neise, and
  St{\"o}cker}]{greiner2012thermodynamics}
Greiner, W., L.~Neise, and H.~St{\"o}cker (2012{\natexlab{b}}).
\newblock \emph{Thermodynamics and statistical mechanics}.
\newblock Springer Science \& Business Media.
\newblock \doi{10.1007/978-1-4612-0827-3}.
\newblock [orig. pub. 1995].

\bibitem[{Greiner and Reinhardt(2008)}]{greiner2008quantum}
Greiner, W. and J.~Reinhardt (2008).
\newblock \emph{Quantum electrodynamics}.
\newblock Springer Science \& Business Media.
\newblock \doi{10.1007/978-3-540-87561-1}.

\bibitem[{Greiner et~al.(2006)Greiner, Schramm, and Stein}]{greiner2006qcd}
Greiner, W., S.~Schramm, and E.~Stein (2006).
\newblock \emph{Quantum chromodynamics}.
\newblock Springer Berlin, Heidelberg.
\newblock \doi{10.1007/978-3-540-48535-3}.
\newblock [orig. pub. 1989].

\bibitem[{Hackebill(2022)}]{Hackebill:2022uxv}
Hackebill, A.~A. (2022).
\newblock \emph{{Magnetic Field Effects on the Physics of Neutron Stars}}.
\newblock Ph.D. thesis, CUNY, Academic Works.
\newblock URL:
  \href{https://academicworks.cuny.edu/gc_etds/5030}{https://academicworks.cuny.edu/gc\_etds/5030}.

\bibitem[{Haro et~al.(2023)Haro, Dickinson, Finkelstein
  et~al.}]{arrabal2023spectroscopic}
Haro, P.~A., M.~Dickinson, S.~L. Finkelstein, et~al. (2023).
\newblock Confirmation and refutation of very luminous galaxies in the early
  universe.
\newblock \emph{Nature}.
\newblock \doi{10.1038/s41586-023-06521-7}.

\bibitem[{Harrison and Scott(2000)}]{Harrison:1999df}
Harrison, P.~F. and W.~G. Scott (2000).
\newblock {CP and T violation in neutrino oscillations and invariance of
  Jarlskog's determinant to matter effects}.
\newblock \emph{Phys. Lett. B}, \textbf{476}, pp. 349--355.
\newblock \doi{10.1016/S0370-2693(00)00153-2}.

\bibitem[{Hegelich et~al.(2014)Hegelich, Mourou, and
  Rafelski}]{Hegelich:2014tda}
Hegelich, B.~M., G.~Mourou, and J.~Rafelski (2014).
\newblock {Probing the quantum vacuum with ultra intense laser pulses}.
\newblock \emph{Eur. Phys. J. ST}, \textbf{223}(6), pp. 1093--1104.
\newblock \doi{10.1140/epjst/e2014-02160-8}.

\bibitem[{Hewett et~al.(2012)}]{Proceedings:2012ulb}
Hewett, J.~L. et~al. (2012).
\newblock {Fundamental Physics at the Intensity Frontier}.
\newblock In \emph{Workshop on Fundamental Physics at the Intensity Frontier}.
\newblock \doi{10.2172/1042577}.

\bibitem[{Holstein and Scherer(2014)}]{Holstein:2013kia}
Holstein, B.~R. and S.~Scherer (2014).
\newblock {Hadron Polarizabilities}.
\newblock \emph{Ann. Rev. Nucl. Part. Sci.}, \textbf{64}, pp. 51--81.
\newblock \doi{10.1146/annurev-nucl-102313-025555}.

\bibitem[{Huber et~al.(2022)}]{Huber:2022lpm}
Huber, P. et~al. (2022).
\newblock {Snowmass Neutrino Frontier Report}.
\newblock In \emph{{Snowmass 2021}}.
\newblock \href{https://arxiv.org/abs/2211.08641}{arXiv:2211.08641}.

\bibitem[{Itzykson and Zuber(1980)}]{Itzykson:1980rh}
Itzykson, C. and J.~B. Zuber (1980).
\newblock \emph{{Quantum Field Theory}}.
\newblock International Series In Pure and Applied Physics. McGraw-Hill, New
  York.
\newblock ISBN 978-0-486-44568-7.

\bibitem[{Jancovici(1969)}]{Jancovici:1969exc}
Jancovici, B. (1969).
\newblock {Radiative correction to the ground-state energy of an electron in an
  intense magnetic field}.
\newblock \emph{Phys. Rev.}, \textbf{187}, pp. 2275--2276.
\newblock \doi{10.1103/PhysRev.187.2275}.

\bibitem[{Jarlskog(1985{\natexlab{a}})}]{Jarlskog:1985cw}
Jarlskog, C. (1985{\natexlab{a}}).
\newblock {A Basis Independent Formulation of the Connection Between Quark Mass
  Matrices, CP Violation and Experiment}.
\newblock \emph{Z. Phys. C}, \textbf{29}, pp. 491--497.
\newblock \doi{10.1007/BF01565198}.

\bibitem[{Jarlskog(1985{\natexlab{b}})}]{Jarlskog:1985ht}
Jarlskog, C. (1985{\natexlab{b}}).
\newblock {Commutator of the Quark Mass Matrices in the Standard Electroweak
  Model and a Measure of Maximal $CP$~Nonconservation}.
\newblock \emph{Phys. Rev. Lett.}, \textbf{55}, p. 1039.
\newblock \doi{10.1103/PhysRevLett.55.1039}.

\bibitem[{Jarlskog(2005)}]{Jarlskog:2004be}
Jarlskog, C. (2005).
\newblock {Invariants of lepton mass matrices and CP and T violation in
  neutrino oscillations}.
\newblock \emph{Phys. Lett. B}, \textbf{609}, pp. 323--329.
\newblock \doi{10.1016/j.physletb.2005.01.057}.

\bibitem[{Jedamzik and Abel(2013)}]{Jedamzik:2013gua}
Jedamzik, K. and T.~Abel (2013).
\newblock {Small-scale primordial magnetic fields and anisotropies in the
  cosmic microwave background radiation}.
\newblock \emph{JCAP}, \textbf{10}, p. 050.
\newblock \doi{10.1088/1475-7516/2013/10/050}.

\bibitem[{Jedamzik and Pogosian(2020)}]{Jedamzik:2020krr}
Jedamzik, K. and L.~Pogosian (2020).
\newblock Relieving the Hubble tension with primordial magnetic fields.
\newblock \emph{Physical Review Letters}, \textbf{125}(18), p. 181302.
\newblock \doi{10.1103/PhysRevLett.125.181302}.

\bibitem[{Jedamzik and Saveliev(2019)}]{Jedamzik:2018itu}
Jedamzik, K. and A.~Saveliev (2019).
\newblock Stringent limit on primordial magnetic fields from the cosmic
  microwave background radiation.
\newblock \emph{Physical review letters}, \textbf{123}(2), p. 021301.
\newblock \doi{10.1103/PhysRevLett.123.021301}.

\bibitem[{Jegerlehner(2017)}]{Jegerlehner:2017gek}
Jegerlehner, F. (2017).
\newblock \emph{{The Anomalous Magnetic Moment of the Muon}}, volume 274.
\newblock Springer, Cham.
\newblock \doi{10.1007/978-3-319-63577-4}.

\bibitem[{Jentschura and Pachucki(1996)}]{Jentschura:1996zz}
Jentschura, U. and K.~Pachucki (1996).
\newblock {Higher-order binding corrections to the Lamb shift of P-2 states}.
\newblock \emph{Phys. Rev. A}, \textbf{54}, pp. 1853--1861.
\newblock \doi{10.1103/PhysRevA.54.1853}.

\bibitem[{Kaluza(1921)}]{Kaluza:1921tu}
Kaluza, T. (1921).
\newblock Zum Unit{\"a}tsproblem der Physik.
\newblock \emph{Sitzungsber. Preuss. Akad. Wiss. Berlin (Math. Phys. )},
  \textbf{1921}, pp. 966--972.
\newblock \doi{10.1142/S0218271818700017}.
\newblock \href{https://arxiv.org/abs/1803.08616}{arXiv:1803.08616}.

\bibitem[{Kaspi and Beloborodov(2017)}]{Kaspi:2017fwg}
Kaspi, V.~M. and A.~Beloborodov (2017).
\newblock {Magnetars}.
\newblock \emph{Ann. Rev. Astron. Astrophys.}, \textbf{55}, pp. 261--301.
\newblock \doi{10.1146/annurev-astro-081915-023329}.

\bibitem[{Klein(1926)}]{Klein:1926tv}
Klein, O. (1926).
\newblock {Quantentheorie und f{\"u}nfdimensionale Relativit{\"a}tstheorie }.
\newblock \emph{Z. Phys.}, \textbf{37}, pp. 895--906.
\newblock \doi{10.1007/BF01397481}.

\bibitem[{Knecht(2004)}]{Knecht:2003kc}
Knecht, M. (2004).
\newblock {The Anomalous magnetic moment of the muon: A Theoretical
  introduction}.
\newblock \emph{Lect. Notes Phys.}, \textbf{629}, pp. 37--84.
\newblock \doi{10.1007/978-3-540-44457-2_2}.

\bibitem[{Kronberg(1994)}]{Kronberg:1993vk}
Kronberg, P.~P. (1994).
\newblock Extragalactic magnetic fields.
\newblock \emph{Reports on Progress in Physics}, \textbf{57}(4), p. 325.
\newblock \doi{10.1088/0034-4885/57/4/001}.

\bibitem[{Labun and Rafelski(2009)}]{Labun:2008re}
Labun, L. and J.~Rafelski (2009).
\newblock {Vacuum Decay Time in Strong External Fields}.
\newblock \emph{Phys. Rev. D}, \textbf{79}, p. 057901.
\newblock \doi{10.1103/PhysRevD.79.057901}.

\bibitem[{Labun and Rafelski(2012)}]{Labun:2012ra}
Labun, L. and J.~Rafelski (2012).
\newblock {Higgs two-gluon decay and the top-quark chromomagnetic moment}.
\newblock \emph{arXiv preprint}.
\newblock \doi{10.48550/arXiv.1210.3150}.

\bibitem[{Labun and Rafelski(2013)}]{Labun:2012fg}
Labun, L. and J.~Rafelski (2013).
\newblock {Top anomalous magnetic moment and the two photon decay of Higgs
  boson}.
\newblock \emph{Phys. Rev. D}, \textbf{88}, p. 071301.
\newblock \doi{10.1103/PhysRevD.88.071301}.

\bibitem[{Larson et~al.(2023)Larson, Finkelstein, Kocevski, Hutchison, Trump,
  Haro, Bromm, Cleri, Dickinson, Fujimoto et~al.}]{CEERSTeam:2023qgy}
Larson, R.~L., S.~L. Finkelstein, D.~D. Kocevski, T.~A. Hutchison, J.~R. Trump,
  P.~A. Haro, V.~Bromm, N.~J. Cleri, M.~Dickinson, S.~Fujimoto, et~al. (2023).
\newblock A CEERS Discovery of an Accreting Supermassive Black Hole 570 Myr
  after the Big Bang: Identifying a Progenitor of Massive $z>6$ Quasars.
\newblock \emph{arXiv preprint}.
\newblock \doi{10.48550/arXiv.2303.08918}.
\newblock [submitted to ApJ].

\bibitem[{Letessier and Rafelski(2023)}]{Letessier:2002ony}
Letessier, J. and J.~Rafelski (2023).
\newblock \emph{Hadrons and Quark–Gluon Plasma}.
\newblock Cambridge Monographs on Particle Physics, Nuclear Physics and
  Cosmology. Cambridge University Press.
\newblock \doi{10.1017/9781009290753}.
\newblock \emph{Open access.} [orig. pub. 2002].

\bibitem[{Lichkunov et~al.(2020)Lichkunov, Popov, and
  Studenikin}]{Lichkunov:2020zzx}
Lichkunov, A., A.~Popov, and A.~Studenikin (2020).
\newblock {Neutrino eigenstates and flavour, spin and spin-flavour oscillations
  in a constant magnetic field}.
\newblock In \emph{{2019 European Physical Society Conference on High Energy
  Physics}}.
\newblock \href{https://arxiv.org/abs/2012.06880}{arXiv:2012.06880}.

\bibitem[{Lim and Marciano(1988)}]{Lim:1987tk}
Lim, C.~S. and W.~J. Marciano (1988).
\newblock {Resonant Spin - Flavor Precession of Solar and Supernova Neutrinos}.
\newblock \emph{Phys. Rev. D}, \textbf{37}, pp. 1368--1373.
\newblock \doi{10.1103/PhysRevD.37.1368}.

\bibitem[{Martin and Glauber(1958)}]{Martin:1958zz}
Martin, P.~C. and R.~J. Glauber (1958).
\newblock {Relativistic Theory of Radiative Orbital Electron Capture}.
\newblock \emph{Phys. Rev.}, \textbf{109}, pp. 1307--1325.
\newblock \doi{10.1103/PhysRev.109.1307}.

\bibitem[{Mart\'\i{}nez-Mirav\'e(2023)}]{Martinez-Mirave:2023fyb}
Mart\'\i{}nez-Mirav\'e, P. (2023).
\newblock \emph{{Neutrino properties from the laboratory and the cosmos}}.
\newblock Ph.D. thesis, Valencia U.
\newblock \doi{10.48550/arXiv.2309.15446}.
\newblock URN/HDL:
  \href{https://roderic.uv.es/handle/10550/89441}{10550/89441}.

\bibitem[{Mathisson(1937)}]{Mathisson:1937zz}
Mathisson, M. (1937).
\newblock {Neue mechanik materieller systemes}.
\newblock \emph{Acta Phys. Polon.}, \textbf{6}, pp. 163--200.

\bibitem[{Mathisson(2010)}]{Mathisson:2010opl}
Mathisson, M. (2010).
\newblock {Republication of: New mechanics of material systems}.
\newblock \emph{Gen. Rel. Grav.}, \textbf{42}(4), pp. 1011--1048.
\newblock \doi{10.1007/s10714-010-0939-y}.

\bibitem[{Melrose(2013)}]{melrose2008quantum}
Melrose, D. (2013).
\newblock \emph{Quantum plasmadynamics: Magnetized plasmas}.
\newblock Springer.
\newblock \doi{10.1007/978-1-4614-4045-1}.

\bibitem[{Mikheev and Smirnov(1986)}]{Mikheev:1986wj}
Mikheev, S.~P. and A.~Y. Smirnov (1986).
\newblock {Resonant amplification of neutrino oscillations in matter and solar
  neutrino spectroscopy}.
\newblock \emph{Nuovo Cim. C}, \textbf{9}, pp. 17--26.
\newblock \doi{10.1007/BF02508049}.

\bibitem[{Mikheyev and Smirnov(1985)}]{Mikheyev:1985zog}
Mikheyev, S.~P. and A.~Y. Smirnov (1985).
\newblock {Resonance Amplification of Oscillations in Matter and Spectroscopy
  of Solar Neutrinos}.
\newblock \emph{Sov. J. Nucl. Phys.}, \textbf{42}, pp. 913--917.

\bibitem[{Morgan(1995)}]{Morgan:1995te}
Morgan, A.~G. (1995).
\newblock {Second order fermions in gauge theories}.
\newblock \emph{Phys. Lett. B}, \textbf{351}, pp. 249--256.
\newblock \doi{10.1016/0370-2693(95)00377-W}.

\bibitem[{Neronov and Vovk(2010)}]{Neronov:2010gir}
Neronov, A. and I.~Vovk (2010).
\newblock Evidence for strong extragalactic magnetic fields from Fermi
  observations of TeV blazars.
\newblock \emph{Science}, \textbf{328}(5974), pp. 73--75.
\newblock \doi{10.1126/science.1184192}.

\bibitem[{Niederle and Nikitin(2006)}]{Niederle:2004bx}
Niederle, J. and A.~G. Nikitin (2006).
\newblock {Relativistic Coulomb problem for particles with arbitrary
  half-integer spin}.
\newblock \emph{J. Phys. A}, \textbf{39}, pp. 10931--10944.
\newblock \doi{10.1088/0305-4470/39/34/023}.

\bibitem[{Nieves(1982)}]{Nieves:1981zt}
Nieves, J.~F. (1982).
\newblock {Electromagnetic Properties of Majorana Neutrinos}.
\newblock \emph{Phys. Rev. D}, \textbf{26}, p. 3152.
\newblock \doi{10.1103/PhysRevD.26.3152}.

\bibitem[{Ohanian(1986)}]{Ohanian:1986wg}
Ohanian, H.~C. (1986).
\newblock What is spin?
\newblock \emph{American Journal of Physics}, \textbf{54}(6), pp. 500--505.
\newblock \doi{10.1119/1.14580}.

\bibitem[{Ohlsson(2012)}]{Ohlsson:2011zz}
Ohlsson, T. (2012).
\newblock \emph{{Relativistic quantum physics: From advanced quantum mechanics
  to introductory quantum field theory}}.
\newblock Cambridge University Press.
\newblock ISBN 978-1-139-21072-0, 978-0-521-76726-2.
\newblock \doi{10.1017/CBO9781139032681}.

\bibitem[{Ortin(2015)}]{Ortin:2015hya}
Ortin, T. (2015).
\newblock \emph{{Gravity and Strings}}.
\newblock Cambridge Monographs on Mathematical Physics. Cambridge University
  Press, 2nd ed. edition.
\newblock ISBN 978-0-521-76813-9, 978-0-521-76813-9, 978-1-316-23579-9.
\newblock \doi{10.1017/CBO9781139019750}.

\bibitem[{Overduin and Wesson(1997)}]{Overduin:1997sri}
Overduin, J.~M. and P.~S. Wesson (1997).
\newblock {Kaluza-Klein gravity}.
\newblock \emph{Phys. Rept.}, \textbf{283}, pp. 303--380.
\newblock \doi{10.1016/S0370-1573(96)00046-4}.

\bibitem[{Pacetti et~al.(2015)Pacetti, Baldini~Ferroli, and
  Tomasi-Gustafsson}]{Pacetti:2014jai}
Pacetti, S., R.~Baldini~Ferroli, and E.~Tomasi-Gustafsson (2015).
\newblock {Proton electromagnetic form factors: Basic notions, present
  achievements and future perspectives}.
\newblock \emph{Phys. Rept.}, \textbf{550-551}, pp. 1--103.
\newblock \doi{10.1016/j.physrep.2014.09.005}.

\bibitem[{Pal(1992)}]{Pal:1991pm}
Pal, P.~B. (1992).
\newblock {Particle physics confronts the solar neutrino problem}.
\newblock \emph{Int. J. Mod. Phys. A}, \textbf{7}, pp. 5387--5460.
\newblock \doi{10.1142/S0217751X92002465}.

\bibitem[{Papapetrou(1951)}]{Papapetrou:1951pa}
Papapetrou, A. (1951).
\newblock {Spinning test particles in general relativity. 1.}
\newblock \emph{Proc. Roy. Soc. Lond. A}, \textbf{209}, pp. 248--258.
\newblock \doi{10.1098/rspa.1951.0200}.

\bibitem[{Pehlivan et~al.(2014)Pehlivan, Balantekin, and
  Kajino}]{Pehlivan:2014zua}
Pehlivan, Y., A.~B. Balantekin, and T.~Kajino (2014).
\newblock {Neutrino Magnetic Moment, CP Violation and Flavor Oscillations in
  Matter}.
\newblock \emph{Phys. Rev. D}, \textbf{90}(6), p. 065011.
\newblock \doi{10.1103/PhysRevD.90.065011}.

\bibitem[{Perrone et~al.(2021)Perrone, Gregori, Reville, Silva, and
  Bingham}]{Perrone:2021srr}
Perrone, L.~M., G.~Gregori, B.~Reville, L.~O. Silva, and R.~Bingham (2021).
\newblock {Neutrino-electron magnetohydrodynamics in an expanding universe}.
\newblock \emph{Phys. Rev. D}, \textbf{104}(12), p. 123013.
\newblock \doi{10.1103/PhysRevD.104.123013}.

\bibitem[{{Pomakov} et~al.(2022){Pomakov}, {O'Sullivan}, {Br{\"u}ggen},
  {Vazza}, {Carretti}, {Heald}, {Horellou}, {Shimwell}, {Shulevski}, and
  {Vernstrom}}]{Pomakov:2022cem}
{Pomakov}, V.~P., S.~P. {O'Sullivan}, M.~{Br{\"u}ggen}, F.~{Vazza},
  E.~{Carretti}, G.~H. {Heald}, C.~{Horellou}, T.~{Shimwell}, A.~{Shulevski},
  and T.~{Vernstrom} (2022).
\newblock {The redshift evolution of extragalactic magnetic fields}.
\newblock \emph{Monthly Notices of the Royal Astronomical Society},
  \textbf{515}(1), pp. 256--270.
\newblock \doi{10.1093/mnras/stac1805}.

\bibitem[{Popov and Studenikin(2019)}]{Popov:2019nkr}
Popov, A. and A.~Studenikin (2019).
\newblock {Neutrino eigenstates and flavour, spin and spin-flavour oscillations
  in a constant magnetic field}.
\newblock \emph{Eur. Phys. J. C}, \textbf{79}(2), p. 144.
\newblock \doi{10.1140/epjc/s10052-019-6657-z}.

\bibitem[{Pshirkov et~al.(2016)Pshirkov, Tinyakov, and
  Urban}]{Pshirkov:2015tua}
Pshirkov, M.~S., P.~G. Tinyakov, and F.~R. Urban (2016).
\newblock {New limits on extragalactic magnetic fields from rotation measures}.
\newblock \emph{Phys. Rev. Lett.}, \textbf{116}(19), p. 191302.
\newblock \doi{10.1103/PhysRevLett.116.191302}.

\bibitem[{Rafelski and Birrell(2014)}]{Rafelski:2013yka}
Rafelski, J. and J.~Birrell (2014).
\newblock {Traveling Through the Universe: Back in Time to the Quark-Gluon
  Plasma Era}.
\newblock \emph{J. Phys. Conf. Ser.}, \textbf{509}, p. 012014.
\newblock \doi{10.1088/1742-6596/509/1/012014}.

\bibitem[{Rafelski et~al.(2023{\natexlab{a}})Rafelski, Birrell, Steinmetz, and
  Yang}]{Rafelski:2023emw}
Rafelski, J., J.~Birrell, A.~Steinmetz, and C.~T. Yang (2023{\natexlab{a}}).
\newblock {A Short Survey of Matter-Antimatter Evolution in the Primordial
  Universe}.
\newblock \emph{Universe}, \textbf{9}(7), p. 309.
\newblock \doi{10.3390/universe9070309}.

\bibitem[{Rafelski et~al.(2023{\natexlab{b}})Rafelski, Evans, and
  Labun}]{Rafelski:2022bsv}
Rafelski, J., S.~Evans, and L.~Labun (2023{\natexlab{b}}).
\newblock {Study of QED singular properties for variable gyromagnetic ratio
  $g\simeq 2$}.
\newblock \emph{Phys. Rev. D}, \textbf{107}.
\newblock \doi{10.1103/PhysRevD.107.076002}.

\bibitem[{Rafelski et~al.(2018)Rafelski, Formanek, and
  Steinmetz}]{Rafelski:2017hce}
Rafelski, J., M.~Formanek, and A.~Steinmetz (2018).
\newblock {Relativistic Dynamics of Point Magnetic Moment}.
\newblock \emph{Eur. Phys. J. C}, \textbf{78}(1), p.~6.
\newblock \doi{10.1140/epjc/s10052-017-5493-2}.

\bibitem[{Rafelski et~al.(1978)Rafelski, Fulcher, and Klein}]{Rafelski:1976ts}
Rafelski, J., L.~P. Fulcher, and A.~Klein (1978).
\newblock {Fermions and Bosons Interacting with Arbitrarily Strong External
  Fields}.
\newblock \emph{Phys. Rept.}, \textbf{38}, pp. 227--361.
\newblock \doi{10.1016/0370-1573(78)90116-3}.

\bibitem[{Rafelski et~al.(2017)Rafelski, Kirsch, M\"uller, Reinhardt, and
  Greiner}]{Rafelski:2016ixr}
Rafelski, J., J.~Kirsch, B.~M\"uller, J.~Reinhardt, and W.~Greiner (2017).
\newblock {Probing QED Vacuum with Heavy Ions}.
\newblock \emph{FIAS Interdisc. Sci. Ser.}, pp. 211--251.
\newblock \doi{10.1007/978-3-319-44165-8_17}.

\bibitem[{Rafelski et~al.(2023{\natexlab{c}})Rafelski, Steinmetz, and
  Yang}]{Rafelski:2023zgp}
Rafelski, J., A.~Steinmetz, and C.~T. Yang (2023{\natexlab{c}}).
\newblock {Dynamic fermion flavor mixing through transition dipole moments}.
\newblock \emph{arXiv preprint}.
\newblock \doi{10.48550/arXiv.2309.15797}.
\newblock [submitted to Int. Journal of Mod. Phys. A].

\bibitem[{Roussy et~al.(2023)}]{Roussy:2022cmp}
Roussy, T.~S. et~al. (2023).
\newblock {An improved bound on the electron\textquoteright{}s electric dipole
  moment}.
\newblock \emph{Science}, \textbf{381}(6653), p. adg4084.
\newblock \doi{10.1126/science.adg4084}.

\bibitem[{Ruffini and Vereshchagin(2013)}]{Ruffini:2012it}
Ruffini, R. and G.~Vereshchagin (2013).
\newblock {Electron-positron plasma in GRBs and in cosmology}.
\newblock \emph{Nuovo Cim. C}, \textbf{036}(s01), pp. 255--266.
\newblock \doi{10.1393/ncc/i2013-11500-0}.

\bibitem[{Ruffini et~al.(2010)Ruffini, Vereshchagin, and Xue}]{Ruffini:2009hg}
Ruffini, R., G.~Vereshchagin, and S.~S. Xue (2010).
\newblock {Electron-positron pairs in physics and astrophysics: from heavy
  nuclei to black holes}.
\newblock \emph{Phys. Rept.}, \textbf{487}, pp. 1--140.
\newblock \doi{10.1016/j.physrep.2009.10.004}.

\bibitem[{Safronova et~al.(2018)Safronova, Budker, DeMille, Kimball,
  Derevianko, and Clark}]{Safronova:2017xyt}
Safronova, M.~S., D.~Budker, D.~DeMille, D.~F.~J. Kimball, A.~Derevianko, and
  C.~W. Clark (2018).
\newblock {Search for New Physics with Atoms and Molecules}.
\newblock \emph{Rev. Mod. Phys.}, \textbf{90}(2), p. 025008.
\newblock \doi{10.1103/RevModPhys.90.025008}.

\bibitem[{Sainte-Marie(2023)}]{Sainte-Marie:2023aqn}
Sainte-Marie, A. (2023).
\newblock \emph{{Strong-field Quantum Electrodynamics in the extremely intense
  light of relativistic plasma mirrors}}.
\newblock Ph.D. thesis, universit\'e Paris-Saclay.
\newblock URL:
  \href{https://theses.hal.science/tel-04057676}{https://theses.hal.science/tel-04057676}.

\bibitem[{Sajjad~Athar et~al.(2022)}]{SajjadAthar:2021prg}
Sajjad~Athar, M. et~al. (2022).
\newblock {Status and perspectives of neutrino physics}.
\newblock \emph{Prog. Part. Nucl. Phys.}, \textbf{124}, p. 103947.
\newblock \doi{10.1016/j.ppnp.2022.103947}.

\bibitem[{Sakurai(1967)}]{sakurai1967advanced}
Sakurai, J.~J. (1967).
\newblock \emph{Advanced quantum mechanics}.
\newblock Pearson Education India.

\bibitem[{Schechter and Valle(1981)}]{Schechter:1981hw}
Schechter, J. and J.~W.~F. Valle (1981).
\newblock {Majorana Neutrinos and Magnetic Fields}.
\newblock \emph{Phys. Rev. D}, \textbf{24}, pp. 1883--1889.
\newblock \doi{10.1103/PhysRevD.25.283}.
\newblock [Erratum: Phys.Rev.D 25, 283 (1982)].

\bibitem[{Schwartz(2014)}]{Schwartz:2014sze}
Schwartz, M.~D. (2014).
\newblock \emph{Quantum Field Theory and the Standard Model}.
\newblock Cambridge University Press.
\newblock \doi{10.1017/9781139540940}.

\bibitem[{Schwinger(1974)}]{schwinger1974spin}
Schwinger, J. (1974).
\newblock Spin precession—a dynamical discussion.
\newblock \emph{American Journal of Physics}, \textbf{42}(6), pp. 510--513.
\newblock \doi{10.1119/1.1987764}.

\bibitem[{Schwinger(1951)}]{Schwinger:1951nm}
Schwinger, J.~S. (1951).
\newblock {On gauge invariance and vacuum polarization}.
\newblock \emph{Phys. Rev.}, \textbf{82}, pp. 664--679.
\newblock \doi{10.1103/PhysRev.82.664}.

\bibitem[{Shrock(1980)}]{Shrock:1980vy}
Shrock, R.~E. (1980).
\newblock {New Tests For, and Bounds On, Neutrino Masses and Lepton Mixing}.
\newblock \emph{Phys. Lett. B}, \textbf{96}, pp. 159--164.
\newblock \doi{10.1016/0370-2693(80)90235-X}.

\bibitem[{Shrock(1982)}]{Shrock:1982sc}
Shrock, R.~E. (1982).
\newblock {Electromagnetic Properties and Decays of Dirac and Majorana
  Neutrinos in a General Class of Gauge Theories}.
\newblock \emph{Nucl. Phys. B}, \textbf{206}, pp. 359--379.
\newblock \doi{10.1016/0550-3213(82)90273-5}.

\bibitem[{Smirnov(2003)}]{Smirnov:2003da}
Smirnov, A.~Y. (2003).
\newblock {The MSW effect and solar neutrinos}.
\newblock In \emph{{10th International Workshop on Neutrino Telescopes}}, pp.
  23--43.
\newblock \doi{10.48550/arXiv.hep-ph/0305106}.

\bibitem[{Steinmetz et~al.(2019)Steinmetz, Formanek, and
  Rafelski}]{Steinmetz:2018ryf}
Steinmetz, A., M.~Formanek, and J.~Rafelski (2019).
\newblock {Magnetic Dipole Moment in Relativistic Quantum Mechanics}.
\newblock \emph{Eur. Phys. J. A}, \textbf{55}(3), p.~40.
\newblock \doi{10.1140/epja/i2019-12715-5}.

\bibitem[{Steinmetz et~al.(2023)Steinmetz, Yang, and
  Rafelski}]{Steinmetz:2023nsc}
Steinmetz, A., C.~T. Yang, and J.~Rafelski (2023).
\newblock {Matter-antimatter origin of cosmic magnetism}.
\newblock \doi{10.48550/arXiv.2308.14818}.
\newblock [Submitted to Phys. Rev. D].

\bibitem[{Studenikin(2016)}]{Studenikin:2016ykv}
Studenikin, A. (2016).
\newblock {Status and perspectives of neutrino magnetic moments}.
\newblock \emph{J. Phys. Conf. Ser.}, \textbf{718}(6), p. 062076.
\newblock \doi{10.1088/1742-6596/718/6/062076}.

\bibitem[{Taylor et~al.(2011)Taylor, Vovk, and Neronov}]{Taylor:2011bn}
Taylor, A.~M., I.~Vovk, and A.~Neronov (2011).
\newblock Extragalactic magnetic fields constraints from simultaneous GeV--TeV
  observations of blazars.
\newblock \emph{Astronomy \& Astrophysics}, \textbf{529}, p. A144.
\newblock \doi{10.1051/0004-6361/201116441}.

\bibitem[{Thaller(2013)}]{Thaller:1992ji}
Thaller, B. (2013).
\newblock \emph{{The Dirac equation}}.
\newblock Springer Science \& Business Media.
\newblock ISBN 978-3-662-02753-0.
\newblock \doi{10.1007/978-3-662-02753-0}.
\newblock [orig. pub. 1992].

\bibitem[{Thomas(1926)}]{Thomas:1926dy}
Thomas, L.~H. (1926).
\newblock {The motion of a spinning electron}.
\newblock \emph{Nature}, \textbf{117}, p. 514.
\newblock \doi{10.1038/117514a0}.

\bibitem[{Tiesinga et~al.(2021)Tiesinga, Mohr, Newell, and
  Taylor}]{Tiesinga:2021myr}
Tiesinga, E., P.~J. Mohr, D.~B. Newell, and B.~N. Taylor (2021).
\newblock {CODATA recommended values of the fundamental physical constants:
  2018}.
\newblock \emph{Rev. Mod. Phys.}, \textbf{93}(2), p. 025010.
\newblock \doi{10.1103/RevModPhys.93.025010}.

\bibitem[{Tsai and Yildiz(1971)}]{Tsai:1971zma}
Tsai, W.~Y. and A.~Yildiz (1971).
\newblock {Motion of charged particles in a homogeneous magnetic field}.
\newblock \emph{Phys. Rev. D}, \textbf{4}, pp. 3643--3648.
\newblock \doi{10.1103/PhysRevD.4.3643}.

\bibitem[{Vachaspati(2021)}]{Vachaspati:2020blt}
Vachaspati, T. (2021).
\newblock {Progress on cosmological magnetic fields}.
\newblock \emph{Rept. Prog. Phys.}, \textbf{84}(7), p. 074901.
\newblock \doi{10.1088/1361-6633/ac03a9}.

\bibitem[{Vaquera-Araujo et~al.(2013)Vaquera-Araujo, Napsuciale, and
  Angeles-Martinez}]{Vaquera-Araujo:2012jlk}
Vaquera-Araujo, C.~A., M.~Napsuciale, and R.~Angeles-Martinez (2013).
\newblock {Renormalization of the QED of Self-Interacting Second Order Spin 1/2
  Fermions}.
\newblock \emph{JHEP}, \textbf{01}, p. 011.
\newblock \doi{10.1007/JHEP01(2013)011}.

\bibitem[{Vazza et~al.(2017)Vazza, Br\"uggen, Gheller, Hackstein, Wittor, and
  Hinz}]{Vazza:2017qge}
Vazza, F., M.~Br\"uggen, C.~Gheller, S.~Hackstein, D.~Wittor, and P.~M. Hinz
  (2017).
\newblock {Simulations of extragalactic magnetic fields and of their
  observables}.
\newblock \emph{Class. Quant. Grav.}, \textbf{34}(23), p. 234001.
\newblock \doi{10.1088/1361-6382/aa8e60}.

\bibitem[{Veltman(1998)}]{Veltman:1997am}
Veltman, M. J.~G. (1998).
\newblock {Two component theory and electron magnetic moment}.
\newblock \emph{Acta Phys. Polon. B}, \textbf{29}, pp. 783--798.
\newblock \doi{10.48550/arXiv.hep-th/9712216}.
\newblock URL:
  \href{https://www.actaphys.uj.edu.pl/R/29/3/783/pdf}{https://www.actaphys.uj.edu.pl/R/29/3/783/pdf}.

\bibitem[{Vernstrom et~al.(2021)Vernstrom, Heald, Vazza, Galvin, West,
  Locatelli, Fornengo, and Pinetti}]{Vernstrom:2021hru}
Vernstrom, T., G.~Heald, F.~Vazza, T.~J. Galvin, J.~L. West, N.~Locatelli,
  N.~Fornengo, and E.~Pinetti (2021).
\newblock {Discovery of magnetic fields along stacked cosmic filaments as
  revealed by radio and X-ray emission}.
\newblock \emph{Monthly Notices of the Royal Astronomical Society},
  \textbf{505}(3), pp. 4178--4196.
\newblock \doi{10.1093/mnras/stab1301}.

\bibitem[{Vryonidou and Zhang(2018)}]{Vryonidou:2018eyv}
Vryonidou, E. and C.~Zhang (2018).
\newblock {Dimension-six electroweak top-loop effects in Higgs production and
  decay}.
\newblock \emph{JHEP}, \textbf{08}, p. 036.
\newblock \doi{10.1007/JHEP08(2018)036}.

\bibitem[{Weinberg(1972)}]{weinberg1972gravitation}
Weinberg, S. (1972).
\newblock \emph{Gravitation and cosmology: principles and applications of the
  general theory of relativity}.
\newblock John Wiley \& Sons.

\bibitem[{Weinberg(2005)}]{Weinberg:1995mt}
Weinberg, S. (2005).
\newblock \emph{{The Quantum theory of fields. Vol. 1: Foundations}}.
\newblock Cambridge University Press.
\newblock ISBN 978-0-521-67053-1, 978-0-511-25204-4.
\newblock \doi{10.1017/CBO9781139644167}.

\bibitem[{Weisskopf(1936)}]{Weisskopf:1936hya}
Weisskopf, V. (1936).
\newblock \emph{The electrodynamics of the vacuum based on the quantum theory
  of the electron}, p. 206–226.
\newblock Cambridge University Press.
\newblock \doi{10.1017/CBO9780511608223.018}.
\newblock [translated and reprinted in 1994].

\bibitem[{Wolfenstein(1978)}]{Wolfenstein:1977ue}
Wolfenstein, L. (1978).
\newblock {Neutrino Oscillations in Matter}.
\newblock \emph{Phys. Rev. D}, \textbf{17}, pp. 2369--2374.
\newblock \doi{10.1103/PhysRevD.17.2369}.

\bibitem[{Workman et~al.(2022)}]{ParticleDataGroup:2022pth}
Workman, R.~L. et~al. (2022).
\newblock {Review of Particle Physics}.
\newblock \emph{PTEP}, \textbf{2022}, p. 083C01.
\newblock \doi{10.1093/ptep/ptac097}.

\bibitem[{Xing(2001)}]{Xing:2000ik}
Xing, Z.~Z. (2001).
\newblock {Commutators of lepton mass matrices, CP violation, and matter
  effects in-medium baseline neutrino experiments}.
\newblock \emph{Phys. Rev. D}, \textbf{63}, p. 073012.
\newblock \doi{10.1103/PhysRevD.63.073012}.

\bibitem[{Xing(2014)}]{Xing:2014wwa}
Xing, Z.~Z. (2014).
\newblock {Neutrino Physics}.
\newblock In \emph{{1st Asia-Europe-Pacific School of High-Energy Physics}},
  pp. 177--217.
\newblock \doi{10.5170/CERN-2014-001.177}.

\bibitem[{Yan et~al.(2023)Yan, Ma, Ling, Cheng, and Huang}]{Yan:2022sxd}
Yan, H., Z.~Ma, C.~Ling, C.~Cheng, and J.~Huang (2023).
\newblock {First Batch of z \ensuremath{\approx} 11\textendash{}20 Candidate
  Objects Revealed by the James Webb Space Telescope Early Release Observations
  on SMACS 0723-73}.
\newblock \emph{Astrophys. J. Lett.}, \textbf{942}(1), p.~L9.
\newblock \doi{10.3847/2041-8213/aca80c}.

\bibitem[{Yang et~al.(2023)Yang, Formanek, Steinmetz, and
  Rafelski}]{Yang:2023aaa}
Yang, C.~T., M.~Formanek, A.~Steinmetz, and J.~Rafelski (2023).
\newblock {Decomposition of Fermi gas into zero and finite temperature
  distributions with examples}.
\newblock \emph{In preparation}.
\newblock [submitted to Int. Journal of Theor. Phys.].

\bibitem[{Zhang(2012)}]{Zhang:2012muc}
Zhang, C. (2012).
\newblock \emph{{Effective field theory for top quark physics}}.
\newblock Ph.D. thesis, Illinois U., Urbana.
\newblock URN/HDL: \href{http://hdl.handle.net/2142/29627}{2142/29627}.

\bibitem[{Zhang and Willenbrock(2011)}]{Zhang:2010dr}
Zhang, C. and S.~Willenbrock (2011).
\newblock {Effective-Field-Theory Approach to Top-Quark Production and Decay}.
\newblock \emph{Phys. Rev. D}, \textbf{83}, p. 034006.
\newblock \doi{10.1103/PhysRevD.83.034006}.

\end{thebibliography}

\appendix

\newgeometry{left=0.8in,right=0.8in,top=1in,bottom=1in}
\chapter{Magnetic dipole moment in relativistic quantum mechanics}
\label{appendixA}
\begin{center}
Steinmetz, A., Formanek, M. \& Rafelski, J. Magnetic dipole moment in relativistic quantum mechanics. European Physical Journal A $\bb{55}$, 40 (2019). \href{https://doi.org/10.1140/epja/i2019-12715-5}{10.1140/epja/i2019-12715-5}
\end{center}

\noindent Copyright \copyright\ 2019 by Springer Nature. Reprinted with kind permission of The European Physical Journal (EPJ). Reprint permission for this thesis granted under \href{https://s100.copyright.com/CustomerAdmin/PLF.jsp?ref=9a7a42d0-4511-4427-8acd-73a16083772c}{License Agreement 5600921273203}. License date: August 2nd, 2023.


\chapter{Strong fields and neutral particle magnetic moment dynamics}
\label{appendixB}
\begin{center}
Formanek, M., Stefan, E., Rafelski, J., Steinmetz, A., Yang, C. T. Strong fields and neutral particle magnetic moment dynamics. Plasma Physics and Controlled Fusion 60, 7 (2018): 074006. \href{https://doi.org/10.1088/1361-6587/aac06a}{10.1088/1361-6587/aac06a}
\end{center}

\noindent Copyright \copyright\ 2018 by IOP Publishing. All rights reserved. Reproduced with permission under \href{https://marketplace.copyright.com/rs-ui-web/mp/license/56222509-d3e0-45fd-8ba6-f519e90d4d18/3a89c359-de4e-466f-93c3-ef1135413aab}{License Agreement 1384591-1}. License date: August 9th, 2023. This is the Accepted Manuscript version of an article accepted for publication in Plasma Physics and Controlled Fusion. IOP Publishing Ltd is not responsible for any errors or omissions in this version of the manuscript or any version derived from it. The Version of Record is available online at \href{https://doi.org/10.1088/1361-6587/aac06a}{10.1088/1361-6587/aac06a}.


\chapter{Relativistic dynamics of point magnetic moment}
\label{appendixC}
\begin{center}
Rafelski, J., Formanek, M., Steinmetz, A. Relativistic dynamics of point magnetic moment. Eur. Phys. J. C $\bb{78}$, 6 (2018). \href{https://doi.org/10.1140/epjc/s10052-017-5493-2}{10.1140/epjc/s10052-017-5493-2}
\end{center}

\noindent Copyright \copyright\ 2018 by Springer Nature. Reprinted with kind permission of The European Physical Journal (EPJ). This article is an open access article distributed under the terms and conditions of the \href{https://creativecommons.org/licenses/by/4.0/}{Creative Commons Attribution 4.0 (CC BY 4.0) License}.


\chapter{Dynamic fermion flavor mixing through transition dipole moments}
\label{appendixD}
\begin{center}
Rafelski, J., Steinmetz, A., Yang, C.T. Dynamic fermion flavor mixing through transition dipole moments. \emph{arXiv preprint}. 2023. \href{https://arxiv.org/abs/2309.15797}{arXiv:2309.15797 [hep-ph]}
\end{center}

\noindent Copyright \copyright\ 2023 by the authors. Contribution to the book edited by Gerhard Buchalla, Dieter L\"ust
and Zhi-Zhong Xing dedicated to memory of Harald Fritzsch. Submitted as a separate article to the International Journal of Modern Physics A (IJMPA) published by World Scientific.


\chapter{A Short Survey of Matter-Antimatter Evolution in the Primordial Universe}
\label{appendixE}
\begin{center}
Rafelski, J., Birrell, J., Steinmetz, A., Yang, C.T. A Short Survey of Matter-Antimatter Evolution in the Primordial Universe. Universe 2023, 9, 309. \href{https://doi.org/10.3390/universe9070309}{10.3390/universe9070309}
\end{center}

\noindent Copyright \copyright\ 2023 by the authors. Licensee MDPI, Basel, Switzerland. This article is an open access article distributed under the terms and conditions of the \href{https://creativecommons.org/licenses/by/4.0/}{Creative Commons Attribution 4.0 (CC BY 4.0) License}.


\chapter{Matter-antimatter origin of cosmic magnetism}
\label{appendixF}
\begin{center}
Steinmetz, A., Yang, C.T. \& Rafelski, J. Matter-antimatter origin of cosmic magnetism. \emph{arXiv preprint}. 2023. \href{https://arxiv.org/abs/2308.14818}{arXiv:2308.14818 [hep-ph]}
\end{center}

\noindent Copyright \copyright\ 2023 by the authors. Submitted to Physical Review D (PRD) published by the American Physical Society (APS).


\end{document}